\documentclass[a4paper,12pt]{book} 
\usepackage[includehead,left=2.6cm,right=2.6cm,top=2.1cm,bottom=2.1cm]{geometry}
\title{\bf\LARGE New approaches to higher-dimensional\\ general relativity\\[20mm]}
\author{{\bf\Large Mark Nicholas Durkee}\\ Churchill College, University of Cambridge \\[10mm]
              \includegraphics[width=3.5cm]{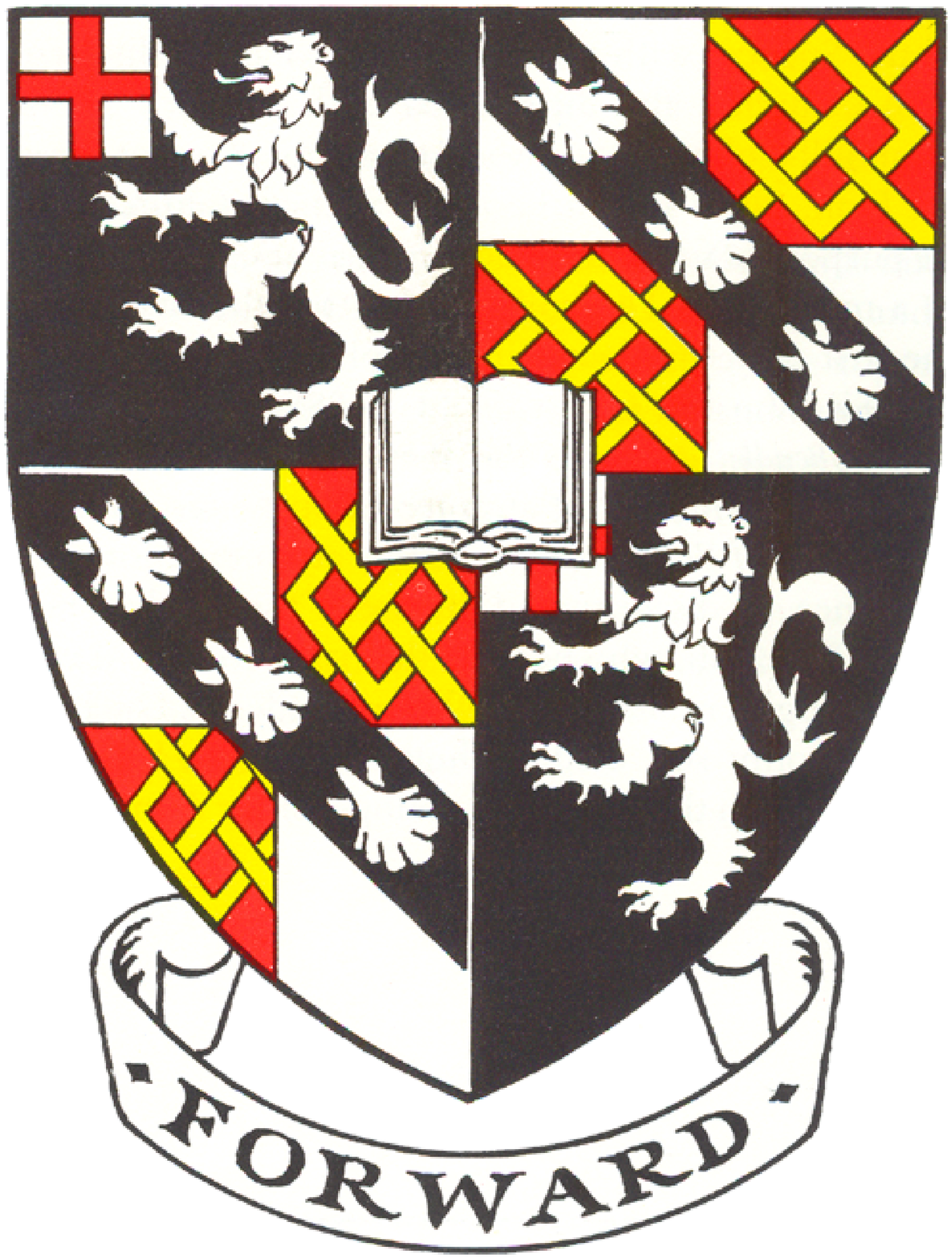} \qquad \qquad 
              \includegraphics[width=3.5cm]{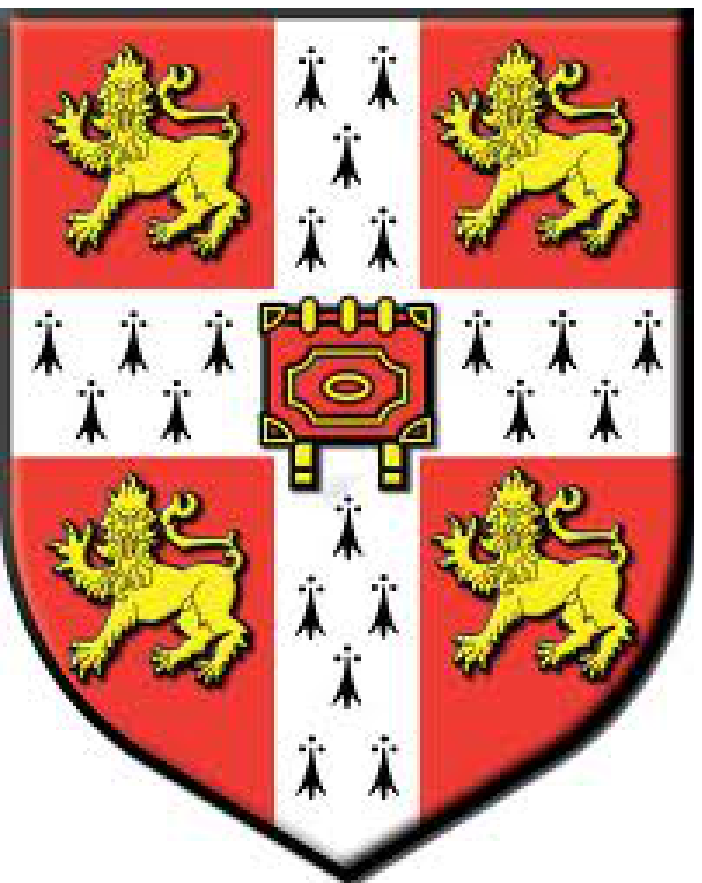} \\[40mm]
        \emph{This dissertation is submitted for the degree of Doctor of Philosophy}\\}
\date{19th April 2011}
\usepackage[dvips]{graphicx} 
\usepackage{amsmath} 
\usepackage{amsfonts}
\usepackage{color} 
\usepackage{epsfig} 
\usepackage{bm}
\usepackage{youngtab}

\newcommand{\al}{\alpha}
\newcommand{\ba}{\beta}

\newcommand{\del}{{\delta}}
\newcommand{\Del}{{\Delta}}
\newcommand{\eps}{{\varepsilon}}
\newcommand{\kap}{{\kappa}}
\newcommand{\la}{{\lambda}}
\newcommand{\La}{{\Lambda}}
\newcommand{\om}{{\omega}}
\newcommand{\Om}{{\Omega}}
\newcommand{\sig}{{\sigma}}
\newcommand{\twomat}[1]{{\left( \begin{array}{cc} #1 \end{array} \right)}}
\newcommand{\threemat}[1]{{\left( \begin{array}{ccc} #1 \end{array} \right)}}
\newcommand{\np}{{\newpage}}

\newcommand{\Mcal}{{\mathcal{M}}}
\newcommand{\Ocal}[1]{{\mathcal{O}^{(#1)}}}
\newcommand{\Hcal}{{\mathcal{H}}}
\newcommand{\Scal}{{\mathcal{S}}}
\newcommand{\Acal}{{\mathcal{A}}}
\newcommand{\Rcal}{{\mathcal{R}}}
\newcommand{\Jcal}{{\mathcal{J}}}
\newcommand{\Bcal}{{\mathcal{B}}}
\newcommand{\Kcal}{{\mathcal{K}}}
\newcommand{\Lcal}{{\mathcal{L}}}
\newcommand{\Dcal}{{\mathcal{D}}}
\newcommand{\Dcalh}{{\mathcal{\hat{D}}}}

\newcommand{\Rbb}{{\mathbb{R}}}
\newcommand{\Cbb}{{\mathbb{C}}}

\newcommand{\half}{{\textstyle{\frac{1}{2}}}}
\newcommand{\nn}{{\nonumber}}
\newcommand{\pd}{{\partial}}
\newcommand{\pard}[1]{\frac{\partial}{\partial #1}}

\newcommand{\eqand}{{\qquad \mathrm{and} \qquad}}
\newcommand{\Veff}{{V_{\mathrm{eff}}}}
\newcommand{\diag}{{\mathrm{diag}}}
\newcommand{\tr}{{\mathrm{tr}}}
\newcommand{\etal}{{\emph{et al.}\ }}

\renewcommand{\Re}{{\mathrm{Re}}}

\newcommand{\estar}{{{}^E\!\star}}
\newcommand{\eref}[1]{{(\ref{#1})}}
\newcommand{\Eref}[1]{{equation (\ref{#1})}}

\newcommand{\proof}{\paragraph*{Proof:}}
\newtheorem{theorem}{Theorem}[chapter]
\newtheorem{lemma}[theorem]{Lemma}
\newtheorem{defn}[theorem]{Definition}

\newtheorem{conjecture}[theorem]{Conjecture}

\newcommand{\dSdx}{{\left(\frac{dS_x}{dx}\right)^2}}
\newcommand{\dSdy}{{\left(\frac{dS_y}{dy}\right)^2}}
\newcommand{\Hxy}{{H(x,y)}}
\newcommand{\Hyx}{{H(y,x)}}
\newcommand{\Hx}{{H(x)}}
\newcommand{\Hy}{{H(y)}}
\newcommand{\Axy}{{A(x,y)}}
\newcommand{\Ayx}{{A(y,x)}}
\newcommand{\Lxy}{{L(x,y)}}
\newcommand{\Gx}{{G(x)}}
\newcommand{\Gy}{{G(y)}}
\newcommand{\Omf}{{\Omega_\phi}}
\newcommand{\Omy}{{\Omega_\psi}}
\newcommand{\lag}{{\mathcal{L}}}
\newcommand{\ham}{{\mathcal{H}}}
\newcommand{\axb}{{\alpha(x)}}
\newcommand{\bxb}{{\beta(x)}}
\newcommand{\ayb}{{\alpha(y)}}
\newcommand{\byb}{{\beta(y)}}
\newcommand{\gx}{{\gamma(x)}}
\newcommand{\gy}{{\gamma(y)}}
\newcommand{\zy}{{\zeta(y)}}
\newcommand{\tzx}{{\xi(x)}}

\newcommand{\tx}{{\theta(x)}}
\newcommand{\cx}{{\chi(x)}}

\newcommand{\Ps}{\Psi}   
\newcommand{\vphi}{\varphi} 
\newcommand{\Phis}{\Phi^\mathrm{S}} 
\newcommand{\Phia}{\Phi^\mathrm{A}} 
\newcommand{\Tf}{T^\Phi}
\newcommand{\Tp}{T^\Psi}
\newcommand{\To}{T^\Om}

\newcommand{\taup}{\tau'}
\newcommand{\rhop}{\rho'}

\newcommand{\kappap}{\kappa'}
\newcommand{\vphip}{\vphi'}

\DeclareTextFontCommand{\textwasy}{\wasyfamily}
\def \wasyfamily{\fontencoding{U}\fontfamily{wasy}\selectfont}
\def \thorn{{\wasyfamily\char105}}
\DeclareTextCommand{\dh}{OT1}{{\wasyfamily\char107}}
\newcommand{\tho}{{\textrm\thorn}}
\newcommand{\eth}{{\textrm{\dh}}}
\newcommand{\thop}{\tho'}

\newcommand{\rhob}{{\bm \rho}}
\newcommand{\taub}{{\bm \tau}}
\newcommand{\kapb}{{\bm \kap}}
\newcommand{\sigb}{{\bm \sig}}

\newcommand{\omb}{{\bm \om}}

\newcommand{\vphib}{{\bm \vphi}}
\newcommand{\Omb}{{\mathbf{\Om}}}
\newcommand{\Psib}{{\mathbf{\Psi}}}
\newcommand{\Phib}{{\mathbf{\Phi}}}

\newcommand{\Fb}{{\mathbf{F}}}
\newcommand{\Id}{{\mathbf{1}}}
\newcommand{\Ob}{{\mathbf{0}}} 

\newcommand{\Qb}{{\mathbf{Q}}}
\newcommand{\Rb}{{\mathbf{R}}}
\newcommand{\Sb}{{\mathbf{S}}}
\newcommand{\Tb}{{\mathbf{T}}}
\newcommand{\vb}{{\mathbf{v}}}
\newcommand{\Xb}{{\mathbf{X}}}
\newcommand{\zb}{{\mathbf{z}}}

\newcommand{\lb}{{\ell}}
\newcommand{\nb}{{n}}
\newcommand{\mb}[1]{{m_{#1}}}
\newcommand{\eb}{{e}}
\newcommand{\M}[1]{{\stackrel{#1}{M}}}   

\newcommand{\mwand}{{multiple WAND}}

\newcommand{\hi}{{\hat{\imath}}}
\newcommand{\hj}{{\hat{\jmath}}}
\newcommand{\hk}{{\hat{k}}}
\newcommand{\hl}{{\hat{l}}}

\newcommand{\nablah}{{\hat{\nabla}}}
\newcommand{\gh}{{\hat{g}}}

\newcommand{\Rh}{{\hat{R}}}
\newcommand{\Ih}{{\hat{I}}}
\newcommand{\Jh}{{\hat{J}}}

\newcommand{\Ah}{{\hat{A}}}
\newcommand{\Bh}{{\hat{B}}}

\newcommand{\alh}{{\hat{\alpha}}}
\newcommand{\betah}{{\hat{\beta}}}
\newcommand{\gammah}{{\hat{\gamma}}}
\newcommand{\delh}{{\hat{\delta}}}

\newcommand{\Ybb}{{\mathbb{Y}}}
\newcommand{\Acalh}{{\mathcal{A}}}
\newcommand{\CP}[1]{{\mathbb{CP}^{#1}}}
\newcommand{\Jcalh}{{\mathcal{J}}}
\newcommand{\Pcalh}{{\mathcal{P}}}

\newcommand{\DelL}{{\Delta_\mathrm{L}}}
\newcommand{\DelLA}{{\Delta_\mathrm{L}^\mathcal{A}}}
\newcommand{\lak}[1]{{\la_{\kap ,m}^\mathrm{#1}}}
\newcommand{\lakz}[1]{{\la_{\kap}^\mathrm{#1}}}

\numberwithin{equation}{chapter}
\newcommand{\url}{\href}

\begin{document}
\maketitle
\frontmatter

\section*{Summary}
This thesis considers various aspects of general relativity in more than four spacetime dimensions.

Firstly, I review the generalization to higher dimensions of the algebraic classification of the Weyl tensor and the Newman-Penrose formalism.  In four dimensions, these techniques have proved useful for studying many aspects of general relativity, and it is hoped that their higher dimensional generalizations will prove equally useful in the future.  Unfortunately, many calculations using the Newman-Penrose formalism can be unnecessarily complicated.  To address this, I describe new work introducing a higher-dimensional generalization of the so-called Geroch-Held-Penrose formalism, which allows for a partially covariant reformulation of general relativity.  This approach provides great simplifications for many calculations involving spacetimes which admit one or two preferred null directions.

The next chapter describes the proof of an important result regarding algebraic classification in higher dimensions.  The classification is based upon the existence of a particular null direction that is aligned with the Weyl tensor of the geometry in some appropriate sense.  In four dimensions, it is known that a null vector field is such a \emph{multiple Weyl aligned null direction} (WAND) if and only if it is tangent to a shearfree null geodesic congruence.  This is not the case in higher dimensions.  However, I have formulated and proved a partial generalization of the result to arbitrary dimension, namely that a spacetime admits a multiple WAND if and only if it admits a geodesic multiple WAND.

Moving onto more physical applications, I describe how the formalism that we have developed can be applied to study certain aspects of the stability of extremal black holes in arbitrary dimension.

The final chapter of the thesis has a rather different flavour.  I give a detailed analysis of the properties of a particular solution to the Einstein equations in five dimensions: the Pomeransky-Sen'kov doubly spinning black ring.  I study geodesic motion around this black ring and demonstrate the separability of the Hamilton-Jacobi equation for null, zero energy geodesics.  I show that this unexpected separability can be understood in terms of a symmetry described by a conformal Killing tensor on a four dimensional spacetime obtained by a Kaluza-Klein reduction of the original black ring spacetime.\newpage

\section*{Declaration}
This dissertation is the result of my own work and includes nothing which is the outcome
of work done in collaboration except where specifically indicated in the text.  The research described was carried out in the Department of Applied Mathematics and Theoretical Physics at the University of Cambridge between October 2007 and December 2010.

Except where reference is made to the work of others, all the results are original and based on the following published works:
\begin{itemize}
  \item {\bf Geodesics and Symmetries of Doubly-Spinning Black Rings}\\
        \emph{Class.\ Quantum Grav., {\bf26} (2009) 085016} \cite{ringgeo}
  \item {\bf Type II Einstein spacetimes in higher dimensions}\\
        \emph{Class.\ Quantum Grav., {\bf26} (2009) 195010} \cite{TypeII}
  \item {\bf A higher-dimensional generalization of the geodesic part of the \\ Goldberg-Sachs theorem} \\
          (with Harvey S. Reall)\\
          \emph{Class.\ Quantum Grav., {\bf26} (2009) 245005} \cite{nongeo}
  \item {\bf Generalization of the Geroch-Held-Penrose formalism to higher \\dimensions} \\
          (with Vojt\v ech Pravda, Alena Pravdov\'a and Harvey S. Reall)\\
         \emph{Class.\ Quantum Grav., {\bf 27} (2010) 215010} \cite{higherghp}
  \item {\bf Perturbations of higher-dimensional spacetimes} \\
          (with Harvey S.\ Reall)\\
           \emph{Class.\ Quantum Grav., {\bf 28} (2011) 035011} \cite{decoupling}
  \item {\bf Perturbations of near-horizon geometries and instabilities of Myers-Perry black holes}\\
          (with Harvey S.\ Reall) \\
          \emph{Accepted by Phys.\ Rev.\ D} \cite{nhperturb}
\end{itemize}

None of the original work contained in this dissertation has been submitted for any other degree, diploma or similar qualification.\\

\noindent Mark Durkee, April 2011
\newpage

\section*{Acknowledgements}
The work that contributes to this thesis was completed between October 2007 and December 2010 in the Department of Applied Mathematics and Theoretical Physics in the University of Cambridge.

I am extremely grateful to Harvey Reall, who has been an excellent supervisor for the last three years, and contributed many important ideas to the work of this thesis.  At various times through the course of my PhD I have also enjoyed useful conversations about aspects of this work with many people, including David Chow, Alan Coley, Oscar Dias, Maciej Dunajski, Pau Figueras, Mahdi Godazgar, David Kubiz{\v n}\'ak, Hari Kunduri, Ricardo Monteiro, Keiju Murata, Marcello Ortaggio, Vojt{\v e}ch Pravda, Alena Pravdov\'a, Donough Regan, Jorge Santos, Jacques Smulevici, Yukinori Yasui and others whom I have forgotten to mention.

My PhD was supported by a studentship from the Science and Technology Facilities Council (STFC), and I am also grateful to Churchill College, DAMTP, the Institute of Mathematics of the Academy of Sciences of the Czech Republic and Dalhoisie University, Canada for travel funding.

\newpage

\setcounter{tocdepth}{1}
\tableofcontents
\listoffigures
\listoftables
\mainmatter
\chapter{Introduction}\label{chap:intro}
In recent years, the study of general relativity in higher dimensions has attracted significant interest in theoretical physics.  As the field develops, it is useful to develop mathematical tools to help answer a variety of physical questions.  This will be the focus of this thesis.

Before moving on to explain what these new approaches are, we begin with some background as to why this study is worthwhile, by placing the study of higher dimensional gravity in a little context.

\section{Historical context}
The last century of progress in understanding the fundamental laws of physics has been based around developing our knowledge of the symmetries that these laws respect.  Prior to the twentieth century, the accepted laws were based on Galileo's principle of relativity.  That is, they do not change over time, and are also invariant under translations and rigid rotations of the three spatial directions.

However, in the early 20th century, Einstein \cite{Einstein:1905} and others understood that this Galilean symmetry was only an approximation to a larger symmetry group, the Lorentz group, acting not on space and time separately, but on a four-dimensional \emph{spacetime}.  Crucially, the Lorentz group encodes a notion of causality, and as a result this new theory of \emph{special relativity} predicts that no information is able to travel faster than the speed of light $c \approx 3\times 10^8 ms^{-1}$.  Galilean symmetry is recovered from special relativity for speeds $v \ll c$, and hence gives a very good approximation for most everyday physics.

Unfortunately, Newton's Law of Gravitation is inherently inconsistent with special relativity, as when a massive body moves, information about the movement is instantaneously transferred across all of space via in the change in its gravitational field.  This violates causality.

This observation motivated the development of general relativity (GR), first written down in full by Einstein in 1916 \cite{Einstein:1916}.  It postulates that the presence of mass causes spacetime to curve, according to a particular set of partial differential equations: the Einstein equations
\begin{equation}\label{eqn:einsteinintro}
  G_{\mu\nu} = \tfrac{8\pi G}{c^4} T_{\mu\nu},
\end{equation}
where the Einstein tensor $G_{\mu\nu}$ encodes some aspects of the curvature of spacetime, and $T_{\mu\nu}$ encodes information about the matter content, with $G$ the Newtonian gravitational constant.  General relativity puts Lorentz symmetry on a different footing; it is now a local symmetry valid over small distances, but broken by curvature on large scales.

This curvature describes the force of gravity, in the sense that test bodies falling freely under gravity follow straight line paths (or geodesics) in this resulting curved manifold.  GR is a deterministic theory; given a consistent set of initial data on some `Cauchy surface', the spacetime is determined uniquely in the causal future of that surface \cite{Bruhat:1958}.

General relativity has been extensively tested observationally, and to date has provided an extremely accurate description of a wide range of phenomena (see e.g.\ Will \cite{Willreview} for an up to date review).

However, there is a serious problem.  General relativity describes gravity, but the other three fundamental forces (electromagnetism and the weak and strong nuclear forces) are best described by a different theory: the standard model of particle physics.  This is a theory of a very different nature, in particular it is quantum mechanical.  It explains phenomena that occur on very small lengthscales, or at very high energies.  Quantum physics is inherently random, with physical observations determined by a probability distribution.  The standard model belongs to a class of physical theories called quantum field theories, which combine the ideas of quantum mechanics with special relativity.  Predictions from the standard model have been extensively tested, for example in particle accelerators, and give remarkably accurate results.

Despite this success, there is a serious problem; general relativity cannot be fitted into this framework.  In particular, if the matter on the right hand side of the Einstein equations (\ref{eqn:einsteinintro}) is quantum mechanical, then it seems reasonable to believe that the left hand side should also be quantized.  However, quantizing general relativity gives a quantum field theory that is \emph{non-renormalizable}, that is it contains infinities that are in some precise sense uncontrolled.  Non-renormalizability is seen as a signal that a physical theory is only valid up a particular energy scale, and there exists some new physics that becomes relevant at higher energies.

The inconsistency between GR and particle physics usually doesn't matter for computational purposes, since the standard model describes the interaction of particles at very small distances (typically subatomic scales), while the force of gravity is only significant at relatively large distances (where quantum fluctuations are `averaged out').  However, from a theoretical point of view it is deeply unsatisfactory that the basic physical forces cannot yet be understood in terms of a self-consistent set of equations.  Also, in certain extreme environments, the effects of gravity and those of particle physics are simultaneously important.  It is commonly thought that there is some `theory of everything' that includes both GR and the standard model as suitable low energy limits.  The search for such a theory of quantum gravity that unifies our understanding of all physical forces has occupied theoretical physicists for many years.

Today, many physicists believe that string theory represents the best possibility for doing this.  Interest in string theory as a candidate theory of quantum gravity was really sparked in 1984 by Green \& Schwarz's discovery \cite{Green:1984sg} that a particular form of string theory allowed for the cancellation of various anomalies.  This theory both contained general relativity (as a low energy limit), and appeared likely to be renormalizable.  The main oddity is that this anomaly cancellation occurs only in ten spacetime dimensions.  Hence, the study of general relativity in higher dimensions is an essential part of better understanding string theory.  Although there are many fundamental questions about the basic nature of string theory that are not yet well understood, it certainly provides a framework in which many difficult questions can be posed in a concrete way.

Even if one does not believe in string theory as a fundamental description of quantum gravity, then studying higher-dimensional general relativity still has the potential to give important new insights into four-dimensional physics.  One particularly exciting aspect of this is the gauge-gravity correspondence \cite{Maldacena:1997re,Witten:1998qj,Aharony:1999ti}.  This conjectures that certain `strongly coupled' four-dimensional gauge theories (with many similarities to those that make up the standard model) are in some precise sense equivalent to theories of gravity in five dimensions.

Although the basic principles behind gauge field theories are well understood, performing accurate computations in the strongly coupled limit is very difficult, with lattice-based computer simulations still lagging behind experiment in terms of precision.  The gauge-gravity correspondence seems to offer a new way to make progress in studying properties of these gauge theories by doing much easier calculations in five-dimensional general relativity.  In this language, five-dimensional black hole spacetimes have particular significance, corresponding to states in the field theory at non-zero temperature \cite{Witten:1998zw}.

Higher-dimensional GR is also interesting from a purely mathematical point of view.  It is fascinating that many familiar results from four-dimensional general relativity turn out to be very specific to four dimensions.  For example, in more than four dimensions there is a far richer set of spacetimes containing black holes \cite{ER:2008}.  We will see many further examples of how special four dimensions is later on in the thesis. 

\section{Review of general relativity}
To fix notation and conventions, we first recall some basic concepts of general relativity in $d$ dimensions.  Spacetime is a differentiable manifold $(\Mcal,g)$, with local distances measured by a line element
\begin{equation}
  ds^2 = g_{\mu\nu} dx^\mu \otimes dx^\nu.
\end{equation}
Summation over indices $\mu,\nu,\ldots = 0,1,\ldots,d-1$ is implied.  The 1-forms $dx^\mu$ provide a local coordinate basis for the co-tangent space of $\Mcal$.  The metric $g_{\mu\nu}$ has signature $(-+\dots +)$, and hence provides an indefinite norm on the tangent space $T(\Mcal)$.  We will raise and lower indices with the metric and its inverse $g^{\mu\nu}$.  Much of the thesis will work with a null frame $\{e_a\}$, which will carry indices $a,b,\ldots = 0,1,2\ldots$.  In this null frame, indices $0,1$ refer to null directions, and indices $i,j,\ldots$ to spacelike ones.

Unless stated otherwise, we will use $\nabla$ to denote the Livi-Civita connection on $(\Mcal,g)$, with the property that $\nabla g = 0$.  The commutator of $\nabla$, acting on an arbitrary vector field $V$, defines the \emph{Riemann curvature tensor} $R_{\mu\nu\rho\sig}$ through
\begin{equation}\label{id:ricci}
  [\nabla_\mu,\nabla_\nu ] V_\rho = R_{\mu\nu\rho\sig}V^\sig.
\end{equation}
The Riemann tensor has $d^2(d^2-1)/12$ independent components, and obeys the symmetries $R_{\mu\nu\rho\sig} = R_{[\mu\nu][\rho\sig]} = R_{\rho\sig\mu\nu}$ and $R_{\mu[\nu\rho\sig]} = 0$, as well as the differential Bianchi identity
\begin{equation}\label{id:bianchi}
  \nabla_{[\mu}R_{\nu\rho]\sig\tau} = 0.
\end{equation}

It is often useful to decompose the Riemann tensor into several parts.  We write
\begin{equation}
  R_{\mu\nu\rho\sig} = C_{\mu\nu\rho\sig}
       + \frac{2}{d-2} \left(R_{\mu[\rho|}g_{\nu|\sig]} - R_{\nu[\rho|}g_{\mu|\sig]}\right)
       + \frac{2R}{(d-1)(d-2)}g_{\mu[\rho|}g_{\nu|\sig]}
\end{equation}
where the \emph{Ricci tensor} and \emph{Ricci scalar} are given by
\begin{equation}
  R_{\mu\nu} \equiv g^{\rho\sig} R_{\mu\rho\nu\sig} \eqand
  R \equiv g^{\mu\nu}R_{\mu\nu},
\end{equation}
and the \emph{Weyl tensor} $C_{\mu\nu\rho\sig}$ is totally traceless.

In general relativity, the spacetime geometry is determined by the Einstein equations
\begin{equation}
  R_{\mu\nu} - \frac{1}{2}R g_{\mu\nu} = 8\pi T_{\mu\nu}
\end{equation}
where $T_{\mu\nu}$ is the \emph{energy-momentum tensor}, defined by the distribution of matter in the spacetime.  We choose natural units where the speed of light $c$ and the $d$-dimensional gravitational constant $G$ are normalized to one.

This thesis will focus on \emph{Einstein spacetimes}, where the only matter allowed is a cosmological constant $\La$ (possibly zero).  The Einstein equation reduces to\footnote{Note that there are (at least) two different conventions for the definition of the cosmological constant in $d$ dimensions, and many references would replace $\La$ by $\tfrac{2\La}{d-2}$ in \eqref{eqn:einstein}.  The two conventions are equivalent in $d=4$ dimensions.}
\begin{equation}\label{eqn:einstein}
  R_{\mu\nu} = \La g_{\mu\nu}.
\end{equation}

The Weyl tensor encodes the information about curvature that is not directly determined by \eqref{eqn:einstein}, and will have particular significance in this work.  One important property of this tensor is that it is \emph{conformally invariant}.  A conformal transformation maps a spacetime $(\Mcal, g)$, to a new spacetime $(\Mcal,\bar{g})$, where the new metric is given by $\bar{g} = \Om^2 g$ for some smooth positive function $\Om: \Mcal \rightarrow \mathbb{R}$.  If $\bar{C}$ be the Weyl tensor for the new spacetime, then the statement of conformal invariance is that $\bar{C}^\mu_{\;\;\nu\rho\sig} = C^\mu_{\;\;\nu\rho\sig}$ (see, e.g.\ \cite{wald}).

\section{Black holes in four dimensions}
Black holes are commonly understood as large astrophysical objects from which nothing, not even light, can escape.  Their existence has been speculated about for many years.  In the eighteenth century, Michell \cite{Michell:1784} and Laplace \cite[App.\ A]{HawkingEllis} both calculated that, given a mass $M$ localized inside a sphere of radius $r_0 = 2GM/c^2$, the escape velocity of a light `particle' obeying Newton's second law of motion would become infinitely large, and hence light would not be able to escape from the body to infinity.

These ideas were not given serious consideration until the mid-twentieth century, when it became apparent that these objects were a feature of general relativity, and the term black hole was coined.  There is now significant astrophysical evidence for their existence, and it is strongly believed that there is a supermassive black hole four million times bigger than our sun at the centre of our galaxy \cite{Ghez:2008ms}.

The first non-trivial exact solution to the Einstein field equations was constructed by Schwarzchild in 1917 \cite{schwarzchild}, described by a metric
\begin{equation}\label{eqn:schwarzchild}
  ds^2 = -\left(1-\tfrac{r_0}{r} \right) dt^2 + \left(1-\tfrac{r_0}{r}\right)^{-1} dr^2 
         + r^2 (d\theta^2 + \sin^2\theta d\phi^2).
\end{equation}
This metric was constructed to represent the gravitational field outside some spherically symmetric massive body; and one can construct `interior' solutions for matter models that can be matched suitably smoothly onto the Schwarzchild solution at any surface of constant $r>r_0$.  A problem seems to occur if the radius of the body is less than $r_0$, as the metric becomes singular.  However, it was later understood \cite{Finkelstein} that this apparent singularity is merely an artifact of the coordinate system that we are using, and one can define a new `advanced Eddington-Finkelstein' coordinate $v=t+r+r_0 \log(r-r_0)$, with respect to which the metric takes the form
\begin{equation}
    ds^2 = 2dv dr -\left(1-\tfrac{r_0}{r} \right) dv^2 + r^2 (d\theta^2 + \sin^2\theta d\phi^2).
\end{equation}
This is manifestly non-singular at $r=r_0$, and hence an observer freely falling from infinity would not observe anything unusual as they pass this surface.  However, once inside this surface, they cannot escape back to asymptotic infinity.  Hence, this metric encapsulates the notion of a black hole, with an \emph{event horizon} at $r=r_0$.  Kruskal \cite{Kruskal} showed how one could introduce further coordinates that revealed the existence of a second asymptotically flat region of this spacetime, causally disconnected from the original spacetime.

There is a true singularity at $r=0$, where the curvature of spacetime (and hence tidal gravitational forces) becomes arbitrarily large.  The extreme region near to $r=0$, where the effects of quantum gravity are thought to be highly significant, is hidden from our view by the presence of the event horizon.

\subsection{The Kerr metric}

The existence of a vacuum solution to the Einstein equations that admits an event horizon is not itself solid evidence for the likely existence of black holes.  However, under reasonable physical assumptions, it is known that a sufficiently massive star will, at the end of its lifecycle, undergo complete gravitational collapse to form a singularity (see e.g.\ \cite{HawkingEllis} for a detailed discussion). 

The end point of this collapse process is thought to be described by the Kerr solution \cite{Kerr}, which is an asymptotically flat, stationary solution to the vacuum Einstein equations.  The exterior region of this spacetime can be described by the metric:
\begin{equation}
  ds^2 = -dt^2 + \frac{2Mr}{\Sigma}(dt-a\sin^2 \theta d\phi)^2 + (r^2 + a^2) \sin^2 \theta d\phi^2 
         + \frac{\Sigma}{\Delta} dr^2 + \Sigma d\theta^2 
\end{equation}
where
\begin{equation}
  \Sigma = r^2 + a^2 \cos^2 \theta \eqand \Delta = r^2 + a^2 -2M r.
\end{equation}
Continuous isometries of spacetimes are generated by Killing vector fields $K$, satisfying $\Lcal_g K = 0$.  The Kerr metric admits two such vector fields, an asymptotically timelike $k \equiv \pard{t}$ which generates time translations, and an asymptotically spacelike $\zeta = \pard{\phi}$ which has closed orbits and generates rotations around the axis of symmetry, giving it an isometry group of $\Rbb\times U(1)$.

Many of the properties of the Kerr geometry generalize to higher dimensions in an interesting way, so here we briefly discuss the key features.

\subsubsection{Ergoregion}
For the static Schwarzchild spacetime, the asymptotically timelike Killing vector field $k$ is timelike everywhere outside the event horizon, and null on the horizon itself.  However, this is not the case for the rotating Kerr black holes, where $k.k = 1-\frac{2Mr}{\Sigma}$ vanishes on the \emph{ergosurface} 
\begin{equation}
  r = r_e(\theta) \equiv M + \sqrt{M^2-a^2\cos^2\theta}.
\end{equation}
The ergosurface lies outside the event horizon, touching it only at $\theta=0,\pi$, and the region between these two surfaces is known as the \emph{ergoregion}.  Although $k$ is spacelike in this region, the signature of the metric remains correct as there exists a vector field $\chi = k+\Om_H \zeta$ that remains timelike everywhere outside the horizon.  The constant $\Om_H$, which has the interpretation of the angular velocity of the horizon, is fixed by the requirement that $\chi$ is null on the horizon. 

The physical interpretation of this is that a massive particle inside the ergoregion cannot follow orbits of $k$, and therefore must co-rotate with the black hole, from the point of view of an observer at infinity.  We will see in Chapter \ref{chap:blackrings} that rotating black holes in higher dimensions also admit ergoregions, but that the properties of this region can become more complicated.  In particular, for black holes of non-spherical topology, the ergoregion does not always have the same topology as the event horizon.

\subsubsection{Algebraic type}
Much of this thesis will be concerned with the algebraic classification of spacetimes.  In this language, the Kerr black hole spacetime is of Petrov Type D.  This is a useful way of understanding various properties of the spacetime.  Chapter \ref{chap:ghp} will give a detailed introduction to algebraic classification.

\subsubsection{Hidden symmetry}\label{sec:kerrkilling}
Finding the geodesics of a $d$-dimensional geometry typically requires solving a set of $d$ coupled, second order ordinary differential equations, which can be most conveniently derived as the Euler-Lagrange equations of the Lagrangian
\begin{equation}
  L = \tfrac{1}{2} g_{\mu\nu}(x) \dot{x}^\mu \dot{x}^\nu,
\end{equation}
where $\dot{ }$ denotes differentiation with respect to some parameter along the geodesic.  However, in the presence of symmetries, the problem considerably simplifies.

Let $p_\mu = \pd L/\pd \dot{x}^\mu$ be the particle momentum conjugate to the particle velocity.  Then, for any Killing vector $K$, $p.K$ is a conserved quantity along the geodesic, as is $\eps = -g^{\mu\nu} p_\mu p_\nu$, which takes values +1 and 0 for timelike and null geodesics respectively.

For the Kerr geometry, there are two independent Killing vector fields $k = \pd / \pd t$ and $\zeta = \pd / \pd \phi$, leading to two conserved quantities $E= - u.k$ and $h = u.m$ which we interpret as the energy and angular momentum of the particle (per unit mass).  This gives us three constants of motion; but in order to render a system of four second-order ODEs integrable we would expect to need four.

Remarkably, Carter \cite{Carter:1968b} found a fourth constant of motion $\Kcal$, which is quadratic in the particle momentum.  This gives sufficient constants to render the geodesic motion integrable.  Later, Walker \& Penrose \cite{Walker:1970} traced the existence of this additional constant of motion to the existence of an additional symmetry, described by a rank-2 \emph{Killing tensor}, satisfying the generalized Killing equation
\begin{equation}\label{eqn:killing}
  \nabla_{(\mu} K_{\nu\rho)} = 0,
\end{equation} 
where the associated constant of motion is $\Kcal = K^{\mu\nu} p_\mu p_\nu$.

In fact, it was later shown that the Killing tensor, as well as the Killing vectors $k$ and $\zeta$, could be constructed from a more fundamental object, a \emph{Killing-Yano} 2-form $f_{\mu\nu}$.  This is a totally antisymmetric tensor, satisfying the Killing-Yano equation
\begin{equation}\label{eqn:ky}
  \nabla_{(\mu} f_{\nu)\rho} = 0.
\end{equation}
Given any solution $f$ to equation \eqref{eqn:ky}, one can construct a solution $K$ to \eqref{eqn:killing} via $K_{\mu\nu} = f_{\mu\rho} f^\rho_{\phantom{\rho}\nu}$ (see, e.g.\ \cite{exact} and references therein).  The Killing and Killing-Yano equations are not invariant under conformal transformations acting on the spacetime.  However, one can define conformally invariant generalizations
\begin{equation}\label{eqn:confkill}
  \nabla_{(\mu} K_{\nu\rho)} = \om_{(\mu} g_{\nu\rho)} \eqand
  \nabla_{(\mu} f_{\nu)\rho} = g_{\mu\nu}\xi_{\rho} - \xi_{(\mu}g_{\nu)\rho} 
\end{equation} 
for 1-forms $\om_\mu = \tfrac{3}{d+2} g^{\nu\rho}\nabla_{(\mu} K_{\nu\rho)}$ and $\xi_\mu = \tfrac{1}{d-1} \nabla^\nu f_{\nu\mu}$.  Solutions to these equations are referred to as \emph{conformal Killing tensors} and \emph{conformal Killing-Yano tensors} respectively. 

These concepts generalize to higher dimensions in a natural way, although they will not play a central role in the work of this thesis.  In Chapter \ref{chap:ghp} we will briefly review how the existence of hidden symmetry is closely linked to the algebraic classification of spacetimes, while in Chapter \ref{chap:blackrings} we will show that black ring spacetimes admit a more limited form of hidden symmetry.

\subsubsection{Stability}
The Kerr black hole is thought to be the stationary end state of the collapse of sufficiently massive stars, under fairly generic conditions.  However, this statement only has physical meaning if the end state is stable against small perturbations.  That is to say; suppose that one starts with (consistent) initial data on some Cauchy surface that is in some appropriate sense close to Kerr initial data.  Is the future development of this Cauchy surface also close to Kerr?  It is strongly believed that the answer to this question is yes.

Much of the evidence for this belief comes from studies of linearized perturbation theory.  In particular, it was shown by Whiting \cite{Whiting:1988vc}, making use of previous work by Teukolsky and others \cite{Teukolsky,Teukolsky:1972,stewperts,Press:1973zz,Stewart:1975vg}, that there are no exponentially growing linearized perturbation modes.  This is interpreted as the absence of an instability.  In Chapter \ref{chap:decoupling} we consider the extent to which is is possible to generalize these methods to higher dimensions.

However, these linearized analyses do not provide a conclusive proof of the full, non-linear, stability of the Kerr family and there are many ongoing attempts to complete this.  This field was re-invigorated by the proof, by Christodoulou \& Klainerman \cite{Christodoulou:1993uv}, of the non-linear stability of Minkowski space.  Significant progress has been made in various aspects of this problem in recent years, see e.g.\ \cite{Dafermos:2010hd,Chrusciel:2010fn} for recent reviews.

\subsubsection{Asymptotic flatness}
For large $r$, the Kerr metric is approximately given by
\begin{equation}
  ds^2 \approx -dt^2 + dr^2 + r^2 (d\theta^2 + \sin^2\theta d\phi^2),
\end{equation}
which is the metric on flat space written in spherical polar coordinates.  This suggests that the spacetime is asymptotically flat.  However, this is a coordinate dependent statement; ideally we would like a notion of asymptotic flatness that is manifestly independent of our choice of coordinates.

The most commonly used definition uses conformal transformations.  For Minkowski space, one can perform such a transformation to obtain a compact space.  For other spacetimes, we say (roughly) that a spacetime is \emph{asymptotically flat} if and only if there exists a suitably regular conformal transformation $g\mapsto \tilde{g} = \Om^2 g$ such that in some neighbourhood of asymptotic infinity, the unphysical spacetime $\tilde{g}$ has the same structure as compactified Minkowski space.  This is indeed the case for the Kerr spacetime.  Such conformal compactifications can be used to define a coordinate $r$ which has many of the properties of familiar radial coordinates; in particular that linearized perturbations fall off as $1/r$ near null infinity.

\section{Black hole uniqueness theorems}
We now move on to talk about vacuum black holes in arbitrary spacetime dimension.  Some physical motivation for this will be given below, but we begin by asking, in a more general setting, why should we expect qualitatively new behaviour in higher dimensions?

In four dimensions, the possible black hole solutions to the vacuum Einstein equations are very tightly constrained.  As we shall see below, it has been established that the Kerr family includes all asymptotically flat, analytic, stationary black hole metrics.  Some aspects of this result have been generalized to higher dimensions, but the results are far less restrictive.  This allows a much richer set of black hole spacetimes to exist in dimensions greater than four.

What do we mean by asymptotic flatness in higher dimensions?  There are difficulties in applying the standard four-dimensional definition, for example because linearized perturbations turn out to fall off as $1/r^{(d-2)/2}$ at null infinity in dimension $d$, and hence in odd dimensions the half-integer powers cause problems with regularity \cite{Hollands:2003ie}.  However, reasonable definitions of asymptotic flatness at null \cite{Hollands:2003ie} and spatial \cite{Tanabe:2009xb} infinity have been made.  As in the Kerr case, when considering exact black hole solutions, one can usually move to a coordinate system in which asymptotic flatness is manifest and hence avoid detailed consideration of these issues.

In an asymptotically flat spacetime $\Mcal$, a \emph{black hole region} is defined as the subset of $\Mcal$ lying outside of the causal past of future null infinity.  This is a global property, requiring knowledge of the entire spacetime including the future of any given Cauchy surface.  It is useful to have an alternative definition that is more local in time.

A result along these lines is given by the notion of an \emph{outermost trapped surface}.  A trapped surface is a closed spacelike surface $\Scal$ in a spacetime with non-positive expansion (i.e.\ an area not increasing with time).  In vacuum, asymptotically flat spacetimes, such surfaces must always lie outside the causal past of future null infinity, i.e.\ a spacetime admitting such surfaces must always contain a black hole \cite{HawkingEllis,wald}.  Furthermore, if one considers takes the union of all trapped surfaces, then the boundary of this region is a non-expanding null surface that can be identified with the event horizon of the black hole.  In the algebraic classification of spacetimes described in Chapter \ref{chap:ghp}, we will see that all black holes must be `algebraically special' on the horizon for this reason.

Black holes, in the sense discussed in this thesis, are time-invariant objects.  This property is captured by the following definition:
\begin{defn}
  An asymptotically flat spacetime is \emph{stationary} if it admits a Killing vector $k$ that is timelike near to asymptotic infinity.  A stationary spacetime is \emph{static} if and only if $k$ is hypersurface orthogonal.
\end{defn}
The fact that this requirement is only imposed near asymptotic infinity is important.  For example, the Kerr spacetime is stationary, but the generator of (asymptotic) time translations $k = \pd/\pd t$ is only timelike outside the ergosphere.

In four dimensions, the topology of (time slices) of the event horizon is constrained by the following:
\begin{theorem}[\cite{Hawking:1972,Chrusciel:1994tr}]
  Let $\Hcal^+$ be the (future) event horizon of a stationary, four-dimensional, vacuum black hole, and $\Sigma$ be any Cauchy surface.  Then $\Hcal^+\cap \Sigma$ is homeomorphic to $S^2$.
\end{theorem}
This has been extended to higher dimensions by Galloway \& Schoen \cite{Galloway:2005}, who show that any black hole horizon (specifically, any marginally trapped outer horizon) is of positive Yamabe type, that is it admits a metric of positive scalar curvature.  In five dimensions, it follows from this (and further work \cite{Hollands:2010qy}) that the horizon topology must be either a connected sum of $S^1\times S^2$ with some number of Lens spaces $L(p,q)$, or the quotient of $S^3$ by some (possibly trivial) finite isometry group.  We will see below that there exist known examples of black hole spacetimes with horizon topologies $S^3$ and $S^1\times S^2$.

In arbitrary dimension, it is also known that a stationary black hole spacetime must be axisymmetric:
\begin{theorem}[\cite{Hollands:2006rj,Moncrief:2008}]\label{thm:rigidity}
  Let $(M,g)$ be a stationary, analytic, asymptotically flat vacuum black hole spacetime with stationary Killing vector field $k$.  Then either $k$ is tangent to the null generators of the horizon (and the black hole is static), or there exists a second Killing vector field $\zeta$ with closed periodic orbits
\end{theorem}
Given such a vector field $\zeta$, there exists some constant $\Om_H$ such that $k+\Om_H \zeta$ is tangent to the null generators of the horizon.  The assumption of analyticity is often seen as undesirable in the context of this `rigidity theorem', and more recent work \cite{Friedrich:1998wq,Alexakis:2009gi} has made some progress in proving this result without needing this assumption.

However, although this axisymmetry result generalizes directly to arbitrary dimension, it is particularly useful in four dimensions, where a spacetime has only one (independent) plane of rotation.  In this case, stationary, axisymmetric, vacuum black hole solutions with $S^2$ horizons are members of the Schwarzchild or Kerr families, characterised uniquely by their mass $M$ and angular momentum $J=Ma$ \cite{Israel:1967wq,Carter:1971zc} .\footnote{These results require various further technical assumptions that have been gradually weakened by various authors over time, see e.g.\ \cite{uniquenessbook} for a review of this progress.}

This is no longer the case in higher dimensions, where the existence of the single $U(1)$ isometry guaranteed by Theorem \ref{thm:rigidity} is far less restrictive.

\section{Black holes in higher dimensions}
Much of the interest in higher-dimensional general relativity has focused on black holes, and in the rest of the introduction we review some of the known results.  We will focus on black holes that are solutions to the vacuum Einstein equations, allowing for a possible cosmological constant (see \cite{ER:2008} for a detailed review).

One physical motivation for this comes from braneworld models; in which, heuristically, our observed universe corresponds to a 4D brane in some bulk spacetime with `large' extra dimensions, of size as large as a millimetre \cite{ArkaniHamed:1998rs}.  It has been argued that in such a scenario, `small' black holes could form.  They could have energy scales as low as a few $TeV$, and would radiate this energy away very rapidly through Hawking radiation \cite{Hawking:1974sw}.  Emparan \etal \cite{Emparan:2000rs} suggest, somewhat speculatively, that this radiation will propagate mainly in the brane directions, and hence could be experimentally observable the Large Hadron Collider.  Due to their small size, such black holes are thought to be well-approximated by asymptotically flat black holes in higher dimensions.

Asymptotically anti-de-Sitter black holes, with a negative cosmological constant, are perhaps of even greater interest, due to the gauge-gravity correspondence \cite{Maldacena:1997re}.  There is a vast recent literature devoted to interpreting certain five-dimensional, asymptotically anti-de-Sitter `bulk' spacetimes in terms of states of four dimensional gauge theories living on the (timelike) boundary of $AdS_5$.  When a black hole is present in the bulk, the dual state in the field theory is at finite temperature (given by the Hawking temperature of the black hole) \cite{Witten:1998zw}.  `Phenomenological' models of this type have led to new ways of describing certain properties of the fluid dynamics of strongly coupled plasmas, for example their viscosity to entropy ratio \cite{Policastro:2001yc,Buchel:2003tz,Son:2007vk} (see e.g.\ \cite{Rangamani:2009xk,Herzog:2009xv,Janik:2010we} for recent reviews).  More recently, dual descriptions of four-dimensional superconductivity have been constructed \cite{Hartnoll:2008vx,Horowitz:2010gk}.

There are also good reasons for wanting to study higher-dimensional black hole spacetimes that are solutions to the Einstein equations for various matter models, in particular those arising from supergravity theories believed to represent a low energy limit of string theory.  Many of the five-dimensional spacetimes used for gauge-gravity calculations are somewhat ad-hoc; i.e.\ they are constructed because they result in interesting 4D physics, rather than because they arise from some more fundamental theory in higher dimensions.  It will be interesting to see if similar clear interpretations can be given for 5D black hole spacetimes arising from more `realistic' matter models.  However, such black holes will generally be beyond the scope of this thesis.\newpage

\subsection{Schwarzschild-Tangherlini black holes}
There are now many known, exact, black hole solutions to the higher-dimensional vacuum Einstein equations.  The first such solution was found by Tangherlini \cite{Tangherlini} in 1963.  He generalized the Schwarzchild solution (\ref{eqn:schwarzchild}) to arbitrary dimension $d$, finding that the metric appears very similar to the 4-dimensional version:
\begin{equation}\label{eqn:tangherlini}
  ds^2 = -\left( 1 - \left(\tfrac{r_0}{r}\right)^{d-3}\right) dt^2 
         + \left( 1 - \left(\tfrac{r_0}{r}\right)^{d-3}\right)^{-1} dr^2
         + r^2 d\Omega_{d-2}^2.
\end{equation}
Here $d\Omega_{d-2}^2$ is the metric on a unit $(d-2)$-sphere, and $r_0>0$ some arbitrary parameter related to the mass of the spacetime.  

Although this metric looks very similar in four and higher dimensions, there is a significant physical difference between the two cases.  In four dimensions, there exist stable, bounded timelike geodesics, corresponding to orbits of massive bodies about the black hole (or other central mass).  However, for $d>4$, no such orbits exist \cite{Tangherlini}.

\subsection{Myers-Perry black holes}
In a $d>4$ dimensional spacetime there are $\lfloor \tfrac{d-1}{2} \rfloor$ independent planes of rotation.  Therefore, one might expect a higher-dimensional generalization of the Kerr black hole to be specified by this number of independent angular momenta $J_i$.  Such a direct generalisation was derived by Myers \& Perry \cite{mp}.  When only one angular momentum is turned on, the solution can be written as
\begin{multline} \label{eqn:singmp}
  ds^2 = -dt^2 + \frac{\mu r^{5-d}}{\Sigma}(dt-a\sin^2 \theta d\phi)^2 
         + (r^2 + a^2) \sin^2 \theta d\phi^2 \\
         + \Sigma \left(\frac{dr^2}{\Delta} + d\theta^2\right) + r^2 \cos^2 \theta d\Omega_{(d-4)}^2 
\end{multline}
where
\begin{equation}
  \Delta = r^2 + a^2 -\frac{\mu}{r^{d-5}}, \eqand \Sigma = r^2 + a^2\cos^2 \theta.
\end{equation}
It has mass and angular momenta
\begin{equation}
  M = \frac{(d-2)\Om_{d-2} \mu}{16\pi G}, \qquad 
  J_1 = \frac{2Ma}{d-2}, \qquad J_i = 0\quad  \forall i>1
\end{equation}
where $\Om_{d-2}$ is the surface area of a unit $(d-2)$-sphere.  Note that this reduces to the Kerr metric when $d=4$.

There is always a coordinate singularity at $r=r_+$, defined by to be the largest value of $r$ such that $\Del(r_+) = 0$.  This corresponds to an event horizon.  However, the equation $\Del(r) = 0$ has a different nature in different dimensions:
\begin{itemize}
  \item For $d=4$, this is a quadratic equation, with roots $r_\pm = \frac{1}{2} (\mu \pm \sqrt{\mu^2 - 4a^2})$.  For fixed mass, this places an upper limit on the value of $a$ that is allowed for a horizon at $r=r_+$ to exist: we must have $|a| \leq \mu/2$.  When this bound is saturated, there is a regular extremal horizon $r=r_+=r_-$.
  \item For $d=5$, the roots are $r_\pm = \pm \sqrt{\mu - a^2}$, and again there is an upper bound on $a$.  However, when the bound is saturated here the roots both lie at $r=0$, and hence there is a naked singularity.  Therefore, singly-spinning Myers-Perry black holes have no extremal limit in five dimensions.
  \item For $d>5$, $\Delta(r)=0$ has a real, positive root for any $r>0$ and hence the black hole can have arbitrarily large angular momentum.  A black hole with an angular momentum per unit mass that is very large is known as an `ultra-spinning' black hole.  Such black holes have `flat' horizons, with a thickness far less than their width.  The presence of these two different lengthscales has a significant impact on the physics of these black holes, as we shall see later when discussing stability.
\end{itemize}

When more than one angular momentum is turned on, the form of the solutions is rather more complicated.  For odd dimension $d=2N+3$, they can be written as \cite{ER:2008}
\begin{equation}
  ds^2 = - dt^2 + \sum_{i=1}^N (r^2+a_i^2)(d\mu_i^2 + \mu_i^2 d\phi_i^2) 
         + \frac{\mu r^2}{\Pi F} \left(dt - \sum_{i=1}^N a_i \mu_i^2 d\phi_i\right)^2 
         + \frac{\Pi F}{\Pi - \mu r^2} dr^2
\end{equation}
where $i=1,\dots, N$ labels the independent planes of rotation,
\begin{equation}
  F(r,\mu_i) = 1 - \sum_{i=1}^N \frac{a_i^2 \mu_i^2}{r^2 + a_i^2}, \qquad
  \Pi(r)     = \prod_{i=1}^N (r^2 + a_i^2)
\end{equation}
and $\mu_i$ are directional cosines with $\sum_i \mu_i^2 = 1$.  The constants $\mu$ and $a_i$ parametrize the mass and angular momenta of the black hole respectively.  An analogous expression can be written down in even dimensions $d=2N+2$, where the metric takes the form
\begin{equation}
  ds^2 = - dt^2 + r^2 d\alpha^2 + \sum_{i=1}^N (r^2+a^2)(d\mu_i^2 + \mu_i^2 d\phi_i^2) 
         + \frac{\mu r^2}{\Pi F} \left(dt - \sum_{i=1}^N a_i \mu_i^2 d\phi_i\right)^2 
         + \frac{\Pi F}{\Pi - \mu r^2} dr^2
\end{equation}
where now $\al^2 + \sum_i \mu_i^2 = 1$.

The spacetime admits $(N+1)$ commuting Killing vectors, $\pd / \pd t$ and $\pd / \pd \phi^i$, and has a $\Rbb \times U(1)^N$ isometry group.  These complicated forms of the metric mean that extracting much information analytically can often be difficult.  However, things are perhaps nicer than might be expected.  All black holes in the Myers-Perry family admit a set of hidden symmetries analogous to those that exist for the Kerr spacetime in four dimensions \cite{Frolov:2007nt}.  To be precise, for any metric in this family, one can construct a conformal Killing-Yano tensor, and this results in a sufficient number of these symmetries to render geodesic motion completely integrable \cite{Page:2006ka}.  These hidden symmetries also exist for the asymptotically $(A)dS$ generalizations of the Myers-Perry metrics, constructed by Refs.\ \cite{Hawking:1998kw,Gibbons:2004,Gibbons:2004js}.

\subsubsection{Cohomogeneity-1 Myers-Perry black holes}
In odd dimensions, a particularly simple class of Myers-Perry metrics are given by setting all of the parameters $a_i$, or equivalently the angular momenta $J_i$, to be equal.  Here, the $U(1)^N$ rotational isometry group is enchanced to $U(N)$, and the metric depends non-trivially on only one coordinate.  We will discuss certain properties of this family in detail in Chapter \ref{chap:nhperts}. 

\subsection{Black Rings}

In five dimensions, there are regular, asymptotically flat, black hole solutions of the vacuum Einstein equations that are qualitatively different from the Myers-Perry family.  This was first demonstrated by Emparan \& Reall's discovery \cite{ER:2001} of a five-dimensional family of \emph{black ring} solutions.  These are (globally) asymptotically flat black hole solutions of the vacuum Einstein equations, with an event horizon of spatial topology $S^1\times S^2$.  This \emph{singly-spinning black ring} has only one non-zero angular momentum; it rotates about the $S^1$ direction but not about the $S^2$.

Unlike topologically spherical black holes, the black ring family does not contain a regular static limit, since there is some lower bound on the allowed angular momentum about the $S^1$ direction.  This condition has a clear physical interpretation; the ring needs enough centrifugal repulsion to balance out the tendency of the black ring to collapse towards its centre.  If this `balance condition' is not satisfied then the spacetime contains a conical singularity.

Assuming that this balance condition holds the solution has two free parameters, one setting an overall lengthscale, and the other parametrizing the `fatness' or `thinness' of the ring.  The properties and structure of this spacetime are described in detail in the review article \cite{ER:2006}, and will be discussed further in Chapter \ref{chap:blackrings} of this thesis.

Do these new black holes violate uniqueness, in the sense of having the same angular momentum and mass as Myers-Perry solutions?  As there is a lower bound on the allowed angular momentum per unit mass of a black ring, and an upper bound on that of the black ring, it is not obvious whether or not this is the case.  However, calculation shows that for a small range of angular momentum per unit mass $J_1/M$, there are both Myers-Perry and (two different) black ring solutions.

The Pomeransky-Sen'kov black ring \cite{Pomeransky} is a \emph{doubly-spinning} generalization of the black ring to include rotation around the $S^2$ as well as around the $S^1$.  Unlike singly spinning black rings, this two parameter family of black rings does admit an extremal limit.  It is described in detail in Chapter \ref{chap:blackrings}.

\subsection{Further solutions}
Another new feature of higher dimensional GR is the existence of a variety of asymptotically flat solutions to the vacuum Einstein equations with multiple black hole event horizons.  In five dimensions, these include black saturn \cite{Elvang:2007sat} (an $S^3$ black hole horizon in the centre of an $S^1\times S^2$ black ring), the black bi-ring \cite{Izumi:2007,Evslin:2007} (an arrangement of concentric, singly spinning black rings rotating in the same plane) and bicycling black rings \cite{Elvang:2007bi} (two singly spinning black rings orthogonal to each other).  In dimensions higher than five, few solutions are known exactly, but it is generally believed that there exists an even richer family of black hole solutions.

All of these solutions admit two commuting spacelike Killing vectors.  In this case, it can be shown (under certain technical assumptions) that the Einstein equations reduce to an integrable system, for which solutions can be found using powerful constructive techniques \cite{ER:weyl}.  The resulting solutions are known as Weyl solutions; and all known exact asymptotically flat, vacuum black hole solutions in five dimensions lie in this class (though not all were originally constructed in this way).

The method works in arbitrary dimension, assuming the existence of $d-3$ commuting angular Killing fields (i.e.\ $\Rbb \times U(1)^{d-3}$ isometry).  However, this number of Killing fields is only consistent with asymptotic flatness in four or five dimensions.  There is also no apparent generalization to asymptotically $AdS$ solutions, since the equations that result are not integrable for solutions with a cosmological constant.
There seems to be no reason to think that there should not be lots of new black holes in these cases; but at present we are lacking a suitable solution generating technique.

Generally, it seems that finding higher-dimensional solutions that are asymptotically anti-de Sitter is significantly harder than the asymptotically flat case.  The generalization of Myers-Perry black holes to include a cosmological constant is known in arbitrary dimension \cite{Hawking:1998kw,Gibbons:2004,Gibbons:2004js}, but attempts to construct an asymptotically $AdS$ black ring have so far proved fruitless.  Approximate results strongly suggest that such solutions exist for all $d\geq 5$ (see, e.g.\ \cite{Caldarelli:2008}).  As for classification, a complete proof of black hole uniqueness in four dimensions has so far proved elusive in the $AdS$ case, and very little is known in higher dimensions.

A potential application of the algebraic classification techniques that will be discussed in Chapter \ref{chap:ghp} is to provide a new approach to discovering and classifying black hole solutions, although it is unclear how likely this is to be successful.  An advantage of these methods though is that including a cosmological constant does not seem to introduce additional difficulties.

\section{Near-horizon geometries}\label{sec:nhintro}
The metrics describing higher-dimensional black holes are often very complicated.  In the case of \emph{extremal} black holes, some useful information about solutions can be extracted without analysis of the full solution.  All such black holes admit a limiting \emph{near-horizon geometry}, which captures certain properties of the full spacetime.  These geometries were introduced in \cite{Reall:2002bh,Chrusciel:2005pa} (certain isolated cases had previously been discussed in four dimensions, e.g.\ \cite{Hajicek:1974}), and have since been used extensively to gain new insights into, for example, the classification problem for higher-dimensional black holes \cite{Kunduri:2008rs,Hollands:2009,Holland:2010bd}.

The existence of these geometries is based around the following result:\footnote{The same result holds for non-degenerate horizons, with the first term in the metric replaced by $-rF(r,x)dv^2$.}
\begin{theorem}[\cite{Moncrief:1983,Friedrich:1998wq,Moncrief:2008}]
  Let $(\Mcal,g)$ be a stationary spacetime in $d$ dimensions, with a degenerate null Killing horizon.  Then, in some neighbourhood of the horizon, one can choose \emph{Gaussian null coordinates} $(v,r,x^A)$ such that the metric takes the form
  \begin{equation}\label{eqn:gaussnullcoords}
    ds^2 = -r^2 F(r,x)dv^2 + 2dv dr + 2r h_A(r,x) dv dx^A + \gamma_{AB}(r,x) dx^A dx^B
  \end{equation}
  where $\pd/\pd v$ is a null Killing vector, $x^A$ are coordinates on spatial slices of the horizon, and the Killing field tangent to the horizon is $\pd/\pd v$.  The null vector field $n=\pd/\pd r$ is tangent to a congruence of null geodesics transverse to the horizon, which is at $r=0$. The functions $F$, $h_A$ and $\gamma_{AB}$ are smooth functions of $r$, with 
  \begin{equation}
    F(r,x)=F(x) + {\cal O}(r), \qquad
    h_A(r,x)=h_A(x) + {\cal O}(r), \qquad 
    \gamma_{AB}(r,x)=\gamma_{AB}(x) + {\cal O}(r)
  \end{equation}
\end{theorem}
Consider a rescaling of coordinates $v \mapsto v/\eps$ and $r \mapsto \eps r$.  We can now take the limit $\eps \rightarrow 0$, to obtain the \emph{near-horizon geometry} of the black hole \cite{Reall:2002bh}, taking the form
\begin{equation}
  ds^2 = - r^2 F(x)dv^2 + 2 dv dr + 2r h_A(x) dv dx^A + \gamma_{AB}(x) dx^A dx^B
\end{equation}
where $F(x)\equiv F(0,x)$ etc.  This metric admits symmetries generated by the Killing vectors $k = \pd/\pd v$ and $X = u\pd/\pd u - v\pd/\pd v$, corresponding to translations $(v \mapsto v + c)$ and rescalings $(v \mapsto v/\eps, r \mapsto \eps r)$ respectively.

The null vector field $\lb\equiv \pd/\pd r$ is non-expanding, non-shearing and non-twisting everywhere.  This implies, by definition, that all near-horizon geometries are \emph{Kundt spacetimes} \cite{Kundt:1961,Podolsky:2008ec,Coley:2009ut}.  In Section \ref{sec:kundt} we will see that all vacuum Kundt spacetimes are algebraically special.

In fact, the near-horizon (NH) geometries of all \emph{known} extremal vacuum black hole solutions have more symmetry than is manifest in the above metric, with the symmetry generated by $k$ and $X$ enhanced to $SO(2,1)$ \cite{Bardeen:1999px,Kunduri:2007vf,Figueras:2008qh,Kunduri:2008rs,Chow:2008dp}.  It is possible to write such NH geometries as a fibration over $AdS_2$ of some $(d-2)$-dimensional real manifold $\Hcal$.

We can think of $\Hcal$ as a (spatial section of) the black hole event horizon, and its metric must therefore be compatible with the horizon topology.  Classification of near-horizon geometries has proved significantly easier than classification of full black hole solutions in higher dimensions.  This allows restrictions to be placed on the existence of certain families of black holes in higher dimensions (assuming that they contain a regular extremal limit).  For example, Kunduri \& Lucietti \cite{Kunduri:2008rs} were able to construct the near-horizon geometries of all extreme vacuum (with possible cosmological constant) black holes in four and five dimensions, assuming a certain amount of rotational symmetry.

More recently, the near-horizon extremal Kerr (NHEK) geometry \cite{Bardeen:1999px} has been given a new interpretation.  Guica \etal \cite{kerrcft} studied quantum states in this geometry, and use their results to propose that Kerr black holes are dual to a chiral conformal field theory in two dimensions; which gives a new approach to understanding, for example, the entropy and temperature of the black hole.  While there are many aspects of this conjecture that are not yet well understood, it certainly serves to emphasize that near-horizon geometries can give insights into a variety of fundamental properties of black holes.  In fact, it has been proposed that similar results may hold for Kerr-AdS spacetimes in higher dimensions \cite{popekerrcft}.

In Chapter \ref{chap:nhperts} we will seek to exploit NH geometries in a different way; as a way of making predictions about the stability of higher dimensional black holes.\newpage

\section{Stability of black holes}\label{sec:higherbhperts}
As interest in higher-dimensional holes has developed, so has interest in their classical stability.  Most of the work so far relies heavily on numerics.  In Chapters \ref{chap:decoupling} and \ref{chap:nhperts} we will take a new approach, and see how much progress can be made analytically.

The standard approach to perturbation theory in GR (see e.g.\ \cite{wald}) begins by making a linearized metric perturbation of the form
\begin{equation}
  g_{\mu\nu} \mapsto g_{\mu\nu} + h_{\mu\nu}.
\end{equation}
Care is needed though, since many choices of $h_{\mu\nu}$ give a perturbed metric that is related to $g_{\mu\nu}$ by an infinitesimal general coordinate transformation.  To eliminate some of this gauge freedom, one can choose a traceless, transverse gauge, fixing $h_{\mu}^{\;\;\mu} = 0$ and $\nabla^\mu h_{\mu\nu} = 0$ (where indices are raised and lowered with $g_{\mu\nu}$).  Given this, the Einstein equations (linearized in $h$), reduce to
\begin{equation}\label{eqn:lich}
  \DelL h_{\mu\nu} = 2\La h_{\mu\nu}
\end{equation}
where $\DelL$ is the Lichnerowicz operator, defined in the case of Einstein spacetimes \eqref{eqn:einstein} by
\begin{equation}
  \DelL h_{\mu\nu} 
      = -\nabla^\rho \nabla_\rho h_{\mu\nu} 
        - 2R_{\mu\phantom{\rho}\nu\phantom{\sig}}^{\phantom{\mu}\rho\phantom{\nu}\sig} h_{\rho\sig}
        + 2\La h_{\mu\nu}.
\end{equation}

In this chapter, we are mainly interested in the classical stability of asymptotically flat, and asymptotically $AdS$ black holes.  However, it is useful to first recall an important result of Gregory \& Laflamme \cite{GL} regarding the stability of black strings and black branes.  They studied a $d$-dimensional spacetime constructed from adding $n=d-D$ flat directions to a $D$-dimensional Schwarzchild black hole, with a metric of the form
\begin{equation}
  ds^2 = -V(r)dt^2 + \frac{dr^2}{V(r)} + r^2 d\Om_{D-2}^2 + \sum_{i=1}^n dz_i dz_i ,\qquad\quad
  V(r) = 1-(r_0/r)^{D-3}
\end{equation}
Consider Fourier mode solutions $h_{\mu\nu} \propto e^{\Om T + i m_i z_i}$ to \eqref{eqn:lich} (in the case $\La=0$), subject to boundary conditions imposing that modes are regular at the event horizon and outgoing at null infinity.  Any mode with $\Re(\Om) > 0$ grows exponentially, and is interpreted as an instability.  Ref.\ \cite{GL} showed numerically that such unstable modes do exist for all such black strings and black branes; corresponding to long wavelength perturbations along the flat directions (i.e.\ those with small $\sum m_i^2$).  This is interpreted as showing that black strings and black branes are classically unstable.  It was speculated that the endpoint of this instability is a chain of localized black holes.  This was investigated in recent numerical work by Lehner \& Pretorius \cite{Lehner:2010pn}.  They showed that the perturbed string evolves first to a sequence of black holes connected by increasingly thin black string sections.  However, the radius of the connecting strings becomes zero in finite asymptotic time, exposing a naked singularity.

The link with asymptotically flat black holes is as follows.  Recall from the introduction that, in six or more dimensions, singly-spinning Myers-Perry black holes can have arbitrarily large angular momentum.  Such \emph{ultra-spinning} black holes have `pancake-like' horizons, with two separate lengthscales corresponding to the thickness of the horizon, and its width.  Emparan \& Myers \cite{Emparan:2003sy} argue that, close to the axis of rotation, such horizons are well approximated by black branes.  Hence, they suffer from the Gregory-Laflamme instability, and are unstable.

Shortly before this, it was established by Ishibashi \& Kodama \cite{Ishibashi:2003} that the higher-dimensional Schwarzschild solution is stable against linearized gravitational perturbations for all $d>4$.  From this, it seems reasonable to conjecture that Myers-Perry black holes will be stable provided they are sufficiently slowly rotating.  If slowly rotating MP black holes are stable, and rapidly rotating ones are unstable, then there must exist some critical value of angular momentum where an instability appears.  Can this value be identified?

A conjecture regarding this can be made by studying the thermodynamics of black hole horizons.  It is known that the area of the black hole horizon(s) in any spacetime is always non-decreasing \cite{HawkingEllis}.  This is reminiscent of the second law of thermodynamics, and for this reason (and others), the entropy of a black hole horizon can be identified as proportional to its area \cite{Bekenstein:1973ur}.  Hence, given two black hole solutions to the Einstein equations (possibly with multiple disconnected horizons), with the same asymptotic mass and angular momenta, the solution with the highest entropy seems to be `thermodynamically preferred'.  Based on this, it seems reasonable to conjecture that when there exist two black hole solutions with the same asymptotic mass and angular momenta, but different entropy, that the solution with lower entropy is likely to be unstable.  Arguments along these lines have been used to make conjectures about the phase space of vacuum black hole solutions in higher dimensions \cite{Elvang:2007hg,Emparan:2007wm}, leading on to the recent development of the so-called \emph{blackfold approach} \cite{Emparan:2009vd,Emparan:2010sx}.

The intuition about links between different kinds of instability was formalised by a conjecture of Gubser \& Mitra \cite{Gubser:2000ec,Gubser:2000mm}, who suggest that a black brane with translational symmetry is classically unstable if and only if it is locally thermodynamically unstable.  This conjecture was proved for a particular class of black brane solutions by Reall \cite{Reall:2001ag}.

These ideas were linked to asymptotically flat black holes by Monteiro \etal \cite{Monteiro:2009tc,Monteiro:2009ke}.  They demonstrate, for several examples of rotating black holes (including singly-spinning black rings), that in the semi-classical approximation the gravitational partition function admits a negative mode precisely when they are locally thermodynamically unstable.  More precisely, for a family of black holes with entropy $S$, labelled by angular momenta $J_i$, one can define the (reduced) Hessian
\begin{equation}
  H_{ij} = \left( \frac{\pd^2 S}{\pd J_i \pd J_j} \right)_M.
\end{equation}
In this paper, they consider a black hole to be \emph{locally thermodynamically unstable} if $H_{ij}$ is not negative definite.  This negative mode appears for all black holes with angular momentum parameter $a$ larger than some critical value $a_0$.  This critical value can be used to give a precise definition of an \emph{ultra-spinning} black hole; i.e.\ all MP black holes with $a>a_0$ are ultraspinning.

This represents further evidence that locally thermodynamically unstable black holes are classically unstable, but does not prove it.  For a proof, one needs to exhibit an explicit linearized instability of a black hole spacetime.  

Dias \etal \cite{Dias:2009iu,Dias:2010maa} made progress towards this goal by studying the case of a singly-spinning, asymptotically flat MP black hole in dimensions $d=7,8,9$.  Rather than working with the black hole spacetime directly, they construct a $d+1$ dimensional black string by adding a single flat direction $dz$, and consider the eigenvalue problem
\begin{equation}
  \DelL h_{\mu\nu} = -k^2 h_{\mu\nu},
\end{equation}
subject to particular boundary conditions, with a Fourier mode ansatz of the form $h_{\mu\nu} \propto e^{ikz} \tilde{h}_{\mu\nu}$.  This is useful because there exist powerful numerical techniques allowing them to find these eigenvalues $k$ with relative ease, for given mass and angular momentum parameter $a$.  Their approach is to start with a particular (small) value of $a$, and find the corresponding eigenvalues $k$.  They then increase $a$ until they find a critical value where $k=0$.  Such a mode is independent of the string direction $z$, and hence can be interpreted as a stationary perturbation mode of the black hole spacetime.  It was argued that this corresponds to the threshold of instability, that is, black holes with larger angular momentum are unstable.

This work motivated the construction of the first explicit example of a linearized instability of an asymptotically flat black hole \cite{Dias:2010eu}, for the cohomogeneity-1 MP black hole.  They demonstrated that, for sufficiently large angular momentum, there exist certain gravitational perturbation modes that grow exponentially with time.  The instabilities found appear at a slightly larger value of angular momentum predicted by the thermodynamic arguments discussed above.

Instabilities of singly-spinning MP black holes have also been found via nonlinear numerical evolution of a perturbed black hole in five \cite{Shibata:2009ad} and higher \cite{Shibata:2010wz} dimensions.  These instabilities are of a qualitatively different nature to those found in \cite{Dias:2010eu}, appearing at a lower value of angular momentum, and breaking more of the symmetry of the original solution.

Despite this recent progress, performing an analysis of the linearized stability of general Myers-Perry black holes seems to be extremely difficult.  Though the principles of doing this are well understood, doing it in practice is not easy.  The equations of motion involved in these perturbations are extremely complicated, which hinders attempts to extract information from them analytically, whilst the large parameter space makes numerical approaches time consuming.  Things are even worse in the case of black rings \cite{ER:2001,ER:2006,Pomeransky}, for which there are physical arguments for various kinds of instabilities \cite{Elvang:2006dd,Dias:2006zv} but little in the way of concrete results.

\section{New results of this thesis}
In this thesis, we will discuss various new results related to some of the questions about higher-dimensional general relativity raised above.  In doing so, we will study powerful mathematical results from four-dimensional general relativity, and investigate the extent to which they can be generalized to higher dimensions.

In Chapter \ref{chap:ghp}, we review the generalization to higher dimensions of the algebraic classification of spacetimes.  In four dimensions, these techniques have proved useful for studying many aspects of general relativity, and we discuss the progress so far in higher-dimensions.  Part of the difficulty with making progress in higher dimensions is that many calculations are extremely complicated.  To ease this difficulty, we will discuss a new approach, a generalization of the four-dimensional Geroch-Held-Penrose (GHP) formalism \cite{ghp}, that simplifies matters in some cases.

The Goldberg-Sachs theorem \cite{Goldberg} is a hugely important theorem in four-dimensional GR.  In Chapter \ref{chap:nongeo} we formulate and prove a partial generalization of the result to arbitrary dimension, as well as discussing what a more complete generalization might look like.

In Chapter \ref{chap:decoupling}, we move on to more physical applications.  We describe how the GHP formalism that we have developed can be applied to construct gauge invariant variables describing perturbations of algebraically special spacetimes.  This opens up a new approach to studying the linearized stability of, for example, Myers-Perry black holes in arbitrary dimension.  In four dimensions, for the Kerr black hole, this approach was exceptionally useful as these gauge invariant variables satisfy a decoupled equation.  Unfortunately, we discover that these gauge invariant variables do not obey a decoupled equation of motion in higher dimensions.

However, the analogous equation does decouple in the near-horizon geometry of any extreme vacuum black hole, and in Chapter \ref{chap:nhperts} we are able to use this equation to conjecture information about instabilities of black holes in arbitrary dimension.  In particular, we show that the equations for linearized perturbations of the near-horizon geometry can be reduced to the equation of motion for a charged, massive scalar field in $AdS_2$.  A generalized Breitenl\"ohner-Freedman stability bound can be defined for such fields.  We conjecture that if there exist perturbation modes that violate this bound, then the full black hole geometry will be unstable, provided that the unstable modes obey a certain symmetry condition.  Although this only allows us to study a limited class of perturbations, it allows progress to be made without resorting to numerics, and offers the possibility of making general statements about stability in arbitrary dimension.  We provide evidence for this conjecture by comparing our results with those obtained by numerical work in a few particular cases, and find good agreement.

The final chapter of the thesis has a rather different flavour, studying properties of a particular solution to the Einstein equations in five dimensions: the Pomeransky-Sen'kov doubly spinning black ring \cite{Pomeransky}.  We will see that the Hamilton-Jacobi equation describing geodesic motion admits separable solutions in the case of null, zero energy geodesics.  Given the very complicated metric describing such black ring spacetimes, this is something of a surprise.  However, we are able to give some insight into this separability by showing that the black ring admits a novel form of hidden symmetry.  While the full spacetime does not admit a conformal Killing tensor, one can make a Kaluza-Klein dimensional reduction to obtain a four-dimensional spacetime that does admit such a tensor.

\chapter[Algebraic classification and null frames]{Algebraic classification and null frames}\label{chap:ghp}
\section{Introduction}\label{sec:cmpp}

When looking to find out more about gravity in higher dimensions, it is natural to try to generalize mathematical methods that have proved powerful in four dimensions.

The algebraic classification of spacetimes, first considered by Petrov \cite{Petrov:1954}, is one example of such a method.  Such classification played a crucial role in understanding various aspects of four-dimensional GR.  For example, Kerr made use of it in order to construct the metric describing a rotating black hole \cite{Kerr}, while the asymptotic behaviour of gravitational radiation can be conveniently understood in this language (see, e.g.\ \cite{stewart}).

The basic idea behind algebraic classification is to divide spacetimes into different types, in order to prove general results about the properties of a precisely defined set of spacetimes.  The schemes discussed below only say useful things about a few particular spacetimes; the reason that they are useful is that these include various important examples, such as the Kerr black hole and pp-waves.

There are at least four distinct approaches to defining such an algebraic classification.  Roughly speaking, the four approaches make use of null vectors, 2-spinors, scalar invariants and bivectors.  In four dimensions, perhaps surprisingly, all of these methods can be used to give different descriptions of the same classification.  In Section \ref{sec:alg4D} we will briefly review these various approaches.

For each technique, it is possible to define (at least one) generalization to higher dimensions.  However, the generalisations are typically not equivalent to each other, and lead to distinct notions of an algebraically special spacetime.  The focus of much of this thesis will be on a null vector based generalization of these classification schemes to higher dimensions, defined in 2001 by Coley, Milson, Pravda \& Pravdov\'a (CMPP) \cite{cmpp,Alignment}.  This will be introduced in Section \ref{sec:cmppclass}, after which we will briefly review some other higher-dimensional classification schemes, including the spinorial de Smet classification \cite{desmet}.

As we shall see below, algebraically special spacetimes are partly characterized by the existence of preferred null directions.  Therefore, it is useful to introduce computational techniques built around one or two particular null directions.  In four dimensions, the Newman-Penrose (NP) \cite{np} and Geroch-Held-Penrose (GHP) \cite{ghp} formalisms are two related examples of such techniques.  Higher-dimensional versions of these approaches will be discussed in detail in Sections \ref{sec:highernp} and \ref{sec:ghp}.  The higher-dimensional generalisation of the NP formalism was developed by various authors (see e.g.\ \cite{Bianchi,Ricci,Coley:2004hu}), while myself, Pravda, Pravdov\'a \& Reall \cite{higherghp} constructed a higher-dimensional version of the GHP formalism.

\section{Algebraic classification in four dimensions}\label{sec:alg4D}
In the case of Einstein spacetimes, all information about the curvature of the spacetime is contained within the Weyl tensor $C_{\mu\nu\rho\sig}$, and the cosmological constant $\La$.  Therefore, algebraic classification of curvature essentially reduces to algebraic classification of the Weyl tensor.

Weyl classification was first considered by Petrov \cite{Petrov:1954}.\footnote{His work was subsequently rederived by various authors throughout the 1950s, see \cite{exact} for a summary of relevant works.}  In this section we review these important results, describing various different approaches to obtaining them, and discussing some applications.

Although one usually refers to algebraic classification of spacetimes, the classification is entirely local, referring to the algebraic structure of the Weyl tensor at a point.  However, unless the point chosen is particularly special for some reason, the algebraic type will usually be the same in any local neighbourhood.  In fact, most spacetimes of interest turn out to be analytic, and hence these local results can be extended globally across the spacetime (see comments in Section \ref{sec:scalarinvs}).

\subsection{2-spinors}
One approach to algebraic classification uses a spinorial representation of the local Lorentz group to construct a `Weyl polynomial,' and then defines a classification according to how this polynomial factorizes.

To see this in more detail, first recall that $SL(2,\Cbb)$ is isomorphic to a double cover of the proper orthochronous Lorentz group $SO(1,3)^\uparrow$, and that this provides a representation of the local Lorentz group acting on 2-spinors $\xi^A$.  This map can be expressed explicitly using the Pauli matrices $\sig^\mu$ as $V^\mu \leftrightarrow V_{A\dot{A}}$ where
\begin{equation}\label{eqn:SL2Cmap}
  V_{A\dot{A}} = i V_\mu \sig^\mu_{A\dot{A}}, \qquad\qquad 
  V^\mu = \tfrac{i}{2} V_{A\dot{A}} \bar{\sig}^{\mu\dot{A} A},
\end{equation}
for $\bar{\sig} = (\sig^0, -\sig^i)$ and 
\begin{equation}
  \sig^0 = \twomat{1 & 0 \\ 0 & 1}, \qquad
  \sig^1 = \twomat{0 & 1 \\ 1 & 0},\qquad
  \sig^2 = \twomat{0 & -i \\ i & 0}, \qquad
  \sig^3 = \twomat{1 & 0 \\ 0 & -1}.
\end{equation}
Our notation is similar to that of \cite{stewart,exact}, and we will use an equals sign $=$ to denote quantities that are equivalent under this map.  In this way, we can define a spinorial counterpart of the Weyl tensor
\begin{equation}
  C_{\mu\nu\rho\sig} = C_{A\dot{A} B\dot{B} C\dot{C} D\dot{D}}.
\end{equation}
Furthermore, it can be shown that this can be expanded as (see, e.g.\ \cite{stewart})
\begin{equation}
  C_{A\dot{A} B\dot{B} C\dot{C} D\dot{D}} = \Psi_{ABCD} \eps_{\dot{A}\dot{B}} \eps_{\dot{C}\dot{D}} 
                                          + \Psi_{\dot{A}\dot{B}\dot{C}\dot{D}} \eps_{AB} \eps_{CD} 
\end{equation}
for some totally symmetric spinor $\Psi_{ABCD}$, which we will refer to as the \emph{Weyl spinor}, where $\eps_{AB}$ is the alternating symbol in two dimensions.  Using this spinor, one can construct the \emph{Weyl polynomial}
\begin{equation}
  C(\xi) = \Psi_{ABCD} \xi^A \xi^B \xi^C \xi^D
\end{equation}
for arbitrary 2-spinors $\xi = (x,y)$.  This is a homogeneous polynomial in two variables $x$ and $y$, and hence it follows from the fundamental theorem of algebra that it has four roots.  Each root defines, up to normalisation, a 2-spinor, and hence there are four such 2-spinors that are somehow inherent to the geometry.

The Petrov classification is defined by considering the multiplicities of these roots, as in Table \ref{tab:petrov}.  We say that a spacetime with all roots distinct is algebraically general (Type I); if at least two of them coincide then a spacetime is algebraically special (Type II, III, IV or D).
\begin{table}[ht]
  \begin{center}
  \begin{tabular}{|c|c|}\hline
     Petrov Type & Multiplicity of roots\\\hline
        I (or G)  &  (1,1,1,1) \\
        II  &  (2,1,1)  \\
        III &  (3,1) \\
        IV (or N)  &  (4) \\
        D  (or II$_{ii}$) & (2,2)\\\hline
  \end{tabular}
    {\it \small\caption[The Petrov classification in four dimensions.]{\label{tab:petrov}The four dimensional Petrov classification expressed in terms of multiplicities of roots of the Weyl polynomial.}}
  \end{center}
\end{table}

This description is specific to four dimensions, as spinorial structures are different in different dimensions.  This suggests that if a spinorial generalization to higher dimensions is possible, then it is likely to be necessary to define this on a dimension by dimension basis.  If we want to be able to write down definitions that work in arbitrary dimension, we can expect to have to use a different method.

\subsection{Vector classification}\label{sec:nullvecclass}
A alternative approach is to work with null vectors.  One motivation for this can be seen from thinking further about the spinorial approach described above.  Under the correspondence \eqref{eqn:SL2Cmap}, null vectors $k$ correspond to rank-1 matrices with zero determinant, which can be expressed as (plus or minus) the outer product of a 2-spinor and its complex conjugate, i.e.\ $k^\mu = \pm \kap^A \bar{\kap}^{\dot{A}}$.  This decomposition is unique up to the sign of $\kap^A$.  Hence, every root of the Weyl polynomial corresponds to a particular null direction in the spacetime; we call this a \emph{principal null direction (PND)}.

By the results above, any spacetime admits exactly four PNDs.  In this language, a spacetime is algebraically special if and only if at least two of the PNDs coincide.  How can this definition be understood in vector language, i.e.\ without making reference to spinors?

To do this, consider a (local) null basis $\{\lb,\nb,m,\bar{m}\}$, where $\lb,\nb$ are real null vectors and $m$ is a complex null vector, with $\lb.n = 1$, $m.\bar{m} = 1$, and all other inner products vanishing.  The metric can be written as
\begin{equation}
  g_{\mu\nu} = 2\lb_{(\mu} \nb_{\nu)} + 2m_{(\mu} \bar{m}_{\nu)} .
\end{equation}
In this basis, one can decompose the Weyl tensor in terms of the complex scalars
\begin{align}\label{eqn:4dweyl}
  \Psi_0 &\equiv C(\lb,m,\lb,m)   &   \Psi_4 &\equiv C(n,\bar{m},n,\bar{m}) \nn\\
  \Psi_1 &\equiv C(\lb,n,\lb,m)   &   \Psi_3 &\equiv C(n,\lb,n,\bar{m}) \nn\\
  \Psi_2 &\equiv -C(\lb,m,n,\bar{m}) \nn\\
         &= \half\left[ C(\lb,n,\lb,n) - C(\lb,n,m,\bar{m})\right].
\end{align}
We say that $\lb$ is a PND if and only if $\Psi_0 = 0$.  This definition depends only on $\lb$, and is equivalent to the definition given above.  The vector $\lb$ is a repeated PND if and only if $\Psi_0=\Psi_1 = 0$; and we say that a spacetime is algebraically special if and only if there exists a choice of $\lb$ such that this is the case.  The complete Petrov classification can be expressed in this form.  To clarify that this classification depends only on $\lb$, equivalent conditions can be given for each type that make this dependence explicit.  The complete classification, given in both of these forms, is given by the statement that, for a particular spacetime (that is not conformally flat), there exists a null vector field $\lb$ such that:
\begin{itemize}
  \item $\Psi_0=\Psi_1 =\Psi_2= \Psi_3 = \Psi_4= 0$ $\Leftrightarrow$ $C_{\mu\nu\rho\sig} = 0$ 
        $\Leftrightarrow$ Spacetime is Type O.
  \item $\Psi_0=\Psi_1 =\Psi_2= \Psi_3 = 0$ $\Leftrightarrow$ $\lb^\sig C_{\mu\nu\rho\sig} = 0$ 
        $\Leftrightarrow$ Spacetime is Type N or O.
  \item $\Psi_0=\Psi_1 =\Psi_2= 0$ $\Leftrightarrow$ $\lb^\rho C_{\mu\nu\rho [\sig} \lb_{\tau]} = 0$ 
        $\Leftrightarrow$ Spacetime is Type III, N or O.
  \item $\Psi_0=\Psi_1 = 0$ $\Leftrightarrow$ $\lb^\mu\lb^\nu C_{\rho \mu\nu [\sig} \lb_{\tau]} = 0$
        $\Leftrightarrow$ Spacetime is Type II, III, N or O.
  \item $\Psi_0 = 0$ $\Leftrightarrow$ $\lb^\mu\lb^\nu \lb_{[\rho} C_{\sig] \mu\nu [\tau} \lb_{\theta]} = 0$ 
        $\Leftrightarrow$ $\lb$ is a PND.
\end{itemize}
Similarly, $\nb$ is a PND iff $\Psi_4=0$, and a repeated PND iff $\Psi_4=\Psi_3 =0$.  Hence, a Type D spacetime is characterized by the existence of a frame in which $\Psi_2$ is the only non-vanishing component of the Weyl tensor.

This null vector language will turn out to be the easiest to generalize to arbitrary dimension, as we shall discuss in detail in Section \ref{sec:cmppclass}.

\subsection{Scalar Invariants}\label{sec:scalarinvs}

So far we have given two distinct methods for working out whether or not a spacetime is algebraically special.  However, both of these methods require several separate steps of working, and the introduction of new structures (e.g.\ the Weyl polynomial and/or the PNDs).  From a computational point of view, it would be nice if there was a more direct condition for checking whether a spacetime is algebraically special.  Such a condition is given by the complex scalar invariants
\begin{equation}
  I \equiv \frac{1}{2} \Psi_{ABCD} \Psi^{ABCD}, \qquad
  J \equiv \frac{1}{6} \Psi_{AB}^{\phantom{AB}CD} \Psi_{CD}^{\phantom{CD}EF} \Psi_{EF}^{\phantom{EF}AB}.
\end{equation}
A spacetime is algebraically special if and only if $I^3 = 27 J^2$ (see e.g.\ \cite{exact}).  It is Type III, N or O if and only if $I=J=0$.  It is possible to express this condition directly in terms of the Weyl tensor, without reference to the Weyl spinor, although the expressions involved are rather more complicated (see e.g.~\cite{Coley:disc}).

The classification can be refined further to fully determine all Petrov types.  Writing
\begin{equation}
  K_{ABCDEF} \equiv \Psi_{PQR(A} \Psi_{BC}^{\quad\; PQ} \Psi^R_{\;\; DEF)}, \qquad
  L_{ABCD} \equiv \Psi_{(AB}^{\quad\; EF} \Psi_{CD)EF},
\end{equation}
it can be shown (see e.g.\ \cite{stewart}) that a non-conformally flat spacetime is Type D if and only if
\begin{equation}
  K_{ABCDEF}K^{ABCDEF}=0
\end{equation}
and Type N if and only if
\begin{equation}
  L_{ABCD}L^{ABCD}=0 .
\end{equation} 
Although computing these invariants explicitly for a given spacetime can be very messy, it is a usually a tractable problem, at least with computer algebra.  From a theoretical point of view, this formulation is useful when dealing with analytic spacetimes.  By the definition of real analyticity, a scalar invariant that is vanishing in some region must vanish everywhere in the spacetime.  Hence, the results above imply that any four dimensional analytic spacetime must have the same algebraic type everywhere (except possibly on some set of zero measure).

Classification using scalar invariants has wider applications than merely giving a different way of understanding the Petrov classification.  Progress in recent years has focused on the use of scalar invariants as providing a continuous characterisation of spacetimes, as opposed to a discrete classification.  An interesting recent result is the following:
\begin{theorem}[Coley \etal \cite{Coley:2009eb}]
  Consider a four-dimensional Lorentzian metric $g$.  Let 
  \begin{equation}
    \mathcal{I} = \{ R, R_{\mu\nu} R^{\mu\nu}, C_{\mu\nu\rho\sig}C^{\mu\nu\rho\sig}, 
                    R_{\mu\nu\rho\sig;\tau}R^{\mu\nu\rho\sig;\tau},\ldots \}
  \end{equation}
  be the set of all scalars constructable from contractions of the Riemann tensor and its derivatives.  Then the metric $g$ is either
  \begin{itemize}
    \item[(i)] determined uniquely by $\mathcal{I}$ \emph{or}
    \item[(ii)] a Kundt metric.
  \end{itemize}
\end{theorem}
Recall that a Kundt spacetime \cite{Kundt:1961} (see also \cite{exact,Coley:2009ut}) is one that admits a shearfree, twistfree, non-expanding, null geodesic congruence $\lb$.\footnote{In fact, many Kundt spacetimes \emph{are} determined uniquely by their scalar invariants, and Ref.\ \cite{Coley:2009eb} gives a precise description of the so-called `degenerate Kundt' spacetimes that are not.}  One important application of this result is to the problem of distinguishing spacetimes.  Clearly if two apparently distinct metrics have differing sets of scalar invariants (e.g.\ an invariant that vanishes identically in one metric but not in the other), then the two metrics must represent genuinely distinct spacetimes.  By this theorem we know that, given a particular pair of spacetimes, it is always possible to show that they are distinct by computing a finite number of elements of $\mathcal{I}$.

\subsection{Bivectors}
Finally, the algebraic classification of the Weyl tensor can also be expressed in terms of the following linear map $\mathrm{C}$, acting on 2-forms (or bivectors) $X = \frac{1}{2}\, X_{[\mu\nu]} dx^{\mu}\wedge dx^{\nu}$ as
\begin{equation}\label{eqn:bivectormap}
  \mathrm{C}: X_{\mu\nu} \mapsto \tfrac{1}{2} C_{\mu\nu}^{\phantom{\mu\nu}\rho\sig} X_{\rho\sig}.
\end{equation}
An algebraic classification can be constructed by considering the eigenvalue structure of this map.

In four dimensions, one can define a duality map ${ }^\sim$ acting on bivectors as
\begin{equation}
  X_{\mu\nu} \mapsto \tilde{X}_{\mu\nu} \equiv \tfrac{1}{2} \eps_{\mu\nu}^{\phantom{\mu\nu}\rho\sig} X_{\rho\sig}.
\end{equation}
Using this, following for example \cite{exact}, we define a complex bivector $X^* = X + i\tilde{X}$ which has the `self-duality' property $(X^*)^\sim = -iX^*$.  We can also construct a self-dual complexified Weyl tensor 
\begin{equation}
  C^*_{\mu\nu\rho\sig} = C_{\mu\nu\rho\sig} 
                       + \frac{i}{2} C_{\mu\nu\tau\theta} \eps^{\tau\theta}_{\phantom{\tau\theta}\rho\sig}.
\end{equation}
The linear map defined by $C^*$ maps the space of self-dual bivectors $X^*$ to itself; and it can be shown \cite{exact} that it contains the same information as the original map \eqref{eqn:bivectormap} (the original map was an endomorphism of a 6-dimensional real vector space, we have converted it into an endomorphism of a 3-dimensional complex vector space).

To make contact with the other forms of classification, we can take a basis
\begin{equation}
  \big\{ 2 \bar{m}\wedge n, 2 m \wedge l, 2 (l\wedge n + m \wedge \bar{m}) \big\}
\end{equation}
of the space of self-dual bivectors.  With respect to this basis, the linear map defined by $C^*$ takes a matrix representation \cite{exact}
\begin{equation}
  \mathbf{Q} 
   = \threemat{\Psi_2 - \tfrac{1}{2}(\Psi_0+\Psi_4) & \tfrac{i}{2} (\Psi_4-\Psi_0)        & \Psi_1-\Psi_3\\
               \tfrac{i}{2} (\Psi_4-\Psi_0)         &\Psi_2 + \tfrac{1}{2}(\Psi_0+\Psi_4) & i(\Psi_1+\Psi_3) \\
               \Psi_1-\Psi_3                        & i(\Psi_1+\Psi_3)                    & -\Psi_2
              }.
\end{equation}
This is a tracefree, symmetric complex matrix, that encodes the 10 independent real Weyl tensor components.  The Petrov classification can then be expressed in terms of this matrix as follows:
\begin{itemize}
  \item A spacetime is Type O iff $\Qb = 0$.
  \item A spacetime is Type N iff $\Qb^2 = 0$ (and it is not Type O).
  \item A spacetime is Type III iff $\Qb^3 = 0$ (and it is not Type N).
  \item A spacetime is Type II iff $\exists\la$ such that $(\Qb+\tfrac{1}{2}\la\Id)^2(\Qb-\la \Id) = 0$ (and it is not Type III).
  \item A spacetime is Type D iff $\exists\la$ such that $(\Qb+\tfrac{1}{2}\la\Id)(\Qb-\la \Id) = 0$ (and it is not Type III).
\end{itemize}
This completes our review of four-dimensional approaches to algebraic classification, we now move on to consider the generalization of these techniques to higher dimensions.

\section{Algebraic classification in higher dimensions}\label{sec:cmppclass} 

In more than four dimensions, we will focus on a particular approach to algebraic classification, which is the natural generalization of the null vector based approach discussed in Section \ref{sec:nullvecclass}.

Coley, Milson, Pravda \& Pravdov\'a (CMPP) \cite{cmpp,Alignment} defined such a classification in arbitrary dimension $d\geq 4$.  In this section we give a detailed account of this approach, and define the notation that will be used in much of the rest of the thesis.

In a $d$-dimensional spacetime we introduce (locally) a frame
\begin{equation}
  \{\lb \equiv \eb_{0}=\eb^{1},
    \nb \equiv \eb_{1} = \eb^{0},
    \mb{i}\equiv\eb_{i} = \eb^{i} \}
\end{equation}
for the tangent space $T(\Mcal)$, where indices $i,j,k,\ldots$ run from $2$ to $d-1$, $\lb$ and $\nb$ are null vector fields and $\mb{i}$ are spacelike vector fields.  We will use $a,b,\ldots$ to denote $d$-dimensional tangent space indices, taking values from $0$ to $d-1$.  We have $\eb_{a}.\eb_{b} = \eta_{ab}$ where
\begin{equation}
  \eta = \left( \begin{array}{ccccc}
                       0 & 1 & 0 & \ldots & 0 \\
                       1 & 0 & 0 & \ldots & 0 \\
                       0 & 0 & 1 & \ldots & 0 \\
                       \vdots & \vdots & \vdots & \ddots & \vdots\\
                       0 & 0 & 0 & \ldots & 1
                     \end{array} \right).
\end{equation}
i.e.\ the only non-vanishing scalar products of basis vectors are $\lb . \nb =1=\eta_{01}$ and $\mb{i} . \mb{j} = \delta_{ij}=\eta_{ij}$.  Although only two of the vectors are null, we will refer to such a basis as a \emph{null frame}.  We will sometimes drop spatial indices $i,j,\dots$ on quantities such as $v_i$, and will use bold font $\vb$ to indicate this.  The Einstein summation convention is used except where explicitly stated otherwise.

\subsection{Changes of basis}

Any tensor $T$ can be expanded with respect to this basis in the obvious way by defining
\begin{equation}
  T_{ab...c} = T(\eb_{a},\eb_{b},\dots,\eb_{c}),
\end{equation}
so, for example, (lowered) indices $0$ correspond to contractions with $\lb$.  The objects $T_{ab\dots c}$ are spacetime scalars, but transform as tensor components under local Lorentz transformations, corresponding to changes in the choice of basis vectors.\footnote{This is if the tensor $T_{\mu\nu...\rho}$ is independent of the choice of null frame. The transformation of tensors constructed from the frame vectors themselves is more complicated, as we shall discuss in Section \ref{sec:ghp}.}

Changes of basis are described by the action of the Lorentz group.  We divide the action of its proper orthochronous component up into the following:
\begin{description}
  \item[Spins:] Rotations of the spatial basis vectors $\mb{i}$: 
                \begin{equation}\label{eqn:spins}
                  \lb \mapsto \lb, \quad\quad
                  \nb \mapsto \nb, \quad\quad
                  \mb{i} \mapsto X_{i j} \mb{j},
                \end{equation}
                where $\Xb:\Mcal\rightarrow SO(d-2)$ is a (position dependent) orthogonal matrix.
  \item[Boosts:] Rescalings of the null basis vectors that preserve the scalar product $\ell \cdot n = 1$:
                \begin{equation}\label{eqn:boosts}
                  \lb \mapsto \la \lb, \quad\quad
                  \nb \mapsto \la^{-1} \nb,\quad\quad
                  \mb{i} \mapsto \mb{i},
                \end{equation}
                where $\lambda$ is an arbitrary non-zero function $\Mcal\rightarrow\Rbb$. We shall say that $\lb$, $\nb$ and $\mb{i}$ have \emph{boost weights} $+1$, $-1$ and $0$ respectively.
  \item[Null Rotations:] Rotation of the rest of the basis about one of the null basis vectors.  A null rotation about $\nb$ takes the form
                \begin{equation}\label{eqn:nullrotn}
                  \lb    \mapsto \lb + z_i \mb{i} - \frac{1}{2}z^2 \nb, \quad \quad 
                  \nb    \mapsto \nb , \quad \quad
                  \mb{i} \mapsto \mb{i} - z_i \lb,
                \end{equation}
                where $z^2 \equiv z_i z_i$, $z_i$ some functions $\Mcal \rightarrow \Rbb^{d-2}$.  An analogous definition can be made for null rotations about $\lb$ (see equation (\ref{eqn:nullrotl} later).
\end{description}
This allows us to make the following definition, first used in this context by CMPP \cite{cmpp}:
\begin{defn}
  A component $T_{a_1 a_2\dots a_m}$ of a tensor $T_{\mu\nu\dots \rho}$ has \emph{boost weight} $b$ if it transforms as
  \begin{equation}
    T_{a_1 a_2\dots a_m} \mapsto \la^b T_{a_1 a_2 \dots a_m}
  \end{equation}
  under boosts of the form \eqref{eqn:boosts}.
\end{defn}
In the following, we will classify components of tensors by their boost weight.  Note that, for tensors that do not depend on the null basis vectors themselves, the boost weight of a component can be read off by subtracting the number of indices 1 from the number of indices 0.  So, for example, components $R_{0i}$ of the Ricci tensor have boost weight $b=+1$, whilst the components $R_{011i}$ of the Riemann tensor have boost weight $b=-1$. 

\subsection{Boost weight decomposition of the Weyl Tensor}
It is most useful to use this classification to make a boost weight decomposition of the Weyl tensor.

In the four-dimensional classification, the complex scalars $\Psi_0$, $\Psi_1$, $\Psi_2$, $\Psi_3$, $\Psi_4$ defined in \eqref{eqn:4dweyl} have boost weights $+2$, $+1$, $0$, $-1$, $-2$ respectively.  Hence, the natural generalization of each of these complex scalars seems to be the collection of components of the Weyl tensor of each boost weight \cite{cmpp}.  Due to the symmetries $C_{abcd} = C_{[ab][cd]}$, the possible boost weights are again $b=2,1,0,-1,-2$.  We define our notation for this decomposition in Table \ref{tab:weyl}.\footnote{Note that there are various different notational conventions in use in the literature, some which differ from others by choices of sign, factors of two etc.}
\begin{table}[ht]
  \begin{center}
  \begin{tabular}{|c|c|c|c|l|c|}
    \hline $b$ & Compt. & Notation & Spin $s$ & Identities & Independent compts. \\\hline
    2 & $C_{0i0j}$& $\Om_{ij}$   & 2 & $\Om_{ij} = \Om_{ji}$, $\Om_{ii}=0$ & $\half d(d-3)$\\\hline
    1 & $C_{0ijk}$& $\Ps_{ijk}$  & 3 & $\Ps_{ijk} = -\Ps_{ikj}$, $\Ps_{[ijk]}=0$ 
                                                            & $\frac{1}{3} (d-1)(d-2)(d-3)$\\
      & $C_{010i}$& $\Ps_{i}$    & 1 & $\Ps_i = \Ps_{kik}$. & \\\hline
    0 & $C_{ijkl}$& $\Phi_{ijkl}$& 4 & $\Phi_{ijkl} = \Phi_{[ij][kl]} = \Phi_{klij}$
                                                          & $\frac{1}{12}(d-1)(d-2)^2(d-3)$\\
      &           &              &   & $\Phi_{i[jkl]}=0$  &                                \\
      & $C_{0i1j}$& $\Phi_{ij}$  & 2 & $\Phi_{(ij)} \equiv \Phis_{ij} = -\half\Phi_{ikjk}$ & \\
      & $C_{01ij}$& $2\Phia_{ij}$& 2 & $\Phia_{ij} \equiv \Phi_{[ij]}$ & $\half (d-2)(d-3)$ \\
      & $C_{0101}$& $\Phi$       & 0 & $\Phi=\Phi_{ii}$ & \\\hline
    -1& $C_{1ijk}$& $\Ps'_{ijk}$ & 3 & $\Ps'_{ijk} = -\Ps'_{ikj}$, $\Ps'_{[ijk]}=0$
                                                             & $\frac{1}{3} (d-1)(d-2)(d-3)$\\
      & $C_{101i}$& $\Ps'_{i}$   & 1 & $\Ps'_i = \Ps'_{kik}$. & \\\hline
    -2& $C_{1i1j}$& $\Om'_{ij}$  & 2 & $\Om'_{ij} = \Om'_{ji}$, $\Om'_{ii}=0$ & $\half d(d-3)$\\
\hline
  \end{tabular}
    {\it\caption[Boost weight decomposition of the Weyl tensor in higher dimensions.]{\small Decomposition of the Weyl tensor by boost weight $b$ for a $d\geq4$ dimensional spacetime.  The various identities given are consequences of the symmetries and tracelessness of the Weyl tensor.  The right hand column shows how many independent components there are of each type, the sum of these numbers gives the total number of independent components of the Weyl tensor for a $d$-dimensional manifold.\label{tab:weyl}}}
  \end{center}
\end{table}

In $d=4$ dimensions, there are exactly two independent components of each boost weight, for example $\Phi_{22}=\Phi_{33} = -\half C_{2323}$ and $\Phi_{23} = -\Phi_{32}$ are the only independent $b=0$ components.  This allows us to express the components in terms of the five complex scalars $\Psi_A$.  However, clearly there are too many components to do this in higher dimensions (c.f.\ the last column of Table \ref{tab:weyl}).

There is also an extra simplification in $d=5$ dimensions, where $\Phi_{ijkl}$ is uniquely fixed in terms of $\Phis_{ij}$ via
\begin{equation}\label{eqn:5dweyl}
  \Phi_{ijkl} \stackrel{d=5}{=}
       2(\del_{il}\Phis_{jk}-\del_{ik}\Phis_{jl}-\del_{jl}\Phis_{ik}+\del_{jk}\Phis_{il})
                 - \Phi (\del_{il}\del_{jk} - \del_{ik} \del_{jl}).
\end{equation}

Note that it is possible to decompose the Weyl tensor further into objects that transform irreducibly under $SO(d-2)$. For example, we could decompose $\Psi_{ijk}$, $\Phi_{ijkl}$ and $\Phi^S_{ij}$ into traceless and pure trace parts.  This may be useful in some contexts (see \cite{bivectors}), but for the applications that will be discussed in this thesis it seems to make things more complicated. 

\subsection{Weyl-aligned null directions}
The higher-dimensional generalization of a principal null direction is given by:
\begin{defn}[\cite{cmpp}]
  A null vector field $\lb$ is a \emph{Weyl-aligned null direction (WAND)} iff all boost weight +2 components of the Weyl tensor vanish everywhere in a frame containing $\lb$.
\end{defn}
In 4 dimensions this definition is equivalent to the statement that $\lb$ is a PND.  Equivalently, $\lb$ is a WAND iff $\Omb=\Ob$.  This definition does not depend on the choice of $\nb$ and $\mb{i}$, since $\Om_{ij}\Om_{ij}$ depends only on $\lb$.

Recall that in four dimensions, all spacetimes with non-vanishing Weyl tensor admit exactly four WANDs (possibly repeated).  This is not the case in higher dimensions: a spacetime may admit no WANDs, a finite number of WANDs, or infinitely many WANDs.  We will see examples of all of these types of behaviour below.

Following the same lines, we can define an algebraically special spacetime as follows:
\begin{defn}\label{def:multwand}
  $\lb$ is a \emph{multiple WAND} iff all boost weight +2 and +1 components of the Weyl tensor vanish everywhere.
\end{defn}
In four dimensions this is equivalent to $\lb$ being a repeated PND.
\begin{defn} \label{def:algspec}
   A spacetime is \emph{algebraically special} if it admits a multiple WAND.
\end{defn}
Note that this notion of being algebraically special is far from the only sensible definition that can be made in higher dimensions.  In fact, most papers on the CMPP classification, including the original papers \cite{cmpp,Alignment}, define a spacetime to be algebraically special if it admits a WAND (not necessarily multiple).  However, the definition that we make here seems to be more useful.  It reduces to the standard definition of algebraically special in 4D, whereas the original definition renders \emph{all} 4D spacetimes algebraically special.  Furthermore, for $d>4$, there exist examples of analytic spacetimes that admit a WAND in some open region, but not in others (see, e.g.\ \cite{brwands,Mahdi}).  More importantly for our purposes, the new results that we derive in Chapters \ref{chap:nongeo}, \ref{chap:decoupling} and \ref{chap:nhperts} will apply to spacetimes that are algebraically special in the sense of Definition \ref{def:algspec}.

To find the algebraic type of a spacetime, one looks first for a choice of (real) null vector $\lb$ that eliminates as many as possible high boost weight Weyl components.  Then we can define:\footnote{We write bold font expressions such as $\Psib=\Ob$ or $\Phib=\Ob$ to indicate that all Weyl components represented by that letter vanish.  So, $\Phib=\Ob$ is the statement that all boost weight 0 components of the Weyl tensor vanish in that basis.}
\begin{defn}
  A spacetime is:
  \begin{itemize}
    \item Type O if its Weyl tensor vanishes everywhere, i.e.\ it is conformally flat. 
    \item Type N if it is not type O and there exists a choice of $\lb$ for which all boost weight 2, 1, 0, -1 Weyl tensor components vanish everywhere (i.e.\ $\Omb=\Psib=\Phib=\Psib'=\Ob$).
    \item Type III if it is not type O or N and there exists a choice of $\lb$ for which all boost weight 2, 1, 0 Weyl tensor components vanish everywhere (i.e.\ $\Omb=\Psib=\Phib=\Ob$).
    \item Type II if it is algebraically special but not type O, N or III (i.e.\ $\Omb=\Psib=\Ob$).
    \item Type I if it admits a WAND, but not a multiple WAND (i.e.\ $\Omb=\Ob$).
    \item Type G if it does not admit a WAND.
  \end{itemize}
\end{defn}
This classification, which depends only on $\lb$, is the {\it primary} classification of the spacetime.  In four dimensions, it is equivalent to the Petrov classification, with the exception of Type G, which does not occur in 4D.  For convenience, we will sometimes say that a null vector field $\lb$ \emph{has the Type III property} if all non-negative boost weight components vanish in a frame containing $\lb$ (and similarly for Type N).

Having fixed $\lb$, one can define a \emph{secondary} classification \cite{cmpp} by choosing $\nb$ so that as many low boost weight components as possible vanish.  For us, the relevant part of this is given by:
\begin{defn}
  A spacetime is \emph{Type D} if it admits two linearly independent multiple WANDs.
\end{defn}
Hence, in a Type D spacetime, one can work in a basis where both $\lb$ and $n$ are multiple WANDS, and hence $\Omb=\Psib=\Psib'=\Omb'=0$.  Recall that the two null vectors are linearly independent if and only if $\lb.n \neq 0$.

\subsection{Determining the CMPP type}
In practical terms, how does one determine the algebraic type of a given spacetime?  One approach to doing this is to start with a convenient choice of null basis $\{\lb,\nb,\mb{i}\}$, and first check whether either $\lb$ or $\nb$ satisfies the WAND condition.\footnote{For algebraically special spacetimes with a lot of symmetry it is often possible to guess correctly which null directions correspond to WANDs.}  If not, then make a null rotation of the form \eqref{eqn:nullrotn} to obtain a new basis $\{\hat{\lb}(\zb),\nb,\hat{m}_i(\zb)\}$, with
\begin{multline}
  \hat{\Om}_{ij}(\zb) = \Om_{ij}-2z_{(j}\Psi_{i)}+2z_k\Psi_{(i|k|j)} 
                        +2Z_{(i|k}\Phi_{|j)k}
                        +z_iz_j\Phi + 4z_kz_{(i}\Phia_{j)k} + z_kz_l \Phi_{kilj}\\
                     + 2z_{(i}Z_{j)k}\Psi'_k + 2z_l Z_{(i|k}\Psi'_{kl|j)} +Z_{ik}Z_{jl}\Om'_{kl} .
\end{multline}
where $Z_{ij}\equiv z_i z_j-\half z^2\del_{ij}$.  At each point in spacetime, the existence of a WAND is therefore equivalent to the question of whether the $(d-2)$ parameters $z_i$ can be chosen to satisfy the $d(d-3)/2$ independent quartic equations $\hat{\Om}_{ij}(\zb) = 0$.

In $d=4$ dimensions, this means that we have two variables and two equations, and hence it is plausible that solutions might always exist.  This is indeed the case, since any 4D spacetime admits WANDs.  However, for $d>4$, there are more equations than free variables, and hence solutions cannot be expected in general.

For $\hat{\lb}$ to be a multiple WAND, the additional condition is that $\hat{\Psi}_{ijk}(\zb)=0$, where
\begin{equation}
  \hat{\Psi}_{ijk}(\zb) =  \Psi_{ijk} + 2z_{[k}\Phi_{i|j]} - 2z_i\Phia_{jk} + z_l \Phi_{lijk}
                        + 2z_iz_{[k} \Psi'_{j]} + 2z_lz_{[k} \Psi'_{j]li} + Z_{il}\Psi'_{ljk}
                      - 2Z_{il} z_{[j}\Om'_{k]l}.
\end{equation}
This corresponds to an additional $\tfrac{1}{3}(d-1)(d-2)(d-3)$ conditions to be satisfied.  In summary, the statement that a spacetime is algebraically special corresponds to the statement that there exists a choice of $\zb$ solving the set of polynomial equations $\hat{\Om}_{ij}(\zb) = 0 = \hat{\Psi}_{ijk}(\zb)$.  Milson \etal \cite{Alignment} discuss how this existence problem can be expressed in the language of alignment varieties, and hence can be approached using tools from algebraic geometry.  However, solving these equations, or proving that solutions do not exist, can be difficult for complicated metrics.

Ref.\ \cite{Alignment} also gives an alternative condition for a null vector to be a WAND, proving that, as in four dimensions:
\begin{lemma}[Milson \etal \cite{Alignment}]
  A null vector field $\lb$ is a WAND if and only if
  \begin{equation}
    \lb^\mu \lb_{[\rho} C_{\sig] \mu\nu [\tau} \lb_{\theta]}\lb^\nu = 0.
  \end{equation}
\end{lemma}
This has the advantage that it does not require the construction of a complete basis to check whether $\lb$ is a WAND. 

This result was later extended by Ortaggio \cite{Ortaggio:2009sb}, who gives a complete characterization of the CMPP classification in terms of a generalization of the Bel-Debever criteria (discussed at the end of Section \ref{sec:nullvecclass}) to higher dimensions, as follows:
\begin{theorem}[Ortaggio \cite{Ortaggio:2009sb}]
  For a $d$-dimensional spacetime with (local) null frame $\{\lb,\nb,\mb{i}\}$:
  \begin{itemize}
    \item $\Omb=\Psib=\Phib=\Psib'=\Ob \Leftrightarrow C_{\mu\nu[\rho\sig} \lb_{\tau]} = 0$.
    \item $\Omb=\Psib=\Phib=\Ob \Leftrightarrow $ $\lb_{[\mu}C_{\nu\rho][\sig\tau}\lb_{\theta]} = 0$ and $C_{\mu\nu\rho[\sig}\lb_{\tau]}\lb^\rho = 0$.
    \item $\Omb=\Psib=\Ob \Leftrightarrow $ $\lb$ is a multiple WAND $\Leftrightarrow $ $\lb_{[\mu}C_{\nu] \rho [\sig\tau}\lb_{\theta]}\lb^\rho = 0$.
  \end{itemize}
\end{theorem} 
These conditions are not identical to the standard ones used in four dimensions, which turn out not to be sufficient to impose the condition that the spacetime is of a particular algebraic type in higher dimensions.

\subsection{Examples of algebraically special spacetimes}\label{sec:algspecexamples}
The algebraic classification of spacetimes is interesting because there are important examples of spacetimes that are algebraically special.  They include the following:
\begin{itemize}
  \item \emph{Schwarzchild-Tangherlini black holes} \cite{Tangherlini} are Type D in all dimensions \cite{cmpp}.  If we write the exterior metric in the form \eqref{eqn:tangherlini}, then the vectors dual to $-dt \pm dr/V$ are tangent to the multiple WANDs.
  \item \emph{Myers-Perry} \cite{mp} and \emph{Kerr-(A)dS} \cite{Hawking:1998kw,Gibbons:2004,Gibbons:2004js} black holes are Type D in all dimensions (see below).
  \item \emph{Vacuum pp-waves} (i.e.\ spacetimes admitting a covaraiantly constant null vector) are Type N in all dimensions.\cite{Coley:2004hu}
  \item \emph{Black string/brane} metrics obtained by adding one or more flat directions to one of the black holes are Type D (this follows, for example, from the results for product spacetimes given in \cite{TypeD}).
  \item \emph{Singly spinning black rings} \cite{ER:2001} are of primary Type II on the horizon, but are not algebraically special in the exterior region (they are of primary Type I or G in different parts of it \cite{brwands}).
\end{itemize}

Why are spherical black holes of algebraic Type D?  Some understanding of this can be obtained from the following two results:
\begin{theorem}[\cite{kerrschild}]
  Let $g$ be a Kerr-Schild metric, i.e.\ one that can be written in the form
  \begin{equation}\label{eqn:kerrschild}
    g_{\mu\nu} = \eta_{\mu\nu} + k_\mu k_\nu
  \end{equation}
  for some conformally flat metric $\eta_{\mu\nu}$ and null vector $k_\mu$.  Then $g$ is algebraically special with multiple WAND tangent to $k$.
\end{theorem}
It is well known that Myers-Perry and Kerr-(A)dS black holes can be written in Kerr-Schild form (indeed, this is how they were originally constructed \cite{mp}), and hence they are algebraically special.  They have multiple WANDs that are expanding everywhere outside the horizon.  Furthermore:\footnote{The proof given for this result in Ref.\ \cite{TypeD} is essentially a `proof by example'.  However, it seems certain that, using for example the results of \cite{Moncrief:2008} on the existence of angular Killing vectors for stationary spacetimes, that this could be made more rigorous.}
\begin{theorem}[\cite{TypeD}] \label{thm:statexp}
  A stationary spacetime admitting an expanding multiple WAND is Type D (or conformally flat).
\end{theorem}
Hence, such black holes are Type D outside the horizon, and hence also on the horizon (by continuity).

In fact, it is reasonably straightforward to show that a spacetime is algebraically special on any null Killing horizon, with a multiple WAND tangent to the null generators of the horizon.\footnote{This can be done, for example, by explicit calculation in Gaussian null coordinates \eqref{eqn:gaussnullcoords} in a neighbourhood of the horizon.}  However, it is well known that the null generators are non-expanding on the horizon, and hence the conditions of Theorem \ref{thm:statexp} fail there.  Hence, there is no inherent reason that a black hole spacetime (e.g.\ the black ring) that is algebraically special only on the horizon should be Type D there.

\section{Aside: Alternative methods of classification}
For comparison, we now briefly discuss some alternative methods of higher-dimensional algebraic classification; namely bivector methods and the De Smet classification.  Neither of these have been as well-developed as the CMPP classification, and we will not make further use of them in the remainder of this thesis.

The existence of multiple distinct methods of algebraic classification in higher dimensions is a disadvantage when it comes to proving general results about a particular classification scheme.  However, there are also advantages, as spacetimes that cannot be usefully analysed using results from one classification might be accessible using another.

\subsection{Bivector methods}
Coley \& Hervik \cite{bivectors} generalized the bivector classification to arbitrary dimension.  The bivector map $\mathrm{C}$ defined by equation \eqref{eqn:bivectormap} is valid in any dimension.  However, note that it is only in four dimensions that Hodge duality provides a map from bivectors to bivectors, and hence the self-duality structure that we then imposed on bivectors cannot be extended to higher dimensions.

Despite this, one can construct a natural bivector classification in arbitrary dimension by classifying the eigenvalue structures (e.g.\ Segre types) of the operator $\mathrm{C}$.  In fact, the authors of Ref.\ \cite{bivectors} chose to describe their classification in terms of the CMPP classification for ease of comparison, and found that even in higher dimensions there are still some links between the bivector and boost weight classifications.  For example, it can be shown that
\begin{lemma}[\cite{bivectors}]
  A spacetime is of CMPP Type III, N or O if and only if the bivector operator is nilpotent.  

  If a spacetime is CMPP Type II, then the bivector operator has at least 3 pairs of matching eigenvalues. 
\end{lemma}
Recent work \cite{Coley:disc} has given a concrete way of computing the eigenvector structure of the bivector operator for a given spacetime, in terms of conditions on a particular series of `discriminants', derived from scalar invariants of various curvature operators.  However, the potential applications of this approach have not yet been explored in great detail.

\subsection{Spinorial methods}

An entirely different approach to a higher-dimensional generalization of the Petrov classification was given by De Smet \cite{desmet}.  His work attempts to generalize the 4D spinorial approach.  However, there are no 2-component spinor representations of the Lorentz group in 5D.  For this reason, de Smet's work uses a particular Dirac spinor representation of the 5D Clifford algebra.  A clear exposition of this approach is given by Godazgar \cite{Godazgar:2010ks}, who also notes that an analogous approach to algebraic classification can be used in four dimensions, but that it gives a different classification scheme to the others discussed above.

Using such a representation $\Gamma^a$, a spinor conterpart of the Weyl tensor can be defined as:
\begin{equation}
  C_{ABCD} = C_{abcd} \Gamma^{ab}_{AB} \Gamma^{cd}_{CD} 
\end{equation}
where $\Gamma^{ab} = \Gamma^{[a}\Gamma^{b]}$.  The motivation behind the particular choice of representation is that it renders $C_{ABCD}$ totally symmetric.  It is not possible to make such a choice in all spacetime dimensions.

The symmetry allows the construction of a Weyl polynomial
\begin{equation}
  C(\psi) = C_{ABCD} \psi^A \psi^B \psi^C \psi^D
\end{equation}
for 4-spinors $\psi^A$.  They have four components, so this is a homogeneous quartic polynomial in 4 variables, which is not guaranteed to factorise.  If it does, then the spacetime is algebraically special in the de Smet classification.

This notion of algebraically special is distinct from the notion of algebraically special in the CMPP classification.  For example, the product of any 4D Petrov Type III spacetime with a flat direction is Type III in the CMPP classification, but algebraically general in the de Smet sense \cite{Godazgar:2010ks}.

The de Smet classification can be refined further, giving a list of possible algebraic types according to the way in which the quartic polynomial factorises.  We use notation where a number represents the degree of a polynomial factor, and underlining a set of factors indicates that they are repeats of each other.  Naively, there are 12 allowed types: 4 (no factorisation, algebraically general), 22, 31, 211, \underline{22}, 1111, 2\underline{11}, 11\underline{11}, \underline{11}\underline{11}, 1\underline{111}, \underline{1111}, 0 (where the last option corresponds to a conformally flat spacetime).

However, the complex spinor $C_{ABCD}$ has 70 independent real components, while the Weyl tensor only has 35 independent components in 5 dimensions.  Godazgar \cite{Godazgar:2010ks} shows how to impose the appropriate reality condition on $C_{ABCD}$ to halve the number of independent components.  After the imposition of this condition, he shows that four of the de Smet types cannot occur, reducing the allowed types to 4, 31, 22, \underline{22}, 211, 1111, \underline{11}\underline{11}, 0. 

Some examples of spacetimes that are algebraically special in this classification include:
\begin{itemize}
  \item Schwarzchild-Tangherlini black holes \cite{Tangherlini} are Type \underline{22}. \cite{DeSmet:2003kt}
  \item Singly-spinning Myers-Perry black holes \cite{mp} are Type \underline{22}. \cite{DeSmet:2003kt}
  \item BMPV black holes \cite{BMPV} are Type 22. \cite{DeSmet:2004if}
  \item Singly-spinning black rings \cite{ER:2001} are Type 4 (algebraically general) \cite{Godazgar:2010ks}.
\end{itemize}

\subsection{Type D spacetimes and hidden symmetry}
In four dimensions, there are strong links between Petrov Type D spacetimes, and the hidden symmetry structures discussed in Section \ref{sec:kerrkilling}.  It is known that:
\begin{theorem}[\cite{Walker:1970,Hughston:1973,exact}]
  In four dimensions, every Petrov Type D vacuum solution admits a conformal Killing tensor.  All Petrov Type D vacuum solutions with the exception of the generalized C-metric admit a rank-2 Killing tensor, and an associated Killing-Yano 2-form.
\end{theorem}
Conversely,
\begin{theorem}[\cite{Collinson:1976,Stephani:1978,exact}]
  A vacuum spacetime admitting a non-degenerate conformal Killing-Yano 2-form is Petrov Type D.
\end{theorem}
These results have been partially generalized to higher dimensions.  It is known that:
\begin{theorem}[\cite{Krtous:2008tb}]\label{thm:KYtypeD}
  A $d$-dimensional vacuum spacetime admitting a closed, non-degenerate conformal Killing-Yano 2-form is Type D in the CMPP classification.
\end{theorem}
However, there is no converse result; it is not known whether all Type D vacuum solutions admit a conformal Killing tensor.  Attempting to prove this in the same way as the four-dimensional result does not work, as it requires the use of the Goldberg-Sachs theorem (which we will discuss in detail later).  

Furthermore, in four dimensions all Type D solutions were constructed explicitly by Kinnersley \cite{Kinnersley}.  In higher dimensions, this has not been done, and it is far from clear that finding all such solutions is likely to be possible.  On this basis, it has been suggested \cite{Mason:2010zzc} that perhaps the natural generalization of the Type D class of metrics to higher dimensions is actually those metrics satisfying the assumptions of Theorem \ref{thm:KYtypeD}.  There is some merit in this suggestion; Krtou{\v s} \etal \cite{Krtous:2008} (generalizing work of Houri \etal \cite{Houri:2008}) are able to explicitly construct all metrics satisfying these conditions.  However, we will see later in the thesis that the more general class of metrics that are algebraically special in the CMPP classification also have useful general properties, which seems to motivate this less restrictive definition.

\section{The Newman-Penrose Formalism} \label{sec:highernp}

So far everything that we have done in this chapter has been algebraic.  We now look to introduce some dynamics, and in particular to do this in a way that is particularly convenient for algebraically special spacetimes.  In four dimensions, such an approach was developed by Newman \& Penrose \cite{np}.  They developed a formalism for studying general relativity that is well-adapted to spacetimes that admit one or more preferred null directions; for example principal null directions.

Working in a frame that includes this null vector often makes calculations simpler than they would otherwise be.  The dynamics comes from writing out the following in the frame basis:
\begin{itemize}
  \item the Bianchi identity (\ref{id:bianchi}),
  \item the Ricci identity (\ref{id:ricci}) as applied to the basis vectors $\{\lb,n,\mb{i}\}$,
  \item the commutators of the frame basis derivatives
        \begin{equation}
          D \equiv \lb.\nabla, \qquad
          \Delta \equiv n.\nabla, \qquad
          \delta \equiv m.\nabla.
        \end{equation}
\end{itemize}
The second of these includes the information from the Einstein equations.

In four dimensions, the Newman-Penrose formalism can be expressed in terms of either spinors or null vectors.  Here, we discuss only the vector version, which has been better studied to date in higher dimensions.  This is part of the reason why the CMPP classification scheme has so far proved more successful than the de Smet classification: it has some dynamics to accompany it.  However, Garc\'ia-Parrado G\'omez-Lobo \& Mart\'in-Garc\'ia \cite{GomezLobo:2009ct} have more recently considered spinor calculus in five dimensions in this context, and it will be interesting to see if their work can generate any useful new results in the future. 

\subsection{Results in four dimensions}
Obviously the aim of the NP formalism is to provide a new approach to solving various problems in general relativity.  In four dimensions, this program proved hugely successful, in part due to the following result:
\begin{theorem}[Goldberg \& Sachs \cite{Goldberg}]
  A null vector field is a principal null direction if and only if it is geodesic and shearfree.
\end{theorem}
This implies immediately that a spacetime is algebraically special if and only if it admits a shearfree null geodesic congruence.  Checking for the existence of such a congruence is, in general, far easier than checking the repeated PND conditions explicitly, as the latter requires computing the Weyl tensor.

Furthermore, it is an easy condition to include in a metric ansatz when searching for new solutions.  The classic example of this approach was the construction of the Kerr metric \cite{Kerr}, which was achieved by searching for axisymmetric, algebraically special solutions of the vacuum Einstein equations.  The Kerr solution is an example of a Type D spacetime, and Kinnersley \cite{Kinnersley} was later able to use the NP formalism to find all Type D vacuum metrics.

The study of gravitational radiation far from an isolated source has been a historically important problem, and one that is gaining increasing relevance today as gravitational wave detectors such as LIGO search for experimental evidence for such radiation.  The NP formalism played an important role in early studies of such radiation.  The classic result is the \emph{peeling theorem} (see, e.g.\ \cite{stewart}).  This states that in an asymptotically flat spacetime, far from some isolated source, the Weyl tensor components can be expanded in terms of some appropriate radial coordinate $r$ (defined in terms of a conformal compactification) as
\begin{equation}
  C_{abcd} \sim \frac{C^{(N)}_{abcd}}{r} + \frac{C^{(III)}_{abcd}}{r^2} 
                + \frac{C^{(II)}_{abcd}}{r^3} + \frac{C^{(I)}_{abcd}}{r^4} + \dots
\end{equation}
where $C^{(II)}$ is a Weyl tensor of Type II etc.  The components falling off as various powers of $r$ can be given fairly general physical interpretations; e.g.\ the terms in $1/r^3$ can be thought of as corresponding to the gravitational field of a massive object, whereas the terms in $1/r$ correspond to transverse gravitational radiation.

The NP formalism also has powerful applications to black hole perturbation theory, as will be discussed in detail in Chapter \ref{chap:decoupling}.

\subsection{Notation}

The four-dimensional NP formalism describes the spin connection associated to the null basis $\{\lb,n,m,\bar{m}\}$ in terms of 12 complex functions $\kap$, $\rho$, $\sig$, $\tau$, $\nu$, $\mu$, $\la$, $\pi$, $\eps$, $\beta$, $\gamma$, $\al$.  There are more components in higher dimensions, so we will need some more general notation; merely increasing the number of Greek letters is clearly not a sensible plan.  Here, we will only discuss the higher-dimensional version in detail, using the notation defined by myself and collaborators in Ref.\ \cite{higherghp}, based around that defined in previous works (e.g.\ \cite{Bianchi,Ricci,TypeD}).

We write the covariant derivatives of the basis vectors themselves as
\begin{equation}
 L_{\mu\nu} = \nabla_\nu \ell_\mu, \qquad 
 N_{\mu\nu} = \nabla_\nu n_\mu, \qquad 
 \M{i}_{\mu\nu} = \nabla_\nu m_{i\mu},
\end{equation}
and then project into the null frame to obtain the scalars $L_{ab}$, $N_{ab}$, $\M{i}_{ab}$. From the orthogonality properties of the basis vectors we have the identities
\begin{equation}\label{eqn:ident1}
  N_{0a} + L_{1a} = 0, \qquad
  \M{i}_{0a} + L_{ia} = 0, \qquad 
  \M{i}_{1a} + N_{ia} = 0, \qquad 
  \M{i}_{ja} + \M{j}_{ia} = 0,
\end{equation}
and
\begin{equation}\label{eqn:ident2}
  L_{0a} = N_{1a} = \M{i}_{ia} = 0.
\end{equation}
The optics of $\lb$ are often particularly important.  In this notation, $\lb$ is tangent to a null geodesic congruence if and only if
\begin{equation}
  \kap_i \equiv L_{i0} = 0,
\end{equation}
and if this is the case we say that $\lb$ is geodesic.  The expansion, shear and twist of the congruence are described by the trace $\rho$, tracefree symmetric part $\sigb$ and antisymmetric part $\omb$ respectively of the matrix $\rhob$, with components
\begin{equation}
   \rho_{ij} \equiv L_{ij}.
\end{equation}
For later convenience, we also define $\tau_i \equiv L_{i1}$.

Finally, we decompose the covariant derivative operator itself in the null frame, writing
\begin{equation}
  D \equiv \lb . \nabla, \quad \Delta \equiv \nb . \nabla \eqand \del_i \equiv \mb{i} . \nabla .
\end{equation}
This approach to the $d>4$ generalization of the 4D Newman-Penrose formalism was developed in Refs.\ \cite{Bianchi, Ricci, VSI}.  The $d>4$ analogues of the 4D NP equations are presented in Ref.\ \cite{Ricci}, the Bianchi identity is written out in Ref.\ \cite{Bianchi} and commutators of the above derivatives are given in Ref.\ \cite{VSI}.  These equations are not presented here explicitly, as in Section \ref{sec:ghp} we will see that there is a more compact way of doing this.

In the non-vacuum case, it is also useful to decompose the Ricci tensor in the frame basis.  The approach to doing this is described in Appendix \ref{app:ghpmatter}.  However, for most of this thesis we will only consider spacetimes that are vacuum, with a possible cosmological constant.

We have chosen much of the notation of this section to resemble as far as possible the standard 4D NP notation, for example $\kap_i$ contains the same information as the complex scalar $\kap$.  However, it is not possible to do this fully.  For example, $\rho_{ij}$ is the $d>4$ analogue of the $d=4$ NP scalars $\rho$ and $\sigma$, and we use $\rho$ without indices to denote the trace of $\rho_{ij}$, which differs from the $d=4$ usage.

\subsection{Results in higher dimensions}
Unfortunately, the NP formalism has not yet led to many important new results in higher dimensions.

In terms of constructing new solutions, perhaps the best attempt was made by Godazgar \& Reall \cite{Mahdi}, who constructed all algebraically special spacetimes in arbitrary dimension that are also \emph{axisymmetric}, in the (relatively strong) sense of admitting an $SO(d-2)$ isometry.  In four dimensions, this class includes the C-metric describing a pair of accelerating black holes.  Unfortunately, Ref.\ \cite{Mahdi} did not find such a metric for $d>4$, so if a higher-dimensional generalization exists it is not algebraically special.

Various papers \cite{TypeD,kerrschild,Pravdova:2008gp} have studied the optical properties of multiple WANDs for various classes of algebraically special spacetimes, partly motivated by attempting to find a higher-dimensional generalization of the Goldberg-Sachs theorem.  We will discuss this further in Chapter \ref{chap:nongeo}.

In the case of asymptotically flat spacetimes, possible higher-dimensional generalizations of the peeling theorem are discussed in Refs.\ \cite{Pravdova:2005ey,Ortaggio:2009zt}.  Pravdova \etal \cite{Pravdova:2005ey} derives the basic peeling properties of the Weyl tensor components in even-dimensional spacetimes, for which a notion of asymptotic flatness at null infinity has been defined by Hollands \& Ishibashi \cite{Hollands:2003ie}.  However, Ortaggio \etal \cite{Ortaggio:2009zt} later showed that such spacetimes, admitting a geodesic multiple WAND with $\det(\rhob)\neq 0$, do not contain gravitational radiation.  Hence, it seems that this formalism may not be a useful way of studying this problem in higher dimensions.

So far, we have reviewed a variety of known results from the literature.  We now move on to discuss the first new results of this thesis.

\section{The Geroch-Held-Penrose Formalism} \label{sec:ghp}

Part of the difficulty of proving general results in the higher-dimensional NP formalism is that the equations involved become rapidly very complicated.  This is in part because lots of redundant information is being carried around. 

The motivation for the formalism was to study spacetimes with one or two preferred null directions, and hence we write out all information relevant to these directions explicitly.  However, in the NP formalism, information that depends on the spacelike components of the spin connection (e.g.\ $\M{i}_{jk}$) is also written out explicitly in all of the equations.  Typically, there is no preferred choice of these spatial directions, and it would be useful to maintain covariance with respect to changes in them.

To do this, we will now construct an alternative formalism that gives a halfway house between covariant calculations, and fully explicit frame basis techniques.  Specifically, we look to retain covariance with respect to boosts \eqref{eqn:boosts} and spins \eqref{eqn:spins} of the null frame.  This was motivated by a similar approach taken by Geroch, Held \& Penrose (GHP) \cite{ghp} in four dimensions, and hence we will refer to this as the \emph{higher-dimensional GHP formalism}.

In four dimensions, the GHP formalism allows for a greatly simplified proof of the Goldberg-Sachs theorem (see, e.g.\ \cite{odonnell,penrind}), and aspects of it were used in the derivation of various classic results, for example Hawking's topology theorem \cite{Hawking:1972}.  In higher dimensions, many existing results from the Newman-Penrose formalism can be derived in a more straightforward manner using our new GHP formalism, for example Lemma \ref{lem:TypeN} in the next section.

Most significantly, the higher-dimensional GHP formalism has allowed the discovery of new results.  The best example of this, to be discussed in Chapter \ref{chap:decoupling}, is its role in understanding the decoupling of linearized perturbations of algebraically special spacetimes. 

\subsection{GHP scalars}

The starting point of the GHP formalism is the following definition:
\begin{defn}
  An object $\Tb$ is a \emph{GHP scalar} of spin $s$ and boost weight $b$ if and only if it transforms as
  \begin{equation}
    T_{i_1...i_s} \mapsto X_{i_1 j_1}...X_{i_s j_s} T_{j_1...j_s}
  \end{equation}
  under spins \eqref{eqn:spins} (with $\mathbf{X}\in SO(d-2)$) and as
  \begin{equation}
    T_{i_1...i_s} \mapsto \la^b T_{i_1...i_s}
  \end{equation}
  under boosts \eqref{eqn:boosts}.
\end{defn}

Note that the outer product of a GHP scalar of spin $s_1$ and boost weight $b_1$ with another of spin $s_2$ and boost weight $b_2$ is a GHP scalar of spin $s_1+s_2$ and boost weight $b_1+b_2$.  The sum of two GHP scalars is a GHP scalar only if $s_1=s_2$ and $b_1=b_2$, in which case the result has spin $s_1$ and boost weight $b_1$.

Not all quantities that appear in the higher-dimensional NP formalism are GHP scalars.  In particular,
\begin{equation}
 \label{eqn:noncov1}
  L_{10} = -N_{00}, \quad L_{11} = -N_{01} \eqand L_{1i} = -N_{0i}
\end{equation}
do not transform covariantly under boosts, while
\begin{equation}
 \label{eqn:noncov2}
  \M{i}_{j0}, \quad \M{i}_{j1} \eqand \M{i}_{jk}
\end{equation}
are not covariant under spins.  However, the remaining quantities \emph{are} GHP scalars, and these quantities are listed in full in Table \ref{tab:weights}.
\begin{table}[ht]
 \begin{center}
   \begin{tabular}{|c|c|c|c|l|}
    \hline Quantity & Notation & Boost weight $b$ & Spin $s$ & Interpretation\\ [1mm]\hline
    $L_{ij}$  & $\rho_{ij}$  & 1  & 2 & expansion, shear and twist of $\lb$\\[1mm]
     $L_{ii}$  & $\rho=\rho_{ii}$  & 1  & 0 & expansion of $\lb$\\[1mm]
    $L_{i0}$  & $\kap_{i}$   & 2  & 1 & non-geodesity of $\lb$\\[1mm]
    $L_{i1}$  & $\tau_{i}$   & 0  & 1 & transport of $\lb$ along $n$\\[1mm]
    $N_{ij}$  & $\rho'_{ij}$ & -1 & 2 & expansion, shear and twist of $n$\\[1mm]
     $N_{ii}$  & $\rho'=\rho'_{ii}$ & -1 & 0 & expansion of $n$\\[1mm]
     $N_{i1}$  & $\kap'_{i}$  & -2 & 1 & non-geodesity of $n$\\[1mm]
    $N_{i0}$  & $\tau'_{i}$  & 0  & 1 & transport of $n$ along $l$\\[1mm]\hline
  \end{tabular}
  {\it\caption{\small\label{tab:weights}GHP scalars constructed from first derivatives of the null basis vectors.}}
 \end{center}
\end{table}

\subsection{GHP derivatives}
If $\Tb$ is a GHP scalar then, in general, $D\Tb$, $\Del \Tb$ and $\del_i \Tb$ are not.  In 4D, GHP \cite{ghp} showed how one can combine this lack of covariance of the NP derivatives with the lack of covariance of the NP scalars (\ref{eqn:noncov1}) and (\ref{eqn:noncov2}) to define new derivative operators that \emph{are} covariant. 
These are straightforward to generalize to higher dimensions as follows:\footnote{The characters `eth' $\eth$ and `thorn' $\tho$ come from the Icelandic alphabet.}

\begin{defn}
  The \emph{GHP derivative operators $\tho$, $\tho'$, $\eth_i$} act on a GHP scalar $\Tb$ of boost weight $b$ and spin $s$ as\emph{
  \begin{eqnarray}
    \tho T_{i_1 i_2...i_s} &\equiv & D T_{i_1 i_2...i_s} - b L_{10} T_{i_1 i_2...i_s} 
                                     + \sum_{r=1}^s \M{k}_{i_r 0} T_{i_1...i_{r-1} k i_{r+1}...i_s},\\
    \tho' T_{i_1 i_2...i_s} &\equiv & \Del T_{i_1 i_2...i_s} - b L_{11} T_{i_1 i_2...i_s} 
                                     + \sum_{r=1}^s \M{k}_{i_r 1} T_{i_1...i_{r-1} k i_{r+1}...i_s},\\
    \eth_j T_{i_1 i_2...i_s} &\equiv & \del_j T_{i_1 i_2...i_s} - b L_{1j} T_{i_1 i_2...i_s} 
                                     + \sum_{r=1}^s \M{k}_{i_r j} T_{i_1...i_{r-1} k i_{r+1}...i_s}.
  \end{eqnarray}}
\end{defn}
So, for example:
\begin{eqnarray}
  \tho \rho_{i j} &=& D \rho_{i j} - L_{1 0} \rho_{i j} + \M{k}_{i 0} \rho_{k j} + \M{k}_{j 0} \rho_{i k}, \\
  \eth_{i} \tau_{j} &=& \delta_{i} \tau_{j} + \M{k}_{j i} \tau_{k},\\
  \tho \Om'_{ij} &=& D\Om'_{ij} + 2L_{10} \Om'_{ij} + 2\M{k}_{(i|0} \Om'_{k|j)}.
\end{eqnarray}
These derivative operators have various useful properties, which are easy to verify by explicit computation:
\begin{enumerate}
  \item They are \emph{GHP covariant}.  That is, if $\Tb$ is a GHP scalar of boost weight $b$ and spin $s$, then $\tho \Tb$, $\tho' \Tb$ and $\eth \Tb$ are all GHP scalars, with boost weights ($b+1$, $b-1$, $b$) and spins ($s$,$s$,$s+1$) respectively.
  \item The \emph{Leibniz rule} holds, that is
        \[ \tho( T_{i_1 i_2...i_s} U_{j_1 j_2...j_t}) = (\tho T_{i_1 i_2...i_s}) U_{j_1 j_2...j_t} 
                                                  + T_{i_1 i_2...i_s} (\tho U_{j_1 j_2...j_t})\]
        for all GHP scalars $\mathbf{T}$ and $\mathbf{U}$, and similarly with $\tho$ replaced by $\tho'$ or $\eth_k$.
  \item They are \emph{metric} for $\del_{ij}$, in the sense that $\tho \del_{ij} = \tho' \del_{ij} = 0$ and
        $\eth_i \del_{jk} = 0$.
\end{enumerate}

\subsection{Priming operation}\label{sec:primingsym}

Following GHP, we have used a prime $'$ to distinguish between certain quantities in the notation introduced above.  This has significance: if we define
\begin{equation}
 \lb'=\nb, \qquad \nb'=\lb, \qquad \mb{i}' = \mb{i},
\end{equation}
then one can interpret the prime as an operator which interchanges $\lb$ and $\nb$.  For example:
\begin{equation}
 (\rho_{i j})' = (\mb{i}^\mu \mb{j}^\nu \nabla_\nu \lb_\mu)' = \mb{i}^\mu \mb{j}^\nu \nabla_\nu \nb_\mu 
               \equiv \rho'_{i j}.
\end{equation}
If a scalar $\Tb$ has boost weight $b$ and spin $s$, then $\Tb'$ has boost weight $-b$ and spin $s$.  Clearly $\Tb'' = \Tb$.

If $\lb$ and $\nb$ are treated symmetrically then use of the prime leads to a significant reduction in the number of independent components e.g.\ of the Bianchi identity.  Note that this is no longer true if the symmetry between $\lb$ and $\nb$ is broken. For example, in an algebraically special spacetime, one can choose $\lb$ to be a multiple WAND.  This is endowing $\lb$ with a property not enjoyed by $\nb$ and hence the priming symmetry is broken and one must write out all of the equations explicitly.  In a Type D spacetime, one can choose both $\lb$ and $\nb$ to be multiple WANDs and the priming symmetry is unbroken.

Note that the action of $'$ on the boost weight 0 components of the Weyl tensor contains one subtlety:
\begin{equation}
  \Phi'_{ij} = (C_{0i1j})' = C_{1 i 0 j}= \Phi_{ji} = \Phis_{ij} - \Phia_{ij}.
\end{equation}
The other boost weight zero Weyl components $\Phi_{ijkl}$ are invariant under the priming operation, as are the boost weight zero Ricci tensor components.

In four dimensions, there are two other discrete symmetries of the system available; complex conjugation and *-symmetry (see \cite{ghp}).  Neither of these extends to an arbitrary number of dimensions in a natural way.

\subsection{Null rotations}

The boosts and spins together generate a $\Rbb \times SO(d-2)$ subgroup of the Lorentz group, under which GHP scalars transform covariantly.  Recall that the full Lorentz group can be recovered by including null rotations of one of the null basis vectors about the other.  Null rotations about $\nb$ take the form \eqref{eqn:nullrotn}, while null rotations about $\lb$ takes the form
\begin{equation} \label{eqn:nullrotl}
  \lb    \mapsto \lb, \quad \quad 
  \nb    \mapsto \nb + z_i \mb{i} - \frac{1}{2}z^2 \lb, \quad \quad
  \mb{i} \mapsto \mb{i} - z_i \lb,
\end{equation}
where $z^2 \equiv z_i z_i$.  Now that we are working in the GHP formalism, we can note that the rotation parameters $z_i$ form $\zb$ a GHP scalar with boost weight $b=-1$ and spin $s=1$ (or $b=1$, $s=1$ in the case of a null rotation \eqref{eqn:nullrotn} about $n$).

Although GHP scalars transform in a simple way under boosts and spins, they do not, in general, transform simply under null rotations.  Consider a null rotation about $\lb$, of the form \eqref{eqn:nullrotl}.  The effect on the various spin coefficients is as follows.  For convenience, we define a boost weight $-2$ GHP scalar $Z_{ij} = z_iz_j - \tfrac{1}{2}\del_{ij} z^2$.\footnote{The NP versions of the following equations have appeared in various places previously.  For example, the spin coefficient rotations are described in \cite{Ricci}, and the Weyl components in \cite{bivectors}.}

The Weyl tensor transforms as:
\begin{eqnarray}
  \Om_{ij}  &\mapsto &\Om_{ij},\\
  \Ps_{i}   &\mapsto &\Ps_{i}+\Om_{ij}z_j,\\
  \Ps_{ijk} &\mapsto &\Ps_{ijk}+2\Om_{i[j}z_{k]},\\
  \Phi      &\mapsto &\Phi + 2z_i\Ps_i + z_i \Om_{ij}z_j,\\
  \Phi_{ij}  &\mapsto& \Phi_{ij}+z_j\Ps_i+z_k\Ps_{ikj}+Z_{jk}\Om_{ik} , \label{eqn:phinullrot}\\
  \Phi_{ijkl}&\mapsto& \Phi_{ijkl} - 2z_{[k}\Ps_{l]ij} - 2z_{[i}\Ps_{j]kl}-2z_jz_{[k}\Om_{l]i}+2z_iz_{[k}\Om_{l]j},\\
  \Ps'_i      &\mapsto& \Ps'_i - z_i\Phi + 3\Phia_{ij}z_j - \Phis_{ij} z_j
                        -2Z_{ij}\Ps_j - Z_{jk} \Ps_{jki} -z_j Z_{ik} \Om_{jk},\ \\
  \Ps'_{ijk}  &\mapsto& \Ps'_{ijk} + 2z_{[k}\Phi_{j]i} + 2z_i\Phia_{jk} + z_l \Phi_{lijk}
                        + 2z_iz_{[k} \Ps_{j]} + 2z_lz_{[k} \Ps_{j]li} + Z_{il}\Ps_{ljk}\nn\\
              &       & + 2Z_{il} z_{[k}\Om_{j]l},\label{eqn:psinullrot}\\
  \Om'_{ij}   &\mapsto& \Om'_{ij}-2z_{(j}\Ps'_{i)}+2z_k\Ps'_{(i|k|j)}
                        +2Z_{(i|k}\Phi_{k|j)}
                        +z_iz_j\Phi-4z_kz_{(i}\Phi^A_{j)k} + z_kz_l \Phi_{kilj}\nn\\
              &       & + 2z_{(i}Z_{j)k}\Ps_k + 2z_l Z_{(i|k}\Ps_{kl|j)} +Z_{ik}Z_{jl}\Om_{kl}.
\end{eqnarray}
and the spin coefficients transform as:
\begin{eqnarray}
  \kap_i    &\mapsto& \kap_i,\label{eqn:kaprot}\\
  \tau_i    &\mapsto& \tau_i + \rho_{ij} z_j - \half z^2 \kap_i,\label{eqn:taurot}\\
  \rho_{ij} &\mapsto& \rho_{ij} - \kap_i z_j,\label{eqn:rhorot}
\end{eqnarray}
and
\begin{eqnarray}
  \kap'_i   &\mapsto& \kap'_i + \rho'_{ij} z_j
                       + Z_{ij}\tau_j - \half z^2 \tau'_i + Z_{ij} \rho_{jk}z_k
                       - \half z^2 Z_{ij}\kap_j + \tho' z_i + z_j \eth_j z_i \nn\\
            &       & - \half z^2 \tho z_i,\label{eqn:kapprot}\\
  \tau'_i   &\mapsto& \tau'_i + Z_{ij}\kap_j + \tho z_i,
                       \label{eqn:tauprot}\\
  \rho'_{ij}&\mapsto& \rho'_{ij} - \tau'_{i} z_j + Z_{ik} \rho_{kj}
                      - Z_{ik} \kap_k z_j + \eth_j z_i - z_j \tho z_i,\label{eqn:rhoprot}
\end{eqnarray}
The analagous equations for null rotations about $\nb$ can be obtained by applying the priming operator to all of the equations above.

\subsection{Newman-Penrose equations for Einstein spacetimes}\label{sec:npeqns}

The curvature tensors can be related to the spin coefficients by evaluating the Ricci identity \eqref{id:ricci} for the basis vectors $V=\lb,\nb,\mb{i}$. The corresponding equations are written out in the higher-dimensional NP formalism in Ref.\ \cite{Ricci}.

In the GHP approach, some of these equations (including all those with $V=\mb{i}$) do not transform as scalars and can be neglected.  In the case of an Einstein spacetime \eqref{eqn:einstein}, the equations that do transform as GHP scalars take the following form:\\
\newcounter{oldeq}
\setcounter{oldeq}{\value{equation}}
\renewcommand{\theequation}{NP\arabic{equation}}
\setcounter{equation}{0}
{\noindent\bf Boost weight +2}
\begin{eqnarray}
  \tho \rho_{ij} - \eth_j \kap_i &=& - \rho_{ik} \rho_{kj} -\kap_i \tau'_j - \tau_i \kap_j  
                                     - \Om_{ij} ,\label{fullsachs}\label{NP1}
\end{eqnarray}
{\noindent\bf Boost weight +1}
\begin{eqnarray}
  \tho \tau_i - \tho' \kap_i &=& \rho_{ij}(-\tau_j + \tau'_j)
                                 - \Psi_i ,\label{R:thotau}\label{NP2}\\[3mm]
  \eth_{[j|} \rho_{i|k]}     &=& \tau_i \rho_{[jk]} + \kap_i \rho'_{[jk]}
                                 - \frac{1}{2} \Psi_{ijk} ,\label{R:ethrho}\label{NP3}
\end{eqnarray}
{\noindent\bf Boost weight 0}
\begin{eqnarray}
  \tho' \rho_{ij} - \eth_j \tau_i &=& - \tau_i \tau_j - \kap_i \kap'_j 
                                      - \rho_{ik}\rho'_{kj}-\Phi_{ij}
                                  - \frac{\La}{d-1}\del_{ij},\label{NP4}
\end{eqnarray}
with another four equations obtained by taking the prime $'$ of these four.  This illustrates the economy of the GHP formalism: not only are the above equations considerably simpler than the corresponding NP equations of Ref.\ \cite{Ricci}, but use of the priming operation enables us to reduce the number of equations by half.  We shall refer to the above equations as `Newman-Penrose equations'; for $d=4$, other names in the literature include `Ricci equations', `spin coefficient equations' and `field equations' (see, e.g.\ \cite{exact,stewart,Ricci,penrind}).
\renewcommand{\theequation}{\arabic{chapter}.\arabic{equation}}
\setcounter{equation}{\value{oldeq}}

Appendix \ref{app:ghpmatter} gives these equations in the more general case of a spacetime with arbitrary matter.  Conversely, Appendix \ref{app:ghpeqns} gives them in an important special case; when the spacetime is an algebraically special Einstein spacetime, for which the symmetry under the priming operation is broken if one chooses $\ell$ to be a multiple WAND.  The symmetry is recovered in the case of a Type D spacetime.

\subsection{Bianchi equations}\label{sec:bianchi}

For an Einstein spacetime, $ R_{\mu\nu} = \Lambda g_{\mu\nu}$, so $\nabla_\rho R_{\mu\nu} = 0$ and hence the differential Bianchi identity $\nabla_{[\tau|} R_{\mu\nu|\rho\sigma]} = 0$ implies that $ \nabla_{[\tau|} C_{\mu\nu|\rho\sigma]} = 0$.

These equations become significantly more complicated in spacetimes with arbitrary matter, the details of how to obtain them in the GHP formalism are given in Appendix \ref{app:ghpmatter}.  The components of this equation are written out in full using the higher-dimensional NP formalism (with different notation) in Ref.\ \cite{Bianchi}. 

In GHP notation, the independent components are equivalent to the following equations:\\
\setcounter{oldeq}{\value{equation}}
\renewcommand{\theequation}{B\arabic{equation}}
\setcounter{equation}{0}
{\noindent\bf Boost weight +2:}
\begin{eqnarray}
  \tho \Ps_{ijk} - 2 \eth_{[j}\Om_{k]i} 
                  &=& (2\Phi_{i[j} \del_{k]l} - 2\del_{il} \Phia_{jk}-\Phi_{iljk})\kap_l \nn\\
                  && -2 (\Ps_{[j|} \del_{il} + \Ps_i\del_{[j|l} + \Ps_{i[j|l} 
                     + \Ps_{[j|il}) \rho_{l|k]} + 2 \Om_{i[j} \tau'_{k]},\label{B1}
\end{eqnarray}
{\bf Boost weight +1:}
\begin{eqnarray}
  - \tho \Phi_{ij} - \eth_{j}\Psi_i + \tho' \Om_{ij} 
                 &=& - (\Psi'_j \del_{ik} - \Psi'_{jik}) \kap_k + (\Phi_{ik} + 2\Phia_{ik} + \Phi \del_{ik}) \rho_{kj} 
                      \nonumber\\
                 &&  + (\Psi_{ijk}-\Psi_i\del_{jk}) \tau'_k - 2(\Psi_{(i}\del_{j)k} + \Psi_{(ij)k}) \tau_k 
                     - \Om_{ik} \rho'_{kj}, \label{B2}\\[3mm]
  -\tho \Phi_{ijkl} + 2 \eth_{[k}\Ps_{l]ij}
                 &=& - 2 \Psi'_{[i|kl} \kap_{|j]} - 2 \Psi'_{[k|ij}\kap_{|l]}\nn\\
                 &&  + 4\Phia_{ij} \rho_{[kl]} -2\Phi_{[k|i}\rho_{j|l]} 
                     + 2\Phi_{[k|j}\rho_{i|l]} + 2 \Phi_{ij[k|m}\rho_{m|l]}\nn\\
                 &&  -2\Ps_{[i|kl}\tau'_{|j]} - 2\Ps_{[k|ij} \tau'_{|l]}
                     - 2\Om_{i[k|} \rho'_{j|l]} + 2\Om_{j[k} \rho'_{i|l]},
                     \label{B3}\\[3mm]
  -\eth_{[j|} \Ps_{i|kl]}
                 &=& 2\Phia_{[jk|} \rho_{i|l]} - 2\Phi_{i[j} \rho_{kl]} 
                     + \Phi_{im[jk|} \rho_{m|l]} - 2\Om_{i[j} \rho'_{kl]},\label{B4}
\end{eqnarray}
{\bf Boost weight 0:}
\begin{eqnarray}
  \tho' \Ps_{ijk} -2 \eth_{[j|}\Phi_{i|k]} 
                 &=& 2(\Ps'_{[j|} \del_{il} - \Ps'_{[j|il}) \rho_{l|k]}
                     + (2 \Phi_{i[j}\del_{k]l} - 2\del_{il}\Phia_{jk} - \Phi_{iljk}) \tau_l \nn\\
                 &&  + 2 (\Ps_i \del_{[j|l} -  \Ps_{i[j|l})\rho'_{l|k]} + 2\Om_{i[j}\kap'_{k]},
                     \label{B5}\\[3mm]
  -2\eth_{[i} \Phia_{jk]} 
                 &=& 2\Ps'_{[i} \rho_{jk]} + \Ps'_{l[ij|} \rho_{l|k]} 
                     - 2\Ps_{[i} \rho'_{jk]} - \Ps_{l[ij|} \rho'_{l|k]},\label{B6}\\[3mm]
  -\eth_{[k|} \Phi_{ij|lm]} 
                 &=& - \Ps'_{i[kl|} \rho_{j|m]} + \Ps'_{j[kl|} \rho_{i|m]} 
                     - 2\Ps'_{[k|ij} \rho_{|lm]}\nn\\
                  && - \Ps_{i[kl|} \rho'_{j|m]} + \Ps_{j[kl|} \rho'_{i|m]} 
                     - 2\Ps_{[k|ij} \rho'_{|lm]}.\label{B7}
\end{eqnarray}
Another five equations are obtained by applying the prime operator to equations (\ref{B1})-(\ref{B5}) above.  The above equations are significantly simpler than those of the NP formalism \cite{Bianchi}. 
Appendix \ref{app:Bianchi} gives these additional equations for the important special case of an algebraically special Einstein spacetime (where symmetry under $'$ is typically broken).

It is sometimes useful to consider the following boost weight +1 equation, constructed from the symmetric part of (\ref{B2}) and a contraction of (\ref{B3}):
\begin{eqnarray}
-\eth_j( \Psi_i \del_{jk} - \Psi_{ijk}) + 2\tho' \Om_{ik}
     &=& -\Om_{ik} \rho' + 2 \Om_{ij} \rho'_{[kj]}
         - 4(\Ps_{(i}\del_{k)j} + \Ps_{(ik)j})\tau_j  \nonumber\\
     &&  + \Phi_{kj} \rho_{ij} - \Phi_{jk} \rho_{ij} + \Phi_{ij} \rho_{kj} - \Phi_{ji} \rho_{jk} \nn\\
     &&  + 2 \Phi_{ij} \rho_{jk} - \Phi_{ik} \rho + \Phi_{ijkl} \rho_{jl} + \Phi \rho_{ik}.
\end{eqnarray}
\renewcommand{\theequation}{\arabic{chapter}.\arabic{equation}}
\setcounter{equation}{\value{oldeq}}
In the case of an algebraically special spacetime, with $\ell$ a multiple WAND, this equation is purely algebraic, see Refs.\ \cite{TypeII,TypeD} and also Chapter \ref{chap:decoupling} for examples of its usefulness. 

\subsection{Commutators of derivatives}\label{sec:comms}

In most respects, the GHP formalism leads to significantly simpler equations than the NP formalism.  One important exception to this statement concerns the commutators of GHP derivatives, which are more complicated than the commutators of the NP derivative operators $D$, $\Delta$ and $\delta_{i}$ (see Ref.\ \cite{VSI} for these commutators).  The GHP commutators contain some information that (in the standard NP formalism) is contained within the NP equations that do not transform as GHP scalars.  These commutators depend on the spin $s$ and boost weight $b$ of the GHP scalar $T_{i_1...i_s}$ that they act on. For an arbitrary Einstein spacetime they read:
\setcounter{oldeq}{\value{equation}}
\renewcommand{\theequation}{C\arabic{equation}}
\setcounter{equation}{0}
\begin{multline}
[\tho, \tho']T_{i_1...i_s} 
         = \left[ (-\tau_j + \tau'_j) \eth_j + 
                    b\left( -\tau_j\tau'_j + \kap_j\kap'_j + \Phi 
                            - \frac{2\La}{d-1}\right)
             \right]T_{i_1...i_s}\\
           + \sum_{r=1}^s \left(\kap_{i_r} \kap'_{j} - \kap'_{i_r} \kap_{j} 
                                  + \tau'_{i_r} \tau_{j} - \tau_{i_r} \tau'_{j} + 2\Phia_{i_r j}
                            \right) T_{i_1...j...i_s}, \label{C1}
\end{multline}
\begin{multline}
[\tho, \eth_i]T_{k_1...k_s}
         = \Bigg[-(\kap_i \tho' + \tau'_i\tho +\rho_{ji}\eth_j)
             + b\left(-\tau'_j\rho_{ji} + \kap_j\rho'_{ji} 
             + \Psi_i \right) \Bigg]T_{k_1...k_s} \\
         + \sum_{r=1}^s \Big[ \kap_{k_r}\rho'_{li} - \rho_{k_r i}\tau'_l
            + \tau'_{k_r} \rho_{li} - \rho'_{k_r i} \kap_l
            - \Psi_{ilk_r} \Big] T_{k_1...l...k_s},
            \label{C2}
\end{multline}
\begin{multline}
[\eth_i,\eth_j]T_{k_1...k_s}
         = \left(2\rho_{[ij]} \tho' + 2\rho'_{[ij]} \tho 
                   + 2b \rho_{l[i|} \rho'_{l|j]} + 2b\Phia_{ij}\right) T_{k_1...k_s}\\\
          + \sum_{r=1}^s \Big[2\rho_{k_r [i|} \rho'_{l|j]} + 2\rho'_{k_r [i|} \rho_{l|j]} 
                                + \Phi_{ijk_r l}
                                + \frac{2\La}{d-1} \del_{[i|k_r}\del_{|j]l} \Big] T_{k_1...l...k_s}.
            \label{C3} 
\end{multline}
\renewcommand{\theequation}{\arabic{chapter}.\arabic{equation}}
\setcounter{equation}{\value{oldeq}}
The 4th commutator $[\tho',\eth_i]$ can be obtained easily by taking the prime of (\ref{C2}).  These equations are given in the case of arbitrary matter in Appendix \ref{app:ghpmatter}.

Again, the equations simplify in the case of an algebraically special Einstein spacetime (although at the cost of breaking the priming symmetry), see Appendix \ref{app:Comm} for more details.

\subsection{Further simplification of equations}

In spacetimes of algebraic type II, III or N, there is a preferred choice for the vector $\lb$ (tangent to the multiple WAND), but not for $\nb$.  For practical calculations, it is often useful to ask if we can make a particular choice of $\nb$ that simplifies the Bianchi and Newman-Penrose equations.  Here we prove the following result, which both gives a convenient choice for doing this, and demonstrates the utility of our new notation. 
\begin{lemma}
  Let $\lb$ be a geodesic multiple WAND in an algebraically special Einstein spacetime, with the property that $\det \rhob \neq 0$.  Then the second null vector $\nb$ can be chosen such that $\taub=\taub'=\Ob$.
\end{lemma}
In fact, in Chapter \ref{chap:nongeo} we will prove that an algebraically special Einstein spacetime must admit a {\it geodesic} multiple WAND, so the first condition of the Lemma is not restrictive.  This Lemma is a useful result for simplifying the GHP equations for some Type II spacetimes.  However, note that when the spacetime is Type D one cannot in general align this choice of $\nb$ with the second multiple WAND.
\paragraph*{Proof:}
Since $\lb$ is a geodesic multiple WAND we have
\begin{equation}
  \Omb=\Psib = \kapb = \Ob.   
\end{equation}
Now, using (\ref{eqn:taurot},\ref{eqn:tauprot}), we see a null rotation about $\lb$ maps $\taub$ and $\taub'$ to
\begin{equation}
  \hat{\taub}  = \taub + \rhob \zb \eqand \hat{\taub}' = \taub'  + \tho \zb.
  \label{eqn:rots}
\end{equation}
When $\det\rhob \neq 0$, we can set $\zb=-\rhob^{-1}\taub$ and hence fix $\hat{\taub}=\Ob$.

Applying $\tho$ to (\ref{eqn:rots}a) gives 
\begin{equation}
  \tho \taub + (\tho \rhob) \zb + \rhob \tho \zb = \Ob.
\end{equation}
Using the Newman-Penrose equations (\ref{fullsachs},\ref{R:thotau}) to eliminate some of the derivatives, and then equation (\ref{eqn:rots}a), this leads to
\begin{equation}
  \tho \zb = -\taub'
\end{equation}
and therefore, by (\ref{eqn:rots}b) we have ${\hat \taub'}=\Ob$.

For spacetimes admitting a multiple WAND with $\det \rhob \neq\Ob$ one can therefore, without loss of generality, choose a gauge with
\begin{equation}
\kapb = \taub = \taub' = \Ob \eqand \Omb = \Ob = \Psib.
\end{equation}
This leads to a considerable simplification of the Newman-Penrose and Bianchi equations. $\Box$

\section{Maxwell fields}\label{sec:max}

Maxwell form fields appear in various higher-dimensional supergravity theories, typically obtained from low energy limits of string theory.  Here we use the GHP formalism to study the linear Maxwell equations for such fields.  One motivation for this, discussed further in Section \ref{sec:algmax}, is the connection in 4D between algebraically special spacetimes, and those admitting an algebraically special Maxwell field.
%
%

We shall study Maxwell test fields (i.e. neglecting gravitational backreaction) with $(p+1)$-form field strength (i.e.\ $p$-form potential) in arbitrary dimension $d\geq 4$, with $1\leq p\leq d-3$.  Note that the energy-momentum tensor is quadratic in the Maxwell field.  Hence, to linear order, we can continue to work with the Newman-Penrose, Bianchi and commutator equations derived for Einstein spacetimes, without including the extra matter terms included in Appendix \ref{app:ghpmatter}.  For $p=1$, our work has some overlap with that of Ortaggio \cite{Ortaggio:2007}.

\subsection{GHP-Maxwell equations in higher-dimensions}
In arbitary dimension $d\geq 4$, the source-free Maxwell equations for a $(p+1)$-form field strength $F_{\nu_1 \ldots \nu_{p+1}}$ (i.e. a $p$-form potential) read
\begin{equation}
  \nabla^{\mu} F_{\mu\nu_1...\nu_p} = 0 \eqand \nabla_{[\nu_1} F_{\nu_2...\nu_{p+2}]} = 0.\label{eqn:max}
\end{equation}
We can convert these into GHP notation as follows.  We define
\begin{align}
  \vphi_{k_1\dots k_p}  &\equiv  F_{0k_1\dots k_p}, &
  f_{k_1\dots k_{p-1}}  &\equiv F_{01k_1\dots k_{p-1}},\nn\\
  F_{k_1\dots k_{p+1}}  &\equiv F_{k_1\dots k_{p+1}}, &
 \vphi'_{k_1\dots k_{p}}&\equiv F_{1k_1\dots k_{p}}, 
\end{align}
so $\vphi_{k_1\dots k_p}$ has $b=1$, $f_{k_1\dots k_{p-1}}$ and 
$F_{k_1\dots k_{p+1}}$ have $b=0$, and $\vphi'_{k_1\dots k_p}$ has $b=-1$. Note that $ f'_{k_1\dots k_{p-1}}=- f_{k_1\dots k_{p-1}}$. The Maxwell equations are equivalent to: 

{\noindent \bf Boost weight +1}
\begin{eqnarray}
   \eth_i \vphi_{i k_1\dots k_{p-1}} + \tho f_{k_1\dots k_{p-1}} 
        &=& \tau'_i \vphi_{i k_1\dots k_{p-1}} - \rho f_{k_1\dots k_{p-1}} 
            + \rho_{[ij]} F_{ijk_1\dots k_{p-1}} \nn\\
        & & - \kap_i \vphi'_{i k_1\dots k_{p-1}}
            + (p-1)\rho_{[k_1| i}f_{i|k_2\dots k_{p-1}]} , 
        \label{max:1}\\
  (p+1) \eth_{[k_1} \vphi_{k_2\dots k_{p+1}]} - \tho F_{k_1\dots k_{p+1}}
        &=& (p+1) \Big( \tau'_{[k_1} \vphi_{k_2\dots k_{p+1}]} + \rho_{i[k_1} F_{|i |k_2 \dots k_{p+1}]}\nn\\
        & & \quad + p \rho_{[k_1 k_2} f_{k_3\dots k_{p+1}]}
            + \kap_{[k_1} \vphi'_{k_2\dots k_{p+1}]} \Big),
        \label{max:2}
\end{eqnarray}
{\noindent \bf Boost weight 0}
\begin{multline}
  2\tho' \vphi_{k_1\dots k_{p}} + \eth_j F_{j k_1\dots k_{p}} 
        - p \eth_{[k_1}f_{k_2\dots k_{p}]} \\
       \qquad\qquad\; = (p \rho'_{[k_1|i} - p\rho'_{i[k_1|} - \rho' \del_{[k_1|i})\vphi_{i|k_2\dots k_p]} 
            + 2\tau_i F_{ik_1\dots k_p} \\
          - 2p\tau_{[k_1} f_{k_2\dots k_p]} 
            + (p \rho_{[k_1|i} + p\rho_{i[k_1|} -\rho\del_{[k_1|i})\vphi'_{i|k_2\dots k_p]},\label{max:3}
\end{multline}
\begin{eqnarray}
  \eth_{[k_1}F_{k_2\dots k_{p+2}]} &=& (p+1)\left(\vphi_{[k_1\dots k_{p}}\rho'_{k_{p+1} k_{p+2}]}
                                       + \vphi'_{[k_1\dots k_{p}}\rho_{k_{p+1} k_{p+2}]}\right), \label{max:4}\\
  \eth_i f_{ik_1\dots k_{p-2}}
        &=& -\rho_{[ij]}\vphi'_{ijk_1\dots k_{p-2}} + \rho'_{[ij]}\vphi_{ijk_1\dots k_{p-2}},
            \quad\quad\mathrm{[for}\; p>1\mathrm{]}, \label{max:5}
\end{eqnarray}
together with the primed equations: (\ref{max:1})$'$, (\ref{max:2})$'$ and (\ref{max:3})$'$.

Note that, in the case $p=1$, the quantity $f$ has no indices, and equation (\ref{max:5}) does not appear. Equation (\ref{max:4}) vanishes identically when $p>d-4$, as is the case in conventional $d=4$, $p=1$ electromagnetism. 

A natural question that arises is whether, given an arbitrary solution of the Maxwell equations, one can always find a vector field $\lb$ that is aligned with it, in the sense that $\vphib=\Ob$.  For $p=1$, a partial answer to this question, in a slightly different context, was given by Milson \cite{Milson:align}.  His results (Propositions 4.4 and 4.5) prove that in even dimension it is always possible to make such a choice, but suggest that this is probably not the case in odd dimension.

\subsection{Hodge duality}
It is well known that the source-free Maxwell equations are invariant under Hodge duality.  That is, if a $(p+1)$-form $F$ satisfies the equations (\ref{eqn:max}), then the $(d-p-1)$-form $\star F$ is also a solution. How can this be seen in our new formalism?

To fix signs, we define the totally antisymmetric symbol $\eps$ with $\eps_{012\dots d-1} = +1$.  This results in a volume form
\begin{eqnarray}
  \epsilon  = \eb^{0}\wedge\eb^{1}\wedge\eb^{2}\wedge\dots\wedge \eb^{d-1} 
                      = -\lb \wedge \nb \wedge \mb{2} \wedge \dots \wedge \mb{d-1}.
\end{eqnarray}

Hodge duality maps the basis components of a $p$-form $A$ to $\star A$ where
\begin{equation}
  (\star A)_{b_1\dots b_{d-p}} 
    \equiv \frac{1}{p!} \eps_{b_1\dots b_{d-p}}^{\phantom{b_1\dots b_{d-p}}a_1\dots a_p}A_{a_1\dots a_p}.
\end{equation}
It is useful to define a Euclidean signature, $(d-2)$-dimensional Hodge duality operator $\estar$ by
\begin{equation}
  (\estar T)_{j_1\dots j_{d-2-r}} \equiv \frac{1}{r!} \eps_{j_1\dots j_{d-2-r}i_1\dots i_r} T_{i_1\dots i_r}
\end{equation}
mapping totally antisymmetric GHP scalars with $r$ spatial indices to totally antisymmetric GHP scalars with $d-2-r$ spatial indices.

Consider the action of Hodge duality on our Maxwell $(p+1)$-form $F$, setting $q=d-2-p$ for convenience, so that
\begin{equation}
  (\star F)_{b_1\dots b_{q+1}} 
    = \frac{1}{(p+1)!} \eps_{b_1\dots b_{q+1}}^{\phantom{b_1\dots b_{q+1}}a_1\dots a_{p+1}} F_{a_1\dots a_{p+1}}.
\end{equation}
Taking components, this implies that
\begin{eqnarray}
  (\star\vphi)_{k_1\dots k_q}   &\equiv& (\star F)_{0k_1\dots k_q} 
                                       = (-1)^{d-p}\left(\estar\vphi\right)_{k_1\dots k_q},\label{eqn:starphi}\\
  (\star f)_{k_1\dots k_{q-1}}  &\equiv& (\star F)_{01k_1\dots k_{q-1}}
                                       = \left(\estar F\right)_{k_1\dots k_{q-1}},\label{eqn:starf}\\
  (\star F)_{k_1\dots k_{q+1}}  &\equiv& (\star F)_{k_1\dots k_{q+1}}
                                       = -\left(\estar f\right)_{k_1\dots k_{q+1}},\label{eqn:starF}\\
  (\star \vphi')_{k_1\dots k_q} &\equiv& (\star F)_{1k_1\dots k_q} 
                                       = (-1)^{d+1-p}\left(\estar\vphi\right)_{k_1\dots k_q}.\label{eqn:starphip}
\end{eqnarray}
Note that applying the Hodge star operation to a primed quantity always introduces an extra minus sign, so it is useful to define $(\estar)' \equiv -(\estar)$ to account for this.

\subsection{Algebraically Special Maxwell Fields}\label{sec:algmax}
We now introduce the notion of an algebraically special Maxwell field:
\begin{defn}
  A Maxwell $(p+1)$-form field $F$ is \emph{algebraically special} if there exists a choice of $\lb$ such that all non-negative boost weight components of $F$ vanish everywhere.  A vector field $\lb$ with this property is \emph{multiply aligned} with $F$.
\end{defn}

Note that, by equations (\ref{eqn:starphi}-\ref{eqn:starF}), the property of being algebraically special is preserved under Hodge duality, that is:
\begin{lemma}
  A Maxwell $(p+1)$-form field $F$ is algebraically special if, and only if, $\star F$ is algebraically special.
\end{lemma}

In four dimensions, the Mariot-Robinson theorem (Theorem 7.4 of Ref. \cite{exact}) states that a null vector field is multiply aligned with a (non-zero) algebraically special Maxwell field if, and only if, is geodesic and shearfree. Therefore, by the Goldberg-Sachs theorem, a vacuum spacetime admits such a Maxwell test field if, and only if, it is algebraically special.  It is natural to ask whether any part of this holds in higher dimensions.  The following result holds: 
\begin{lemma}\label{lem:algmax}
  Let $\lb$ be a null vector field in a $d$-dimensional spacetime, multiply aligned with a non-zero Maxwell $(p+1)$-form field $F$, with $0<p<d-2$.  Then
  \begin{itemize}
    \item[(i)] $\lb$ is tangent to a null geodesic congruence.
    \item[(ii)] $\rho_{(ij)}$ has $p$ eigenvalues whose sum is $\rho/2$ (hence the remaining $d-2-p$ eigenvalues must also sum to $\rho/2$).
  \end{itemize}
\end{lemma}
\proof
(i) Choose a null frame in which $\lb$ is one of the basis vectors. Equations (\ref{max:1}) and (\ref{max:2}) reduce to
\begin{equation}
  \kap_i \vphi'_{i k_1\dots k_{p-2}} = 0 =  \kap_{[k_1} \vphi'_{k_2\dots k_p]}.
\end{equation}
If $\kapb\neq \Ob$, then we can use spins to move to a frame where $\kap_i = \kap\del_{i2}$ and immediately show that this implies $\vphi'_{k_1\dots k_p}=0$, and hence the Maxwell field vanishes.  Hence, if the Maxwell field is non-vanishing, $\kapb = \Ob$ and $\lb$ is geodesic, which completes the proof of (i).

(ii) Let $\Sb$ denote the symmetric part of $\rhob$. The Maxwell equation (\ref{max:3}) reduces to
\begin{equation}
  0 = (2p S_{[k_1|i} - \rho \del_{[k_1|i})\vphi'_{i|k_2\dots k_p]}.
\end{equation}
Working in a basis where $\Sb$ is diagonal with eigenvalues $s_i$, this implies
\begin{equation}
  \left[ \sum_{r=1}^p s_{k_r} - \frac{\rho}{2}\right] \vphi'_{k_1\dots k_p} = 0,
\end{equation}
where we drop the summation convention for the remainder of this proof.
The Maxwell field is non-vanishing, so we can shuffle indices to set $\vphi'_{23\dots p+1} \neq 0$; which implies that
\begin{equation}
  \sum_{i=2}^{p+1} s_{i} = \frac{\rho}{2},
\end{equation}
which gives the required result. $\Box$
%
%
%
%

Note that this result is consistent with Hodge duality. In four dimensions, it reduces to the statement that a null vector field multiply aligned with a Maxwell field must be geodesic and shearfree. 

In the case $p=1$ one can prove a slightly stronger result:\footnote{Note that part of this result was first proved in \cite{Ortaggio:2007}.}
\begin{lemma}
  Let $\lb$ be a null vector field in a $d$-dimensional spacetime, multiply aligned with a Maxwell $2$-form field.  Then $\lb$ is geodesic, and the symmetric and anti-symmetric parts of the optical matrix $\rhob$ have the following properties:
  \begin{enumerate}
    \item $\rho_{(ij)}$ has an eigenvalue $\rho/2$, with corresponding eigenvector $\vphi'_i$ (the $b=-1$ part of the Maxwell field)
    \item $\rho_{[ij]} =  \vphi'_{[i} \omega_{j]}$ for some $\omega_i$.
  \end{enumerate}
\end{lemma}
\proof
The geodesity property was proved in Lemma \ref{lem:algmax}.  Now the Maxwell equations (\ref{max:3}-\ref{max:5}) reduce to:  
\begin{eqnarray}
  0 &=& (\rho_{(ki)} - \half \rho \del_{ki})\vphi'_{i}, \label{max:3N}\\
  0 &=& \rho_{[k_1 k_2} \vphi'_{k_3]}.                \label{max:4N}
\end{eqnarray}
These are equivalent to statements 1 and 2 respectively. $\Box$

There is an important difference between $d=4$ and $d>4$ in the above results. As mentioned above, for $d=4$,  $\lb$ is multiply aligned with a Maxwell (test) field if, and only if, it is multiply aligned with the Weyl tensor (in vacuum). The results above demonstrate that this is not true for $d>4$.

For example, consider the Schwarzschild solution, for which the multiple WANDs are geodesic and shearfree, i.e., choosing $\lb$ to be a multiple WAND, all eigenvalues of $\rho_{(ij)}$ are equal to $\rho/(d-2)$.  Then, for $\lb$ also to be multiply aligned with an algebraically special Maxwell $(p+1)$-form field we would need, from Lemma \ref{lem:algmax},  $p\rho/(d-2) = \rho/2$ and hence $d=2(p+1)$.  Therefore only in an even number  $d=2(p+1)$ of dimensions is it possible for a null vector field to be multiply aligned simultaneously with the Weyl tensor and with a $(p+1)$-form Maxwell field in the Schwarzschild spacetime.  This shows that, for a general higher-dimensional spacetime, we cannot expect any relation between vectors multiply aligned with a $(p+1)$-form Maxwell field and vectors multiply aligned with the Weyl tensor, except possibly when $d=2(p+1)$.

\section{Codimension-2 hypersurfaces}

The GHP formalism is particularly useful for spacetimes admitting a preferred pair of null directions.  One example, discussed for $d=4$ by GHP \cite{ghp} (see also \cite{penrind}), is when one is interested in a codimension-2 spacelike surface $\Scal$.   There is a unique (up to a sign) choice of null directions that lie orthogonal to $\Scal$.  Choosing $\lb$ and $\nb$ to lie in those directions implies that $\Scal$ is spanned by the spacelike vectors $\mb{i}$. 

Projections onto the surface are given by
\begin{equation}
  h^\mu_{\phantom{\mu}\nu} = \sum_{i=2}^{d-1} \mb{i}^\mu \mb{i}_\nu, 
\end{equation}
and $h_{\mu\nu}$ is the induced metric on $\Scal$.  Note that $\eth_i$, when acting on boost weight 0 quantities (which are those invariant under the rescaling of $\lb$ and $\nb$), is simply the metric covariant derivative on $\Scal$:
\begin{equation}
  \eth_i h_{jk} = \del_i h_{jk} + \M{l}_{ji} h_{lk} + \M{l}_{ki} h_{jl} = \M{k}_{ji} + \M{j}_{ki} = 0.
\end{equation}

Consider the commutator (\ref{C3m}) (from Appendix \ref{app:ghpmatter}), acting on a boost weight zero GHP scalar $V_k$.  This takes the form
\begin{multline}\label{eqn:Scomm}
  [\eth_i,\eth_j]V_{k} =
                \Big[2\rho_{k [i|} \rho'_{l|j]} + 2\rho'_{k [i|} \rho_{l|j]} 
                   + \Phi_{ijkl} + \frac{2}{d-2} (\del_{[i|k}\phi_{|j]l} - \del_{[i|l}\phi_{|j]k})\\
                   - 2\del_{[i|k}\del_{|j]l}\frac{2\phi+\phi_{mm}}{(d-1)(d-2)} \Big] V_{l}.
\end{multline}
We have used $\rho_{[ij]} = \rho'_{[ij]} = 0$, which follows from Frobenius' theorem.

The terms on the RHS give us the induced Riemann tensor on $\Scal$, in terms of the null vector fields that define the embedding of the surface, and the curvature of the spacetime in which it is embedded.  To see this, we can compare (\ref{eqn:Scomm}) with the $(d-2)$-dimensional Ricci identity
\begin{equation}
  (\nabla_i \nabla_j - \nabla_j \nabla_i) V_k = {}^{(d-2)}\! R_{ijkl}V_l 
\end{equation}
to obtain
\begin{equation}\label{eqn:hypersurfacecurv}
  {}^{(d-2)}\! R_{ijkl} = 2\rho_{k [i|} \rho'_{l|j]} + 2\rho'_{k [i|} \rho_{l|j]} 
                   + \Phi_{ijkl} + \frac{2}{d-2} (\del_{[i|k}\phi_{|j]l} - \del_{[i|l}\phi_{|j]k})
                   - 2\del_{[i|k}\del_{|j]l}\frac{2\phi+\phi_{mm}}{(d-1)(d-2)} .
\end{equation}
This approach to dealing with $(d-2)$-dimensional surfaces has an important advantage over approaches that require a particular choice of basis on the surface in that it is always guaranteed to be well defined across the whole surface \cite{penrind}.  For example, in even dimensions, if $\Scal$ has the topology $S^{d-2}$ then it is well known that there is no continuous, globally valid choice of vector basis $\{\mb{i}\}$ that can be made on $\Scal$.  The GHP approach does not require the introduction of such an explicit basis, and therefore does not suffer from this problem.

Further examples of the use of the higher-dimensional GHP formalism will be discussed in the rest of the thesis, in particular in Chapter \ref{chap:decoupling}.

\chapter[Geodesity of multiple WANDs]{Geodesity of multiple WANDs}\label{chap:nongeo}
\section{Introduction}
Recall that, in four dimensions, a key result in the early development of the Newman-Penrose formalism was the following (re-written here in the language used in higher-dimensions):
\begin{theorem}[Goldberg \& Sachs \cite{Goldberg}]
  In a four-dimensional Einstein spacetime, a null vector field is a multiple WAND if and only if it is tangent to a shearfree null geodesic congruence.
\end{theorem}

In this chapter we investigate the generalization of this result to higher dimensions.  It has been known for some time that the theorem does not generalize in an obvious way.  A geodesic multiple WAND need not be shear-free (this occurs for example in Myers-Perry black holes \cite{TypeD,Frolov}), and a multiple WAND need not be geodesic \cite{TypeII,Mahdi,TypeD}.  The simplest example of the latter behaviour is a product spacetime, for example $dS_3 \times S^2$, where {\it any} null vector field tangent to $dS_3$ is a multiple WAND irrespective of whether or not it is geodesic \cite{Mahdi}.  However, in this example there also exist geodesic multiple WANDs.  The main result of this chapter is a proof that this always happens, at least for Einstein spacetimes:
\begin{theorem}\label{thm:geo}
  An Einstein spacetime admits a multiple WAND if, and only if, it admits a geodesic multiple WAND.
\end{theorem}
The `if' part of this theorem is trivial. To prove the `only if' part, we shall assume that the multiple WAND is non-geodesic and prove that there exists another multiple WAND that is geodesic.  As a first step, we will prove that
\begin{lemma}\label{lem:typeD}
  An Einstein spacetime that admits a non-geodesic multiple WAND is Type D (or conformally flat).
\end{lemma}

We then go on to show that the properties of spacetimes admitting non-geodesic multiple WANDs are further restricted, in particular that:
\begin{theorem}\label{thm:submfd}
  An Einstein spacetime that admits a non-geodesic multiple WAND is foliated by totally umbilic, constant curvature, Lorentzian, submanifolds of dimension three or greater, and any null vector field tangent to the leaves of the foliation is a multiple WAND. 
\end{theorem}
A submanifold is `totally umbilic' if and only if its extrinsic curvature is proportional to its induced metric, i.e.\ $K_{\mu\nu\rho} = \xi_\mu h_{\nu\rho}$, for some $\xi_\mu$ orthogonal to the submanifold, where $h_{\mu\nu}$ is the projection onto the submanifold. This property is useful because:
\begin{lemma}\label{lem:umbilic}
  A Lorentzian submanifold is totally umbilic if, and only if, it is ``totally null geodesic'', i.e., any null geodesic of the submanifold is also a geodesic of the full spacetime. 
\end{lemma}
Hence any \emph{geodesic} null vector field in the constant curvature submanifolds of Theorem \ref{thm:submfd} is a geodesic multiple WAND of the full spacetime, so Theorem \ref{thm:geo} follows as a direct corollary of these two results.  Note that these results also imply immediately that in a Type D spacetime, one can choose \emph{both} of the multiple WANDs to be geodesic.

For the special case of five dimensions, as well as Theorems \ref{thm:geo} and \ref{thm:submfd}, we have the stronger result:
\begin{theorem}\label{thm:list}
  A five-dimensional Einstein spacetime admits a non-geodesic multiple \\WAND if, and only if, it is locally isometric to one of the following:
  \begin{enumerate}
    \item Minkowski, de Sitter, or anti-de Sitter spacetime 
    \item A direct product $dS_3 \times S^2$ or $AdS_3 \times H^2$ 
    \item A spacetime with metric 
              \begin{equation*}
            ds^2 = r^2 d\tilde{s}_3^2 + \frac{dr^2}{U(r)} + U(r) dz^2, \qquad  U(r) = k - \frac{m}{r^2} -  \frac{\La}{4}r^2, 
                      \end{equation*}
where $m \ne 0$, $k \in \{1,0,-1\}$, $d\tilde{s}_3^2$ is the metric of a 3D Lorentzian space of constant curvature (i.e. 3D Minkowski or (anti-)de Sitter) with Ricci scalar $6k$, and the coordinate $r$ takes values such that $U(r)>0$.
 \end{enumerate}
\end{theorem}
Note that (ii) and (iii) are Type D.  Both admit 3D Lorentzian submanifolds of constant curvature, in agreement with Theorem \ref{thm:submfd}. Solution (iii) is an analytically continued version of the 5D Schwarzschild solution\footnote{It is a higher-dimensional generalization of the 4D B-metrics.} (generalized to allow for a cosmological constant and planar or hyperbolic symmetry).  Special cases of (iii) are the Kaluza-Klein bubble \cite{witten1} and the anti-de Sitter soliton \cite{adssoliton}.

In more than five dimensions, there are many Einstein spacetimes that admit non-geodesic multiple WANDs. A large class of examples can be obtained as follows. Consider a 6D static axisymmetric solution (which need not admit a WAND)
\begin{equation}
 ds^2 = - A(r,z)^2 dt^2 + B(r,z)^2 (dr^2 + dz^2) + C(r,z)^2 d\Omega^2,
\end{equation}
where $d\Omega^2$ is the metric on a unit $S^3$.  There are many solutions of the Einstein equation of this form, although the general solution is not known (except in the algebraically special case \cite{Mahdi}).  Now set $t=i\tau$ and analytically continue $d\Omega^2$ to the metric on 3D de Sitter space.  This gives an Einstein metric for which any null vector field tangent to the $dS_3$ is a multiple WAND.  This shows that there exist many six-dimensional Einstein spacetimes admitting non-geodesic multiple WANDs.  Obviously similar constructions work in higher dimensions too.

This chapter is organized as follows.  In Section \ref{sec:typeD}, we prove that an Einstein spacetime admitting a non-geodesic multiple WAND must be Type D (or conformally flat).  This is the starting point for the proof of Theorem \ref{thm:submfd} in Section \ref{sec:submfd}, which also contains the proof of Lemma \ref{lem:umbilic}.  In Section \ref{sec:list}, we restrict to five dimensions in order to prove Theorem \ref{thm:list}, and make some additional remarks about the six-dimensional case.  Most of our results are obtained from the Bianchi identity, whose components were written out in Section \ref{sec:bianchi}.  As many of our equations in this chapter will not be GHP invariant (since the preferred submanifolds discussed in Theorem \ref{thm:submfd} break the GHP invariance), we will rewrite some of these equations in Newman-Penrose notation as we go along.

Recall that the vector $\lb$ is non-geodesic if, and only if, $\kapb \neq \Ob$.  We shall work in an open subset of spacetime in which $\kapb \neq \Ob$.  Most work on algebraically special solutions assumes that spacetime is analytic, and in an analytic spacetime we expect that our results can be extended from this open subset to the rest of the spacetime.  In smooth but non-analytic spacetimes, the algebraic type can differ in disjoint open subsets of spacetime, even in 4D, and all of the results here should be understood as holding in some open subset of the spacetime which is algebraically special.\np

\section{Non-geodesity implies Type D}\label{sec:typeD}
Prior to the main work of this chapter, the following result was known: 
\begin{lemma}\label{lem:oldnongeo}
  In an Einstein spacetime that is not conformally flat, a multiple WAND $\lb$ is always geodesic if any of the following conditions on boost weight 0 components of the Weyl tensor hold:
  \begin{enumerate}
    \item[(i)] $\Phia_{ij}$ is non-vanishing.
    \item[(ii)] None of the eigenvalues of $\Phis_{ij}$ are $-\Phi$.
    \item[(iii)] $\Phi_{ijkl}$ vanishes identically.
  \end{enumerate}
\end{lemma}
This lemma combines two known results from the literature.  CMPP \cite{Bianchi} proved that all multiple WANDs with the Type III or Type N property are geodesic, and then Pravda \etal \cite{TypeD} derived the further restrictions described above in the Type II case (the lemma was first published in this form in my paper \cite{TypeII}).

The authors of \cite{TypeD} interpreted this as a statement that, `generically', a multiple WAND in a vacuum spacetime is geodesic.  The intention of our work here was to provide a concrete statement of exactly when WANDs can be non-geodesic.  For completeness, we begin by reviewing the proof of this lemma:

\paragraph*{Proof of Lemma \ref{lem:oldnongeo}}
Suppose that the spacetime admits a non-geodesic multiple WAND $\lb$, i.e. $\kapb\neq\Ob$.  We aim to show that none of the conditions (i)-(iii) hold.  The Bianchi equation \eref{B1} reduces to
\begin{equation} \label{B1:nongeo}
  0 = (2\Phi_{i[j}\del_{k]l} - 2\del_{il}\Phia_{jk} - \Phi_{iljk})\kap_{l}.
\end{equation} 
Contraction of \eref{B1:nongeo} on $ik$ gives
\begin{equation}\label{eqn:sim1}
  (\Phis_{ij}+3\Phia_{ij})\kap_j = -\Phi \kap_i.
\end{equation}
and further contraction with $\kap_i$ implies
\begin{equation}\label{eqn:phikapkap}
  \Phis_{ij} \kap_i \kap_j = - \Phi \kap_i \kap_i.
\end{equation}
Contracting \eref{B1:nongeo} with $\kap_i\kap_j$, and using \eref{eqn:phikapkap} gives
\begin{equation}\label{eqn:sim2}
  (\Phis_{ij}-3\Phia_{ij})\kap_j = -\Phi \kap_i.
\end{equation}
Taking $\eref{eqn:sim1}+\eref{eqn:sim2}$ implies that $\kapb$ is an eigenvector of $\Phis_{ij}$ with eigenvalue $-\Phi$, so (ii) fails.  Meanwhile $\eref{eqn:sim1}-\eref{eqn:sim2}$ implies that $\Phia_{ij}\kap_j = 0$.  Contracting \eref{B1:nongeo} with $\kap_i$, and using these last two results implies that $\kap_i\kap_i \Phia_{jk}=0$, and hence $\Phia_{ij}=0$ and (i) fails.

So either (iii) fails (in which case we're done), or $\Phi_{ijkl}=0$, and the spacetime is either Type III or Type N.

In the Type III case, consider equation \eref{B2}, which reduces to
\begin{equation}
  \Psi'_j \kap_i = \Psi'_{jik} \kap_k.
\end{equation}
Contracting with $\kap_i$ gives $\Psi'_j\kap_i\kap_i=0$, and hence $\Psi'_i=0$.  Inserting this back into \eref{B2} implies $\Psi'_{ijk}\kap_k = 0$, and hence $\Psi'_{ijk}\kap_i = 0$.  Now consider \eref{B3}, which gives
\begin{equation}
  0 = - 2 \Psi'_{[i|kl} \kap_{|j]} - 2 \Psi'_{[k|ij}\kap_{|l]},
\end{equation}
and contracting with $\kap_j$ implies $\Psi'_{ikl} \kap_j\kap_j=0$, and hence if $\kap_i\neq0$ the spacetime is not Type III.

It remains to consider the Type N case.  Here, \eref{B5} reads
\begin{equation}
  2\Om'_{i[j}\kap_{k]} = 0.
\end{equation}
Contracting this on $ik$ implies that $\Om'_{ij}\kap_i=0$, while contracting with $\kap_k$ gives $\Om'_{ij}\kap_k\kap_k = \Om'_{ik}\kap_k \kap_j$.  Combining these results implies $\Om'_{ij}\kap_k\kap_k=0$, and hence multiple WANDs in Type N spacetimes must also be geodesic.  This suffices to prove the result.$\Box$

Now we move on to prove Lemma \ref{lem:typeD}, namely that all spacetimes admitting a non-geodesic multiple WAND are either Type D or Type O.

\paragraph*{Proof of Lemma \ref{lem:typeD}}
Assume we have an Einstein spacetime with a non-geodesic \mwand\ $\lb$.  By Lemma \ref{lem:oldnongeo}, we know that $\Phi_{ij}$ is symmetric, and has an eigenvalue $-\Phi$ with associated eigenvector $\kap_i$, that is
\begin{equation}  \label{eqn:Li0evec}
 \Phi_{ij} \kap_j = -\Phi \kap_i.
\end{equation}  
We know that $\kapb \neq \Ob$.  Since $\kapb$ transforms as a vector under rotations of the spatial basis vectors $\mb{i}$, we can choose these basis vectors so that
\begin{equation}
 \kap_2 \ne 0, \qquad \kap_{\hi}=0,
\end{equation}
where $\hi$, $\hj$ etc take values $3,4, \ldots , (d-1)$.\footnote{This particular choice breaks GHP covariance, so we will now need to use explicit Newman-Penrose forms of the Bianchi identities etc.  One could simplify the calculation slightly by introducing GHP-like derivative operators $\hat{\tho}$ etc that are only covariant under spins that preserve the $\mb{2}$ direction, but we will see that the proof as it is will not be complicated enough to justify this additional machinery.}  From equation \eref{eqn:Li0evec}, we have
\begin{equation}
 \Phi_{22} = -\Phi, \qquad \Phi_{2\hi} = 0.
\end{equation}
Equation \eref{B1:nongeo} implies that
\begin{equation}
\label{eqn:C2i2j}
 \Phi_{2\hi 2 \hj} = \Phi_{\hi \hj} \qquad \Phi_{2 \hi \hj \hk} = 0.
\end{equation}
The Bianchi equation \eref{B4} reads
\begin{eqnarray}\label{B4:nongeo} 
  0 &=& -2\Phi_{i[j}\rho_{kl]} + \Phi_{im[jk|}\rho_{m|l]}
\end{eqnarray}
and setting $ijkl=22 \hi \hj$ gives
\begin{equation}
\label{PhiL}
 \Phi \rho_{[\hi \hj]} - \rho_{\hk [\hi} \Phi_{\hj] \hk} = 0.
\end{equation}
Now consider \eref{B2}, which reads
\begin{eqnarray}\label{B2:nongeo}
  D \Phi_{ij} &=& -(\Phi_{ik} + \Phi \del_{ik}) \rho_{kj} - 2\Phi_{(i|k}\M{k}_{|j)0}
                                   + (\Psi'_j \del_{ik} - \Psi'_{jik}) \kap_{k}.
\end{eqnarray}
First look at the antisymmetric part. The $2\hi$ component gives
\begin{equation}
\label{B32i}
 \left( \Psi'_{22\hi} - \Psi'_{\hi} \right) \kap_2 
          = \left( \Phi_{\hi \hj} + \Phi \delta_{\hi\hj} \right) \rho_{\hj 2},
\end{equation}
and, using (\ref{PhiL}), the $\hi\hj$ component gives
\begin{equation}
 \Psi'_{2\hi\hj}=0.
\end{equation}
Now look at the symmetric part of \eref{B2:nongeo}. 
Setting $i=2$, $j=\hi$ gives
\begin{equation}
 \left( \Psi'_{22\hi} + \Psi'_{\hi} \right) \kap_2 
      = \left( \Phi_{\hi \hj} + \Phi \delta_{\hi\hj} \right) \rho_{\hj 2} 
        - 2 \Phi_{\hi \hj} \M{2}_{\hj 0} - 2 \Phi \M{2}_{\hi 0}.
\end{equation}
Subtracting this from (\ref{B32i}) gives
\begin{equation}
  -\Psi'_{\hi} \kap_2 = \Phi_{\hi \hj} \M{2}_{\hj 0} +  \Phi \M{2}_{\hi 0}.
  \label{eqn:psii}
\end{equation}
Now we shall show that the basis vectors $n$, $\mb{i}$ can be chosen to make the negative boost weight Weyl components vanish. Consider moving to a new basis $\{\bar{\lb},\bar{\nb},\bar{m}_i \}$ by performing a null rotation about $\lb$:
\begin{equation}
  \bar{\lb}= \lb, \qquad 
  \bar{\nb} = \nb - z_i \mb{i} - \frac{1}{2} z^2 \lb, \qquad
  \bar{m}_i = \mb{i} + z_i \lb,
\end{equation}
where $z_i$ are some smooth functions, and $z^2\equiv z_i z_i$.  In the new basis we have
\begin{equation}
 \bar{\M{2}}_{\hi 0} = \M{2}_{\hi 0} - z_{\hi} \kap_2.
\end{equation}
We can always choose $z_{\hi}$ so that the RHS vanishes. Hence we can always choose our basis so that (this equation is trivial for $i=2$)
\begin{equation}
\label{basischoice}
 \M{2}_{i 0}=0. 
\end{equation}
We shall assume this henceforth, and drop the bars.  We now have, from \eref{eqn:psii}, that $\Psi'_{\hi} = 0$.  Now, the $\hi\hj$ component of the symmetric part of \eref{B2:nongeo} gives
\begin{equation} \label{eqn:DPhi1}
 D\Phi_{\hi \hj} = - \Phi \rho_{(\hi\hj)} - \rho_{\hk (\hi} \Phi_{\hj) \hk} 
                   - \Psi'_{(\hi\hj)2} \kap_2 + \M{\hi}_{\hk 0} \Phi_{\hk \hj} 
                   + \M{\hj}_{\hk 0} \Phi_{\hk \hi}.
\end{equation}

Also, \eref{B3} reduces to
\begin{eqnarray}\label{B3:nongeo} 
  -D \Phi_{ijkl} &=& - 2\Phi_{i[k|} \rho_{j|l]} + 2\Phi_{j[k|} \rho_{i|l]} 
                  + 2 \Phi_{ij[k|m} \rho_{m|l]} + 2 \Phi_{[i|mkl} \M{m}_{|j]0} 
                  + 2 \Phi_{ij[k|m} \M{m}_{|l]0}\nn\\ 
              & & - 2 \Psi'_{[i|kl} \kap_{|j]} - 2 \Psi'_{[k|ij} \kap_{|l]},
\end{eqnarray}
and taking the $2\hi 2 \hj$ component gives
\begin{equation}\label{eqn:DPhi2}
  D\Phi_{\hi \hj} = - \Phi \rho_{(\hi\hj)} - \rho_{\hk (\hi} \Phi_{\hj) \hk} 
                    + 2 \Psi_{(\hi\hj)2} \kap_2 
                    + \M{\hi}_{\hk 0} \Phi_{\hk \hj} + \M{\hj}_{\hk 0} \Phi_{\hk \hi}.
\end{equation}
Comparing \eqref{eqn:DPhi1} and \eqref{eqn:DPhi2} reveals that $\Psi'_{(\hi\hj)2} =0$. However, we also have that $\Psi'_{2\hi \hj}=0$, so the identity $\Psi'_{[ijk]}=0$ implies that $\Psi'_{[\hi\hj]2}=0$. Combining these results, we learn that
\begin{equation}
 \Psi'_{\hi \hj 2}=0.
\end{equation}
Using (\ref{basischoice}), the $2\hi\hj\hk$ component of \eref{B3:nongeo} reduces to $\kap_2 \Psi'_{\hi\hj\hk}=0$, and hence
\begin{equation}
 \Psi'_{\hi\hj\hk}=0.
\end{equation}
Now we have $0=\Psi'_{\hi}=\Psi'_{2\hi 2} + \Psi'_{\hj \hi \hj} = \Psi'_{2\hi 2}$. Hence all components of $\Psi'_{ijk}$ vanish, and therefore so must $\Psi'_i$:
\begin{equation}
 \Psi'_{ijk} = \Psi'_i = 0.
\end{equation}
Next, consider the following equation, constructed from \eref{B5} and its primed version:
\begin{eqnarray}\label{B5p:nongeo}
  -D \Psi'_{ijk} &=& (2 \Phi_{i[j} \del_{k]l} - \Phi_{iljk}) (\tau_{l}-\tau'_{l})  
                    + 2 \Psi'_{[j} \rho_{k]i} + \Psi'_{ijk} L_{10} + \Psi'_{sjk}\rho_{si} \nn\\
                & & + 2\Psi'_{i[j|l} \M{l}_{|k]0} + \Psi'_{ljk}\M{l}_{i0} - 2\Om'_{i[j}\kap_{k]}
\end{eqnarray}
Setting $i=j=2$ and $k=\hi$ gives
\begin{equation} \label{Psi2i}
  \Om'_{2\hi} \kap_2 = \left( \Phi_{\hi\hj} + \Phi \delta_{\hi\hj} \right) 
                                               \left( \tau_\hj - \tau'_\hj \right).
\end{equation}
and setting $ijk=\hi\hj 2$ gives $\Om'_{\hi \hj} = 0$.

The $\hi\hj\hk$ component gives
\begin{equation}
 0 = \left( 2 \Phi_{\hi [ \hj} \delta _{\hk] \hl} 
    - \Phi_{\hi\hl\hj\hk} \right) \left( \tau_\hl - \tau'_\hl \right).
\end{equation}
Contracting on $\hi$ and $\hj$, using 
\begin{equation}\label{eqn:trace}
  \Phi_{\hi\hl\hi\hk}=\Phi_{i\hl i\hk} - \Phi_{2\hl 2 \hk} 
                     = -3 \Phi_{\hl\hk} \eqand \Phi_{\hi\hi} = \Phi_{ii} - \Phi_{22} = 2\Phi
\end{equation}
gives
\begin{equation}
 0 =  \left( \Phi_{\hk\hl} + \Phi \delta_{\hk\hl} \right) \left( \tau_\hl - \tau'_\hl \right).
\end{equation}
Substituting this into (\ref{Psi2i}) gives $\Om'_{2\hi}=0$, and hence it only remains to show that $\Om'_{22}=0$.  From the Bianchi equations \eref{B5} and \eref{B5}$'$ we obtain an equation
\begin{equation}\label{eqn:oldB1}
  D \Psi'_i - \del_i \Phi = -(\Phi_{ij}+\Phi\del_{ij}) (\tau_{j}+\tau'_{j})
            - \Psi'_j ( \del_{ji} L_{10} + 2 \rho_{ji} + \M{j}_{i0}) + \Om'_{ij} \kap_{j}
\end{equation}
 Setting $i=2$ gives
\begin{equation}\label{eqn:Om22}
 \delta_2 \Phi = - \Om'_{22} \kap_2.
\end{equation}

Now consider \eref{B5}, which reduces to
\begin{eqnarray}\label{B5:nongeo} 
  -2 \del_{[j}\Phi_{k]i} &=& (2 \Phi_{i[j}\del_{k]l} - \Phi_{iljk}) \tau_{l}
                             - 2\Phi_{il}\M{l}_{[jk]} - 2\Phi_{l[j|} \M{l}_{i|k]}\nn\\
                         & & + (2\Psi'_{[j|} \del_{il} - 2 \Psi'_{[j|il}) \rho_{l|k]}
\end{eqnarray}
Setting $ijk=\hi 2\hk$, and tracing on $\hi$ and $\hk$ gives
\begin{equation}
 \delta_2 \Phi = - \frac{1}{2} \left( \Phi \M{2}_{\hi\hi} + \Phi_{\hi\hj} \M{2}_{\hi\hj} \right).
\end{equation}
However, we can compare this to a different result obtained from equation \eref{B7}, which reads
\begin{eqnarray}\label{B7:nongeo}
  -\del_{[k|} \Phi_{ij|lm] } &=&  - \Psi'_{i[ kl|} \rho_{j|m ]} + \Psi'_{j[ kl|} \rho_{i|m ]}
                                  - 2\Psi'_{[k| ij} \rho_{|lm]} \nn \\
                             & &  + 2\Phi_{ij[ k|n} \M{n}_{|lm]}
                                  + \Phi_{in[ kl|} \M{n}_{j|m]} - \Phi_{jn[ kl|} \M{n}_{i|m]}
\end{eqnarray}
Setting $m=2$ and $ijkl=\hi\hj\hk\hl$, tracing on $\hi$ and $\hk$ and then tracing on $\hj$ and $\hl$ gives
\begin{equation}
 \delta_2 \Phi = - \frac{2}{3} \left( \Phi \M{2}_{\hi\hi} + \Phi_{\hi\hj} \M{2}_{\hi\hj} \right).
\end{equation}
Hence we can conclude that $\delta_2 \Phi=0$, and hence, by \eref{eqn:Om22}, $\Om'_{22}=0$.  Therefore
\begin{equation}
 \Om'_{ij}=0.
\end{equation}
Therefore, all of the components of the Weyl tensor of non-zero boost weight vanish.

It remains to exclude the possibility that the spacetime is Type III or Type N.  To see that this cannot occur, suppose that there exists a different null vector $\bar{\lb}$ along which the Weyl tensor has negative boost order (that is all Weyl components of non-negative boost weight vanish).  There are two possibilities; either $\bar{\lb}.\nb = 0$ or $\bar{\lb}.\nb \neq 0$.  In the former case, this implies that $\bar{\lb}\parallel\nb$, and hence the boost weight zero components of the Weyl tensor must have vanished in our original frame, implying that the Weyl tensor of the spacetime vanishes identically.  Alternatively, if $\bar{\lb}.\nb \neq 0$ then we can rescale such that $\bar{\lb}.\nb = 1$ and hence work in a null frame containing both $\bar{\lb}$ and $\nb$ in which the Weyl tensor again vanishes identically.  This implies that the spacetime must be either Type D or Type O, as required.$\Box$

\section{Foliation by submanifolds}\label{sec:submfd}

Having established that the spacetime in question is Type D, we now go on to show that it is foliated by a particular family of submanifolds:

\paragraph*{Proof of Theorem \ref{thm:submfd}}
Assume that we have an Einstein spacetime admitting a non-geodesic multiple WAND $\lb$.  From Lemma \ref{lem:typeD}, we can use a basis in which the Type D condition is satisfied. Consider a new basis defined by a null rotation about $\nb$:
\begin{equation} \label{newbasis}
 \hat{\lb} = \lb - z_i \mb{i} - \frac{1}{2} z^2 n, \qquad 
 \hat{\nb}=\nb, \qquad 
 \hat{m}_i = \mb{i} + z_i \lb,
\end{equation}
where $z_i$ are arbitrary smooth functions and $z^2 \equiv z_i z_i$. Using the Type D property, in the new basis we have
\begin{equation}
 \hat{\Psi}_{ijk} = \Phi_{iljk} z_l - 2\Phi_{i[j} z_{k]},
\end{equation}
\begin{equation}
 \hat{\Om}_{ij} = \hat{\Psi}_{ijk} z_k + z_i \left( \Phi_{jk} z_k + \Phi z_j \right).
\end{equation}
Now choose the functions $z_i$ so that $\hat{\Psi}_{ijk}=0$, i.e.,
\begin{equation}
\label{zdef}
 \Phi_{iljk} z_l - 2\Phi_{i[j} z_{k]}=0.
\end{equation}
This equation certainly admits non-vanishing solutions $z_i$ because \Eref{B1:nongeo} shows that $z_i = \kap_i$ is a solution and, by our assumption that $\lb$ is non-geodesic, this solution is non-vanishing. Tracing on $i$ and $k$ reveals that $z_i$ is an eigenvector of $\Phi_{ij}$ with eigenvalue $-\Phi$:
\begin{equation}
\label{Phiz}
 \Phi_{ij} z_j = -\Phi z_i.
\end{equation}
The previous two equations imply that $\hat{\Om}_{ij}=0$. Hence for any change of basis defined by $z_i$ satisfying (\ref{zdef}), all positive boost weight Weyl components vanish  in the new basis (\ref{newbasis}), and hence $\hat\lb$ is a multiple WAND. Since $\hat\nb=\nb$ is also a multiple WAND, the negative boost weight Weyl components still vanish, and hence the Type D condition is still satisfied in the new basis.  Note that, when working in a fixed basis that satisfies the Type D condition, the priming symmetry discussed in Section \ref{sec:primingsym} holds.

The LHS of (\ref{zdef}) defines a linear map on $z_i$ at any point in spacetime. We know that the kernel $K$ of this map is non-empty.  Let $n$ be the dimension of $K$ at some point $p$ in the spacetime.  By smoothness there must be a neighbourhood of $p$ in which the dimension also equals $n$, and assuming analyticity, we can extend this to all points in the spacetime, except possibly some set of zero measure where the dimension of $K$ differs.

In this neighbourhood, there exist $n$ linearly independent solutions $z_i$ of (\ref{zdef}), and hence a $n$-parameter family of multiple WANDs at any point. This family obviously contains $\lb$.

The $n$ solutions $z_i$ define a $n$-dimensional distribution spanned by vector fields of the form $z_i \mb{i}$. By rotating the spatial basis, we can divide it into a set $\{ \mb{I} \}$ that spans this distribution and a set $\{ \mb{\alpha} \}$ that is orthogonal to it. Here, indices $I,J,\ldots$ take values $2,3, \ldots, (n+1)$ and indices $\alpha,\beta,\ldots$ take values $(n+2),(n+3),\ldots, (d-1)$. By definition, the general solution of equation (\ref{zdef}) is
\begin{equation}
 z_{\alpha} = 0,
\end{equation}
with $z_I$ arbitrary functions. From equation (\ref{Phiz}), it follows that
\begin{equation}
\label{eqn:PhiIJ}
 \Phi_{IJ} = -\Phi \delta_{IJ}, \qquad \Phi_{I\alpha}=0.
\end{equation}
The vectors $\mb{\alpha}$ can be chosen to diagonalize $\Phi_{\alpha\beta}$.  Note that we do {\it not} know that all eigenvalues of $\Phi_{\alpha\beta}$ differ from $-\Phi$.

In this basis, equation (\ref{zdef}) reduces to
\begin{equation}
\label{eqn:Weylcpts}
 \Phi_{IJKL} = -2\Phi \delta_{I[K} \delta_{L]J}, \qquad \Phi_{I\alpha J\beta} = \delta_{IJ} \Phi_{\alpha\beta}, \qquad 
 \Phi_{IJK \alpha} = \Phi_{IJ \alpha\beta} = \Phi_{I \alpha \beta\gamma}=0.
\end{equation}
We shall now use the Bianchi identities to deduce constraints on the form of $L_{ab}$, $N_{ab}$ and $\M{\al}_{Ia}$. The following will be useful:
\begin{lemma}\label{lem:useful}
  If $X_\alpha$ obeys $\Phi_{\alpha \beta\gamma \delta} X_{\delta} - 2 \Phi_{\gamma[ \alpha} X_{\beta]} =0$ everywhere then $X_\alpha = 0$ everywhere.
\end{lemma}
\noindent {\it Proof}. Extend $X_\alpha$ to $X_i$ by defining $X_I=0$. Tracing on $\alpha$ and $\gamma$ gives $\Phi_{\beta \delta} X_\delta = -\Phi X_{\delta}$. One can now check that all components of $\Phi_{ijkl} X_l - 2\Phi_{k[i} X_{j]}$ vanish everywhere, and therefore $X_i$ lies in the kernel $K$ described above.  But the directions $\mb{\alpha}$ were defined to be those orthogonal to the kernel, and hence it follows that $X_\alpha = 0$. $\Box$

Using this Lemma, we now note that Equations \eref{B1}, and \eref{B1}$'$ imply that
\begin{equation}
\label{eqn:Lalpha0}
 \kap_\al = \kap'_\al = 0.
\end{equation}
Similarly, equation \eref{B5p:nongeo} says that $(\tau_i - \tau'_i)$ obeys (\ref{zdef}) everywhere and hence
\begin{equation}\label{eqn:Nalpha0}
 \tau_\al = \tau'_\al.
\end{equation}
Setting $ijkl=\gamma\alpha \beta I$ in \eref{B4} gives
\begin{equation}
 \Phi_{\alpha \beta\gamma \delta} \rho_{\delta I} - 2 \Phi_{\gamma[ \alpha} \rho_{\beta]I} =0
\end{equation}
so from Lemma \ref{lem:useful} (treating $I$ as fixed) we obtain
\begin{equation}
 \label{eqn:LalphaI0}
 \rho_{\alpha I}=0.
\end{equation}
Similarly, from \eref{B4}$'$ we obtain
\begin{equation}
\label{eqn:NalphaI0}
 \rho'_{\alpha I} =0.
\end{equation}
Setting $ijkl=\alpha\beta\gamma I$ in \eref{B3:nongeo} (which was obtained from \eref{B3}), and using \eref{eqn:LalphaI0} gives
\begin{equation}
 \Phi_{\alpha \beta\gamma \delta} \stackrel{I}M_{\delta 0} - 2 \Phi_{\gamma[ \alpha} \stackrel{I}{M}_{\beta]0} =0,
\end{equation}
so Lemma \ref{lem:useful} gives
\begin{equation}
\label{eqn:MalphaI0}
 \stackrel{I}M_{\alpha 0} =0.
\end{equation}
Similarly, working from \eref{B3}$'$ we obtain 
\begin{equation}
\label{eqn:MalphaI1}
 \stackrel{I}M_{\alpha 1} =0.
\end{equation}
Next, setting $ijklm=I\beta\delta J \gamma$ in \eref{B7:nongeo} we obtain
\begin{equation}
 2 \delta_{[\gamma} \Phi_{\delta] \beta} \delta_{IJ} 
     = - 2 \Phi_{\beta [\gamma} \M{I}_{\delta] J} - \Phi_{\beta \alpha \gamma\delta} \M{\al}_{IJ} 
       + 2 \delta_{IJ} \left( \Phi_{\alpha \beta} \M{\al}_{[\gamma \delta]} 
       + \Phi_{\alpha [\gamma|} \M{\al}_{\beta| \delta]} \right).
 \end{equation}
However, setting $ijk=\beta\gamma\delta$ in \eref{B5:nongeo} gives
 \begin{equation}
   -2 \delta_{[\gamma} \Phi_{\delta] \beta} 
       = 2 \Phi_{\beta[\gamma} \tau_{\delta]} - \Phi_{\beta\alpha \gamma \delta} \tau_{\alpha} 
        - 2 \left( \Phi_{\alpha \beta} \M{\al}_{[\gamma \delta]} + \Phi_{\alpha [\gamma|} \M{\al}_{\beta| \delta]} \right).
\end{equation}
Combining these two equations gives
\begin{equation}
 \Phi_{\beta\alpha \gamma\delta} X_{\alpha IJ} - 2 \Phi_{\beta [\gamma} X_{\delta] IJ} = 0,
\end{equation}
where $X_{\alpha IJ} = \tau_{\alpha} \delta_{IJ} + \M{\alpha}_{IJ}$.  Hence, using Lemma \ref{lem:useful}, and \eref{eqn:Nalpha0}, we have
\begin{equation}\label{eqn:MalphaIJ}
 \stackrel{\alpha}{M}_{IJ} = - \tau_{\alpha} \delta_{IJ} = -\tau'_\al \del_{IJ}.
\end{equation}
A convenient way of summarizing the above results is to define indices $A,B,\ldots$ to take values $0,1,2,\ldots (n+1)$. Using equations \eref{eqn:Weylcpts} and the definition of $\Phi_{ij}$, we find that
\begin{equation}
\label{eqn:submfdWeyl}
 \Phi_{ABCD} = -2\Phi \eta_{A[C} \eta_{D]B},
\end{equation}
where $\eta_{AB}$ is the Minkowski metric ($\eta_{01}=\eta_{10}=1$, $\eta_{IJ} = \delta_{IJ}$).
We also have (using the Type D condition and equations \eref{eqn:Weylcpts})
\begin{equation}
\label{eqn:submfdWeyl2}
 \Phi_{ABC\alpha} = \Phi_{AB\alpha\beta} = \Phi_{A\alpha\beta\gamma}=0, \qquad \Phi_{A\alpha B\beta}= \eta_{AB} \Phi_{\alpha\beta}.
\end{equation}
Equations (\ref{eqn:Lalpha0}, \ref{eqn:Nalpha0}, \ref{eqn:LalphaI0}, \ref{eqn:NalphaI0}, \ref{eqn:MalphaI0}, \ref{eqn:MalphaI1}, \ref{eqn:MalphaIJ}) are equivalent to
\begin{equation}
\label{eqn:MalphaAB}
 \M{\alpha}_{AB} = -\tau_{\alpha} \eta_{AB}.
\end{equation}
Using this, we have
\begin{equation}
\label{eqn:integrable}
 [e_{A},e_{B}]_\alpha \equiv 2 \M{\alpha}_{[AB]}=0.
\end{equation}
Hence the distribution spanned by $\{e_A \} = \{\lb,\nb,\mb{I} \}$ is integrable, in the sense of Frobenius' Theorem, and hence it is tangent to $(n+2)$-dimensional submanifolds of spacetime.  From equations \eref{eqn:submfdWeyl} and \eref{eqn:submfdWeyl2}, it follows that {\it any} null vector tangent to these submanifolds is a multiple WAND.

Now the extrinsic curvature tensor of one of the submanifolds is defined by
\begin{equation}
 K(X,Y) = (\nabla_X Y)^\perp,
\end{equation}
where $X$ and $Y$ are vector fields tangent to the submanifold, and $\perp$ is the projection perpendicular to the submanifold. The non-vanishing components are
\begin{equation}\label{eqn:umbilic}
 K_{\al AB} = - \M{\alpha}_{AB} = \tau_{\al} \eta_{AB},
\end{equation}
where we used \eref{eqn:MalphaAB}. Hence the submanifolds are totally umbilic.

Let $\Scal$ be one of the submanifolds. Calculating the Riemann tensor $\tilde{R}_{abcd}$ of $\Scal$ gives
\begin{equation}
\label{eqn:RiemS1}
   \tilde{R}_{abcd} = h_{a}^{\phantom{a}a'} h_{b}^{\phantom{b}b'} h_{c}^{\phantom{c}c'} h_{d}^{\phantom{d}d'} \left[R_{a'b'c'd'} + 2\M{\alpha}_{c'[a'|} \M{\alpha}_{d'|b']} \right],
\end{equation}
where
\begin{equation}
  h_{ab} = \eta_{AB} e^{A}_a e^{B}_b = \eta_{ab} - \del_{\al a} \del_{\al b}
\end{equation}
is the projection operator onto $\Scal$. Using $e_{A}$ as a basis on $\Scal$, we have (using the relation between the Riemann and Weyl tensors in $d$ dimensions, as well as the Einstein equation)
\begin{equation} \label{eqn:RiemS2}
  \tilde{R}_{ABCD} = \Phi_{ABCD} + \frac{2\Lambda}{d-1}  \eta_{A[C} \eta_{D]B} 
                     + 2 \M{\alpha}_{[C|A} \M{\alpha}_{|D]B}
\end{equation}
Using equations \eref{eqn:submfdWeyl} and \eref{eqn:MalphaAB} now gives
\begin{equation} \label{eqn:projcurv}
 \tilde{R}_{ABCD} = 2 {\cal R} \eta_{A[C} \eta_{D]B},
\end{equation}
where
\begin{equation}
  {\cal R} = \frac{\Lambda}{d-1} - \Phi + \tau_\al \tau_\al .
\end{equation}
Furthermore, equation \eref{eqn:projcurv} is the statement that $\Scal$ has constant curvature, since the $(n+2)$-dimensional Bianchi identity implies that ${\cal R}$ is constant on $\Scal$.  This completes the proof of Theorem \ref{thm:submfd}. $\Box$

Note that from the $IJ$ components of equations \eref{B2:nongeo}, its primed version, and the $IJK$ components of \eref{B5:nongeo} we also have
\begin{equation}
 D\Phi = \Delta\Phi = \delta_I \Phi = 0, \label{eqn:constphi}
\end{equation}
so $\Phi$ is constant on any of the constant curvature submanifolds.

One further comment is useful here.  Ref.\ \cite{TypeD} gives an example of a class of 7-dimensional spacetimes that admit non-geodesic WANDs.  For this example, it can be shown that the foliation is by 3-dimensional Lorentzian submanifolds of constant curvature, and that the arbitrary function appearing in this solution can be eliminated by a change of coordinates.

Recall that the reason we are interested in the totally umbilic condition \eref{eqn:umbilic} was that a submanifold with this condition is \emph{totally null geodesic}.  For completeness, we now give a proof of the lemma that establishes this:
\paragraph*{Proof of lemma \ref{lem:umbilic}}
Let $\Scal$ be a Lorentzian submanifold of spacetime. Consider an affinely parametrized null geodesic of $\Scal$ with tangent vector $U$, i.e., we have $U \cdot \hat{\nabla} U= 0$, where $\hat{\nabla}$ is the Levi-Civita connection in $\Scal$. This is equivalent to $(U \cdot \nabla U)^\parallel = 0$, where $\parallel$ denotes the projection tangential to $\Scal$. Now, from the definition of the extrinsic curvature $K$, we have
\begin{equation}
 (U \cdot \nabla U)^\perp = K(U,U).
\end{equation}
If $\Scal$ is totally umbilic then the RHS vanishes because $U$ is null. Therefore all components of $U \cdot \nabla U$ vanish so $\Scal$ is totally null geodesic.

Conversely, if the manifold is totally null geodesic then pick a point $p$ on $\Scal$, let $U$ be an arbitrary null vector tangent to $\Scal$ at $p$, and consider the geodesic in $\Scal$ that has tangent vector $U$ at $p$. By assumption, this is a geodesic of the full spacetime, so the RHS of the above equation must vanish. But $p$ and $U$ are arbitrary, so $K(U,U)$ must vanish for any null $U$ tangent to $\Scal$, which implies that $\Scal$ is totally umbilic.$\Box$

\section{Results in five dimensions} \label{sec:list}

In five dimensions, Theorem \ref{thm:list} gives an explicit list of spacetimes that can admit non-geodesic multiple WANDs.  Here we prove this theorem.
\paragraph*{Proof of Theorem \ref{thm:list}}
In a 5-dimensional spacetime, the boost weight zero components of the Weyl tensor are all determined by $\Phi_{ij}$, via equation \eqref{eqn:5dweyl}.  Assume that we have a non-geodesic multiple WAND $\lb$.  From Lemma \ref{lem:typeD}, we know that we can choose a basis $\{\lb,n,m_i \}$ so that the Type D condition is satisfied. Following Ref.\ \cite{TypeD}, we can substitute equation \eref{eqn:5dweyl} into \eref{B1:nongeo}, to learn that the eigenvalues of $\Phi_{ij}$ must be $-\Phi,\Phi,\Phi$. Therefore we can choose the spatial basis vectors $\mb{i}$ so that
\begin{equation}
  \Phi_{ij} = \diag (-\Phi,\Phi,\Phi)  \label{eqn:nongeophi}.
\end{equation}
and
\begin{equation}
 \kap_2 \ne 0, \qquad \kap_{3}=\kap_{4}=0.
\end{equation}
The Weyl tensor is fully determined by the single scalar $\Phi$.  If $\Phi=0$ then the Weyl tensor vanishes, in which case the spacetime is Minkowski or (anti-) de Sitter, i.e., case (i) of the theorem. Henceforth we assume $ \Phi \ne 0$.
Since there is only one eigenvalue $-\Phi$, we know immediately (from \eref{eqn:PhiIJ}) that the constant curvature submanifolds of Theorem \ref{thm:submfd} must be 3-dimensional.

The Weyl tensor is sufficiently constrained that we can now solve completely the full set of Bianchi equations.  The results of Section \ref{sec:submfd} still apply here, and we will make use of them below.  Indices $\al,\beta$ take values $3,4$, consistent with previous sections.

Firstly, the $\al$ component of \eref{eqn:oldB1}, combined with (\ref{eqn:Nalpha0},\ref{eqn:MalphaIJ}), gives
\begin{equation}
  \tau_\al = \tau'_\al =  \M{2}_{\al 2} = \frac{1}{4\Phi}\del_\al \Phi.
\end{equation}
Also, the $\al\ba$ components of \eref{B2} and its primed version, along with the $\al\ba2$ component of \eref{B5:nongeo} give, after using \eref{eqn:constphi}, that
\begin{equation}
  \rho_{\al\ba} = 0,\quad \rho'_{\al\ba} = 0 \eqand \M{2}_{\al\ba}=0
\end{equation}
respectively.

This now leaves us with the following results, for some unknown $\rho_{2i}$, $\rho'_{2i}$, $\kap_2$, $\kap'_2$, $\tau_2$, $\tau'_2$, $\Phi$:
\begin{equation}\label{eqn:phiresults}
  D\Phi = 0, \quad \quad \Delta \Phi = 0,\quad\quad \del_2 \Phi = 0,
\end{equation}
\begin{equation}\label{eqn:Lresults}
 \rho_{ij} = \left( \begin{array}{ccc}
                   \rho_{22} & \rho_{23} & \rho_{24}\\
                   0 & 0 & 0\\
                   0 & 0 & 0
                  \end{array}\right) , \quad
  \kap_i = \left( \begin{array}{c}
                   \kap_2\\ 0\\ 0
                  \end{array}\right), \quad
  \tau_{i} =   \left( \begin{array}{c}
                   \tau_{2}\\ \del_3\Phi /(4\Phi) \\ \del_4\Phi/(4\Phi)
                  \end{array} \right),
\end{equation}
\begin{equation}\label{eqn:Nresults}
 \rho'_{ij} = \left( \begin{array}{ccc}
                   \rho'_{22} & \rho'_{23} & \rho'_{24}\\
                   0 & 0 & 0\\
                   0 & 0 & 0
                  \end{array}\right) , \quad
  \kap'_{i} = \left( \begin{array}{c}
                   \kap'_{2}\\ 0\\ 0
                  \end{array}\right), \quad
  \tau'_{i} =  \left( \begin{array}{c}
                    \tau'_{2}\\ \del_3 \Phi /(4\Phi) \\ \del_4\Phi/(4\Phi)
                  \end{array} \right),
\end{equation}
\begin{equation}\label{eqn:M2results}
 \M{2}_{ij} =  \left( \begin{array}{ccc}
                   0 & 0 & 0\\
                   \del_3\Phi/(4\Phi) & 0 & 0\\
                   \del_4\Phi/(4\Phi) & 0 & 0
                  \end{array}\right) , \quad
  \M{2}_{i1} = \left( \begin{array}{c}
                   0\\ 0\\ 0
                  \end{array}\right), \quad
  \M{2}_{i0} = \left( \begin{array}{c}
                   0\\ 0\\ 0
                  \end{array}\right).
\end{equation}
Furthermore, inserting (\ref{eqn:phiresults}-\ref{eqn:M2results}) into the full set of Bianchi equations ((\ref{B1}-\ref{B7}), along with their primed versions), we find that this is sufficient to satisfy all of them, with no further restrictions.  This is to be expected, as we have given all information that is invariant under both boosts, and the subgroup of spins that preserve the $\mb{2}$ direction.

Next we shall show that the 5D spacetime must be a {\it warped product}.  We shall do this by showing that it is conformal to a product spacetime.  To this end, let $h^\mu{}_\nu$ denote the tensor that projects onto the 3D submanifolds, i.e.,
\begin{equation}
 h^\mu{}_\nu = \lb^\mu n_\nu + n^\mu \lb_\nu + m_{2}^\mu m_{2\, \nu}.
\end{equation}
 Now define
\begin{equation}
  H_{\mu\nu\rho} \equiv \nabla_\rho h_{\mu\nu}.
\end{equation}
Using the above results, the only non-vanishing null frame components of this are
\begin{equation}
  H_{301}=H_{310}=H_{322}= \frac{\del_3 \Phi}{4\Phi} \eqand
  H_{401}=H_{410}=H_{422}= \frac{\del_4 \Phi}{4\Phi}
\end{equation}
as well as those related to these components by the symmetry in the first two indices. Now consider a conformally related spacetime, with metric
\begin{equation}
 \tilde{g} = |\Phi|^{1/2} g.
\end{equation}
Let $\tilde{\nabla}$ denote the Levi-Civita connection in the new spacetime. Using the relation between $\tilde{\nabla}$ and $\nabla$, and the above results for $H$, we find that
\begin{equation}
 \tilde{\nabla}_\rho h^\mu{}_\nu = 0.
\end{equation}
However, this is the necessary and sufficient condition for the new spacetime to be {\it decomposable} \cite{exact}.  That is, there exist coordinates $(x^A,y^\alpha)$ so that the metric takes the form
\begin{equation}
 d\tilde{s}^2 = \tilde{g}_{AB}(x)\, dx^A dx^B 
                + \tilde{g}_{\alpha\beta}(y)\, dy^\alpha dy^\beta,
\end{equation}
where $A,B=0,1,2$ and $\al,\ba=3,4$ are coordinate indices only for this equation, and the remainder of this section. The 3D submanifolds are surfaces of constant $y^\alpha$. These are orthogonal to 2D submanifolds of constant $x^A$.
In this coordinate chart, equations \eref{eqn:phiresults} reduce to $\Phi=\Phi(y)$. We now see that the physical metric is a warped product:
\begin{equation}
   ds^2 = |\Phi(y)|^{-1/2} \tilde{g}_{AB}(x) dx^A dx^B + g_{\alpha\beta}(y) dy^\alpha dy^\beta,
\end{equation}
where $g_{\alpha\beta}(y) = |\Phi(y)|^{-1/2} \tilde{g}_{\alpha\beta}(y)$. The surfaces of constant $y^\alpha$ are constant curvature, so $\tilde{g}_{AB}(x)$ is the metric of 3D Minkowski or (anti-)de Sitter spacetime. We now see that the symmetries of the constant curvature submanifolds extend to symmetries of the full spacetime. Hence we can apply Birkhoff's theorem to deduce that the 5D spacetime must be isometric to either (ii) (if $\Phi$ is constant) or (iii) in the statement of the theorem. $\Box$

\subsection{Comments on the six-dimensional case}

Consider now a 6D Einstein spacetime admitting a non-geodesic multiple WAND.  Let us use the same notation as we did in the proof of Proposition \ref{lem:typeD}. We already know some of the components of $\Phi_{ijkl}$ from equation \eref{eqn:C2i2j}. The remaining components $\Phi_{\hi\hj\hk\hl}$ have the symmetries of the Riemann tensor in 3D, hence they are completely determined by their trace $\Phi_{\hi\hj\hk\hj}=-3\Phi_{\hi\hk}$ (see \eref{eqn:trace}). This gives
\begin{equation}\label{eqn:6weyl}
 \Phi_{\hi\hj\hk\hl} = -6 ( \del_{\hi[\hk} \Phi_{\hl]\hj} - \del_{\hj[\hk} \Phi_{\hl]\hi} ) 
                       + 6\Phi \del_{\hi [\hk} \del_{\hl]\hj} .
\end{equation}
Together with Proposition \ref{lem:typeD}, this implies that the Weyl tensor is fully determined by $\Phi_{ij}$. We can substitute this into equation \eref{B1:nongeo} to learn that the constant curvature submanifolds of Theorem \ref{thm:submfd} have dimension three {\it unless} (i) the eigenvalues of $\Phi_{ij}$ are $-\Phi,-\Phi,3\Phi/2,3\Phi/2$ with $\Phi \ne 0$, in which case they have dimension four; or (ii) $\Phi_{ij}=0$, in which case the spacetime is type O (i.e. Minkowski or (anti-)de Sitter spacetime).

The case of the foliation by four-dimensional submanifolds can be analyzed using a similar method to the proof of Theorem \ref{thm:list}. The result is that the spacetime must be either a direct product $dS_4 \times S^2$ or $AdS_4 \times H^2$, or a spacetime with metric 
\begin{equation}
  ds^2 = r^2 d\tilde{s}_4^2 + \frac{dr^2}{U(r)} + U(r) dz^2, \qquad 
  U(r) = k - \frac{m}{r^3} -  \frac{\La}{5}r^2, 
\end{equation}
where $m \ne 0$, $k \in \{1,0,-1\}$, $d\tilde{s}_4^2$ is the metric of a 4D Lorentzian space of constant curvature (i.e. 4D Minkowski or (anti-)de Sitter) with Ricci scalar $12k$, and the coordinate $r$ takes values such that $U(r)>0$. These solutions are the 6D analogues of cases (ii) and (iii) of Theorem \ref{thm:list}. 

Now consider the case in which the constant curvature submanifolds are three dimensional. In this case, we might hope to prove that the distribution orthogonal to these submanifolds is integrable.  Here, coordinates could be introduced so that the metric takes the form
\begin{equation} \label{eqn:6dnongeowand}
  ds^2 =  F(x,y)^2 g_{AB}(x) dx^A dx^B + g_{\alpha\beta}(x,y) dy^\alpha dy^\beta,
\end{equation}
where $A,B$ range from $0$ to $2$ and $\alpha,\beta$ range from $3$ to $5$ and the surfaces of constant $y^\alpha$ are the constant curvature submanifolds. The constant curvature condition implies that the coordinates $x^A$ can be chosen so that
\begin{equation}
  F(x,y)^2 g_{AB}(x) = \frac{\eta_{AB}}{\big(a(y) \eta_{CD}x^C x^D + b_C(y) x^C + c(y)\big)^2},
\end{equation}
for some $a(y)$, $b_C(y)$ and $c(y)$, where $\eta_{AB}$ is the 3d Minkowski metric. Note that it is not obvious that the symmetries of the constant curvature submanifolds extend to symmetries of the spacetime.

Using the Bianchi identity, we are able to prove that the distribution orthogonal to the constant curvature submanifolds is indeed integrable {\it except} when $\Phi_{ij}$ has eigenvalues $0,0,\phi,-\phi$ for some scalar $\phi \ne 0$ (this implies $\Phi=0$). We have not made any progress in analyzing this exceptional case so we shall not give further details here. In more than six dimensions, it seems likely that the distribution orthogonal to the submanifolds of constant curvature will be non-integrable except in special cases.

\section{Discussion}
Theorem \ref{thm:geo} is a useful result when attempting to prove general results about algebraically special spacetimes, as one can work with a geodesic multiple WAND without loss of generality.

However, much of the power of the four-dimensional Goldberg-Sachs theorem is that it allows one to check whether a null vector field is tangent to a repeated principal null direction without explicitly calculating the Weyl tensor.  From a computational point of view, this is a huge advantage.  Is there a full generalisation of such a result to higher dimensions?  Given our results on geodesity, it seems that an analogous result in higher dimensions would have to apply only to \emph{geodesic} multiple WANDs.

To be specific, to be a true generalization of the original Goldberg-Sachs theorem, a result would need to take the form: \emph{A null vector field $\lb$ is a geodesic multiple WAND if and only if it is tangent to a null geodesic congruence with particular optical properties.}

\subsection{Progress towards a full theorem}
It seems fair to say that, to date, progress on this issue has been rather limited.  There are a series of partial results, mainly attempting to derive restrictions on the optics of null vector fields, with the assumption that they are tangent to multiple WANDs.

One example of such a partial result is the following:
\begin{lemma}\label{lem:TypeN}
  Let $\lb$ be a multiple WAND of Type N alignment in an Einstein spacetime.  Then the optical matrix $\rhob$ takes the form
  \begin{equation}
 \label{eqn:rhotypeN}
    \rhob = \frac{1}{2}\left(\begin{array}{c|c}
                    \!\!\begin{array}{cc}
                      \rho    & a\\
                      -a & \rho
                    \end{array} & \Ob \\ 
                    \hline
                    \Ob & \Ob
                  \end{array}\right)
  \end{equation}
  (in a frame where its symmetric part is diagonalized), for some $\rho$, $a$. If $\rho = 0$ then $a=0$ and the spacetime is Kundt (i.e. $\rhob =\Ob$).
\end{lemma}
This result was previously obtained in \cite{Bianchi}, but the proof given here is significantly simpler; this also provides a nice example of the utility of the GHP formalism.
\paragraph*{Proof of Lemma \ref{lem:TypeN}:}
By Lemma \ref{lem:oldnongeo}, all multiple WANDs in Type N spacetimes are geodesic.  Hence $\lb$ is geodesic ($\kapb=\Ob$).  For Type N spacetimes, by definition, the only non-vanishing Weyl components are $\Omb'$. The Bianchi equations imply that
\begin{eqnarray}
  \tho \Om'_{ij}       &=& -\Om'_{ik} \rho_{kj}, \label{A13:N} \\
  \Om'_{i[j}\rho_{kl]} &=& 0 \label{A15:N},\\
  \Om'_{i[k|}\rho_{j|l]} &=& \Om'_{j[k|}\rho_{i|l]} \label{A14:N}.
\end{eqnarray}
Let $\Sb$ and $\omb$ denote the symmetric and antisymmetric parts of $\rhob$ respectively.
Tracing (\ref{A15:N}) on $i$ and $k$ gives
\begin{equation}\label{eqn:psiA}
  \Omb' \omb + \omb \Omb' = 0.
\end{equation}
Similarly, tracing (\ref{A14:N}) on $i$ and $k$ gives
\begin{equation}\label{eqn:psirho}
  \Omb' \rhob + \rhob \Omb' = (\tr \rhob) \Omb'
\end{equation}
and, using (\ref{eqn:psiA}), this gives
\begin{equation}\label{eqn:psiS}
  \Omb' \Sb + \Sb \Omb' = (\tr \Sb) \Omb'.
\end{equation}
Now we take the antisymmetric part of (\ref{A13:N}) to obtain
\begin{equation}
  0 = -[\Omb',\Sb] - ( \Omb' \omb + \omb \Omb' ),
\end{equation}
and after applying (\ref{eqn:psiA}) this tells us that $[\Omb',\Sb]=0$, and hence $\Omb'$ and $\Sb$ are simultaneously diagonalizable, via rotations of the $\mb{i}$. Work in a basis where $\Omb'$ and $\Sb$ are diagonal.  Let $N$ be the number of eigenvalues of $\Omb'$ that do not vanish everywhere in the spacetime, then we can shuffle the $\mb{i}$ so that
\begin{equation}
  \Omb' = \diag(\Om'_{(2)},...,\Om'_{(N+1)},0,...,0) \eqand \Sb = \diag(s_{(2)},...,s_{(d-1)}),
\end{equation}
with all the $\Om'_{(\al)}$ non-zero (where from now on in this section, indices $\al,\beta,...$ range over $2,...,N+1$ and $I,J,...$ range over $N+2,...,d-1$).  As the spacetime is Type N not Type O, we must have $N\geq1$. Putting this into (\ref{eqn:psiS}) gives (with no summation),
\begin{equation}
  \Om'_{(i)} s_{(i)} = \frac{1}{2} \Om'_{(i)} (\tr \Sb)
\end{equation}
for all $i$ and hence
\begin{equation}
  s_{(\al)} = \frac{\tr \Sb}{2} \quad \mathrm{for} \quad \al=2,...,N+1.
\end{equation}
Also, the $\al I$ component of (\ref{eqn:psiA}) implies that $\om_{I\al} = 0 = \om_{\al I}$, so $\rhob$ is block diagonal with blocks of size $N$ and $d-2-N$. Finally, taking the $ijkl=I\al J \beta$ component of the Bianchi equation (\ref{A14:N}) gives $\Om'_{\al\beta} \rho_{IJ} = 0$ and hence $\rho_{IJ}=0$.

In summary, we have shown so far that (recall $\tr \Sb=\rho$)
\begin{equation}
  \rhob = \left(\begin{array}{c|c}
                  \frac{\rho}{2} \Id_N + \omb_N & 0 \\ 
                  \hline
                  0 & 0
                \end{array}\right)
\end{equation}
where $\Id_N$ is the $N\times N$ identity matrix, and $\omb_N$ is antisymmetric.
Taking the trace tells us that $\rho=N \rho/2$ hence either (i) $N=2$ or (ii) $\rho=0$.

In case (i), we have proved that $\rhob$ must take the form \eqref{eqn:rhotypeN} for some $a$.

In case (ii), $\Sb=\Ob$.  The trace of equation (\ref{Sachs}) gives $\tho( \tr \Sb) = -\tr(\Sb^2)-\tr(\omb^2)$
and hence we see that $\tr(\omb^2) = -\om_{ij} \om_{ij}=0$, so $\omb=\Ob$ and the spacetime is Kundt. (In fact $\Sb=\Ob$ implies $\omb=\Ob$ for all Einstein spacetimes \cite{Ricci}.) $\Box$

Type N is the simplest of the algebraic types to analyse in this way, and obtaining similar results for Type II or Type III spacetimes is more difficult.  Various partial results exist, see e.g.\ \cite{Bianchi,TypeD,TypeII}; and it seems that a natural decomposition into $2\times 2$ blocks often occurs.  This perhaps gives a hint about why four dimensions is so special in this context; this is the dimension where there is always exactly one of these $2\times 2$ blocks.

To try and build more intuition, it is interesting to consider other special cases.  For example, the following result gives a clear example of a decomposition into multiple $2\times 2$ blocks:
\begin{lemma}[Ortaggio \etal \cite{Ortaggio:2009bz}]\label{lem:ksoptics}
  Let $g$ be an Einstein metric of the Kerr-Schild form
  \begin{equation}
    g_{\mu\nu} = \eta_{\mu\nu} + H \lb_\mu \lb_\nu
  \end{equation}
  where the function $H$ is chosen such that $\lb$ is tangent to \emph{affinely parametrized null geodesics}, and hence we can choose coordinates such that $\lb=\pd/\pd r$.  Then, for some non-negative integers $p$,$q$ with  $p+q\leq d-2$, one can choose a real spatial basis $\mb{i}$ such that the optical matrix of $\lb$ takes the block diagonal form
  \begin{equation}
    \rhob = \mathrm{blockdiag}\big( \underbrace{\Rb_{(1)},\Rb_{(2)},\ldots, 
                                                \Rb_{(p)}}_{2p},
                                    \underbrace{\tfrac{1}{r},\ldots,\tfrac{1}{r}}_{q},
                                    \underbrace{0,\ldots,0}_{d-2-q-2p}
                              \big)
  \end{equation}
  where the $\Rb_{(\al)}$ are shearfree $2\times 2$ blocks of the form
  \begin{equation}
    \Rb_{(\al)} = \frac{1}{r^2 + c_{(\al)}^2} \twomat{ r  & c_{(\al)} \\ - c_{(\al)} & r}
  \end{equation}
  for some $c_{(\al)}$ (not dependent on $r$).  Furthermore, $\M{i}_{j0}=0$ and hence this block diagonal form is parallelly propagated along $\lb$.
\end{lemma}

\subsubsection{Kundt spacetimes}\label{sec:kundt}
Going the other way, i.e.\ proving that a null vector field is a multiple WAND given certain optical properties, seems to be more difficult.  One example of where this can be done is for \emph{Kundt spacetimes}.  These are spacetimes admitting a null geodesic congruence with vanishing expansion, rotation and shear.  This important class of exact solutions to the Einstein equations in four dimensions was first discussed by Kundt \cite{Kundt:1961}, typically this null vector can be thought of as the direction of propagation of a plane fronted gravitational wave.  The definition extends naturally to higher dimensions.  In our notation, this means that a spacetime is Kundt if and only if there exists a choice of $\lb$ such that $\rhob=\Ob=\kapb$.

It is known \cite{Ricci} that any Kundt spacetime, with matter such that it admits a choice of basis for which $R_{00} = R_{0i} = 0$, is algebraically special in the sense of Definition \ref{def:algspec}.  Using the GHP formalism we can both prove this in a more convenient manner, and in fact generalize the result slightly:
\begin{theorem}\label{thm:kundt}
  Let $\lb$ be a non-expanding, non-twisting, non-shearing null geodesic congruence in a Kundt spacetime $(\Mcal,g)$.  Then $\lb$ is a WAND.

  If $(\Mcal,g)$ has matter such that the vector field $R(\lb) \equiv R^\mu_{\phantom{\mu}\nu} \lb^\nu$ is null then it is algebraically special with multiple WAND $\lb$.
\end{theorem}
The conditions of this theorem include Einstein spacetimes, as well as any matter for which $\lb$ is an eigenvector of the Ricci tensor (e.g.\ aligned null radiation).
\proof
In GHP notation, the Kundt property is equivalent to the statement that $\kapb=\Ob$ and $\rhob=\Ob$.  Consider the Newman-Penrose equations (for spacetimes with arbitrary matter) given in Appendix \ref{app:ghpmatter}.  Equation \eqref{NP1m} reads
\begin{equation}
  0 = -\Om_{ij} - \tfrac{1}{d-2} \om \del_{ij},
\end{equation}
and taking the trace and tracefree parts of this implies that $\om\equiv R_{00} = 0$ and $\Om_{ij}=0$.  Hence $\lb$ is a WAND.

Now, the NP equation \eqref{NP3m} implies that
\begin{equation}
  \Psi_{ijk} = \tfrac{2}{d-2}\del_{i[j} \psi_{k]}
\end{equation}
where $\psi_k \equiv R_{0k}$.  We now need to use the assumption on the matter content of the spacetime.  Note that 
\begin{equation}
  \psi_i \psi_i = \lb^\mu\lb^\rho m_i^\nu m_i^\sig R_{\mu\nu}R_{\rho\sig} = R(\lb)_\mu R(\lb)^\mu - 2\om \phi
\end{equation}
where $\phi \equiv R_{01}$.  But we have already shown that $\om=0$, and hence the assumption $R(\lb).R(\lb)=0$ is sufficient to give $\psi_i=0$ and hence $\Psi_{ijk}=0$.  Hence, $\lb$ is a multiple WAND.$\Box$

In fact, it is also known that all Einstein spacetimes admitting a shearfree null geodesic congruence are algebraically special.  The spacetimes in this class that are not Kundt are known as Robinson-Trautman spacetimes, and are characterized by an optical matrix with $\rho_{(ij)} = \frac{\rho}{d-2}\del_{ij}$ with non-zero expansion $\rho$.  Ref.\ \cite{RobTraut} constructed a canonical form for all such metrics, and showed that they are algebraically special.

In the shearing case, progress has proved significantly more difficult.

\subsection{Non-existence of a full theorem?}
Perhaps the difficulty in proving a more complete version of such a theorem is that no such theorem exists?  One way of gaining intuition for whether two sets of conditions are likely to be equivalent is to count components.

A null vector field is a multiple WAND if and only if $\Omb=\Psib=\Ob$.  Naively, it appears that this is a condition on $[\Omb] + [\Psib]$ independent real components, where 
\begin{equation}
  [\Omb] = \tfrac{1}{2}d(d-3), \qquad 
  [\Psib] = \tfrac{1}{3}(d-1)(d-2)(d-3)
\end{equation}
are the number of boost weight +2 and +1 components of the Weyl tensor in $d$ dimensions.

Algebraic conditions on the optics of $\lb$ are given by the optical matrix $\rhob$, which can be split into conditions on its expansion $\rho$, shear $\sigb$ and twist $\omb$.  They have the following number of independent components:
\begin{equation}
  [\rho] = 1, \qquad
  [\sigb] = \tfrac{1}{2} d (d-3), \qquad
  [\omb] = \tfrac{1}{2} (d-2)(d-3), \qquad
  [\kapb] = d-2.
\end{equation}
In four dimensions, $[\Omb] = [\Psib] = 2$, and $[\kapb] = [\sigb] = 2$ (and hence each of these objects can be encoded in terms of a single complex scalar).  Naively, the fact that there are four independent components on `each side' of the Goldberg-Sachs theorem makes it plausible that an equivalence between the two sets of conditions is potentially possible.

What happens if we try to perform a similar counting argument in higher dimensions?  We know that it will not be sufficient to only specify information about the shear, as multiple WANDs are sometimes, but not always, shearing in higher dimensions.  If we allow ourselves to specify information about all parts of $\rhob$, then the `optical side' of the Goldberg-Sachs theorem involves $[\kapb] + [\rho] + [\sigb] + [\omb]$ components.  A brief calculation shows that
\begin{equation}
  \big([\Omb] + [\Psib]\big) - \big([\kapb] + [\rho] + [\sigb] + [\omb]\big) \geq 1 \qquad 
  \forall d\geq 5.
\end{equation}
Given this, is it tempting to speculate that the multiple WAND condition is more restrictive than any possible condition on the optics of a null direction, and hence that a two-way Goldberg-Sachs theorem probably doesn't exist in higher dimensions.

However, we have missed some important constraints; namely how the canonical form of the optical matrix $\rhob$ varies under parallel transport along $\lb$.  In, for example, the Kerr-Schild case discussed above it is crucial that the canonical form of the matrix is preserved under parallel propagation along $\lb$.  If $\mb{i}$ and $\mb{j}$ are vectors spanning the $2\times 2$ block $\Rb_{(\al)}$ in Lemma \ref{lem:ksoptics}, then this requires $\tho \mb{i},\tho\mb{j} \in \mathrm{span}\{\mb{i},\mb{j}\}$.  However, this can only introduce ${\cal O}(d^2)$ further conditions, while $\Psib=\Ob$ gives ${\cal O}(d^3)$ conditions, so for sufficiently large dimension $d$ there will still be more conditions on one side of the possible equivalence than the other.

Note that in the case of Kundt or Robinson-Trautman spacetimes, where it \emph{can} be deduced from optical properties that a vacuum spacetime is algebraically special \cite{Ricci,RobTraut}, this issue of parallel propagation does not occur, as there are not multiple distinct blocks.  Of course these classes include all algebraically special vacuum metrics with a twistfree repeated PND in four dimensions.

If such a theorem does not exist in higher dimensions, how could this be demonstrated conclusively?  A clear counterexample would take the form of two null vector fields (probably in two distinct vacuum spacetimes), with optical properties that appear to be identical (i.e.\ two optical matrices of the same canonical form, preserved properly under parallel transport), only one of which is a multiple WAND.  Further investigation is needed to clear this issue up; it is not clear whether or not the problem is tractable in the immediate future. 

\subsection{Other approaches}
Perhaps the difficulty with making further progress on a Goldberg-Sachs theorem in higher dimensions is that we are using the `wrong' definition of algebraically special.

Recent work \cite{Mason:2010zzc,TaghaviChabert:2010bm} proposes an alternative generalization of the notion of a Type II spacetime in higher dimensions.  Their definition is more restrictive than that of the CMPP classification; in CMPP language it corresponds to the vanishing of all positive boost weight components, as well as particular boost weight 0 and -1 components.  In five dimensions, Taghavi-Chabert \cite{TaghaviChabert:2010bm} was able to use this definition to prove that a particular optical structure (associated to the algebraically special property) is integrable.  However, it is also demonstrated that the direct converse to this result does not hold, with the black ring admitting an appropriate integrable optical structure but not being algebraically special.

Separately, other recent work \cite{bivectors,lode} suggests ways of refining the CMPP classification, using the decomposition of components of different boost weights into irreducible components under the action of spins $\Xb\in SO(d-2)$.  It is possible that more progress could be made towards a Goldberg-Sachs theorem by using this refinement to make a more (or less) restrictive definition of algebraically special.  This idea has not yet been investigated in detail.

\chapter[Decoupling of perturbations]{Decoupling perturbations of algebraically special spacetimes}\label{chap:decoupling}
\section{Introduction}
The previous two chapters have focused on developing the theory behind the higher-dimensional GHP formalism.  Here, we move on to study an application.  We look to develop a new approach to analysing the stability of higher-dimensional black holes, motivated by successful approaches in four dimensions.  In particular, we will consider linearized scalar field, electromagnetic and gravitational perturbations of algebraically special spacetimes, and define a new set of gauge invariant variables which can be used to describe the perturbations.

Furthermore, we attempt to find a decoupled equation describing these perturbations.  This is not fully successful; in higher dimensions we will only be able to achieve decoupling for Kundt spacetimes.  The usefulness of this will be discussed in Chapter \ref{chap:nhperts}.  Despite the failure of decoupling for black hole spacetimes in higher dimensions, the new gauge invariant quantities may provide a useful new approach to studying linearized perturbations numerically.

In Section \ref{sec:higherbhperts} we reviewed some existing results for analysing the linearized stability of black holes in higher dimensions.  The existing methods have two particular difficulties.  Firstly, they require the numerical solution of highly complicated, coupled differential equations.  Furthermore, after solving these equations, it must be checked that the solutions found do not correspond to `pure gauge' modes that do not represent physical perturbations of the spacetime.

The cases where a stability analysis has proved tractable concern black holes with a large isometry group, occurring for example in the Schwarzschild solution, or particularly symmetric examples of Myers-Perry black holes with many of the angular momenta coinciding.

However, in four dimensions, there is an alternative approach to studying black hole stability, due to Teukolsky \cite{Teukolsky:1972,Teukolsky:1973}.  This exploits the algebraically special nature of black hole solutions.  It is this approach, using the 4D Newman-Penrose formalism, which originally rendered tractable the study of perturbations of the Kerr solution, and hence it is natural to ask whether this method can be applied in higher dimensions.

Consider a 4D spacetime with null tetrad $(\lb,n,m,\bar{m})$.  Recall that the Weyl tensor is encoded in the Newman-Penrose scalars $\Psi_0,\ldots, \Psi_4$.  Now, consider a linearized perturbation of such a spacetime.  Let $\Psi_A^{(0)}$ denote the unperturbed value of $\Psi_A$, and let $\Psi_A^{(1)}$ denote the perturbation.  In general, there is gauge freedom in this perturbation, corresponding to the possibility of infinitesimal coordinate transformations and infinitesimal changes of tetrad.  However, it can be shown \cite{stewperts} that $\Psi_0^{(1)}$ is gauge invariant if and only if $\lb$ is a repeated principal null direction of the background spacetime.  Therefore, for perturbations of algebraically special spacetimes, there exists a local, gauge-invariant quantity, linear in the metric perturbation.

This gauge-invariant quantity seems likely to be useful when studying perturbations.  However, something much more surprising happens.  In a general spacetime, the linearized equations of motion will lead to coupled equations for the quantities $\Psi_A^{(1)}$.  Remarkably, in an algebraically special spacetime, Teukolsky \cite{Teukolsky:1972} showed that one can decouple these equations to obtain a single, second order, wave equation for $\Psi_0^{(1)}$.  In fact, this \emph{Teukolsky equation} can be generalized to describe perturbations of other kinds; for example electromagnetic test fields in the background of an algebraically special spacetime.

If the background is Type D, then we can choose both $\lb$ and $n$ to be repeated principal null directions, and both $\Psi^{(1)}_0$ and $\Psi^{(1)}_4$ are gauge invariant and both satisfy decoupled equations of motion.

Stewart \& Walker \cite{stewperts} used the GHP formalism to gain a fuller understanding of why Teukolsky's approach is successful, as well as giving a far simpler derivation of the Teukolsky equation.  It is their approach that we will follow when looking for the higher-dimensional generalization.

In Chapter \ref{chap:ghp} we described the development of a higher-dimensional generalization of the GHP formalism; based around a particular choice of two null vectors $\lb$ and $n$.  Recall that the appropriate generalization of $\Psi_0$ is a $(d-2) \times (d-2)$ traceless symmetric matrix $\Om_{ij}\equiv C_{0i0j}$, while the analogue of $\Psi_4$ is another such matrix, $\Om'_{ij} \equiv C_{1i1j}$.  These quantities transform as scalars under general coordinate transformations.  Note that the number of independent components of $\Omb$ (or $\Omb'$) is the same as the number of physical degrees of freedom of the gravitational field.

Just as for $d=4$, we find that $\Omb^{(1)}$ is invariant under infinitesimal coordinate transformations and infinitesimal changes of basis if (and only if) $\lb$ is a multiple WAND.  If the background is is Type D (or O), we can choose both $\lb$ and $n$ to be multiple WANDs, and find that both $\Omb^{(1)}$ and $\Omb'^{(1)}$ are gauge invariant.  This gauge invariance implies that, irrespective of decoupling, these quantities are natural objects to consider when studying gravitational perturbations of higher-dimensional algebraically special solutions.

We will study linearized gravitational perturbations of algebraically special spacetimes satisfying the vacuum Einstein equation (allowing for a cosmological constant).  We find that $\Omb^{(1)}$ satisfies a decoupled equation in an algebraically special vacuum spacetime with $d>4$ if, and only if, $\ell$ is geodesic and free of expansion, rotation and shear.  We also analyze the simpler case of a Maxwell field and find that exactly the same condition is required for decoupling in this case.

Recall that a spacetime admitting a null geodesic congruence with vanishing expansion, rotation and shear is known as a Kundt spacetime.  In Theorem \ref{thm:kundt} we used the GHP formalism to show that any such spacetime is algebraically special (in vacuum).  Hence our result is that electromagnetic and gravitational perturbations can be decoupled in this way if, and only if, the spacetime is Kundt.

In four dimensions, decoupling requires only that $\lb$ be geodesic and shearfree.  By the Goldberg-Sachs theorem, we know that such an $\lb$ can be found in any algebraically special spacetime; so this is not a restrictive condition.  By contrast, the condition that we have found in higher dimensions is far more restrictive.

Sometimes both $\Omb^{(1)}$ and $\Omb'^{(1)}$ satisfy decoupled equations; this occurs if $\lb$ and $n$ are both geodesic with vanishing expansion, rotation and shear.  We will refer to such a spacetime as \emph{doubly Kundt}; and note that a doubly Kundt spacetime must be Type D (or O).

Unfortunately, black hole spacetimes are not Kundt and therefore decoupling does not occur in higher-dimensional black hole spacetimes.\footnote{There is no contradiction with the results of Ishibashi \& Kodama \cite{Ishibashi:2003}, since that reference studies perturbations by exploiting the spherical symmetry of the Schwarzschild solution rather than its Type D property, i.e., $\Omb^{(1)}$ is not used to describe the perturbation.  The quantities that satisfy decoupled equations are \emph{non-local} in the metric perturbation.}  Obviously is it disappointing that decoupling does not occur for the Myers-Perry solution.\footnote{Nevertheless, we emphasize that the quantities $\Omb^{(1)}$, $\Omb^{'(1)}$ should be useful in studies of Myers-Perry perturbations because of their locality and gauge invariance.}  However, as we noted in Section \ref{sec:nhgeom}, the \emph{near-horizon geometries} of extreme vacuum black holes \emph{are} Kundt spacetimes.  Therefore our decoupled equation is ideal for studying perturbations of near-horizon geometries.  In Chapter \ref{chap:nhperts} we will apply our techniques to this problem, and discuss what information this can give us about the stability of higher-dimensional black holes.

The current chapter is organized as follows.  In Section \ref{sec:gaugeinv} we investigate the existence of gauge invariant quantities, and show that $\Omb^{(1)}$ is gauge invariant if and only if the background spacetime is algebraically special.  We then move on to consider decoupling of perturbations.  As a warm-up exercise, in Section \ref{sec:maxdecoupling} we consider the decoupling of Maxwell perturbations, as this simpler example illustrates the approach that we use in the gravitational case in Section \ref{sec:gravdecoupling}.  Finally, in Section \ref{sec:decouplingdiscussion} we discuss the possible applications of our results, leading into Chapter \ref{chap:nhperts} where we will use our results to study perturbations of near-horizon geometries.

\section{Gauge-invariant variables}\label{sec:gaugeinv}

We are interested in linearized perturbations of spacetimes.  For a quantity $\Xb$, we shall write $\Xb=\Xb^{(0)}+\Xb^{(1)}$ where $\Xb^{(0)}$ is the value in the background spacetime and $\Xb^{(1)}$ is the perturbation.  Following Ref.\ \cite{stewperts}, we look to find variables that are gauge invariant under both infinitesimal coordinate transformations and infinitesimal changes of basis.

Let $\Xb$ be a spacetime scalar.  Then, under an infinitesimal coordinate transformation with parameters $\xi^\mu$, we have $ \Xb^{(1)} \rightarrow \Xb^{(1)} + \xi .\, \partial \Xb^{(0)}$.  Hence $\Xb^{(1)}$ is invariant under infinitesimal coordinate transformations if, and only if, $\Xb^{(0)}$ is constant.

In four dimensions, as discussed above, $\Psi^{(1)}_0$ is a gauge invariant quantity describing gravitational perturbations in four-dimensional algebraically special spacetimes.  Hence, we are motivated to consider $\Xb = \Omb$.  Is this gauge invariant?  It turns out that the result is the same as the four-dimensional one:
\begin{lemma}\label{lem:Ominv}
  $\Omb^{(1)}$ is a gauge invariant quantity if and only if $\lb$ is a multiple WAND of the background spacetime (or equivalently, if and only if $\Psib^{(0)} = \Ob = \Omb^{(0)}$).
\end{lemma}
\proof
First we consider infinitesimal basis transformations.  Consider an infinitesimal spin of the form \eqref{eqn:spins}. If $\Omb^{(0)}$ is non-vanishing then this will induce a change in $\Omb^{(1)}$.  Hence we must have $\Omb^{(0)}=\Ob$ for $\Omb^{(1)}$ to be gauge invariant.

Next consider an infinitesimal null rotation \eqref{eqn:nullrotn} about $\nb$.  From equation \eqref{eqn:nullrotn}, the change in $\Omb^{(1)}$ is, to linear order in the infinitesimal parameters $z^i$,
\begin{equation}
  \Om^{(1)}_{ij}
     \mapsto \Om^{(1)}_{ij}-2 z_k (\Psi^{(0)}_{(i}\del_{j)k}+\Psi^{(0)}_{(ij)k}) .
\end{equation}
For invariance, we need 
\begin{equation}
  \Psi^{(0)}_{(i}\del_{j)k}+\Psi^{(0)}_{(ij)k} = 0.
\end{equation}
Taking the trace on $j$ and $k$ gives $\Psi^{(0)}_i=0$.  We then use the relation $\Psi_{ijk} = \tfrac{2}{3}(\Psi_{(ij)k} - \Psi_{(ik)j})$ to deduce that $\Psi^{(0)}_{ijk}=0$.  So we conclude that invariance of $\Omb^{(1)}$ under infinitesimal basis transformations implies that
\begin{equation}
 \Omb^{(0)} = \Ob = \Psib^{(0)}.
\end{equation}
It is easy to see that these conditions are both necessary and sufficient for $\Omb^{(1)}$ to be invariant under infinitesimal basis transformations.  These conditions are equivalent to the statement that $\lb$ is a multiple WAND of the background geometry. 

Finally, since $\Omb^{(0)}=\Ob$, it follows that $\Omb^{(1)}$ is invariant under infinitesimal coordinate transformations. $\Box$

Similarly, $\Omb'^{(1)}$ is gauge invariant if, and only if, $\nb$ is a multiple WAND. Hence both quantities are gauge invariant if, and only if, the spacetime is Type D.\footnote{Note that further gauge invariant quantities exist for higher dimensional spacetimes satisfying additional restrictions, see Lemma \ref{lem:otherinvs} later.}

Now consider a Maxwell field. We shall consider only a test field, i.e.\ we neglect gravitational backreaction and treat the Maxwell field as an infinitesimal quantity that vanishes in the background.  It follows that all components are invariant to first order under infinitesimal coordinate transformations and infinitesimal basis transformations.  Note that, since we are treating the Maxwell field as infinitesimal, and working to first order, there is no distinction between Maxwell theory and Maxwell theory with a Chern-Simons term.

So far we have discussed only \emph{infinitesimal} basis transformations.  However, sometimes one might want to consider finite transformations.  For example, consider a Type D spacetime.  Then $\lb$ and $n$ are fixed (up to scaling) in the background by the requirement of being multiple WANDs.  But there is no preferred way of choosing the spatial basis vector $\mb{i}$. Different choices are related by {\it finite} spins.  $\Omb^{(1)}$ and $\Omb'^{(1)}$ are not invariant under finite spins.  Exactly the same issue arises in 4D, where $\Psi_0^{(1)}$ and $\Psi_4^{(1)}$ pick up complex phases under finite spins. 

Physical quantities should not care about the choice of spatial basis vectors so such quantities must be related to GHP scalars with zero spin.  For example, in an asymptotically flat 4D spacetime, the energy flux in ingoing and outgoing gravitational waves is related to the spin-0 GHP scalars $|\Psi_0^{(1)}|^2$ and $|\Psi_4^{(1)}|^2$, respectively (for appropriate choices of $\ell$ and $n$, see \cite{Teukolsky:1972}).  For $d>4$, the analogous quantities are $\Om^{(1)}_{ij} \Om^{(1)}_{ij}$ and $\Om'^{(1)}_{ij} \Om'^{(1)}_{ij}$, although as discussed in Section \ref{sec:highernp}, they probably do not carry the same physical interpretation in higher dimensions.

We can also define additional invariant quantities such as $\Phi_{ij}^{(0)} \Om^{(1)}_{ij}$ (which vanishes identically in 4D because $\Phis_{ij} = \tfrac{1}{2} \Phi \delta_{ij}$ for all spacetimes).

\section{Decoupling of electromagnetic perturbations}\label{sec:maxdecoupling}
%

The highest boost weight components of the Maxwell $(p+1)-$form field strength are denoted by a GHP scalar $\vphib$ of boost weight 1 and spin $p$.  In 4D (where $p=1$) the quantity analogous to $\vphib$ satisfies a decoupled equation of motion in an algebraically special background.  We shall investigate the conditions under which $\vphib$ satisfies a decoupled equation of motion in $d>4$ dimensions.  The motivation for doing this is mainly that the Maxwell field illustrates the arguments that we shall also employ in the gravitational case, but the equations are considerably simpler.  For this reason, we restrict to the simplest case $p=1$.

In this section, we show how, in a particular class of background Einstein spacetimes, we can construct decoupled 2nd order differential equations for a Maxwell test field.  We show that this decoupling is possible if and only if the background spacetime is Kundt, that is it admits a geodesic null vector field that is not shearing, twisting or expanding.

In particular, we will show that the dynamics of a Maxwell test field on the background of a Kundt spacetime can be described by the following equation:
\begin{equation}\label{eqn:maxpert}
    \left(2\tho'\tho + \eth_j\eth_j + \rho'\tho -4\tau_j\eth_j 
                                   + \Phi-\tfrac{2d-3}{d-1}\La\right)\vphi_i 
             + (- 2\tau_i\eth_j + 2\tau_j \eth_i + 2\Phis_{ij} + 4\Phia_{ij})\vphi_j = 0.  
\end{equation}
We also show that analogous decoupled equations cannot be constructed for spacetimes that are not Kundt, and discuss briefly whether any alternative progress can be made.

It is interesting to compare this to the equation of motion for a massive scalar field $\Psi$:
\begin{equation}
  (\nabla_\mu \nabla^\mu - \mu^2) \Psi = 0.
\end{equation}
When written out in GHP form in a general background, this equation is
\begin{equation} \label{eqn:scalarperts}
  (2\tho'\tho + \eth_i\eth_i + \rho'\tho  -2\tau_i\eth_i + \rho\tho'  - \mu^2)\Psi = 0.
\end{equation}
To compare this with the decoupled Maxwell equation, one must specialize to a Kundt spacetime, for which $\rho=0$. Note that $\tau'_i$ does not appear in either equation.
\subsection{Derivation of results}

In the case of a 2-form field strength $F_{\mu\nu}$, the GHP Maxwell equations (\ref{max:1}-\ref{max:4}) reduce to:
\begin{eqnarray}
  \eth_i \vphi_i + \tho f\label{max1}
         &=& \tau'_i \vphi_i + \rho_{ij} F_{ij} - \rho f - \kap_i \vphi'_i \\
  2 \eth_{[i} \vphi_{j]} - \tho F_{ij} \label{max2}
         &=& 2\tau'_{[i} \vphi_{j]} + 2f \rho_{[ij]}
             + 2F_{[i|k} \rho_{k|j]} + 2\kap_{[i}\vphi'_{j]}\\
  2\tho' \vphi_i + \eth_j F_{ji} - \eth_i f
         &=& (2\rho'_{[ij]}-\rho' \del_{ij}) \vphi_j
                      - 2F_{ij} \tau_j - 2 f\tau_i
                      + (2\rho_{(ij)}-\rho \del_{ij}) \vphi'_j \label{max3}\\
  \eth_{[i} F_{jk]} \label{max4}
         &=& \vphi_{[i} \rho'_{jk]} + \vphi'_{[i} \rho_{jk]}
\end{eqnarray}
A further three equations can be obtained by priming equations (\ref{max1}),(\ref{max2}) and (\ref{max3}).  We will often make use of the combination $ \del_{ij} (\ref{max1})- (\ref{max2}) $:
\begin{multline}\label{max1-2}
  \tho(F_{ij}+\del_{ij} f)
         = 2 \eth_{[i} \vphi_{j]} - \del_{ij}\eth_k \vphi_k - 2\tau'_{[i} \vphi_{j]} - 2f \rho_{[ij]}
             - 2F_{[i|k} \rho_{k|j]} - 2\kap_{[i}\vphi'_{j]} \\
           + \del_{ij}(\tau'_k \vphi_k + \rho_{kl} F_{kl} - \rho f - \kap_k \vphi'_k) 
\end{multline}
Now consider the combination $\tho(\ref{max3}) + \eth_j (\ref{max1-2})$.  This gives
\begin{eqnarray}
  0 &=& (2\tho' \tho + \eth_j \eth_j) \vphi_i + 2[\tho,\tho']\vphi_i - 
         [\tho,\eth_j] (F_{ij}+f\del_{ij}) + [\eth_i,\eth_j] \vphi_j \nn \\
    &&  + \tho \big(-(2\rho'_{[ij]}-\rho' \del_{ij}) \vphi_j
                      + 2(F_{ij}+f\del_{ij}) \tau_j
                      - (2\rho_{(ij)}-\rho \del_{ij}) \vphi'_j\big)\label{eqn:maxmaster}\\
    &&  + \eth_i \big(-\rho_{jk} F_{jk} + \rho f - \tau'_j \vphi_j 
                     + \kap_j \vphi'_j\big)+ \eth_j \big(2\tau'_{[i} \vphi_{j]} + 2f \rho_{[ij]}
        + 2F_{[i|k} \rho_{k|j]} + 2\kap_{[i}\vphi'_{j]}\big).\nn
\end{eqnarray}

This involves second derivatives of $\vphib$, as well as of the boost weight 0 quantities $F_{ij}$ and $f$. However, the latter occur in the form of a commutator $[\tho,\eth_j](F_{ij} + f\delta_{ij})$ and can therefore be eliminated using \eqref{C2}.  Now we consider first derivatives of Maxwell components other than $\vphib$.  We need to eliminate these from the equation if it is to decouple.

First consider terms involving $\tho$:
\begin{itemize}
  \item $\tho$ acts on $f$ and $F_{ij}$ through the combination $\tho(F_{ij} + f\delta_{ij})$, which we eliminate using equation \eqref{max1-2}.
  \item Terms involving $\tho \vphi'_i$ are eliminated using equation \eqref{max3}$'$.
  \item Terms in which $\tho$ acts on $\rhob$, $\taub$ and $\rhob'$ are eliminated using the Newman-Penrose equations (\ref{NP1}), (\ref{NP2}) and (\ref{NP4})$'$ respectively.
\end{itemize}

The resulting equation is very long: 
\begin{multline}\label{eqn:maxfull}
  \Big[(2 \tho \thop + \eth_{j} \eth_{j} + \rhop \tho + \rho \thop - 2\taup_{j} \eth_{j}
        - 2 \tau_{j}\eth_{j}) \vphi_{i}  \\
       +( - \rhop_{ij} \tho - 2   \tau_{i} \eth_{j} + \rhop_{ji} \tho
      - \rho_{ij} \thop + [\eth_{i}, \eth_{j}]  + 2 \tau_{j}\eth_{i} - \rho_{ji} \thop )\vphi_{j} \\
       - \kap_{i} \kappap_{j} \vphi_{j} 
       - 2 \vphi_{j} \rho_{k i} \rhop_{jk} + \vphi_{j} \rho_{k j} \rhop_{i k} 
       - 2 \vphi_{j} \tau_{j} \taup_{i} + 2 \vphi_{i} \tau_{j} \taup_{j} 
       - \vphi_{j} \rho_{i k} \rhop_{jk}  + \vphi_{j} \rho\rhop_{ji} \\
       + 2 \vphi_{j} \tau_{i} \taup_{j} 
       + \kap_{j} \kappap_{i} \vphi_{j}
       - \kap_{j} \kappap_{j} \vphi_{i} - \vphi_{i} \rho_{k j} \rhop_{jk} - 2\Phia_{ij} \vphi_{j} - \Phi \vphi_{i} 
       -  \tfrac{d-2}{d-1}\Lambda \vphi_{i}
\Big] \\
+ \Big[ \kap_{j} \thop (F_{ij}+f\del_{ij}) +\rho_{ji}  \eth_{j} f - \rho_{k i}  \eth_{j} F_{jk} 
        + 2  \rho_{ij} \eth_{j} f  + \rho_{k j}\eth_{j} F_{i k} - \rho_{jk} \eth_{i} F_{jk}
        +  \rho_{jk}\eth_{j} F_{i k}  \\
       - f \eth_{j} \rho_{ji}  - F_{jk} \eth_{j} \rho_{k i} + f \eth_{j} \rho_{ij}  + F_{ij} \eth_{k} \rho_{jk}  
        + f \eth_{i} \rho   - F_{jk} \eth_{i} \rho_{jk}  
        + 2 f \thop \kap_{i} + 2 F_{ij} \thop \kap_{j} \\
        - 5 f \rho_{ij} \tau_{j} - 2 f \Psi_{i} - 4 F_{ij} \rho_{jk} \tau_{k} - F_{ij} \Psi_{j}
         - F_{jk} \kap_{i} \rhop_{jk} + F_{jk} \kap_{j} \rhop_{i k} - F_{jk} \Psi_{j k i} 
        - F_{ij} \kap_{k} \rhop_{jk} \\
        + F_{ij} \kap_{j} \rhop + f \rho_{ji} \tau_{j} - 3 F_{jk} \rho_{ji} \tau_{k} 
        - f \rho \tau_{i} + 2 F_{jk} \rho_{jk} \tau_{i} - F_{jk} \rho_{ij} \tau_{k} + F_{ij} \rho \tau_{j}
\Big] \\
+ \Big[\kap_{j}  \eth_{i} \vphip_{j} + \kap_{i}\eth_{j} \vphip_{j}  - \kap_{j}  \eth_{j} \vphip_{i} 
       + 2 \rho_{k i} \rho_{jk} \vphip_{j}  + \kap_{j} \tau_{j} \vphip_{i} + \rho_{i k} \rho_{k j}\vphip_{j} \\
         + \vphip_{j} \rho_{i k} \rho_{jk} - \kap_{j} \tau_{i} \vphip_{j}- \kap_{i} \tau_{j} \vphip_{j}  
        - \rho_{ji} \rho\vphip_{j} - \rho_{jk} \rho_{k j} \vphip_{i} + 2 \Om_{ij} \vphip_{j} 
\Big]= 0
\end{multline}
The only terms above involving derivatives of Maxwell components other than $\vphib$ are of the (schematic) form $\kapb \tho' \Fb$, $\kapb\tho'f$ $\kapb \eth \vphib'$, $\rhob\eth \Fb$ and $\rhob \eth f$.  We need to eliminate all of these from our equations if we are to obtain a decoupled equation for $\vphib$.  Consider the first three, which are
\begin{equation}
  \kappa_j \thop \left( F_{i j} + f \delta_{i j} \right) + 2\kap_j \eth_{[i}\vphi'_{j]} + \kap_i \eth_j\vphi'_j
        = 2\kappa_j \thop \left( F_{i j} + f \delta_{i j} \right) + \dots
\end{equation}
where we have used \eqref{max1-2}$'$ to eliminate the $\kapb \eth \vphib'$ terms in favour of $\kapb \thop \Fb$, $\kapb \thop f$ and some other terms not involving derivatives.

Now, the Maxwell equations cannot be used to eliminate the terms of the form $\kapb \thop (\Fb+f)$ without re-introducing 1-derivative terms of the form $\kapb \eth \vphib'$.  Hence the only way in which the $\kapb \thop (\Fb+f)$ terms can be eliminated is if $\kapb=\Ob$, and therefore the vector field $\lb$ must be geodesic for decoupling to be possible.  We assume henceforth that this is the case.

Now examine the $\rhob \eth \Fb$ and $\rhob\eth f$ terms above. These are:
\begin{equation} \label{rhoethF}
  \rho_{ji} \eth_{j}f - \rho_{ki} \eth_{j}F_{jk} + 2 \rho_{ij} \eth_{j}f + \rho_{kj} \eth_{j}F_{ik} 
  - \rho_{jk} \eth_{i}F_{jk} + \rho_{jk} \eth_{j}F_{ik}
\end{equation}
To achieve decoupling, we need to eliminate these terms from the equation without introducing any 1-derivative terms (unless the derivative acts on $\vphib$). It is convenient to decompose $\eth_i F_{jk}$ into parts that transform irreducibly under $SO(d-2)$:
\begin{equation}
 \eth_i F_{jk} = {\cal F}_{ijk} + \frac{2}{d-3} \delta_{i[j} \eth_{|l} F_{l|k]},
 \end{equation}
where ${\cal F}_{ijk}$ is traceless and can be decomposed further into objects transforming irreducibly according to the Young tableaux {\tiny\yng(1,1,1)} and {\tiny\yng(2,1)}.  The quantity $\eth_i f$ transforms in the same way as $\eth_j F_{ji}$, i.e. as a vector ({\tiny\yng(1)}) under $SO(d-2)$.  The latter can be eliminated in favour of the former using equation \eqref{max3}, which gives $\eth_j F_{j i} = \eth_i f + \ldots$, where the ellipsis denotes terms in which derivatives act only on $\vphib$. The contribution of the `vector' terms to \eqref{rhoethF} is then
\begin{equation}
\label{ethFterms}
 \frac{2}{d-3} \left( \rho_{ji} + (d-3) \rho_{ij} - \rho \delta_{ij} \right) \eth_j f
\end{equation}
We can substitute our decomposition of $\eth F$ into the Maxwell equations. There are no Maxwell equations that can be used to eliminate $\eth_i f$ without reintroducing new derivative terms of the form $\rhob \tho \vphib'$. Hence the only way in which the Maxwell equation will decouple is if the expression in brackets in \eqref{ethFterms} vanishes. The symmetric and antisymmetric parts of the resulting equation give
\begin{equation}
\label{zeroshearrot}
 \sigma_{ij} = 0 = (d-4) \omega_{ij},
\end{equation}
where $\sigb$ and $\omb$ are the shear and rotation of $\ell$ respectively.  Hence a necessary condition for decoupling is that $\ell$ be shearfree and, for $d>4$, rotation free (and hence hypersurface orthogonal since $\ell$ is geodesic).  We now assume $d>4$, so we set $\sigb=\omb=\Ob$ henceforth, and therefore have
\begin{equation}
  \rho_{ij} = \frac{\rho}{d-2} \delta_{ij}.
\end{equation}
A spacetime admitting a null geodesic congruence with vanishing rotation and shear is called a Robinson-Trautman spacetime if $\rho \ne 0$ and a Kundt spacetime if $\rho =  0$.  We saw in Chapter \ref{chap:nongeo}that an Einstein spacetime of either of these types is algebraically special, with the vector field $\ell$ aligned with the congruence being a multiple WAND.  Therefore we can take $\Omb = \Ob = \Psib$.  Note that (\ref{NP3}) now implies $\eth_i \rho=0$.

It is now guaranteed that we can use equation \eqref{max3} to eliminate `vector' terms of the form $\eth_j F_{ij}$ or $\eth_i f$ from \eqref{eqn:maxfull}.  Upon doing so, we find that the terms involving ${\cal F}_{ijk}$ all drop out.  The commutators $[\tho,\thop]$ and $[\eth_i, \eth_j]$ can be used to tidy up the equation, giving
\begin{multline}\label{maxwellexpanding}
  0 =  \left[2 \thop\tho + \eth_{j} \eth_{j} + \rhop \tho + \tfrac{d+2}{d-2} \rho \thop - 4 \tau_j \eth_{j}
         \right]\vphi_{i} 
          + 2( \tau_j \eth_{i} - \tau_i \eth_{j})\vphi_{j} \\
    + \left[ 3 \Phi_{i j} - \Phi_{j i} - \frac{2\rho}{d-2} \rhop_{[ij]}+  \left( \Phi  + \frac{\rho \rhop}{d-2}  
        - \frac{2d-3}{d-1} \Lambda  \right) \delta_{ij} \right] \vphi_j \\ 
    + \frac{d-4}{d-2} \rho \left[\tau_j \left( F_{ij} - F \delta_{ij} \right)  + \frac{\rho}{d-2} \vphip_i\right] .
\end{multline}
The only term involving $\vphib'$ is the final one, so for $\vphib'$ to decouple we need $(d-4) \rho = 0$. This also ensures that the terms involving $F_{ij}$ and $f$ drop out of the equation. Hence decoupling requires $\rho=0$ (since $d>4$), which implies $\rho_{ij}=0$, so $\lb$ must be free of expansion as well as shear and rotation.  That is, the spacetime must be Kundt.  The equation reduces to
\begin{multline}
  \left[2 \thop\tho + \eth_{j} \eth_{j} + \rhop \tho - 4 \tau_j \eth_{j} + \Phi 
         - \tfrac{2d-3}{d-1} \La 
  \right]\vphi_{i} 
          + ( 2\tau_j \eth_{i} - 2\tau_i \eth_{j} + 3 \Phi_{i j} - \Phi_{j i})\vphi_{j} = 0.
\end{multline}
which is equivalent to \eqref{eqn:maxpert}.

To summarize, for $d>4$, $\vphib$ satisfies a second-order decoupled equation if, and only if, $\lb$ is geodesic with vanishing expansion, rotation and shear.  The existence of such a choice of $\lb$ implies, by definition, that the spacetime is Kundt.

Note the presence of factors of $(d-4)$ in several of our equations above. When $d=4$, it is not necessary for the rotation $\omb$ of $\ell$ to vanish in equation \eqref{zeroshearrot}, or for the expansion $\rho$ to vanish in equation (\ref{maxwellexpanding}). Indeed, in 4D, all that is required is that $\lb$ be geodesic and shearfree, which is equivalent (by the Goldberg-Sachs theorem) to the spacetime being algebraically special. 

It is clear that $\vphib'$ will satisfy a second-order decoupled equation (the prime of the above equation) if, and only if, $\nb$ is geodesic with vanishing expansion, rotation and shear. Hence $\vphib$ and $\vphib'$ will both satisfy second order decoupled equations if, and only if, $\kapb = \kapb' = \rhob = \rhob' = 0$.  A natural name for a spacetime admitting such null vector fields seems to be:
\begin{defn}\label{def:doubly}
  A spacetime is \emph{doubly Kundt} if and only if it admits a pair of non-expanding, non-shearing, non-twisting geodesic null vector fields $\lb$ and $\nb$ with $\lb.\nb \neq 0$.
\end{defn}

\subsection{The Schwarzchild Solution}
Consider the special case of the higher-dimensional Schwarzschild solution, which is not Kundt.  This solution has $\rhob = \frac{\rho}{d-2} \Id$ and $\taub=\Ob$ (a consequence of spherical symmetry).  The latter implies that the terms in $\Fb$ and $f$ drop out of equation \eqref{maxwellexpanding}, leaving us with an equation of the form
\begin{equation}
 ( \Box \vphi)_i +  \frac{(d-4)}{(d-2)^2} \rho^2 \vphi'_i = 0,
\end{equation}
where $\Box$ is a second order differential operator.  The second term remains an obstruction to decoupling.  For the Schwarzschild solution, the two multiple WANDs have identical properties so we can take the prime of the equation to obtain
\begin{equation}
 ( \Box' \vphi')_i +  \frac{(d-4)}{(d-2)^2} \rho'^2 \vphi_i = 0,
\end{equation}
and hence
\begin{equation}
  \left[ \Box' \left( \frac{1}{\rho^2} \Box \vphi \right) \right]_i - \frac{(d-4)^2}{(d-2)^4} \rho'^2 \vphi_i = 0.
\end{equation}
So in fact $\vphib$ does satisfy a decoupled equation but it is fourth order in derivatives. Note that we had to make use of several special properties of the Schwarzschild solution to obtain this result. It would be interesting to investigate more generally the circumstances under which one can obtain a decoupled equation of higher order for $\vphib$. 

\section{Decoupling of gravitational perturbations}\label{sec:gravdecoupling}

\subsection{Introduction and main result}

We now move on to gravitational perturbations.  In Lemma \ref{lem:Ominv} we found a set of gauge invariant quantities $\Omb^{(1)}$ under the assumption that $\lb$ was a multiple WAND of the background spacetime.  Hence, we shall consider gravitational perturbations of an algebraically special Einstein spacetime, for which we can take $\lb$ to be a multiple WAND.  Hence $\Omb$ and $\Psib$ vanish in the background, so we can treat them as first order quantities: $\Omb = \Omb^{(1)}$, $\Psib=\Psib^{(1)}$.  Therefore we shall not bother including a superscript ${}^{(1)}$ on $\Omb$ or $\Psib$ below.

The final result will be similar to that of the electromagnetic perturbations; we will find that we can only achieve decoupling when the spacetime is Kundt.  We will show that gravitational perturbations of such a Kundt spacetime are described by
\begin{multline}\label{eqn:gravperts}
  \left(2\tho'\tho+ \eth_k \eth_k + \rho'\tho - 6\tau_k\eth_k  + 4\Phi 
        - \tfrac{2d}{d-1} \La \right) \Om_{ij}\\
  + 4\left(\tau_k\eth_{(i|}- \tau_{(i|}\eth_k + \Phis_{(i|k} + 4\Phia_{(i|k}\right) \Om_{k|j)} 
  + 2\Phi_{ikjl} \Om_{kl} = 0,
\end{multline}
where all quantities except $\Omb$ are evaluated in the background geometry (e.g. $\Phi$ denotes $\Phi^{(0)}$ etc.).

In a doubly Kundt spacetime, $\Omb'$ also will satisfy a decoupled equation, which is given by taking the prime of the above equation.

\subsection{Derivation of main result}

We follow as closely as possible the 4D approach of Stewart \& Walker \cite{stewperts}.  Many of the equations in this section were checked using the computer algebra package Cadabra \cite{Cadabra1,Cadabra2}.  We start by obtaining an equation in which second derivatives act only on $\Omega_{ij}$.  Consider the equations
\begin{multline}\label{eqn:A}
  0   = - \eth_k \Om_{ij} - \del_{jk} \eth_l \Om_{il} + \eth_j \Om_{ik} -\tho (\Ps_i \del_{jk} + \Ps_{ijk}) 
          + \del_{jk}(\Phi_{li} - 2\Phi_{il} - \Phi \del_{il}) \kap_l  \\
        + (-2\Phi_{i[k|} \del_{j]l} + 2\del_{il} \Phia_{kj} + \Phi_{ilkj})\kap_l
          + \del_{jk} \left[ - \Psi_i \rho - \rho_{il} \Psi_l 
          - (\Psi_{mil}+\Psi_{iml})\rho_{lm} \right]\\
        + 2 (\Ps_{[k|} \del_{il} + \Ps_i\del_{[k|l} + \Ps_{i[k|l} + \Ps_{[k|il}) \rho_{l|j]}  
          + (\Om_{il} \tau'_l \del_{jk} - \Om_{ik} \tau'_{j} + \Om_{ij} \tau'_{k})
\end{multline}
and
\begin{multline}\label{eqn:B}
  0 = -2\tho' \Om_{ij} + \eth_k( \Psi_i \del_{jk} + \Psi_{ijk})
                    + \left( -\Om_{ij} \rho' + 2 \Om_{ik} \rho'_{[jk]} \right)
                    - 4(\Psi_{(i}\del_{j)k} + \Psi_{(ij)k})\tau_k \\
       + \Phi_{jk} \rho_{ik} - \Phi_{kj} \rho_{ik} + \Phi_{ik} \rho_{jk} - \Phi_{ki} \rho_{kj}
                   + 2 \Phi_{ik} \rho_{kj} - \Phi_{ij} \rho + \Phi_{ikjl} \rho_{kl} + \Phi \rho_{ij}.
\end{multline}
Equation \eqref{eqn:A} is obtained by taking various linear combinations and contractions of the Bianchi equations (\ref{B1}), while equation \eqref{eqn:B} is constructed from the symmetric part of (\ref{B2}) and a contraction of (\ref{B3}).  These equations are {\it exact}: no decomposition into background and perturbation has been performed at this stage.

Now we consider the linear combination $\eth_k \eqref{eqn:A}+\tho \eqref{eqn:B}$.  This contains second derivatives acting on $\Omb$ and on $\Psib$.  However, the point of taking this particular combination is that the second derivatives of $\Psib$ occur in the combination $- [\tho,\eth_k ] ( \Psi_{ijk} + \Psi_i \del_{jk} )$ and therefore can be eliminated in favour of terms involving one or zero derivatives of $\Psib$ using the formula (\ref{C2}) for the commutator $[\tho,\eth_k]$.

We can also symmetrize the entire equation on $ij$ without losing any useful information, as the antisymmetric terms do not contain any second derivatives of $\Omb$.  This reduces the equation to
\begin{eqnarray}
 0 &=& - (2\tho'\tho+ \eth_k\eth_k) \Om_{ij} - 2[\tho,\tho']\Om_{ij} - [\eth_{(i|},\eth_k]\Om_{k|j)}
       + \tho(\To_{ijkl} \rho'_{kl}) - \eth_l (\To_{ijkl}\tau'_k) \nn\\
   & & +[\tho, \eth_k]\Tp_{ijk} - 4\tho ( \Tp_{ijl}\tau_l ) 
       - 2\eth_{(i|} ( \Tp_{|j)lk}\rho_{kl}) + 2\eth_l(\Tp_{(j|lk} \rho_{k|i)}) - 2\eth_l(\Tp_{ijk}\rho_{kl})\nn\\
   & & +\tho(\Tf_{ikjl}\rho_{kl}) - \eth_l (\Tf_{ikjl}\kap_k) \label{eqn:star2}\label{master}
\end{eqnarray}
where
\begin{eqnarray}
  \Tf_{ikjl} & \equiv & \Phi_{(i|k|j)l} + \Phi \del_{(i|k}\del_{|j)l} - \Phis_{ij}\del_{kl} 
                        + (2\Phi_{(i|l}-\Phi_{l(i|})\del_{k|j)}
                        + (2\Phi_{(i|k}-\Phi_{k(i|})\del_{l|j)},\nn\\
  \Tp_{ijk}  & \equiv & \Psi_{(ij)k} + \Psi_{(i}\del_{j)k},\nn\\
  \To_{ijkl} & \equiv & -\Om_{ij}\del_{kl} + \Om_{(i|l}\del_{k|j)} - \Om_{(i|k} \del_{l|j)}.
\end{eqnarray}
Note that these quantities satisfy the following relations:
\begin{equation}\label{eqn:Tfid}
  \Tf_{ijkl} = \Tf_{(i|j|k)l} = \Tf_{i(j|k|l)}, \qquad
  \Tf_{ijil} = 0 \eqand 
  \Tf_{ijkj} = -(d-2)\Phis_{ik} + \Phi \del_{ik},
\end{equation}
\begin{equation}\label{eqn:Tpid}
  \Tp_{ijk} = \Tp_{(ij)k}, \quad \Tp_{iik} = 0 \eqand \Tp_{iji} = \half d\Psi_j
\end{equation}
\begin{equation}\label{eqn:Toid}
  \To_{ijkl} = \To_{(ij)kl},\quad 
  \To_{ij(kl)} = -\Om_{ij}\del_{kl} \quad 
  \To_{iikl} = 0 \eqand 
  \To_{ijkk} = -(d-2)\Om_{ij}.
\end{equation}
In this notation, the parts of \eqref{eqn:A} and \eqref{eqn:B} symmetric on $ij$ become
\begin{equation}\label{eqn:Asimp}
  \tho \Tp_{ijk} - \eth_l \To_{ijlk} = - \To_{ijlk}\tau'_l + 2\Tp_{(i|kl} \rho_{l|j)} 
                   - 2\Tp_{ijl}\rho_{lk} - 2\Tp_{l(i|m}\rho_{ml} \del_{k|j)} - \Tf_{ikjl}\kap_l
\end{equation}
and
\begin{equation}\label{eqn:Bsimp}
  -\eth_k \Tp_{ijk} + 2\tho'\Om_{ik} = \Tf_{ikjl}\rho_{jl} - 4\Tp_{ijk}\tau_k 
                                       + \To_{ijkl}\rho'_{kl}.
\end{equation}
Next we perform the following steps:
\begin{enumerate}
  \item Use the commutator (\ref{C2}) to eliminate the terms $[\tho, \eth_k]\Tp_{ijk}$ from \eqref{master} (note that this introduces a new kind of term, of the schematic form $\kapb \tho' \Psib$).
  \item Expand out the brackets using the Leibniz rule for GHP derivatives.
  \item Eliminate the term $\tho\Tp_{ijk}$ using equation \eqref{eqn:Asimp}.
  \item Use the NP equations (\ref{NP1}), (\ref{NP2}) and (\ref{NP4})$'$ to eliminate terms 
in which $\tho$ acts on $\rhob$, $\taub$ and $\rhob'$ respectively.
  \item Take a linear combination of the Bianchi equations (\ref{B2},\ref{B3},\ref{B4}) to get an equation
  \begin{eqnarray}\label{eqn:thophi}
    \tho \Tf_{ikjl} &=& \tho' \To_{ijkl} + \eth_{(i|}\Psi_{l|j)k} - \eth_l\Psi_{(ij)k} 
                        - \del_{(i|k}\del_{|j)l}\eth_m \Psi_m
                        + \del_{kl}\eth_{(i}\Psi_{j)} \nn\\ 
                    & & + (-2\eth_l\Psi_{(i|} + \eth_{(i|} \Psi_l)\del_{k|j)}
                        + (-2\eth_k\Psi_{(i|} + \eth_{(i|} \Psi_k)\del_{l|j)} + \dots,
  \end{eqnarray}
  where the ellipsis indicates terms that involve no derivatives.  Use this to eliminate $\tho \Tf_{ikjl}$ from \eqref{master}.
  \item Use a combination of (\ref{B5}) and (\ref{B7}) to show that
      \begin{equation}
        \eth_l \Tf_{ijkl} = 3\tho'\Tp_{ijk} + 3\Tf_{ikjl}\tau_l + \dots
      \end{equation}
where the ellipsis denotes first order terms not involving any derivatives.  Use this to eliminate $\eth_l \Tf_{ijkl}$ from \eqref{master}.
\end{enumerate}
The resulting equation is very long so we shall not write it out in full. It has the schematic form
\begin{multline}\label{eqn:schematic}
  \big(\tho'\tho + \eth . \eth + [\tho,\tho'] + [\eth,\eth] + \rhob'\tho + \rhob\tho' 
   + \taub\eth + \taub'\eth + \taub\taub' + \rhob\rhob' + \Phib \big) \Omb \\ 
      + \kapb \tho' \Psib + \rhob \eth \Psib  + (\taub\rhob + \taub'\rho + \kapb \rho' + \tho'\kapb 
      + \eth\rhob )\Psib \\
   + ( \taub \kapb + \taub'\kapb + \rhob^2)\Phib + ( \kapb \rhob )\Psib' =0
\end{multline}
Here, we neglect terms that are of quadratic order or higher when we decompose quantities into a background piece and a perturbation.  Recall that $\Omb$ and $\Psib$ are first order quantities.
Note that the only terms containing derivatives of Weyl components other than $\Omb$ are of the schematic form $\kapb \tho' \Psib$ and $\rhob \eth \Psib$.  For decoupling to occur, these must vanish for any possible perturbation.  We shall now examine the circumstances under which we can eliminate these terms.

The detailed form of the $\kapb \tho' \Psib$ terms is
\begin{equation}
 4 \kap_k \tho' ( \Psi_{(ij)k} + \Psi_{(i} \delta_{j) k})
\end{equation}
If $\kapb^{(0)} \ne \Ob$ then there is nothing we can do to eliminate these terms.  The only Bianchi equation containing $\tho'\Psib$ is \eqref{B5}, and using this again would reintroduce the 1-derivative terms that we have eliminated above.  Hence the only way for these terms to drop out is for $\kapb$ to vanish in the background.  Hence $\kapb^{(0)}=\Ob$ is a necessary condition for decoupling.  Henceforth we assume $\kapb$ is a first-order quantity, in which case the above terms become second order terms and can be neglected.

Recall that $\kapb^{(0)}=\Ob$ is equivalent to the statement that $\ell$ is geodesic in the background.  By Theorem \ref{thm:geo}, this places no further restrictions on the spacetimes that can be analysed.

Having set $\kapb^{(0)}=\Ob$, the only remaining terms involving derivatives of Weyl components other than $\Omb$ are of the form $\rhob \eth \Psib$.  The detailed form of these terms is:
\begin{multline} \label{eqn:rhoethPsi}
  4 \rho_{(i|k} \eth_k \Psi_{|j)} 
  + \rho_{kl} \left[ 2\eth_l \Psi_{(ij)k} + \eth_{(i|}\Psi_{l|j)k} - \eth_{(i}\Psi_{j)kl}
                     + 2\eth_k \Psi_{(ij)l} - \eth_{(i|} \Psi_{k|j)l} \right]\\
  + \rho_{k(i|} \left[ -\eth_l \Psi_{|j)lk} - \eth_l \Psi_{l|j)k} 
                       - \eth_{|j)} \Psi_k + 2\eth_k \Psi_{|j)} \right]
\end{multline}
For decoupling we need to eliminate these terms in favour of terms in which derivatives act only on $\Omb$.

Certain combinations of terms of the form $\eth \Psib$ can be eliminated using Bianchi equations.  In order to understand precisely what kinds of terms can be so eliminated, we can decompose $\eth_i \Psi_{jkl}$ into parts that transform irreducibly under $SO(d-2)$.  If we do the same for the Bianchi equations at our disposal (or combinations of them such as \eqref{eqn:B}) then we will see which irreducible parts of $\eth \Psib$ can be eliminated from the above equation.

Decomposing into tracefree and trace parts gives, for $d>4$:
\begin{equation}
 \eth_i \Psi_{jkl} = V_{ijkl} + 2 \delta_{i[k|} W_{j|l]} + \delta_{ij} X_{kl} 
                     + 2 \delta_{j[k|} Y_{i|l]} + 2 \delta_{i[k|} \delta_{j|l]} Z,
\end{equation}
where $V_{ijkl}$ is traceless and satisfies $V_{i[jkl]}=V_{ij(kl)}=0$. The other terms are given by
\begin{equation}
  W_{[ij]} = \frac{1}{2} X_{ij} 
           = \frac{1}{d(d-4)} \left( - (d-3) \eth_k \Psi_{[ij]k} + \eth_{[i} \Psi_{j]} \right), 
\end{equation}
\begin{equation}
  Y_{[ij]} =  \frac{1}{d(d-4)} \left( 3 \eth_k \Psi_{[ij]k} -(d-1) \eth_{[i} \Psi_{j]} \right),
\end{equation}
\begin{equation}
  W_{(ij)} = \frac{1}{(d-2)(d-4)} \left( -(d-3) \eth_k \Psi_{(ij)k} 
             + \eth_{(i} \Psi_{j)} - \eth_k \Psi_k \delta_{ij} \right),
\end{equation}
\begin{equation}
  Y_{(ij)} = \frac{1}{(d-2)(d-4)} \left( \eth_k \Psi_{(ij)k} - (d-3)  \eth_{(i} \Psi_{j)} 
                                         + \eth_k \Psi_k \delta_{ij} \right),
\end{equation}
\begin{equation}
 Z = \frac{1}{(d-2)(d-3)} \eth_k \Psi_k.
\end{equation}
Note that $W_{(ij)}$ and $Y_{(ij)}$ are traceless and $X_{(ij)}=0$.

The traceless part $V_{ijkl}$ can be decomposed further into parts that transform irreducibly under $SO(d-2)$.  The relevant irreducible representations correspond to Young tableaux with 4 boxes. However, it turns out that we will not need to discuss these. As well as these quantities, we have two independent quantities transforming as {\tiny$\yng(1,1)$}, namely $W_{[ij]}$ and $Y_{[ij]}$, two quantities transforming as {\tiny$\yng(2)$}, namely $W_{(ij)}$ and $Y_{(ij)}$, and a singlet $Z$.

Consider first the singlet $Z$. The contribution of this to equation \eqref{eqn:rhoethPsi} is 
\begin{equation}
 4 (d-3) \sigma_{ij} Z = \frac{4}{d-2} \sig_{ij}  \eth_k \Psi_k,
\end{equation}
where the shear $\sigb$ is the traceless symmetric part of $\rhob$.  In order to achieve decoupling, we would need to add to \eqref{eqn:rhoethPsi} a combination of Bianchi components containing a singlet term that cancelled this, and did not introduce any 1-derivative terms (e.g. $\tho \Phib$ terms) that we have already eliminated.  However, there is no such combination.  For example, the singlet drops out of equation \eqref{eqn:B}.  Therefore, the only way to eliminate the singlet term from our equation, as required for decoupling, is to set $\sigb^{(0)}=0$.  This is the condition that, in the background geometry, the shear of the multiple WAND $\lb$ must vanish.  Henceforth we assume that this is the case.

Next consider the traceless symmetric tensors $W_{(ij)}$ and $Y_{(ij)}$ that arise in the above decomposition of $\eth_i \Psi_{jkl}$.  The contribution of these to \eqref{eqn:rhoethPsi} is:
\begin{equation}\label{eqn:WY}
  - 5 \rho (W_{(ij)} + Y_{(ij)} ) 
  + \tfrac{1}{2}(d-10) \left(W_{(ik)} \om_{jk} + W_{(jk)} \om_{ik} \right)
  - \tfrac{3}{2}(d-2)\left( Y_{(ik)} \omega_{jk} + Y_{(jk)} \omega_{ik} \right),
\end{equation}
where $\om_{ij} \equiv \rho_{[ij]}$.  

Now consider again the Bianchi equations.  The only combination of equations involving $W_{(ij)}$ and $Y_{(ij)}$ that does not introduce any 1-derivative terms that we have already eliminated is \eqref{eqn:Bsimp}, which gives an expression for 
\begin{equation}
  \eth_k \Tp_{ijk} \equiv -(d-2)\left(W_{(ij)}+Y_{(ij)}\right).
\end{equation}
We can use this to eliminate, say, $Y_{(ij)}$ from \eqref{eqn:WY}, via the expression $Y_{(ij)} = - W_{(ij)} + \ldots$, where the ellipsis denotes terms in which derivatives act only on $\Omb$. Equation \eqref{eqn:WY} then reduces to
\begin{equation}
 2 ( d - 4) \left( W_{(ik)} \om_{jk} + W_{(jk)} \om_{ik} \right) + \ldots.
\end{equation}
Since we have no independent equation that will allow us to eliminate $W_{(ij)}$, we conclude that in order for the $\eth\Psib$ terms to decouple we must have $\omb=\Ob$ in the background, i.e., the multiple WAND $\lb$ must be shearfree \emph{and} rotation free.  Note the factor of $d-4$: for $d=4$, vanishing rotation is {\it not} necessary for decoupling.\footnote{For $d=4$, $\Psi_{ijk} = -2 \delta_{i[j} \Psi_{k]}$, so the irreducible parts of $\eth_{i} \Psi_{jkl}$ are just the trace, tracefree symmetric and antisymmetric parts of $\eth_i \Psi_j$. Considering the trace gives $\sigb=\Ob$ as for $d>4$. The tracefree symmetric part can be eliminated with \eqref{eqn:B}. The antisymmetric part simply drops out of \eqref{eqn:rhoethPsi}, using the fact that all $2 \times 2$ antisymmetric matrices commute.}

Having set $\sigb^{(0)} = \omb^{(0)} = \Ob$, we find that that the 1-derivative terms \eqref{eqn:rhoethPsi} reduce to
\begin{equation}
 \frac{5 \rho}{d-2} \left( \eth_k \Psi_{(ij)k} + \eth_{(i} \Psi_{j)} \right)
    =  \frac{5 \rho}{d-2} \eth_k \Tp_{ijk},
\end{equation}
where $\rho\equiv \rho_{ii}$.  These terms can be eliminated from \eqref{eqn:schematic} with equation \eqref{eqn:Bsimp}.

In the resulting equation, we now use (\ref{NP3}) to argue that $\eth_i \rho$ is a first order quantity. It appears only when multiplied by $\Psib$, so such terms are second order and can be dropped. The only Weyl components that are now acted on by derivatives are $\Omb$, and the equation has been reduced to the schematic form
\begin{equation}\label{eqn:schematic2}
  (\tho'\tho + \eth .\eth + [\tho,\tho'] + [\eth,\eth] + \rho'\tho + \rho\tho' + \taub\eth 
   + \taub'\eth + \rho\rho' + \taub\taub' + \Phib ) \Omb + \rho^2 \Phib + \taub\rho\Psib=0
\end{equation}
At this point, we can also simplify the form of the terms involving $\Omb$, by using the commutators (\ref{C1},\ref{C3}) to eliminate the terms of the form $[\tho,\tho']\Omb$ and $[\eth,\eth]\Omb$ respectively, in favour of terms that involve at most first derivatives of $\Omb$. 

The terms of the form $\Phib\rho$ are simplified by noting that equation \eqref{eqn:B}, evaluated in the background geometry implies that
\begin{equation}
\label{Phiback}
 \rho^{(0)} \Phi^{(0)}_{ij} = \frac{1}{d-2} \rho^{(0)} \Phi^{(0)} \delta_{ij}.
\end{equation}
Equation \eqref{eqn:schematic2} now reduces to something sufficiently simple to write out explicitly:
\begin{multline}\label{eqn:expstar}
  \left(2\tho'\tho+ \eth_k \eth_k + \rho'\tho + \tfrac{d+6}{d-2}\rho\tho' 
        + \tfrac{2}{d-2}\rho\rho' - 6\tau_k\eth_k + 4\Phi - \tfrac{2d}{d-1} \La \right) \Om_{ij}\\
    + \left(4\tau_k\eth_{(i|}- 4\tau_{(i|}\eth_k + \tfrac{2}{d-2}\rho(\rho'_{k(i|} - \rho'_{(i|k})
             +  4\Phis_{(i|k} + 16\Phia_{(i|k}\right) \Om_{k|j)} 
    + 2\Phi_{ikjl} \Om_{kl} \\
    + \frac{2\rho^2}{d-2}\left(\Phis_{ij} - \tfrac{1}{d-2}\Phi\del_{ij}\right)
    + 2\rho \tau_k  \left(\Psi_{(ij)k} - \Psi_{(i}\del_{j)k}
                                               +\tfrac{2}{d-2}\del_{ij}\Psi_k \right)=0. 
\end{multline}
This equation is the analogue of equation \eqref{maxwellexpanding} for the Maxwell field. Note that \eqref{Phiback} implies that $\rho (\Phis_{ij} - \tfrac{1}{d-2} \Phi \del_{ij} )$ is a first order quantity.

To achieve decoupling we have to eliminate the terms not involving $\Om_{ij}$, i.e., those on the final line of this equation.  For $d=4$, this is automatic since the particular combination of $\Phi$ terms appearing in this equation vanishes identically (i.e. $\Phis_{ij} = \frac{1}{2} \Phi \delta_{ij}$ if $d=4$), as does the particular combination of $\Psib$ terms.  For $d>4$, the only way of eliminating the $\Phib$ terms above is to set $\rho^{(0)}=0$, i.e., take $\rho$ to be first order.  All terms on the final line above are then of higher order and can be neglected.

Hence we see that, for $d>4$, decoupling requires that
\begin{equation}
 \kapb^{(0)} = \Ob = \rhob^{(0)},
\end{equation}
that is the multiple WAND must be geodesic and free of expansion, rotation and shear.  The existence of such a vector field implies, by definition, that the spacetime is Kundt.  This is a necessary condition for decoupling; it is also sufficient since we now have an equation in which the only perturbed Weyl components that appear are $\Omb$.

The resulting decoupled equation is:
\begin{multline}
    \left(2\tho'\tho+ \eth_k \eth_k + \rho'\tho - 6\tau_k\eth_k  + 4\Phi 
          - \tfrac{2d}{d-1} \La \right) \Om_{ij}\\
    + 4\left(\tau_k\eth_{(i|}- \tau_{(i|}\eth_k + \Phis_{(i|k} + 4\Phia_{(i|k}\right) \Om_{k|j)} 
    + 2\Phi_{ikjl} \Om_{kl} = 0.
\end{multline}
We remind the reader that $\Omb$ is a first order quantity, so quantities multiplying $\Omb$ (e.g.\ $\Phi$, $\taub$) must be evaluated in the background geometry.

\subsection{Comment on the expanding case}

Just as we did for Maxwell perturbations, it is interesting to consider what happens if $\lb$ is geodesic with vanishing rotation and shear, but non-vanishing expansion (i.e.\ the spacetime is Robinson-Trautman).  Under these circumstances, we have equation \eqref{eqn:expstar}, a perturbation equation for a gauge invariant quantity $\Omb$.  However, it contains two terms that obstruct the decoupling of the equation.  It is interesting to ask how these terms are consistent with gauge invariance.  The answer is supplied by:
\begin{lemma}\label{lem:otherinvs}
  Let $\lb$ be an expanding, non-twisting, non-shearing geodesic multiple WAND for an Einstein spacetime of dimension $d>4$.  Then
  \begin{equation}\label{eqn:gaugeinv1}
    \left(\Phis_{ij} - \tfrac{1}{d-2}\Phi\del_{ij}\right)^{(1)}
  \end{equation}
  is a gauge invariant quantity. If $\taub^{(0)} \neq \Ob$, then
  \begin{equation}\label{eqn:gaugeinv2}
    \Psi^{(1)}_i \eqand \tau^{(0)}_k \Psi^{(1)}_{ijk}
  \end{equation}
 also are gauge invariant quantities.
\end{lemma}
The Schwarzschild black hole in arbitrary dimension is an example of a spacetime admitting such a multiple WAND (although in this case, $\taub^{(0)}=\Ob$).  In four dimensions, \eqref{eqn:gaugeinv1} vanishes identically in all spacetimes, while the quantities \eqref{eqn:gaugeinv2} are not gauge invariant. 
\proof
From equation \eqref{Phiback} we have
\begin{equation} \label{Phiback2}
  \Phi_{ij}^{(0)} = \tfrac{1}{d-2}\Phi^{(0)}\del_{ij}
\end{equation}
in any such spacetime.  Hence we see immediately that \eqref{eqn:gaugeinv1} is invariant under infinitesimal coordinate transformations, and also under infinitesimal spins. Furthermore, Ref. \cite{RobTraut} showed that all such spacetimes are of algebraic Type D so we can choose our basis so that all Weyl tensor components with non-zero boost weight vanish.  Under an infinitesimal null rotation about $\lb$, equation \eqref{eqn:phinullrot} implies that, to first order in $z_i$,
\begin{equation}
   \Phis_{ij} \mapsto \Phis_{ij}+z_{(i}\Psi_{j)} - z_k\Psi_{(ij)k}, 
\end{equation}
but $\Psi$ is a first order quantity and hence $\Phi_{ij}^{\mathrm{S}(1)}$ and $\Phi^{(1)}$ are both invariant in a Type D background.  An identical argument applies to null rotations about $\nb$, and hence \eqref{eqn:gaugeinv1} is a gauge invariant quantity.

For an algebraically special spacetime, $\Psi_{ijk}$ and $\Psi_i$ both vanish in the background, and so, to first order, they are invariant under infinitesimal spins and infinitesimal coordinate transformations.  They are also invariant under infinitesimal null rotations about $\lb$, as these can only introduce terms involving $\Om_{ij}$ which also vanishes in the background.  We now consider the effect of an infinitesimal null rotation about $\nb$.  Taking the prime of (\ref{eqn:psinullrot}) implies that, to linear order,
\begin{equation}\label{eqn:psirot1}
  \Psi_{ijk}  \mapsto \Psi_{ijk} + \tfrac{2}{d-2} \Phi^{(0)} \delta_{i[j} z_{k]}  
                      + z_l \Phi_{lijk}^{(0)}
\end{equation}
and 
\begin{equation}\label{eqn:psirot2}
 \Psi_{j}  \mapsto \Psi_{j} - \tfrac{d-1}{d-2} \Phi^{(0)} z_j,
\end{equation}
where we have used \eqref{Phiback2}.  We will show that the quantities \eqref{eqn:gaugeinv2} are invariant under this transformation if $\tau_i^{(0)} \ne 0$.

Take a double trace of the Bianchi equation (\ref{B7}) for the background spacetimes.  This implies that $(d-4)\eth_k\Phi^{(0)} = 0$, and hence, for $d>4$, $\eth_k\Phi^{(0)} = 0$.  The trace of (\ref{B5}) gives
\begin{equation}\label{eqn:ethphi}
  \eth_j\Phi^{(0)} = \tfrac{d-1}{d-3} \tau_j^{(0)} \Phi^{(0)} ,
\end{equation}
and hence $\Phi^{(0)}=0$ if $\tau_i^{(0)} \ne 0$.  From \eqref{Phiback2} we then have $\Phi_{ij}^{(0)} = 0$. Putting these results back into (B5) implies that $\Phi_{ijkl}^{(0)}\tau_l^{(0)} = 0$.
Inserting these results into (\ref{eqn:psirot1},\ref{eqn:psirot2}) implies that, although $\Psi_{ijk}$ is not invariant under infinitesimal null rotations about $\nb$, both $\tau_k \Psi_{ijk}$ and $\Psi_i$ are invariant, and hence both of these are new gauge invariant quantities, provided that $d>4$ and $\tau_i^{(0)} \neq 0$.\, $\Box$

\section{Discussion}\label{sec:decouplingdiscussion}

To summarize, we have shown that, for linearized gravitational perturbations of an algebraically special spacetime, there exist local quantities $\Omb^{(1)}$, linear in the perturbation, that are invariant under infinitesimal coordinate transformations and infinitesimal changes of basis.  For perturbations of a Type D background, e.g.\ a Myers-Perry black hole, both $\Omb^{(1)}$ and $\Omb'^{(1)}$ are gauge invariant.  Irrespective of decoupling, the locality and gauge invariance of these quantities should make them useful in studies of gravitational perturbations.

Furthermore, $\Omb^{(1)}$ satisfies a decoupled equation of motion in $d>4$ dimensions if, and only if, the background is Kundt.  Therefore the decoupling property which is satisfied in the Kerr spacetime does not extend to the Myers-Perry spacetimes in an obvious way.

When decoupling does occur, an important question is whether a solution of the decoupled equation uniquely characterizes the gravitational perturbation.  If one has two solutions with the same $\Omb^{(1)}$ then do they describe the same metric perturbation?  This is equivalent to the question of whether there exist non-trivial linearized gravitational perturbations with $\Omb^{(1)}=\Ob$.

In four dimensions, for perturbations of a Kerr black hole, this problem was addressed by Wald \cite{waldtypeDpert}.  He showed that a well-behaved solution with $\Psi^{(1)}_0=0$ must also have $\Psi^{(1)}_4=0$.  A null rotation about $n$ can then be used to set $\Psi^{(1)}_1=0$ and a null rotation about $\ell$ can be used to set $\Psi^{(1)}_3=0$.  It follows that the perturbation must preserve the Type D condition to linear order.  Since Type D solutions are specified by just a few constants \cite{Kinnersley}, it is natural to expect that there will be only a finite number of solutions satisfying these conditions.  Wald showed that the only well-behaved solutions correspond simply to perturbations in the mass or angular momentum of the Kerr solution; i.e.\ the end-point of such perturbations is another member of the Kerr family.

For $d>4$, even if one can show that $\Omb^{(1)} = \Ob$ implies that $\Omb'^{(1)}=\Ob$ then it is no longer true that one can use null rotations to set $\Psib^{(1)} = \Psib'^{(1)} = \Ob$.  This is because a null rotation about $n$ contains fewer parameters than the number of independent components of $\Psib^{(1)}$ (whereas for $d=4$ both have two degrees of freedom).  So for $d>4$ it seems likely that the perturbations overlooked by our decoupled equation are more general than perturbations preserving the Type D condition.  Nevertheless, since $\Omb^{(1)}$ has the same number of degrees of freedom as the gravitational field, it seems reasonable to expect that our decoupled equation of motion captures `nearly all' of the information about linearized metric perturbations.

Sometimes it might not be enough to know the solution for $\Omb^{(1)}$, one might need to know the metric perturbation explicitly.  Wald \cite{Wald:1978vm} gives a systematic procedure for constructing solutions of the linearized Einstein equation (in a certain gauge), given the existence of a decoupled equation of motion for a quantity linear in the metric perturbation.  It seems likely that this procedure can be applied in the present case to generate solutions of the higher-dimensional linearized Einstein equation whenever $\Omb^{(1)}$ satisfies a decoupled equation.

However, this will not be necessary in the next chapter, where we will move on to discuss an application of our decoupled equation.  Recall from Section \ref{sec:nhintro} that all extreme vacuum black hole solutions have near-horizon geometries, and that these near horizon geometries are Kundt spacetimes.  Hence, we can apply our formalism to study their perturbations, and will discuss in detail how to do this for a large class of spacetimes (including the near-horizon geometries of all known extreme vacuum black holes).

\chapter[Perturbations of near-horizon geometries]{Perturbations of near-horizon geometries and instabilities of Myers-Perry black holes}\label{chap:nhperts}
\section{Introduction}

In Chapter \ref{chap:decoupling}, we developed a new approach to studying perturbations of Kundt spacetimes, and observed that this could be applied to study perturbations of the near-horizon geometries of extreme vacuum black holes.  The purpose of this chapter is twofold.  Firstly, we study in detail perturbations of vacuum near-horizon geometries, using the decoupled equations (\ref{eqn:maxpert},\ref{eqn:gravperts}) for electromagnetic and gravitational perturbations.

Secondly, we ask whether one can learn anything about stability of an extreme black hole solution from a study of perturbations of its near-horizon geometry?  Clearly, we will not be able to deduce that the full black hole is stable just by looking at its near-horizon geometry.  So, a more precise question is: \emph{if the near-horizon geometry is unstable then does this imply that the full black hole is also unstable?}

If true, this would give a fairly simple way of predicting instabilities of extreme black holes because perturbations of a near-horizon geometry can be studied using the decoupled equations of Chapter \ref{chap:decoupling}.  Furthermore, if an extreme black hole is unstable then it seems likely that near-extreme elements of the same family of black holes will also be unstable.

We should clarify what we mean by an instability of a near-horizon geometry.  As we observed in the introduction, the near horizon geometries of all known extreme vacuum black holes take the form of a compact space $\Hcal$ fibred over $AdS_2$.  We will show that the decoupled equations describing scalar field, electromagnetic and gravitational perturbations can be separated, and hence reduced to a an equation for a massive, charged, scalar field in $AdS_2$ with a homogeneous electric field, and a mass determined by the eigenvalues of a self-adjoint operator on $\Hcal$.  We argue that one can define an `effective Breitenl\"ohner-Freedman bound' \cite{BF}.  We shall say that the near-horizon geometry is unstable if there is some mode that violates this BF bound.  This reduction can be regarded as a Kaluza-Klein compactification with internal space $\Hcal$; it is non-trivial because the `KK gauge fields' arising from rotation of the BH are non-vanishing in the background spacetime.

Some motivation for believing that an instability of a near-horizon geometry implies an instability of the full black hole comes from studies of charged scalar fields in the background of an extreme Reissner-Nordstr\"om-AdS black hole.  Numerical results \cite{Hartnoll:2008kx} suggest that the scalar field becomes unstable in the black hole geometry when the near-horizon $AdS_2$ BF bound is violated.  In fact instability can occur even for an uncharged scalar field.  In this case, we shall present a proof (in Section \ref{sec:nhinstab}) that instability of the near-horizon geometry does imply instability of the full black hole.

Returning to gravitational perturbations, are any useful results known already?  Consider the four-dimensional near-horizon extreme Kerr (NHEK) geometry \cite{Bardeen:1999px}.  It has $\Hcal=S^2$ (with an inhomogeneous `squashed' metric).  Linearized gravitational perturbations of NHEK were studied in Refs.\ \cite{Amsel:2009ev,Dias:2009ex}.  After KK reduction to $AdS_2$, they found that certain non-axisymmetric modes violate the effective BF bound.  In this sense, the NHEK geometry is unstable against linearized gravitational perturbations.  But, as discussed in the introduction, the full Kerr solution is believed to be stable to such perturbations. 

Naively, this seems to suggest that perhaps instability of the near-horizon geometry does not imply instability of the full black hole.  However, we believe that there is a connection.  We shall argue that instability of the near-horizon geometry does imply instability of the full black hole, but only \emph{if the unstable mode respects certain symmetries}.  In the Kerr example, the symmetry in question is axisymmetry.  Axisymmetric perturbations of NHEK do respect the BF bound \cite{Dias:2009ex}, and hence the stability of such modes is consistent with the stability of the full black hole.

Before attempting to understand why an instability of the near-horizon geometry implies an instability of the full black hole when certain symmetries are respected, we will start by gathering some more data.  We will consider the most symmetric rotating black hole solutions: Myers-Perry (MP) black holes \cite{mp} in an odd number of dimensions, with equal angular momenta.  Recall that such black holes are cohomogeneity-1, i.e.\ they depend non-trivially on only a single coordinate.  The Killing field tangent to the horizon generators has the form $k + \Omega_H \zeta$ where $k$ is the generator of asymptotic time translations, $\zeta$ is an angular Killing field with closed orbits, and $\Omega_H$ is the angular velocity of the black hole.

In the extremal limit, such a black hole has a near-horizon geometry for which $\Hcal=S^{d-2}$ (with a homogeneous metric).  After KK reduction to $AdS_2$, we find that there exist modes that violate the effective BF bound, but most of these violate the symmetry generated by $\zeta$.  These are the analogue of the non-axisymmetric modes in NHEK.  What about modes that preserve the symmetry generated by $\zeta$?  For $d=5$, we find that such modes always respect the BF bound, just as for NHEK.  However, for $d \ge 7$, we find that some of these modes violate the BF bound.

How does this compare with the stability properties of the full extreme black hole solution?  The only known results are for gravitational perturbations of \emph{non-extreme} cohomogeneity-1 MP solutions \cite{Kunduri:2006,Murata:2008,Dias:2010eu}.  However, it is natural to expect that a reliable guide to the stability of an extreme black hole should be the stability of black holes that are very close to extremality. For modes that are invariant under the symmetry generated by $\zeta$, it turns out that, in the cases for which data exists, for any mode that is unstable in the near-horizon geometry, there is a corresponding unstable mode of the full black hole solution close to extremality.  This leads us to predict that \emph{all} cohomogeneity-1 MP black holes with $d \ge 7$ are unstable sufficiently close to extremality. 

Are these isolated examples of a more general result?  If so, under what circumstances does an instability of the near-horizon geometry imply an instability of the full black hole?  In the cases discussed above, the relevant instabilities of the near-horizon geometries are those preserving particular rotational symmetries.  To investigate this in more generality, consider a stationary black hole with $n$ commuting angular Killing fields $\partial/\partial \phi^I$ and a metric of the form
\begin{equation} \label{eqn:ADM}
 ds^2 = - N(x)^2 dt^2 + g_{IJ}(x)\left(d\phi^I + N^I(x) dt\right)\left(d\phi^J + N^J(x) dt\right) 
        + g_{AB}(x) dx^A dx^B
\end{equation}
where $1 \le I,J \le n$, $\phi^I \sim \phi^I + 2\pi$, and the metric depends only on the coordinates $x^A$.  In any number of dimensions, Theorem \ref{thm:rigidity} guarantees the existence of at least one rotational Killing vector for any stationary black hole solution, so we know that $n\geq 1$.  All \emph{known} exact black hole solutions in $d>4$ dimensions (e.g.\ Myers-Perry black holes, black rings) have more symmetry than this; they have multiple rotational symmetries.

Our conjecture for the circumstances under which a near-horizon geometry instability implies an instability of the full black hole is the following:
\begin{conjecture}\label{conj:perts}
  Consider linearized gravitational perturbations of the near-horizon geometry of an extreme vacuum black hole with metric \eqref{eqn:ADM}.  These can be Fourier decomposed into modes with $\phi^I$ dependence $e^{im_I \phi^I}$.  A sufficient condition for instability of the full black hole geometry is that the near-horizon geometry is unstable against some perturbation mode satisfying
  \begin{equation}\label{eqn:axicondition}
    m_I N^I(x) = 0
  \end{equation}
\end{conjecture}
For most MP black holes, or doubly-spinning black rings, the functions $N^I(x)$ are linearly independent, and hence this condition implies $m_I = 0$ for all $I$.  However, for MP solutions with enhanced symmetry this condition is less restrictive, e.g.\ in the cohomogeneity-1 case it implies only that $\Sigma_I m_I = 0$, which is equivalent to the perturbation being invariant under the symmetry generated by $\zeta$.

The purpose of this chapter is to build evidence in favour of this conjecture.  In the first part of the chapter we shall proceed phenomenologically.  In Section \ref{sec:decoupling}, we will show how to use the formalism developed in Chapter \ref{chap:decoupling} to study scalar field, electromagnetic and gravitational perturbations of the near-horizon geometry of an extreme black hole.

We then apply Conjecture \ref{conj:perts} to the case of cohomogeneity-1 Myers-Perry-AdS black holes, i.e.\ odd dimensional Myers-Perry black hole solutions with all angular momenta equal.  If we find an instability satisfying (\ref{eqn:axicondition}) then we shall predict an instability of the full extreme black hole solution.  If the extreme black hole is unstable then it seems natural to expect that near-extremal black holes will also be unstable.  Hence, in some cases, this prediction can be tested by comparing to the results of Dias \etal \cite{Dias:2010eu}, where it was found that certain cohomogeneity-1 Myers-Perry solutions become unstable near extremality, with the instability respecting (\ref{eqn:axicondition}).  We will find that our predictions are in good agreement with the results of \cite{Dias:2010eu}, which gives us confidence to make further predictions concerning the stability of extreme cohomogeneity-1 Myers-Perry black holes \cite{mp}, including asymptotically AdS solutions \cite{Hawking:1998kw,Gibbons:2004js}.

Finally, in Section \ref{sec:kerrcft}, we discuss briefly whether our results have any relevance to the conjectured Kerr-CFT correspondence \cite{kerrcft}.  It has been suggested that this extends to extreme Myers-Perry black holes \cite{popekerrcft}. Consider the asymptotic behaviour of perturbations in $d=5$ (where we do not predict an instability).  Following standard AdS/CFT rules we can determine operator dimensions in the dual CFT using our results for gravitational perturbations of the near-horizon geometry.  We find that all operators dual to perturbations respecting (\ref{eqn:axicondition}) have integer conformal dimensions.  This seems somewhat surprising, and perhaps hints at the existence of some symmetry protecting these operator dimensions. 

In Section \ref{sec:nhinstab} we return to more general questions, and explain how the precise form of the conjecture was arrived at.  Further motivation will come from considering the toy model of a scalar field.  We argue that an instability of the scalar field in the near-horizon geometry implies an instability in the full black hole spacetime if the condition \eqref{eqn:axicondition} holds.  For gravitational perturbations, we do not have a complete argument but the results discussed above, and further evidence that we shall discuss, suggests that an argument similar to the scalar field case should also apply.

\section{Decoupling and near-horizon geometries}\label{sec:decoupling}
\subsection{Near-horizon geometries} \label{sec:nhgeom}

Recall from Chapter \ref{chap:decoupling} that decoupling of $\vphib$ and $\Omb$ occurs only for Kundt spacetimes.  Although such spacetimes do not describe black holes, some Kundt spacetimes are closely related to black holes, since any extremal black hole admits a near-horizon geometry and any near-horizon geometry is a Kundt spacetime.

Consider an extreme black hole, i.e.\ one with a degenerate Killing horizon.  In the introduction, we noted that the near horizon geometries of all \emph{known} extreme vacuum black holes take the form of a fibration of some manifold $\Hcal$ over $AdS_2$.  More explicitly, they an be written as \cite{Bardeen:1999px,Kunduri:2007vf,Figueras:2008qh,Kunduri:2008rs,Chow:2008dp}
\begin{equation}\label{nhgeom}
 ds^2 = L(y)^2 \left( - R^2 dT^2 + \frac{dR^2}{R^2} \right) 
        + g_{IJ}(y) \left( d\phi^I - k^I R dT \right) \left( d\phi^J - k^J R dT \right) 
        + g_{AB}(y) dy^A dy^B.
\end{equation}
where $\pd/\pd \phi^I$, $I=1, \ldots, n$ are the rotational Killing vector fields of the black hole and $k^I$ are constants. The metric in the first set of round brackets is the metric of $AdS_2$ (in Poincar\'e coordinates). The coordinates $\phi^I$ have period $2\pi$.  The metric depends non-trivially only on the $d-n-2$ coordinates $y^A$.

A calculation (see Appendix \ref{app:nhframe}) reveals that the vector fields $\lb$ and $\nb$ dual to $-dT \pm dR/R^2$ are tangent to affinely parametrized null geodesics with vanishing expansion, rotation and shear, and hence these spacetimes are \emph{doubly Kundt} spacetimes, in the sense of Definition \ref{def:doubly}.  Such a spacetime is of algebraic Type D.  If we consider perturbing such a spacetime then the perturbations in both $\Omb$ and $\Omb'$ are gauge invariant and satisfy the decoupled equation (\ref{eqn:gravperts}).

\subsection{Decomposition of perturbations}\label{sec:decomposition}

The metric \eqref{nhgeom} takes a Kaluza-Klein form.  There is an `internal' compact space $\Hcal$, parametrized by $(\phi^I,y^A)$, corresponding to a spatial cross-section of the black hole horizon.  More precisely, $\Hcal$ denotes a surface of constant $T$ and $R$ in \eqref{nhgeom}, with geometry
\begin{equation}
  d\hat{s}^2 = g_{IJ}(y) d\phi^I d\phi^J + g_{AB}(y) dy^A dy^B .
\end{equation}
Additionally, there is a non-compact $AdS_2$ space parametrized by the Poincar\'e type coordinates $T$ and $R$.  Mixing between these two spaces is described by the terms $-k^I R dT$, which can be thought of as `Kaluza-Klein gauge fields' associated to a $U(1)^n$ gauge group.  These preserve the symmetries of $AdS_2$ because the associated field strengths $k^I dT \wedge dR$ are proportional to the volume form of $AdS_2$ (they describe homogeneous electric fields).

Our strategy will be to decompose perturbations as scalar fields in $AdS_2$, with the effective mass of these scalar fields given by eigenvalues of some operator on $\Hcal$.  This is more complicated than a standard (linearized) Kaluza-Klein reduction because the `KK gauge fields' are non-vanishing in the background geometry.  Fields with non-vanishing $\phi^I$ dependence will be charged with respect to the $AdS_2$ gauge fields.  We give more details of this decomposition below.

\subsubsection{Scalar fields}
It is instructive to consider first the example of a complex scalar field $\Psi(T,R,\phi^I,y^A)$ satisfying the Klein-Gordon equation\footnote{Alternatively, we could have started with the GHP version \eqref{eqn:scalarperts} of this equation, but in this case it does not make things any simpler.}
\begin{equation}\label{eqn:kg}
 \left( \nabla^2 - M^2  \right) \Psi= 0.
\end{equation}
We start with a separable ansatz
\begin{equation}
  \Psi(T,R,\phi,y) = \chi_0(T,R) Y(\phi,y)
\end{equation}
and Fourier decompose $Y$ along the periodic directions $\phi^I$:
\begin{equation}
  Y(\phi,y) = e^{i m_I \phi^I} \Ybb(y)
\end{equation}
The Klein-Gordon equation \eqref{eqn:kg} separates, and we see that the function $\chi_0(T,R)$ satisfies the equation of a massive charged scalar field in $AdS_2$ with a homogeneous electric field.  More precisely, we write the $AdS_2$ metric and gauge field $A_2$ as
\begin{equation} \label{ads2}
  ds^2 = -R^2 dT^2 + \frac{dR^2}{R^2}, \qquad A_2 = -R \, dT,
\end{equation}
and introduce a gauge-covariant derivative for a scalar with charge $q$:
\begin{equation}\label{eqn:ads2deriv}
  D \equiv \nabla_2 - i q A_2,
\end{equation}
where $\nabla_2$ is the Levi-Civita associated to the $AdS_2$ metric.  The scalar $\chi_0$ satisfies the equation of an $AdS_2$ scalar with charge $q$ and squared mass $\mu^2 = \la + q^2$:
\begin{equation}
 \left( D^2 - \la - q^2\right) \chi_0(T,R) = 0
\end{equation}
where the charge $q$ is given by\footnote{We are considering $AdS_2$ with a single gauge field $A= -R dT$. We could consider $AdS_2$ with multiple gauge fields, as is natural from the KK perspective, $A^I = -k^I R dT$. We would then obtain an $AdS_2$ scalar with charge $m_I$ with respect to $A^I$.  However, for fields of higher spin, it turns out to be more useful to consider a single gauge field.  The motivation for taking the separation constant to be $\la = \mu^2 - q^2$ rather than $\mu^2$ itself will also become apparent when we consider higher spin fields.}
\begin{equation}
 q = m_I k^I.
\end{equation}
The separation constant $\la$ is given by the eigenvalue equation
\begin{equation}\label{eqn:Oscalar}
 \Ocal{0} Y  \equiv  - \nablah_\mu\big( L(y)^2 \nablah^\mu Y\big) +L(y)^2  (M^2-q^2) Y= \la Y,
\end{equation}
where $\nablah$ is the Levi-Civita connection on $\Hcal$ and $\mu,\nu$ denote indices on $\Hcal$, raised and lowered with the metric on $\Hcal$.

The operator $\Ocal{0}$ is self-adjoint with respect to the inner product
\begin{equation}
 (Y_1,Y_2) = \int_\Hcal \bar{Y}_1 Y_2\,\, \mathrm{d(vol)}
\end{equation}
defined on the compact manifold $\Hcal$.  This self-adjointness guarantees that $\la$ is real, and furthermore that the harmonics $Y$ form a complete set and hence any solution $\Psi$ can be expanded as a sum of separable solutions of the above form.  Note also that $\Ocal{0}$ commutes with the Lie derivative with respect to $\pd/\pd{\phi^I}$ and hence eigenfunctions of $\Ocal{0}$ may be assumed to have the $\phi^I$ dependence assumed above.

\subsubsection{Gravitational perturbations}
The same procedure works for the linearized gravitational field.  As things are more complicated here, we give the full details in Appendix \ref{app:nhframe} and merely summarize the argument here.  We are looking to separate the decoupled equation \eqref{eqn:gravperts}, and start with a separable ansatz
\begin{equation}
 \Omega_{ij} = {\rm Re} \left[ \chi_2(T,R) Y_{ij} (\phi,y) \right].
\end{equation}
Since we are choosing our null basis vectors $\lb$ and $n$ to be tangent to the null geodesic congruences with vanishing expansion, rotation and shear, i.e., to $-RdT \pm dR/R$, it follows that the spatial basis vectors $m_i$ span $\Hcal$.  Therefore, we can regard $Y_{ij}$ as the components of a symmetric traceless tensor $Y_{\mu\nu}$ on $\Hcal$.  For the remainder of this chapter $\mu,\nu,\ldots$ will represent indices on $\Hcal$, with indices raised and lowered with $\gh$.  We take a Fourier decomposition of this tensor as above, that is we assume that
\begin{equation}
 \Lcal_I Y_{\mu\nu} = i m_I Y_{\mu\nu},
\end{equation}
where $\Lcal_I$ is the Lie derivative with respect to $\pd/\pd \phi^I$.  We can again perform a separation of the perturbation equation for $\Omb$, and show that it reduces to the equation of a massive charged scalar in $AdS_2$, satisfying
\begin{equation}
  \left( D^2 - q^2- \lambda \right) \chi_2 = 0.
\end{equation}
Here the charge is given by
\begin{equation}
 q = m_I k^I + 2i
\end{equation}
and the separation constant $\la$ by the eigenvalue equation
\begin{equation}
 (\Ocal{2} Y)_{\mu\nu} = \la Y_{\mu\nu}
\end{equation}
for an operator
\begin{multline}
  (\Ocal{2} Y)_{\mu\nu}
   =  -\frac{1}{L^4}\nablah^\rho \left(L^6 \nablah_\rho Y_{\mu\nu}\right)
      + \left(6-(k^I m_I)^2 - \tfrac{4}{L^2} k_\mu k^\mu - 2(d-4)\La L^2 \right)Y_{\mu\nu} \\
      + 2L^2\left(\hat{R}_{(\mu|\rho} 
             + \hat{R}\hat{g}_{(\mu|\rho}\right)Y^\rho_{\;\;\;|\nu)}
      - 2L^2\hat{R}_{\mu\phantom{\rho}\nu}^{\phantom{\mu}\rho\phantom{\nu}\sig}
                                          Y_{\rho\sig} \\
   + \Big[ - (dk)_{(\mu|\rho} - \tfrac{2}{L^2}\left(d(L^2)\wedge k\right)_{(\mu|\rho}\\
            + 2 \left( k-d(L^2)\right)_{(\mu|}\nablah_\rho 
            - 2 \left( k-d(L^2)\right)_{\rho} \nablah_{(\mu|}
            \Big]
                        Y^\rho_{\;\;\;|\nu)}.
 \label{eqn:Ograv}
\end{multline}
In this expression, $\hat{R}_{\mu\nu\rho\sigma}$ is the Riemann tensor on $\Hcal$ (with $\hat{R}_{\mu\nu}$ and $\hat{R}$ the Ricci tensor and Ricci scalar), indices are raised and lowered with the metric on $\Hcal$, $k$ is the Killing vector field on ${\cal H}$ defined by
\begin{equation}
 k = k^I \frac{\partial}{\partial \phi^I}
\end{equation}
and $(dk)_{\mu\nu} = 2 \hat{\nabla}_{[\mu} k_{\nu]}$.  We have written $\Ocal{2}$ in a covariant way, so that it can be evaluated using any basis on $\Hcal$, not limited to the particular one that we used above.  The explicit $m_I$ dependence enters only via $k^I m_I$, which can be determined from
\begin{equation}
\label{kdotm}
 \Lcal_k Y_{\mu\nu} = i k^I m_I Y_{\mu\nu}
\end{equation}

As in the scalar case, we can show that the separation constant $\la$ is real by showing that $\Ocal{2}$ is self-adjoint.  To do this, we define an inner product between traceless, symmetric, square integrable 2-tensors on $\Hcal$ by
\begin{equation}\label{eqn:gravinner}
  (Y_1,Y_2) \equiv \int_\Hcal L^4 \bar{Y}_1^{\mu\nu} Y_{2\mu\nu} \; \mathrm{d(vol)},
\end{equation}
and find that it can be shown, by integrating by parts, that $\Ocal{2}$ is self-adjoint with respect to this, which implies that its eigenvalues $\la$ are real.

The function $\chi_2(T,R)$ satisfies the equation of a charged scalar in $AdS_2$ where the mass $\mu$ is given by
\begin{equation}
 \mu^2 = q^2 + \la
\end{equation}
Note that $q$ is {\it complex}.  This has been observed previously for gravitational perturbations of the NHEK geometry \cite{Amsel:2009ev,Dias:2009ex}. Self-adjointness implies that $\la$ is real and hence $\mu^2$ also is complex but the combination $\mu^2 - q^2$ is always real.

Note that the use of the gauge-invariant quantity $\Omb$ to describe metric perturbations implies that we will not be able to study certain non-generic perturbations that preserve the algebraically special property of the background geometry and hence have $\Omb$.  In particular, perturbations that deform the near-horizon geometry into another near-horizon geometry will be missed.

\subsubsection{Electromagnetic Perturbations}

Finally, we can also analyse the behaviour of Maxwell fields in a similar manner. In a Kundt background, these satisfy a decoupled equation in terms of $\vphib$.  Similarly to previous cases, we write
\begin{equation}
  \vphi_i (T,R,\phi^I,y^A) = \mathrm{Re} \big[ \chi_1 (T,R) Y_i(\phi^I,y^A) \big]
\end{equation}
The decoupled equation for $\vphi_i$ can be separated to give the equation of a charged scalar in $AdS_2$:
\begin{equation}
  (D^2-\la-q^2)\chi_1 = 0
\end{equation}
where the charge is
\begin{equation}
 q =  k^I m_I + i,
\end{equation}
the mass $\mu$ is given by $\mu^2 = q^2 + \la$, and $\la$ is given by
\begin{equation}
  (\Ocal{1} Y)_\mu = \la Y_\mu 
\end{equation}
where
\begin{multline}
  (\Ocal{1} Y)_\mu 
   =  -\frac{1}{L^2}\nablah^\rho \left(L^4 \nablah_\rho Y_\mu \right)
      + \left(2-(k^I m_I)^2 - \tfrac{5}{4L^2} k_\mu k^\mu - \tfrac{d-6}{2}\La L^2 \right)Y_\mu \\
      + L^2(\hat{R}_{\mu\nu} + \tfrac{1}{2} \hat{R}\hat{g}_{\mu\nu})Y^\nu
   + \left( - \tfrac{1}{2} (dk)_{\mu\nu} 
            + 2 \left( k-d(L^2)\right)_{[\mu} \nablah_{\nu]}
            - \tfrac{1}{L^2} (dL^2)_{[\mu} k_{\nu]} 
     \right) Y^\nu .
 \label{eqn:Omax}
\end{multline}
This is again self-adjoint, this time with respect to the inner product
\begin{equation}\label{eqn:maxinner}
  (Y_1,Y_2) \equiv \int_\Hcal L^2 \bar{Y}_1^{\mu} Y_{2\mu} \; \mathrm{d(vol)},
\end{equation}
and hence the eigenvalues $\la$ are real.

\subsection{Behaviour of solutions}\label{sec:behaviour}

We've seen that for a scalar field, linearized gravitational field, or Maxwell field, we can reduce the equation of motion to that of a massive, charged, scalar field $\chi_b(T,R)$ in $AdS_2$ with a homogeneous electric field (\ref{ads2}). Solutions of this equation of motion were considered in Refs.\ \cite{Strominger:1998yg,Amsel:2009ev,Dias:2009ex}. At large $R$, they behave as $\chi_b \sim R^{-\Del_\pm}$ where
\begin{equation}\label{eqn:modefreq}
  \Del_\pm = \frac{1}{2} \pm \sqrt{ \mu^2 - q^2 + \frac{1}{4} }
\end{equation}
Therefore solutions grow or decay as real powers of $R$ if the `effective BF bound' is respected:
\begin{equation}\label{eqn:ads2stabbound}
  \mu^2 - q^2 \ge -\frac{1}{4}.
\end{equation}
If this bound is violated then solutions oscillate at infinity. 

In the uncharged case ($q=0$), it is known that boundary conditions can be imposed that lead to stable, causal, dynamics when the bound is respected \cite{BF,Ishibashi:2004wx}. If the bound is violated then no choice of boundary conditions leads to stable, causal, dynamics \cite{Ishibashi:2004wx}. Motivated by this, we make the following definition for the remainder of the paper:
\begin{defn} 
  A near-horizon geometry is \emph{unstable} against linearized gravitational (or scalar field or Maxwell) perturbations if expanding in harmonics on $\Hcal$ gives a massive, charged, scalar field in $AdS_2$ that violates the bound (\ref{eqn:ads2stabbound}).
\end{defn}

This is just introducing some terminology, we are not claiming anything about the dynamics of a scalar field in $AdS_2$ when (\ref{eqn:ads2stabbound}) is violated.  Of course, it would be interesting to see if the arguments of Ishibashi \& Wald \cite{Ishibashi:2004wx} could be extended to the charged case to show that violation of (\ref{eqn:ads2stabbound}) implies that there exists no choice of boundary conditions for which the scalar field has stable dynamics.  However, such considerations are not actually relevant to this paper, as we are interested in the question of whether violation of (\ref{eqn:ads2stabbound}) implies instability of the full black hole geometry rather than just its near-horizon geometry. 

In fact, the results of Refs.\ \cite{Amsel:2009ev,Dias:2009ex} show that it probably doesn't make sense to consider perturbations of the near-horizon geometry as a spacetime in its own right since there will be a large backreaction when one goes beyond linearized theory.

We showed above that $\mu^2-q^2 = \la$, the eigenvalue of a self-adjoint operator $\Ocal{b}$.  Hence, our condition for instability of the near-horizon geometry is the existence of an eigenvalue $\la < -1/4$.  This means that the question of stability has been reduced to studying the spectrum of these operators on $\Hcal$.  In the next section we shall study the spectrum of these operators for the case of extreme cohomogeneity-1 MP black holes.

\section{Cohomogeneity-1 extreme MP black holes} \label{sec:nhmp}

\subsection{Metric and near-horizon limit}

We shall now illustrate the methods described above with an example.  Consider a Myers-Perry-(AdS) black hole \cite{mp,Hawking:1998kw,Gibbons:2004js} in odd dimension $d=2N+3$, with all angular momentum parameters set to be equal, $a_i=a$.  Such a black hole has enhanced rotational symmetry; the $U(1)^{N+1}$ is enlarged to $U(N+1)$, i.e.\ the symmetry is that of a homogeneously squashed $S^{d-2} = S^{2N+1}$.  The metric is cohomogeneity-1, that is it depends non-trivially on a single coordinate.  This makes the study of gravitational perturbations of this class of black holes more tractable than the general case, and certain types of perturbation of the full black hole geometry have been studied previously \cite{Kunduri:2006,Murata:2008,Dias:2010eu}.

The metric for the full black hole solution can be written in the form \cite{Kunduri:2006}
\begin{equation}\label{eqn:mpmetric}
  ds^2 = -\frac{V(r)}{h(r)^2} dv^2 +  \frac{2drdv}{h(r)} + r^2 h(r)^2 (d\hat{\psi} + \Acal - \Om(r)dv)^2 
         + r^2\hat{g}_{\al\beta} dx^\al dx^\beta
\end{equation}
where $(v,r,\hat{\psi},x^\al)$ are ingoing Eddington-Finkelstein type coordinates, $\hat{\psi}$ has period $2\pi$,
\begin{equation}
  V(r) = 1 + \frac{r^2}{l^2} 
         + \left(\frac{r_0}{r}\right)^{2N} \left(-1 + \frac{a^2}{l^2} + \frac{a^2}{r^2}
           \right),
\end{equation}
\begin{equation}
  h(r) = \sqrt{1 + \frac{a^2}{r^2}\left(\frac{r_0}{r}\right)^{2N}}\eqand
  \Om(r) = \frac{a}{r^2 h(r)^2}\left(\frac{r_0}{r}\right)^{2N}.
\end{equation}
The solution is parameterized by three quantities with the dimensions of length: $r_0$, $a$ (which determines the ratio of angular momentum to mass), and $l$ (the AdS radius). We are writing the $S^{2N+1}$ as a $U(1)$ fibration over $\CP{N}$, with $\hat{g}_{\al\beta}$ the Fubini-Study metric on $\CP{N}$ (normalized to have Ricci tensor $2(N+1) \hat{g}_{\al\beta}$) and $\Acal = \Acal_\al dx^\al$ satisfying $d\Acal = 2\Jcal$, where $\Jcal$ is the K\"ahler form on $\CP{N}$.  The metric satisfies the vacuum Einstein equation
\begin{equation}
  R_{ab} = -\frac{d-1}{l^2} g_{ab} \equiv \La g_{ab},
\end{equation}
and is asymptotically $AdS_d$ with radius $l$. The limit $l\rightarrow \infty$ gives the asymptotically flat MP solution.

The event horizon lies at $r=r_+$, with $V(r_+)=0$.  This family of black holes admits an extremal limit, i.e. there exists a value of $a$ for which $V'(r_+)=0$.  In this case, the solution is uniquely labelled by $l$ and $r_+$, with
\begin{equation}
  r_0^N = r_+^{N+2} \sqrt{N+1} \left(\frac{1}{r_+^2} + \frac{1}{l^2} \right), \qquad
  a^2   = \frac{r_+^2 l^2}{N+1} \left(\frac{(N+1)r_+^2 + Nl^2}{(r_+^2+l^2)^2}\right). 
\end{equation}
To obtain the near-horizon limit, we define new coordinates $\tilde{r},\tilde{v}, \tilde{\psi}$ by 
\begin{equation}
  r = r_+ + \eps \tilde{r}, \quad v = \frac{\tilde{v}}{\eps} \eqand \hat{\psi} = \tilde{\psi} + \Om(r_+) v,
\end{equation}
and then take the limit $\eps\rightarrow 0$, to obtain a metric
\begin{equation}\label{eqn:ds2exact}
  ds^2 =  -\frac{V''(r_+)\tilde{r}^2}{2h(r_+)^2} d\tilde{v}^2 + \frac{2d\tilde{r}d\tilde{v}}{h(r_+)}
         + r_+^2 h(r_+)^2 \left(d\tilde{\psi} + \Acal - \Om'(r_+)\tilde{r}d\tilde{v}\right)^2 
         + r_+^2\hat{g}_{\al\beta} dx^\al dx^\beta.
\end{equation}
Finally, to simplify this, and recover a form of the metric more similar to that used in the discussion above, we define new coordinates $(T,R,\psi,x^\al)$ by
\begin{equation}
  T = \frac{V''(r_+)}{2h(r_+)} \tilde{v} + \frac{1}{\tilde{r}},\qquad
  R = \tilde{r}, \qquad
  \psi = \tilde{\psi} - \frac{2h(r_+) \Om'(r_+)}{V''(r_+)}\log(\tilde{r})
\end{equation}
and define constants
\begin{eqnarray}
\label{Ldef}
 \frac{1}{L^2} &=& \frac{V''(r_+)}{2} 
                = 2(N+1) \left( \frac{N}{r_+^2} + \frac{N+2}{l^2} \right)\label{eqn:Ldef} \\ 
  B^2 &=& r_+^2 h(r_+)^2 
       = (N+1) r_+^2 \left(1+ \frac{r_+^2}{l^2}\right), \label{eqn:Bdef}\\
  \Om &=& \frac{2h(r_+) \Om'(r_+)}{V''(r_+)} 
       =  \frac{-1}{(N+1)\big(1+(N+2)(r_+/l)^2\big)}\sqrt{\frac{Nl^2 + (N+1)r_+^2}{l^2+r_+^2}} , \\
  \frac{1}{E}  &=& \frac{B\Om}{2L^2} 
                  = -\Big(1+\frac{r_+^2}{l^2}\Big) 
                    \sqrt{(N+1)\left(\tfrac{N+1}{l^2} + \tfrac{N}{r_+^2}\right)} \label{eqn:Edef}.
\end{eqnarray}
This gives a simple form for the near-horizon metric:
\begin{equation}\label{eqn:metric}
  ds^2 = L^2(-R^2 dT^2 + \frac{dR^2}{R^2}) + B^2 \left(d\psi + \Acal - \Om R dT\right)^2 
         + r_+^2 \hat{g}_{\al\beta} dx^\al dx^\beta.
\end{equation}
As expected, this metric takes the form of a $(d-2)$-dimensional manifold $\Hcal$ fibred over $AdS_2$.  Here, $\Hcal$ is a homogeneously squashed $(d-2)$-sphere, with metric
\begin{equation}
  ds_{d-2}^2 = B^2 \left(d\psi + \Acal\right)^2 + r_+^2 \hat{g}_{\al\beta} dx^\al dx^\beta,
\end{equation}
where $\hat{g}$ is the metric on $\CP{N}$ as above and $\psi$ has period $2\pi$.

We are writing the metric in a form that makes manifest its enhanced symmetry, rather than in the form \eqref{nhgeom} (which makes manifest only the Killing directions $\pd/\pd\phi^I$).  Since we know that the near-horizon geometry of a general extreme MP solution \emph{can} be written in the form \eqref{nhgeom} \cite{Figueras:2008qh} it follows that there must be a coordinate transformation that would allow us to bring our metric to this form.  However, it is not necessary to perform such a transformation since the operators $\Ocal{b}$ on $\Hcal$ are defined in a covariant way.  We can read off the vector $k$ by looking at the cross-terms proportional to $\pd/\pd T$ in the inverse metric:
\begin{equation}
  \frac{1}{2L^2} \left(-2\Om \pard{\psi} \pard{T} \right) = 
  \frac{1}{2L^2} \left( -2k^I \pard{\phi^I} \pard{T} \right)
\end{equation}
and hence
\begin{equation}\label{eqn:nhmpk}
  k \equiv k^I \pard{\phi^I} =  \Om \left(\pard{\psi}\right).
\end{equation}
We can Fourier decompose our perturbation in the $\psi$ direction, i.e. assume dependence $e^{im\psi}$ so that eigenfunctions $Y$ on $\Hcal$ obey $\Lcal_k Y = i\Om m Y$.  Equation \eqref{kdotm} now enables us to read off 
\begin{equation}
  k^I m_I = \Omega m
\end{equation}
For these black holes, the condition (\ref{eqn:axicondition}) reduces to $m=0$.  However, we will obtain results for general $m$.  We will determine the spectrum of our operators $\Ocal{b}$ by expanding them in harmonics on $\CP{N}$, with metric $\gh_{\al\beta}$ (where $\al,\beta,\ldots$ are indices on $\CP{N}$, raised and lowered with $\gh$).  From the $\CP{N}$ perspective, $m$ acts like a charge which couples to the `gauge field' $\Acal$ (see \cite{Kunduri:2006}).  We therefore define a charged covariant derivative on $\CP{N}$
\begin{equation}\label{eqn:Dhatdef}
  \Dcalh_\al = \hat{D}_\al - im \Acalh_\al
\end{equation}
where $\hat{D}$ is the Levi-Civita connection on $\CP{N}$.

\subsection{Scalar field perturbations}\label{sec:scalarfield}

As a simple first example, we show how to deal with massive scalar field perturbations. The operator $\Ocal{0}$ defined by (\ref{eqn:Oscalar}) reduces to
\begin{equation}\label{eqn:massivescalarperts}
   \Ocal{0} Y =  -\frac{2Nm^2L^4}{r_+^4}Y - \frac{L^2}{r_+^2} \Dcalh^2 Y + L^2M^2 Y,
\end{equation}
acting on functions $Y(\psi,x) = e^{im\psi} \Ybb(x)$.  We shall assume that the $AdS_d$ BF bound is respected, i.e. that
\begin{equation} \label{eqn:adsdbf}
  M^2 \geq -\frac{(d-1)^2}{4l^2} = -\frac{(N+1)^2}{l^2}.
\end{equation}
Scalar eigenfunctions of the charged covariant Laplacian $\Dcalh^2$ on $\CP{N}$ were studied in \cite{Hoxha:2000}.  For each integer $m$, there exist $\CP{N}$ scalars $\Ybb(x)$ satisfying 
\begin{equation} \label{eqn:scalarYbb}
  (\Dcalh^2 + \lak{S}) \Ybb = 0,
\end{equation}
for eigenvalues
\begin{equation}\label{eqn:scalarevals}
  \lak{S} = 4\kap(\kap+N) + 2|m|(2\kap+N) \qquad \kap = 0,1,2,\dots.
\end{equation}
Hence, the eigenvalues of $\Ocal{0}$ are
\begin{equation}
  \la = \frac{\left(4\kap(\kap+N) + 2|m|(2\kap+N)\right)L^2}{r_+^2} -\frac{2Nm^2L^4}{r_+^4} + M^2L^2.
\end{equation}
Therefore, for large $|m|$, $\la$ becomes arbitrarily negative and the BF bound \eqref{eqn:ads2stabbound} is always violated.  However, for the axisymmetric modes $m=0$ that are relevant for our conjecture, the eigenvalues are given by 
\begin{equation}
  \frac{\la}{L^2} = \frac{4\kap(\kap+N)}{r_+^2} + M^2 
\end{equation}
for non-negative integers $\kap$ (recall that $L$ is defined by (\ref{Ldef})). 

Consider first asymptotically flat black holes $l\rightarrow \infty$ and $M^2 \ge 0$.  Here, we manifestly have $\la \geq 0$, and hence the $AdS_2$ BF bound is not violated.

This is not always the case for asymptotically $AdS$ black holes. Clearly there is no problem if $M^2 \ge 0$. However, if $M^2<0$ then it is possible for the $AdS_2$ BF bound to be violated even if the $AdS_d$ BF bound is respected \cite{Dias:2010ma}.  Consider for example the case in which the $AdS_d$ bound is saturated. Then a mode labelled by $\kap$ violates the $AdS_2$ BF bound if
\begin{equation}
  \frac{r_+^2}{l^2} > \frac{4\kap(\kap+N) + N(N+1)}{4N(N+1)},
\end{equation}
that is, for sufficiently large black holes.  In this case, our conjecture predicts that the scalar field should be unstable in the full black hole geometry. This issue was investigated numerically in Ref.~\cite{Dias:2010ma}.  It was found that the full black hole is indeed unstable, and there exists a new nonlinear family of `hairy' rotating black holes.  In Section~\ref{sec:nhinstab} we shall prove analytically that the full black hole solution must be unstable.

\subsection{Gravitational perturbations of asymptotically flat BHs}

We now consider the more complicated case of gravitational perturbations.  The calculations here are significantly more involved.  In this section we will merely give the results for different classes of perturbation mode, reserving the details of the calculations for Appendix \ref{app:technical}.

Our approach to determining the eigenvectors $Y_{\mu\nu}$ of $\Ocal{2}$ is to decompose $Y_{\mu\nu}$ into parts parallel and perpendicular to $\CP{N}$ and then expand each part in terms of harmonics on $\CP{N}$, assuming dependence $e^{im\psi}$ along the $S^1$ fibre.  By `harmonics', we mean eigenfunctions of the charged $\CP{N}$ Laplacian $\hat{{\cal D}}^2$.  They can be divided into scalar, vector, and (traceless) tensor types \cite{Kunduri:2006,Martin:2008pf,Dias:2010eu} where vector and tensor harmonics are transverse with respect to the derivatives $\Dcalh_\al$ and $\Jcalh_\al{}^\beta \Dcalh_\beta$.  See Ref.~\cite{Martin:2008pf} for detailed discussion of this decomposition.  The orthogonality properties of these different types of harmonic implies that eigenfunctions of $\Ocal{2}$ must each be built from $\CP{N}$ harmonics of a particular type (scalar, vector or tensor) and with the same eigenvalue of $\Dcalh^2$.

The modes that are relevant to our conjecture are those that are $\psi$ independent, i.e.\ those with $m=0$.  Therefore, we only list our results in this case, although in Appendix \ref{app:technical} we derive all of these results for general $m$.  It turns out that, as in the scalar field case, the coefficient of $m^2$ in these eigenvalues is always negative, and hence for sufficiently large $|m|$ there are instabilities in every sector of perturbations of the near-horizon geometry.

We begin with the asymptotically flat case, corresponding to $l\rightarrow \infty$.

\subsubsection{Tensor modes}\label{sec:tensorevals}

These eigenfunctions $Y_{\mu\nu}$ have components only in the direction of $\CP{N}$, and are proportional to a transverse, traceless, tensor harmonic on $\CP{N}$. Such harmonics exist only for $N>1$ ($d>5$).  Tensor perturbations of the full black hole geometry were considered in Ref.~\cite{Kunduri:2006}.  In the asymptotically flat case, no evidence of any instability was found near extremality. Hence, if our conjecture holds, we would not expect to find any unstable modes satisfying \eqref{eqn:axicondition} (i.e. $m=0$) in this sector.

We find that the eigenvalues $\la$ of $\Ocal{2}$ are given by 
\begin{equation}\label{eqn:tensorevalsflat}
  \la = \frac{2\kap(\kap+N) + 2N(1-\sig)}{N(N+1)},
\end{equation}
where $\kap=0,1,2,\ldots$, and the parameter $\sig = \mp 1$ separates two different classes of tensor harmonic which are respectively Hermitian, or anti-Hermitian, on $\CP{N}$ (more details are given in Appendix \ref{sec:gravcalcs}).

The eigenvalues $\lambda$ are manifestly non-negative.  Hence the effective BF bound $\lambda \ge -1/4$ is respected and there is no instability of the near-horizon geometry in this sector.  Hence our conjecture is consistent with the results of Ref.~\cite{Kunduri:2006}.

\subsubsection{Vector modes}

Next, we move on to study vector-type perturbation modes. Again, these exist only for $N>1$ ($d>5$). These have not been previously studied in the literature, so we have no numerical results for the full black hole geometry to compare our results to. 

For vector-type perturbations $Y_{\mu\nu}$ is written as a linear combination of three different types of term built from a $\CP{N}$ vector harmonic and its derivatives, so it is determined by the three coefficients in this expansion. Acting with $\Ocal{2}$ has the same effect as acting with a certain $3\times 3$ matrix on these coefficients. Hence finding the eigenvalues of $\Ocal{2}$ for vector type perturbations reduces to finding the eigenvalues of a $3\times 3$ matrix. The elements of this matrix involve the eigenvalue of the vector harmonic on $\CP{N}$, which is labelled by a non-negative integer $\kap$ (and the integer $m$). Perhaps surprisingly, the eigenvalues of $\Ocal{2}$ turn out to be rational (given here for $m=0$):
\begin{equation}\label{eqn:gravvecevals}
  \la = \frac{2(N+(\kap+1)^2)}{N(N+1)}, \quad
  \frac{2(\kap+2)(\kap+N+1)}{N(N+1)}, \quad
  \frac{2\left(N^2 + (\kap+2)^2 + N(2\kap+5)\right)}{N(N+1)}.
\end{equation}
These are all manifestly positive, so there is no violation of the generalized $AdS_2$ BF bound in this sector.

\subsubsection{Scalar modes}\label{sec:gravscalarevalsflat}

The most complicated sector is that of scalar type gravitational perturbations.  In this case, $Y_{\mu\nu}$ is written as a linear combination of six terms, each of which is constructed from $\CP{N}$ scalar harmonics and their derivatives. 

Harmonics are again labelled by an integer $\kappa \geq 0$, as well as $m$. Acting with $\Ocal{2}$ has the effect of acting with a $6 \times 6$ matrix. Hence determining the eigenvalues of $\Ocal{2}$ is equivalent to determining the eigenvalues of a $6 \times 6$ matrix. For the special cases $\kappa=0,1$ some combinations of derivatives of the $\CP{N}$ harmonics vanish, which a corresponding reduction in the size of the matrix. There is also a reduction in size for the special case of $N=1$ (i.e.\ $d=5$) for which the matrix is generically $5 \times 5$.  In all cases, we again find that the eigenvalues of $\Ocal{2}$ are rational. 

For $\kap=0$, there is just one eigenvalue
\begin{equation}\label{eqn:gravkap0}
  \la = \frac{2(2N+1)}{N} 
\end{equation}
which is manifestly positive, and hence there is no instability here.

For $\kap=1$, the eigenvalues $\la$ correspond to the eigenvalues of a $4\times 4$ matrix ($3 \times 3$ for $N=1$). They are
\begin{equation}
  \la  = \frac{2}{N}, \quad \frac{2(N+1)}{N}, \quad \frac{2(N+2)}{N},\quad \frac{4(N+2)}{N},
\end{equation}
which are again all positive.  The second eigenvalue does not appear for $N=1$.

Things get more interesting for $\kap\geq 2$, where we have a $6 \times 6$ matrix ($5\times 5$ for $N=1$).  The eigenvalues are given by
\begin{multline}\label{eqn:gravscalarevecs}
  \la = \frac{2(\kap-1)(\kap-N-1)}{N(N+1)}, \quad
                     \frac{2\kap(\kap-1)}{N(N+1)}, \quad 
                   \frac{2\kap(\kap+N)}{N(N+1)}, \\
                   2+\frac{2\kap(\kap+N)}{N(N+1)}, \quad
                   \frac{2(\kap+N)(\kap+N+1)}{N(N+1)},\quad
                   \frac{2(\kap+1+N)(\kap+2N+1)}{N(N+1)},
\end{multline}
with the fourth of these absent for $N=1$.

Five of these eigenvalues are manifestly non-negative, so in order to check for an instability of the near-horizon geometry, we need only to analyse whether there exist $\kap$, $N$ such that
\begin{equation}\label{eqn:minevalflat}
  \frac{2(\kap-1)(\kap-N-1)}{N(N+1)} < -\frac{1}{4}.
\end{equation}
We list the values of the left hand side explicitly in Table \ref{tab:evals1} for all $\kap=2,\dots 10$, in dimensions $d=5,7,\dots,23$. 

\begin{table}[htb]
\begin{center}
  \begin{tabular}{|c|c||c|c|c|c|c|c|c|c|c|}
\hline
       &   & $\kap$&       &       &       &        &   &   &   &  \\
   $d$ &$N$&     2 &     3 &     4 &     5 &     6  & 7 & 8 & 9 & 10\\ \hline\hline
     5 & 1 &  0.00 & 2.00  &  6.00 & 12.00 & 20.00 & 30.00 & 42.00 & 56.00 & 72.00\\
     7 & 2 & \bf-0.33 & 0.00  &  1.00 & 2.67  & 5.00  & 8.00  & 11.70 & 16.00 & 21.00\\
     9 & 3 & \bf-0.33 & \bf-0.33 &  0.00 & 0.67 & 1.67  & 3.00  & 4.67  & 6.67 & 9.00\\
    11 & 4 & \bf-0.30 & \bf-0.40 & \bf-0.30 & 0.00  & 0.50  & 1.20  & 2.10  & 3.20 & 4.50\\
    13 & 5 & \bf-0.27 & \bf-0.40 & \bf-0.40 & \bf-0.27 & 0.00  & 0.40  & 0.93  & 1.60 & 2.40\\
    15 & 6 & -0.24 & \bf-0.38 & \bf-0.43 & \bf-0.38 & -0.24 & 0.00  & 0.33  & 0.76 & 1.29\\
    17 & 7 & -0.21 & \bf-0.36 & \bf-0.43 & \bf-0.43 & \bf-0.36 & -0.21 & 0.00  & 0.29 & 0.64\\
    19 & 8 & -0.19 & \bf-0.33 & \bf-0.42 & \bf-0.44 & \bf-0.42 & \bf-0.33 & -0.19 & 0.00 & 0.25\\
    21 & 9 & -0.18 & \bf-0.31 & \bf-0.40 & \bf-0.44 & \bf-0.44 & \bf-0.40 & \bf-0.31 & -0.18& 0.00\\
    23 &10 & -0.16 & \bf-0.29 & \bf-0.38 & \bf-0.44 & \bf-0.46 & \bf-0.44 & \bf-0.38 & \bf-0.29& -0.16\\\hline
  \end{tabular}
  \caption[Smallest eigenvalue of the operator $\Ocal{2}$ for asymptotically flat extreme cohomogeneity-1 Myers-Perry black holes in odd dimensions.]{\it\small\label{tab:evals1} Smallest eigenvalue of $\Ocal{2}$ for $m=0$, in the case of asymptotically flat extremal cohomogeneity-1 Myers-Perry black holes in dimensions $d=5,7,\dots 23$, for modes $\kap=2,\dots 10$.  The BF bound is $-1/4$, eigenvalues violating this bound, and indicating an instability of the near horizon geometry, are shown in bold (NB: all of these values are rational numbers determined by \eqref{eqn:gravscalarevecs}, we give decimal approximations here for readability purposes.)}
\end{center}
\end{table}

In dimension $d=5$ there are no modes that violate the effective BF bound, and we conclude that there are no unstable scalar modes of the near horizon geometry that satisfy the condition (\ref{eqn:axicondition}).  Therefore we do not predict any instability of the full black hole in this case.  This is consistent with a study of linearized perturbations of the full black hole \cite{Murata:2008}, which did not find any evidence of instability near extremality.

Our main result in this section is that for $d \ge 7$ there is always at least one mode that violates the effective BF bound and hence the near-horizon geometry is unstable.  Since this mode respects (\ref{eqn:axicondition}), our conjecture predicts that the full black hole solutions should be unstable.  Perturbations of the full non-extreme black hole were studied in Ref.~\cite{Dias:2010eu}.  For $d = 9$ it was found that $\kap=2$ scalar perturbations are unstable near extremality, in agreement with our conjecture.  However no instability was found for the cases $d=7$, $\kap=2$ or $d=9$, $\kap=3$ for which we predict that one should be present.

The reason for this discrepancy is that the results of Ref.~\cite{Dias:2010eu} do not get close enough to extremality to see the instability that we predict.  J.~E.~Santos has kindly repeated the numerical analysis of Ref.~\cite{Dias:2010eu} for black holes that are very close to extremality.  He finds instabilities that were missed in the analysis of Ref.~\cite{Dias:2010eu}.  Let $a_{\rm ext}$ denotes the value of $a$ at which the black hole becomes extreme. Table \ref{tab:jorge} gives the critical value of $1-a/a_{\rm ext}$ below which the black hole is unstable.\footnote{There is no instability of the black hole for $\kap=1$ but Ref.~\cite{Dias:2010eu} showed that there is an instability of the corresponding black {\it string} close to extremality. For completeness, we give Santos' results for the critical values of $1-a/a_{\mathrm{ext}}$ for this instability: $ 4.116 \times 10^{-2}$, $ 8.347 \times 10^{-2}$, $ 1.351 \times 10^{-1}$, $ 1.517 \times 10^{-1}$ for $d=7,9,13,15$ respectively.}  There is indeed an instability very near extremality for $d=7$, $\kap=2$ and $d=9$, $\kap=3$, for $d=13$ with $\kap = 2,3,4,5$ and for $d=15$ with $\kap = 3,4,5$, all in perfect agreement with our conjecture.  He also finds that there are cases for which we do not predict an instability but nevertheless one exists (e.g. $d=7$, $\kap=3$), which emphasizes that our conjecture supplies a sufficient, but not necessary, condition for instability. 

\begin{table}
\begin{center}
\small
\begin{tabular}{|c||c|c|c|c|c|c|}
\hline
     $d$          & $\kappa = 2$ & $\kappa = 3$ & $\kappa = 4$ & $\kappa = 5$ & $\kappa = 6$ & $\kappa = 7$ \\
\hline\hline
$ 7$       & $ 2.34 \times 10^{-5}$  & $ 2.51 \times 10^{-7}$  &  &  &  & \\
\hline
$ 9$       & $ 2.12 \times 10^{-3}$  & $ 2.94 \times 10^{-7}$  & $ 8.02 \times 10^{-9}$  &  &  & \\
\hline
$ 13$      & $ 1.50 \times 10^{-2}$  & $ 1.36 \times 10^{-3}$  & $ 2.11 \times 10^{-5}$  
               & $ 1.056 \times 10^{-6}$  & $ 7.35 \times 10^{-8}$  & \\
\hline
$ 15$      & $ 2.23 \times 10^{-2}$  & $ 3.46 \times 10^{-3}$  & $ 2.87 \times 10^{-4}$  
               & $ 5.05 \times 10^{-6}$  & $ 7.57 \times 10^{-7}$  & $ 6.10 \times 10^{-8}$  \\
\hline
\end{tabular}
\caption[Critical values below which cohomogeneity-1, Myers-Perry black holes become unstable to various perturbation modes.] {\label{tab:jorge}
{\it\small Critical values $1-a/a_{\mathrm{ext}}$ below which an asymptotically flat, cohomogeneity-1, Myers-Perry black hole becomes unstable against scalar-type gravitational perturbations with the given $\kappa$.  These numerical results were obtained by J.~E.~Santos using the methods described in \cite{Dias:2010eu}.}}
\end{center}
\end{table}
In general dimension $d=2N+3$, straightforward algebra shows that a violation of the effective BF bound occurs if
\begin{equation}
  1 + \tfrac{N}{2} - \tfrac{1}{2} \sqrt{\tfrac{N(N-1)}{2}} < \kap 
          < 1 + \tfrac{N}{2} + \tfrac{1}{2} \sqrt{\tfrac{N(N-1)}{2}}. 
\end{equation}
This proves that for any $N\geq 2$, there is at least one integer value of $\kap$ for which the effective BF bound is violated.

\subsection{Gravitational perturbations of asymptotically $AdS$ BHs}

We now move on to consider gravitational perturbations of cohomogeneity-1 Myers-Perry-AdS black holes. Ref.~\cite{Kunduri:2006} demonstrated that such black holes suffer a `super-radiant' instability near extremality.  This instability corresponds to perturbations with $m\neq 0$, which are excluded from the scope of our conjecture. We shall consider eigenfunctions of $\Ocal{2}$ with $m=0$ to see if any new instability appears. Once again, we consider separately eigenfunctions of $\Ocal{2}$ constructed from tensor, vector and scalar harmonics on $\CP{N}$.

\subsubsection{Tensor modes}

The eigenvalues $\la$ of $\Ocal{2}$ are given by 
\begin{equation}\label{eqn:tensorevalsads}
  \frac{\la}{L^2}
    = 4(1-\sig) \left(\frac{N}{r_+^2}+\frac{N+1}{l^2}\right) + \frac{4\kap(\kap+N)}{r_+^2} ,
\end{equation}
where again $\sig=\pm 1$.  This is manifestly non-negative.  Hence the BF bound is respected so we do not predict any instability. This is in agreement with Ref.~\cite{Kunduri:2006}, which proved that $m=0$ tensor perturbations are stable in the full black hole geometry.

\subsubsection{Vector modes}

In contrast with the asymptotically flat case, we are unable to give a simple explicit form for the eigenvalues of $\Ocal{2}$ corresponding to vector modes.  However, we can still prove that for all $N$, for any value of the dimensionless ratio $r_+/l$, the eigenvalues are all non-negative, and hence the effective $AdS_2$ BF bound is respected.  Hence, we do not predict any instability in this sector.  The proof is given in Appendix \ref{sec:gravcalcs}.

\subsubsection{Scalar modes}\label{sec:gravscalarevalsads}

The analysis proceeds in the same way as in the asymptotically flat case. 

For $\kap=0$, there is a single eigenvalue
\begin{equation}
  \la = L^2\left( \frac{4}{E^2} + 4(N+1)\frac{B^2}{r_+^4}\right)
\end{equation}
which is manifestly positive, and hence there is no instability.

For $\kap=1$, the eigenvalues $\la$ correspond to the eigenvalues of a $4\times 4$ matrix, and these cannot be found explicitly in a convenient way.  However, plotting these eigenvalues against the dimensionless parameter $r_+/l$ shows immediately that all these eigenvalues lie above the BF bound, and hence there is no instability in this sector.

For $\kappa=2,3,4,\ldots$, the problem reduces to finding eigenvalues of a $6 \times 6$ matrix parametrized by $r_+/l$.  For each $\kap=2,3,4,\ldots$, there are six real eigenvalues of $\Ocal{2}$.

\begin{figure}
\begin{center}
\begin{minipage}[c]{\textwidth}
  \includegraphics[width=0.49\textwidth]{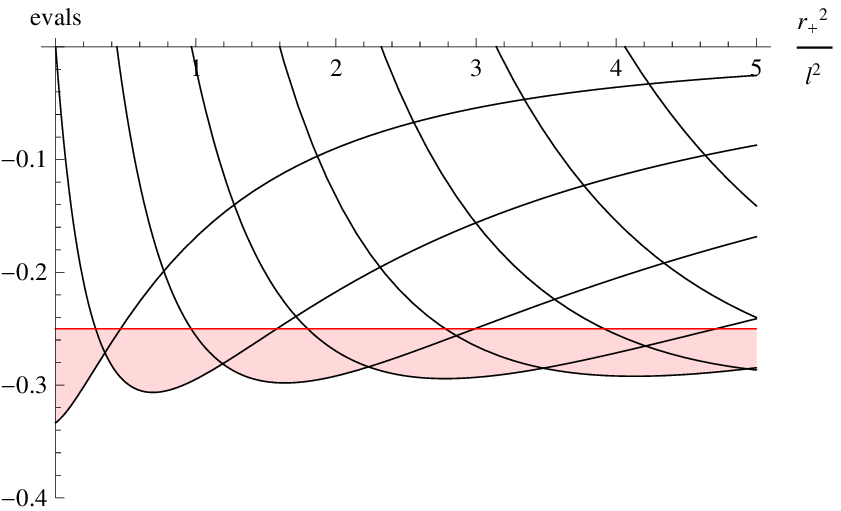}
  \includegraphics[width=0.49\textwidth]{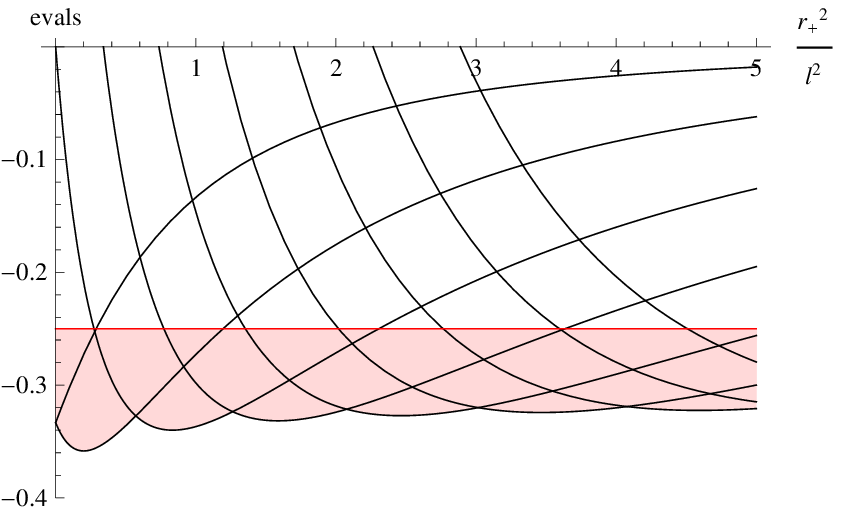}
  \caption[Eigenvalues of the operator $\Ocal{2}$ describing gravitational perturbations, plotted against the size of the black hole in $AdS$ units, in seven and nine dimensions.]{\it\small\label{fig:evalsN=23}Lowest eigenvalues of $\Ocal{2}$ plotted against the size of the $AdS$ black hole ($r_+^2/l^2$), in dimensions $d=7$ (left) and $d=9$ (right). The shaded region corresponds to violation of the effective BF bound. The separate curves shown correspond to $\kap=2,3,4,5,6$, moving from left to right as $\kap$ is increased (the curves that are negative for $r_+/l \rightarrow 0$ are $\kap=2$ on the left and $\kap=2,3$ on the right).  In both cases, there is some mode that violates the BF bound for any black hole size.} 
\end{minipage}
\end{center}
\end{figure}

Our results are easiest to understand for $d \ge 7$ ($N \ge 2$).  Consider first the case $N=2$.  The lowest eigenvalue for each value of $\kap$ is plotted in Figure  \ref{fig:evalsN=23}. We find that there is a violation of the effective BF bound by the lowest $\kap=2$ eigenvalue for sufficiently small $r_+/l$. This makes sense: the eigenvalues here are continuously connected to the eigenvalues in the asymptotically flat case as $r_+/l \rightarrow 0$, and we saw that there is an instability with $\kap=2$ in the asymptotically flat case. Modes with higher $\kap$ are unstable for ranges of $r_+/l$ corresponding to larger black holes.  The ranges for successive values of $\kap$ overlap, and in fact for any $r_+/l$, there exists some $\kap$ corresponding to an unstable mode.  For $N=3$ ($d=9$) the results are similar and are also shown in Figure \ref{fig:evalsN=23}.

We can perform similar studies for higher dimensions $d=11,13,\ldots$, and find results that are qualitatively similar to those for $d=7,9$ (although note that modes with small $\kap$ become stable for small $AdS$ black holes in larger dimensions, however instabilities for higher $\kap$ ensure that such black holes remain unstable).
Therefore our conjecture predicts that all extreme, cohomogeneity-1 MP-AdS black holes with $d \ge 7$ should be unstable against scalar-type gravitational perturbations with $m=0$. We emphasize that this is distinct from the previously discovered superradiant instability.
\begin{figure}
\begin{center}
\begin{minipage}[c]{\textwidth}
  \includegraphics[width=0.49\textwidth]{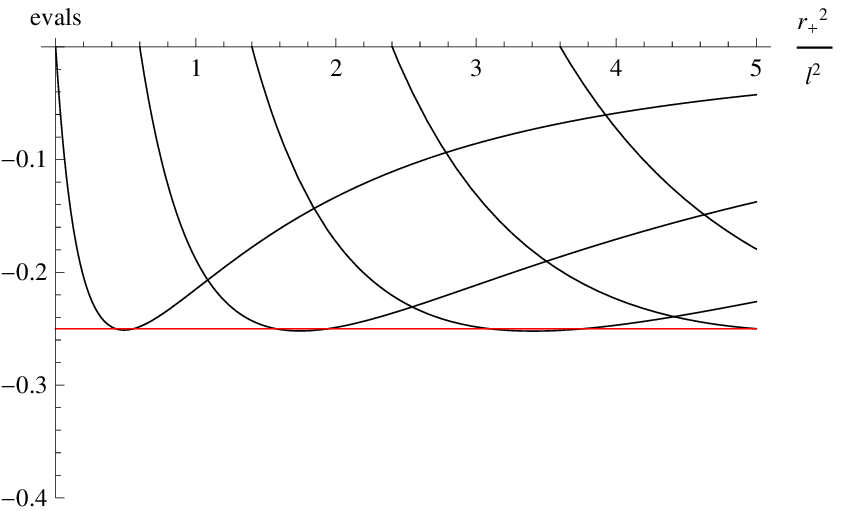}
  \includegraphics[width=0.49\textwidth]{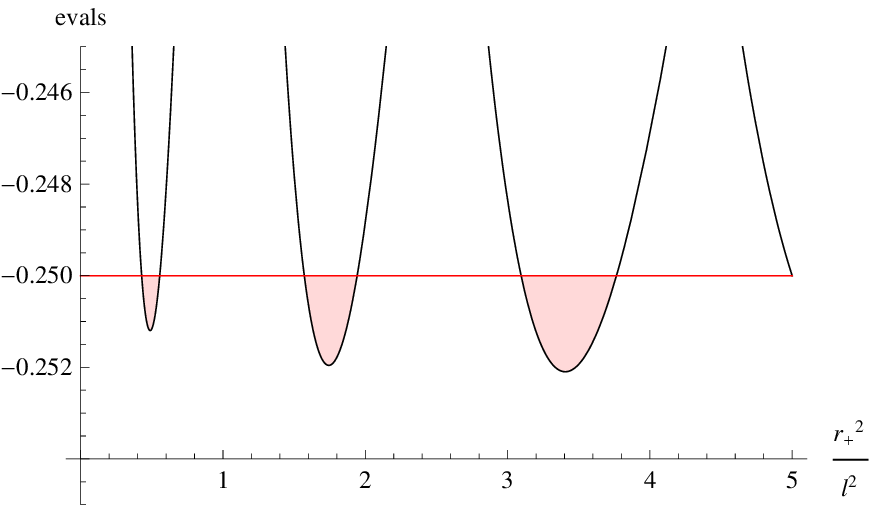}
  \caption[Eigenvalues of the operator $\Ocal{2}$ describing gravitational perturbations, plotted against the size of the black hole in $AdS$ units, in five dimensions.]{\it\small\label{fig:evalsN=1}Eigenvalues of $\Ocal{2}$ plotted against the size of the $AdS$ black hole ($r_+^2/l^2$), in dimension $d=5$ (the right hand graph is a zoomed version of the left hand one).  The separate curves shown correspond to $\kap=2,3,4,5,6$, moving from left to right as $\kap$ is increased.  We find that the generalized BF bound $\la \geq-1/4$, shown by the horizontal line, is violated by a small amount in various small ranges of black hole size.  As $\kap$ is increased, the violation of the BF bound occurs for increasingly large black holes.} 
\end{minipage}
\end{center}
\end{figure}

For $d=5$ ($N=1$), we plot the lowest eigenvalue of $\Ocal{2}$ with given $\kappa$ in Figure \ref{fig:evalsN=1}.
For $\kap=2$, there is a violation of the effective BF bound for $0.43<r_+^2/l^2<0.56$.  The violation is small: by less than $1\%$.  Modes with higher $\kap$ are also unstable in particular small ranges of the black hole size, but for increasingly large black holes as $\kap$ increases.  We do not have a good explanation for why these unstable modes are found only in these small ranges. The fact that the bound is violated only by a small amount may imply that the instability appears much closer to extremality than anything in Table \ref{tab:jorge} so confirming our conjecture in this case may require a numerical study of the full, \emph{extremal} black hole solution.  As we can only give a sufficient condition for instability, not a necessary one, it might be the case that the full extremal black hole solution is unstable against $m=0$ perturbations for any $r_+/l$ above a certain lower bound (we know that an instability is not present for the asymptotically flat case $r_+/l \rightarrow 0$).\newpage 

\subsection{Electromagnetic Perturbations}

Recall that an instability of the near-horizon geometry under electromagnetic perturbations  corresponds to an eigenvalue of $\Ocal{1}$ being less than $-1/4$, and the eigenvectors of $\Ocal{1}$ are vectors $Y_\mu$ on $S^{2N+1}$. Just as in the gravitational case, we can decompose these into parts parallel and perpendicular to $\CP{N}$ and then decompose these parts into scalar and vector harmonics on $\CP{N}$. As things are simpler here, we can consider both asymptotically flat and asymptotically $AdS$ black holes together.  We find no evidence of any instability in either of these cases. Once again, we restrict attention to modes with $m=0$ since these are the ones relevant to our conjecture.

\subsubsection{Vector modes}\label{sec:emvectors}
For eigenvectors $Y_\mu$ built from vector harmonics on $\CP{N}$, we find eigenvalues
\begin{equation}
 \la = 4\big(\kap^2+(N+3)\kap + 2(N+1)\big)\frac{L^2}{r_+^2},
\end{equation}
where $\kap$ is a non-negative integer.  These are all positive so there is no instability.

\subsubsection{Scalar modes}\label{sec:emscalars}

For $Y_\mu$ built from scalar harmonics on $\CP{N}$ (labelled by a non-negative integer $\kap$), there are two cases to consider separately.  For $\kap=0$, there is a positive single eigenvalue:
\begin{equation}\label{eqn:emevals0}
  \la = 4N(N+1)L^2\left(\frac{1}{l^2} + \frac{1}{r_+^2}\right) 
\end{equation}
For $\kap\geq 1$, there are three eigenvalues for each $\kap$, given by
\begin{equation}
  \la = \frac{2\kap(\kap+N)}{(N+1)\left(N+(N+2)\tfrac{r_+^2}{l^2}\right)},
\end{equation}
and
\begin{equation}
  \la = \frac{\tfrac{2\kap(\kap+N)}{N+1}+N + (N+1)\tfrac{r_+^2}{l^2}
              \pm \sqrt{4\kap(\kap+N)\left(1 + \left(\tfrac{N+2}{N+1}\right) \tfrac{r_+^2}{l^2}
                                     \right)
                        + \left(N+(N+1)\tfrac{r_+^2}{l^2}\right)^2
                       }
             }
             {N+(N+2)\tfrac{r_+^2}{l^2}}.
\end{equation}
Two of these are positive, but the third can sometimes be negative.  In order to check whether the effective $AdS_2$ BF bound $\la \geq -1/4$ is violated, we plotted this eigenvalue against the $AdS_d$ black hole size $r_+/l$, finding that there is no violation of the effective BF bound for any $N$ or $\kap$.

In the asymptotically flat case, these eigenvalues are again very simple, reducing to
\begin{equation}
  \frac{2\kap(\kap+N)}{N(N+1)},\quad
  \frac{2(\kap+N)(\kap+N+1)}{N(N+1)}, \quad
  \frac{2\kap(\kap-1)}{N(N+1)}.
\end{equation}

\subsection{Dual operators and conformal dimensions}\label{sec:kerrcft}

It has been conjectured that there exists a CFT dual to the NHEK geometry \cite{kerrcft}. Assuming that CFT operator dimensions are related to the decay rate of fields in $AdS_2$ in the usual way, then equation (\ref{eqn:modefreq}) gives the operator dimensions. In general, these turn out to be complex, which may be a problem for the Kerr-CFT conjecture. However, the results of Refs.\ \cite{Amsel:2009ev,Dias:2009ex} show that operators dual to {\it axisymmetric} gravitational perturbations are particularly simple, with integer dimensions $\Delta_+ = l+1 $ where $l=2,3,\ldots$ labels the harmonic on $\Hcal = S^2$. 

It has been suggested that the Kerr-CFT conjecture can be extended to the Myers-Perry black holes \cite{popekerrcft} so it is interesting to use our results to compute operator dimensions for this case too. Consider a cohomogeneity-1 extreme Myers-Perry black hole. The operator $\Ocal{2}$ governing gravitational perturbations of the near-horizon geometry appears very complicated.  It is striking that its eigenvalues are all rational numbers (for asymptotically flat black holes\footnote{In the asymptotically AdS case, the eigenvalues generically are all irrational but this case seems less interesting for the present discussion since there always is a superradiant instability \cite{Kunduri:2006}.}). 

For $d>5$ we have seen that our conjecture predicts an instability so presumably a CFT dual does not exist (or is also unstable).  So consider the case $d=5$ ($N=1$). In this case, only scalar-type gravitational perturbations exist.  Again, if $m \ne 0$ then there are complex operator dimensions but for the modes with $m=0$ that are relevant to our conjecture the operator dimensions are real and particularly simple.  For $\kappa=0$ we have $\Delta_+ = 3$.  The $\kappa=1$ harmonics give
\begin{equation}
 \Delta_+ = 2, 3, 4.
\end{equation}
For $\kap=2,3,4, \ldots$, we find
\begin{equation}
 \Delta_+ = \kappa-1,\, \kappa, \, \kappa+1, \, \kappa+2,\, \kappa+3.
\end{equation}
Hence if there is a CFT description that obeys the usual AdS/CFT rules then the $m=0$ gravitational perturbations give rise to five infinite families of operators with {\it integer} dimensions, just as for NHEK.\footnote{A massless scalar field would give operators with $\Delta_+ = \kap+1$ for $\kap=0,1,2\ldots$. 
For $N>1$, if we ignore the instability and calculate $\Delta_+$ formally for gravitational perturbations (for stable modes) then the results are generically irrational.}  This result hints that some symmetry is protecting the dimensions of operators dual to $m=0$ gravitational perturbations.  Note that the operator of lowest dimension is marginal (in 1D): $\Delta_+ = 1$. 
                                                             
\section{Instabilities from near-horizon geometries}\label{sec:nhinstab}

Does an instability of the near-horizon geometry imply the existence of an instability of the full spacetime?  We conjectured in the introduction that this was the case for a particular class of perturbation modes and explained how extreme Kerr is consistent with the conjecture. In Section \ref{sec:nhmp} we have shown that our conjecture predicts an instability for certain Myers-Perry black holes, and this prediction is confirmed by studies of perturbations of the full black hole geometry. 

In this section, we will present some ideas that explain why our conjecture appears to work. In the case of a scalar field, we shall sketch a proof of the conjecture.  We shall present some evidence suggesting that the method of proof in the scalar field case might also generalize to gravitational perturbations.\footnote{The material in this section provides motivation for much of the rest of work already described in this chapter.  However, the results of this section were largely derived by my supervisor Harvey Reall, and appear in our paper \cite{nhperturb}.}

\subsection{Scalar field instabilities}

Consider an uncharged, scalar field $\Psi$ of mass $M$ in the extreme planar Reissner-Nordstr\"om-AdS black hole background in arbitrary dimension $d\geq 4$.  This has a near-horizon geometry of the form $AdS_2 \times \Rbb^{d-2}$.  So, in the language described above, we have $\Hcal=\Rbb^{d-2}$ here.

As before, we can Fourier analyze on $\Rbb^{d-2}$ to reduce the scalar field equation of motion to that of a massive scalar in $AdS_2$.  The BF bound (\ref{eqn:ads2stabbound}) associated to the $AdS_2$ is more restrictive than that associated to the asymptotic $AdS_d$ geometry.  Numerical work \cite{Hartnoll:2008kx,Denef:2009tp} suggests that if the scalar field violates the $AdS_2$ BF bound then the scalar field is unstable in the full black hole geometry (even when the $AdS_d$ BF bound is respected).  Moreover, it has been proved \cite{Dias:2010ma} that if the $AdS_2$ BF bound is satisfied then the scalar field is stable in full black hole geometry, i.e., stability of the near-horizon geometry implies stability of the full black hole.  Here we will prove that {\it instability} of the near-horizon geometry implies instability of the full black hole, in agreement with our conjecture.

Consider an extreme static black hole with geometry
\begin{equation}\label{eqn:staticgeom}
  ds^2 = -f(r) dt^2 + f(r)^{-1} dr^2 + r^2 d\Sigma_k^2.
\end{equation}
where $d\Sigma_k^2$ is the metric on a unit sphere if $k=1$, a unit hyperboloid if $k=-1$ and flat if $k=0$.  This metric encompasses the Schwarzschild(-AdS) and Reissner-Nordstr\"om(-AdS) black holes with various horizon topologies.

As the black hole is extreme, we can assume that it has a degenerate horizon at $r=r_+$, and hence that
\begin{equation}
  f(r) = \frac{(r-r_+)^2}{L^2} + \mathcal{O} (r-r_+)^3.
\end{equation}
The near horizon geometry is then $AdS_2 \times \Sigma_k$ where the $AdS_2$ has radius $L$.

In the full spacetime, the equation of motion of a scalar field $\Psi$ of mass $M$ can be written
\begin{equation} \label{Aeq}
 -\frac{\partial^2 \Psi}{\partial t^2} = {\cal B} \Psi,
\end{equation}
where
\begin{equation}
 {\cal B}\Phi \equiv f \left[ - \frac{1}{r^{d-2}} \pd_r \left( r^{d-2} f \pd_r \Psi \right) 
                        + \frac{1}{r^2} \hat{\nabla}^2 \Psi + M^2 \Psi \right],
\end{equation}
with $\hat{\nabla}$ the connection on $\Sigma_k$.  Now define the following inner product between functions defined on a surface of constant $t$ outside the horizon:
\begin{equation}
 ( \Psi_1, \Psi_2 ) = \int_{r_+}^\infty dr \, d\Sigma_k \, r^{d-2} f^{-1} \Psi_1 \Psi_2 .
\end{equation}
We impose boundary conditions that the functions of interest must decay sufficiently fast for this integral to converge at $r=\infty$, and they must vanish at least as fast as $(r-r_+)$ as $r \rightarrow r_+$ in order that the integral converges at $r=r_+$.  Now, if our functions decay fast enough at infinity, then ${\cal B}$ is self-adjoint with respect to this inner product.\footnote{Note that this is different to the self-adjointness of operators discussed in Section \ref{sec:decomposition}; we are integrating over the exterior region of the full spacetime, not just over the manifold $\Hcal$.}

We can estimate the lowest eigenvalue $\lambda_0$ of $\Bcal$ using the Rayleigh-Ritz method, noting that
\begin{equation} \label{rayleighritz}
 \lambda_0 \le \frac{ (\Psi, {\cal B} \Psi )}{(\Psi,\Psi)},
\end{equation}
for any function $\Psi$ satisfying the boundary conditions.

Suppose that $\lambda_0$ is negative, with $\Psi_0$ the associated eigenfunction. Then (\ref{Aeq}) has solutions 
\begin{equation}
  \Psi(t,r,x) = e^{\pm \sqrt{-\lambda_0 } t} \Psi_0.
\end{equation}
From the form of ${\cal B}$, it is easy to show that near $r=r_+$, the eigenfunction behaves as \begin{equation}
  \Psi_0 \sim \exp \left(-\frac{\sqrt{-\lambda_0} L^2}{r-r_+}\right) .
\end{equation}
Transforming to ingoing Eddington-Finkelstein coordinates ($dv = dt + dr/f$ so $t \sim v + L^2/(r-r_+)$ near $r=r_+$) reveals that the solution $e^{+ \sqrt{-\lambda_0 } t} \Psi_0$ is regular at the future horizon.  This grows exponentially with time, and hence represents an instability of the scalar field in the black hole background.

The idea now is to show that violation of the $AdS_2$ BF bound (\ref{eqn:ads2stabbound}) implies the existence of a trial function $\Psi$ with $(\Psi, {\cal B} \Psi) < 0$. This implies that $\lambda_0$ must be negative, hence the scalar field is unstable and the conjecture is proved.

To see how this works, consider the case of a 4D extreme Reissner-Nordstr\"om-AdS black hole, for which
\begin{equation}
 f(r) = \left( 1-\frac{r_+}{r} \right)^2 \left(k + \frac{3 r_+^2  + 2 r r_+ + r^2}{\ell^2} \right), 
\end{equation}
where $\ell$ is the $AdS_4$ radius.  This has a near-horizon geometry with
\begin{equation}
 \frac{1}{L^2} = \frac{6}{\ell^2} + \frac{k}{r_+^2}.
\end{equation}
Consider the following trial function (motivated by a similar example in Ref.\ \cite{Dias:2010ma}) 
\begin{equation} \label{trial}
 \Psi(r) = \frac{(r-r_+) \ell^{9/2}}{r^4 (r-r_+ + \epsilon \ell)^{3/2}},
\end{equation}
with $\epsilon >0$. This satisfies the boundary conditions required for self-adjointness of ${\cal B}$. As $\epsilon \rightarrow 0$, this gives
\begin{equation}
 (\Psi, {\cal B} \Psi ) 
    \equiv \int_{r_+}^\infty dr \, d\Sigma_k \, r^{2} \left( f (\partial_r \Psi)^2 
                                                                   + \mu^2 \Psi^2 \right) 
    = V_k \left( M^2 + \frac{1}{4 L^2} \right) \frac{\ell^9}{r_+^6} \log \left( \epsilon^{-1} \right)
      + {\ldots}
\end{equation}
where the ellipsis denotes terms subleading in $\epsilon$, and $V_k$ is the volume of $\Sigma_k$.  The $AdS_2$ BF bound states that the quantity in brackets on the RHS should be non-negative.\footnote{
More precisely: this is the BF bound for modes which are homogeneous on $\Sigma_k$.}
From the above expression we see that ${\cal B}$ admits a negative eigenvalue if this bound is violated.  Hence there is an instability of the scalar field when the $AdS_2$ BF bound is violated. 
The argument generalizes easily to $d>4$.\footnote{
Ref.\ \cite{Dias:2010ma} proved that {\it stability} of the near-horizon geometry implies stability of the full black hole for $k=-1,0$. Combining this with our result, we learn that, for these cases, the full black hole is stable if, and only if, its near-horizon geometry is stable. For $k=1$, stability of the near-horizon geometry is not sufficient to guarantee stability of the full black hole because the $AdS_2$ BF bound can be less restrictive than the $AdS_d$ bound.}

A similar example is the cohomogeneity-1 Myers-Perry-AdS black hole discussed in Section \ref{sec:nhmp}. For large black holes, the $AdS_2$ BF bound is more restrictive than that of $AdS_d$. 
Hence a scalar field can violate the $AdS_2$ BF bound but respect the $AdS_d$ BF bound.  Ref.\ \cite{Dias:2010ma} studied the case of a scalar field invariant under $\partial/\partial \psi$ (i.e.\ those modes corresponding to $m=0$ in Section \ref{sec:scalarfield}) and presented numerical evidence that such a scalar field is indeed unstable if its mass lies between the two BF bounds. Furthermore, it was proved that the scalar field (with $m=0$) is stable if it respects both bounds.

This example also can be understood using the argument above.  Even though the black hole is rotating, the fact that the scalar field is invariant under $\partial/\partial \psi$ implies that its equation of motion takes the form (\ref{Aeq}). The only difference is the form of ${\cal B}$:
\begin{equation}
 {\cal B}\Psi = \frac{V(r)}{h(r)^2}  
   \left[ - \frac{1}{r^{d-2}} \pard{r} \left( r^{d-2} V(r) \frac{\pd \Psi}{\pd r} \right) 
          - \frac{1}{r^2} \hat{\nabla}^2 \Psi + \mu^2 \Psi \right],
\end{equation}
where $\hat{\nabla}$ is the connection of the metric on $\CP{N}$. $\Bcal$ is self-adjoint with respect to the scalar product
\begin{equation}
 (\Psi_1,\Psi_2) = 2 \pi \int_{r_+}^\infty dr \, d\hat{\Sigma}_N\, r^{d-2} \frac{h(r)^2}{V(r)} \Psi_1\Psi_2,
\end{equation}
where $d\hat{\Sigma}_N$ is the volume element on $\CP{N}$.  Consider, for simplicity, the five-dimensional case (where $N=1$).  We can use the trial function (\ref{trial}) with the modification $r^4 \mapsto r^5$ (to improve the convergence at $r=\infty$).  The result is the same: $(\Psi, {\cal B} \Psi)$ is proportional to $\log (\epsilon^{-1})$ with a coefficient of proportionality that is negative if, and only if, the $AdS_2$ BF bound is violated.  Hence we have proved that the scalar field is unstable in the extreme black hole geometry if it violates the $AdS_2$ BF bound, in agreement with our conjecture.

Now recall from the introduction to this chapter that for the extreme Kerr black hole, we know that instability of the near-horizon geometry does not always imply instability of the full black hole. Even for a scalar field, there exist modes that violate the effective BF bound in the near-horizon geometry \cite{Bardeen:1999px}. So how does the above argument fail for Kerr? The key step above was to impose a symmetry condition on the scalar field that makes its equation of motion take the form  (\ref{Aeq}), in which first time derivatives are absent. For Kerr, eliminating first time derivatives implies that the scalar field must be axisymmetric, and axisymmetric modes do respect our conjecture. 

More generally, if we consider an extreme black hole with metric (\ref{eqn:ADM}) then the necessary and sufficient condition for the equation of motion of a massive scalar field to reduce to (\ref{Aeq}) is 
\begin{equation}
 N^I(x) \frac{\partial}{\partial \phi^I} \Psi = 0.
\end{equation} 
if we Fourier analyze $\Psi \propto e^{i m_I \phi^I}$ for integers $m_I$ then this equation reduces the axisymmetry condition (\ref{eqn:axicondition}) in the conjecture that we made in the introduction. If this condition is satisfied then the argument we have sketched above should apply. This explains why our conjecture should work for scalar fields.

\subsection{Gravitational perturbations}

We have sketched an argument that explains why a scalar field instability in the near-horizon geometry of an extreme black hole implies an instability of the full black hole, provided the scalar field satisfies the symmetry condition (\ref{eqn:axicondition}).  We are really interested in linearized gravitational perturbations.  If we attempt to repeat the same argument, we would need to convert the equations governing gravitational perturbations to something of the form 
\begin{equation} \label{Aeq2}
 -\frac{\partial^2 \Psi_\alpha}{\partial t^2} = {\cal A}_\alpha{}^\beta \Psi_\beta
\end{equation}
with $\Psi_\alpha$ a vector encoding the perturbation, and $ {\cal A}_\alpha{}^\beta$ a matrix operator self-adjoint with respect to a suitable inner product.  Can this be done?  For axisymmetric (i.e.\ $m=0$) metric perturbations of the Kerr black hole, in a certain gauge, it can indeed: a variational formula analogous to (\ref{rayleighritz}) is given in Chandrasekhar \cite[\S 114]{chandra}.  Hence the extreme Kerr black hole should obey our conjecture and, as we discussed in the introduction, it does.

What about higher dimensions? Can we bring the equations governing gravitational perturbations of, for example, a Myers-Perry black hole to the form (\ref{Aeq2}), provided the perturbation satisfies the symmetry condition (\ref{eqn:axicondition})? Evidence that this is indeed possible comes from recent work \cite{Dias:2010eu} on instabilities of cohomogeneity-1 MP black holes.  This work considered metric perturbations satisfying (\ref{eqn:axicondition}).  In the cases for which an instability was found, the time dependence was $e^{-i \omega t}$ where $\omega$ has positive imaginary part. In general, one would expect $\omega$ to be complex but it turned out that unstable modes had purely imaginary $\omega$. This would be explained if perturbations were governed by an equation of the form (\ref{Aeq2}) (with $ {\cal A}_\alpha{}^\beta$ self-adjoint), which predicts that $\omega^2$ should be real.

Perturbations of Myers-Perry black holes with a single non-vanishing angular momentum have also been considered \cite{Dias:2009iu}. Again, perturbations satisfying (\ref{eqn:axicondition}) were considered ((\ref{eqn:axicondition}) reduces to $m_1=0$ in this case). The critical mode associated to the onset of instability was identified. This mode has zero frequency, which suggests that unstable modes should have purely imaginary frequency (if unstable modes had $\omega$ with a non-zero real part then there is no reason why the mode at the threshold of instability should have $\omega=0$ rather than $\omega$ some non-zero real number).  Again, this suggests that a formula of the form (\ref{Aeq2}) exists for this situation.

In these two examples, it appears that the condition (\ref{eqn:axicondition}) is indeed sufficient to obtain an equation of the form (\ref{Aeq2}) governing gravitational perturbations (in a certain gauge).  This is encouraging evidence that it will indeed be possible to demonstrate that an instability of the near-horizon geometry of an extreme black hole will imply instability of the full black hole provided this symmetry condition is respected.

\chapter{Hidden symmetries of black rings} \label{chap:blackrings}
\section{Introduction}\label{sec:br1intro}
The final chapter of this thesis starts out along a slightly different track.  Rather than deriving general results about Einstein spacetimes in higher dimensions as we have done in the previous chapters, we will discuss the properties of a particular spacetime in five dimensions, namely the \emph{doubly-spinning black ring}.

This doubly-spinning black ring is an asymptotically flat solution to the vacuum field equations, discovered by Pomeransky and Sen'kov \cite{Pomeransky} using solution generating techniques for higher-dimensional Weyl solutions \cite{ER:weyl}.  It is a generalisation of the original Emparan-Reall black ring \cite{ER:2001} with rotation around the $S^2$ as well as the $S^1$.  The solution is rather more complicated than \cite{ER:2001}, but reduces to the balanced version of that solution in a particular limit.

Various authors have studied properties of this solution in the past.  Kunduri \etal \cite{Kunduri:2007vf} studied the extremal limit and associated near-horizon geometry, while Elvang \& Rodriguez \cite{Elvang:2007bi} studied its phase structure, asymptotics and horizon.  There is a more general version of the solution, corresponding to an `unbalanced' ring with conical singularities, which is explicitly presented in \cite{Morisawa}.  Much of the current literature on this spacetime is reviewed in \cite{ER:2008}, we give some brief details of its properties in Section \ref{sec:br2metric}.

More recently, and after the work of this chapter was complete, Chrusciel \etal \cite{Chrusciel:2009} produced an extensive rigorous study of various properties; and in particular constructed an analytic extension of the spacetime through its event horizon, as well as explicitly exhibiting the regularity of the metric on both the ergosurface and the axes of rotation.  In later work, Chrusciel \& Szybka \cite{Chrusciel:2010jg} proved stable causality of the domain of outer communications.

As the black ring is rotating, there is an ergosurface.  The topology of this ergosurface changes as the black ring parameters vary.  For a ring with sufficiently small rotation about the $S^2$, the topology is $S^1\times S^2$, as in the singly spinning case.  However, for more rapid $S^2$ rotation the ergosurface has topology $S^3 \cup S^3$, consisting of a small sphere around the centre of the ring, and a large sphere enclosing the entire ring.  There is a critical case, where a topology change occurs: the surface `pinches' on an $S^1$ (see Section \ref{sec:ergo}).  After the work of this section was mostly complete, a similar analysis \cite{Elvang:2008erg} appeared, as part of a paper discussing the properties of ergoregions in various higher-dimensional solutions.

At face value, the doubly-spinning black ring metric seems to be extremely complicated.  However, we will see that it admits some expected symmetry that makes it possible to study certain properties analytically.  Consider the Hamilton-Jacobi (HJ) equation, describing geodesic motion in the spacetime.  We will see in Section \ref{sec:br3geo} that, for both the singly-spinning and doubly-spinning rings, this equation admits separable solutions in the case of null, zero energy geodesics.  These null, zero energy geodesics can only exist inside the ergoregion, and correspond to massless particles coming out of the white hole horizon in the past, and falling into the black hole horizon in finite parameter time in the future.  On the other hand, the Klein-Gordon equation is not separable in ring-like coordinates, even if we restrict to looking for massless, time-independent solutions.  We will briefly discuss the reasons for this in Section \ref{sec:kg}.

In Section \ref{sec:br4coord}, we will see that the existence of these geodesics allows us to construct new coordinate systems for the black ring that are valid across the event horizon.  In the singly spinning case, it is possible to construct a new set of coordinates $(v,x,y,\tilde{\phi},\tilde{\psi})$ such that $v$, $\tilde{\phi}$, $\tilde{\psi}$ are constant along one of these geodesics.  These coordinate systems are regular at the future black hole horizon, and a particular subset of them cover the entire horizon.  The coordinate change given in \cite{ER:2006} is included in this family of coordinate systems, and hence this allows us to understand its geometric significance.

In the doubly-spinning case, the best approach is to use coordinates $(v,x,y,\tilde{\phi},\tilde{\psi})$ where only $\tilde{\phi}$ and $\tilde{\psi}$ are constant along the geodesics, and a change of coordinate $v$ is made that simply makes the metric regular at the horizon (rather than demanding that it is constant along the geodesics).  Using this approach, we are able to present explicitly a form for the doubly-spinning metric that is valid across the horizon.

We then see in Section \ref{sec:br5sym} that the null, zero energy separability of the Hamilton-Jacobi equation is related to the existence of a conformal Killing tensor in a 4-dimensional spacetime obtained by a spacelike Kaluza-Klein reduction of the black ring spacetime in the ergoregion, reducing along the asymptotically timelike Killing vector field.  Furthermore, a pair of conformal Killing-Yano tensors exist for the 4-dimensional spacetime if, and only if, the associated ring is singly-spinning.\np

\section{The Doubly-Spinning Black Ring Spacetime} \label{sec:br2metric}
Here, we briefly describe some properties of the doubly-spinning black ring spacetime, in order to set up notation, and gather together some results that will be useful in what follows.  We also explore some interesting properties of the ergoregion, which will be relevant later when we move on to consider geodesics.

\subsection{Form of the metric} \label{sec:met}
The doubly rotating ring solution can be written in the form
\begin{multline} \label{eqn:brmetric}
  ds^2 = -\frac{\Hyx}{\Hxy}(dt+\Omega)^2 \\ 
        + \frac{R^2 \Hxy}{(x-y)^2 (1-\nu)^2} \left[\frac{dx^2}{\Gx}-\frac{dy^2}{\Gy} 
                          +\frac{\Ayx d\phi^2-2\Lxy d\phi d\psi- \Axy d\psi^2}{\Hxy \Hyx}
                                              \right].
\end{multline}
The coordinates lie in ranges $-\infty<y\leq -1$, $-1\leq x \leq 1$ and $-\infty < t < \infty$, with $\psi$ and $\phi$ $2\pi$-periodic.  Varying the parameters $\la$ and $\nu$ changes the shape, mass and angular momentum of the ring.  They are required to lie in the ranges $0\leq\nu<1$ and $2\sqrt{\nu}\leq \la < 1+\nu$.

The functions $G$, $H$, $A$ and $L$ are moderately complicated polynomials, and are
given by
\begin{eqnarray*}
  \Gx  &=&(1-x^2)(1+\la x + \nu x^2),\\
  \Hxy &=& 1+\la^2-\nu^2+2 \la \nu(1-x^2)y+2 x \la(1-y^2 \nu^2)+x^2 y^2 \nu (1-\la^2-\nu^2),\\
  \Lxy &=& \la \sqrt{\nu} (x-y)(1-x^2)(1-y^2) \big[1+\la^2-\nu^2+2(x+y)\la \nu \\
       & & \quad\quad\quad\quad\quad\quad\quad\quad\quad\quad\quad\quad\quad - xy\nu(1-\la^2-\nu^2)\big],\\
  \Axy &=& \Gx(1-y^2)\left[((1-\nu)^2-\la^2)(1+\nu)+y\la(1-\la^2+2\nu-3\nu^2)\right] \\
       & & + \Gy\big[ 2\la^2+x \la((1-\nu)^2+\la^2)+x^2((1-\nu)^2-\la^2)(1+\nu) \\
       & & \quad \qquad + x^3 \la(1-\la^2-3\nu^2+2\nu^3)+x^4\nu(1-\nu)(1-\la^2-\nu^2)\big]
\end{eqnarray*}
The rotation is described by the 1-form $\Om = \Om_\psi(x,y) d\psi + \Om_\phi(x,y) d\phi$, where
\begin{equation}
  \Om_\psi = -\frac{R\la\sqrt{2((1+\nu)^2-\la^2)}}{\Hyx} \frac{1+y}{1-\la+\nu}\left(1+\la-\nu+x^2 y \nu (1-\la-\nu)+2\nu x(1-y)\right)
\end{equation}
and
\begin{equation}
  \Om_\phi = - \frac{R\la\sqrt{2((1+\nu)^2-\la^2)}}{\Hyx} (1-x^2)y\sqrt{\nu}.
\end{equation}

The form of the metric we use here is slightly different, although entirely equivalent, to that presented elsewhere in the literature.  Relative to \cite{Pomeransky}, the $\phi$ and $\psi$ coordinates have been exchanged, to be consistent with the singly spinning solution as presented in the review \cite{ER:2006}, and the functions $F(x,y)$ and $J(x,y)$ have been replaced with $A(x,y)$ and $L(x,y)$ defined such that
\begin{equation}
  F(x,y) = \frac{R^2 A(x,y)}{(1-\nu)^2 (x-y)^2} \eqand J(x,y) = \frac{R^2 L(x,y)}{(1-\nu)^2 (x-y)^2}.
\end{equation}
The length-scale parameter $R$ is related to their $k$ by $R^2=2k^2$.

It is useful at this stage to think a little bit more carefully about the properties of the metric functions $\Axy$ and $\Lxy$.  Is it immediately apparent from the definition of $\Axy$ that we can write it in the form
\begin{equation}
 \Axy = \Gx \ayb + \Gy \bxb 
\end{equation}
for some $\alpha(\xi)$ and $\beta(\xi)$.  Note that there is a freedom in our choice of these functions; we can add an arbitrary multiple of $G(\xi)$ to one and subtract it from the other without affecting $A(x,y)$ itself.  It turns out that the most convenient way of doing this is to pick 
\begin{equation}
  \alpha (\xi) = \nu (1-\xi^2) \left[- (1+\la^2) - \nu(1-\nu) + \la \xi (2-3\nu) - (1-\la^2)\xi^2  \right] 
\end{equation}
and
\begin{equation}
  \beta(\xi) = (1+\la^2) + \la \xi (1+(1-\nu)^2)
               - \nu\xi^2 (2\la^2+\nu(1-\nu)) - \la \nu^2\xi^3 (3-2\nu) - \nu^2\xi^4 (1-\la^2 + \nu(1-\nu)) .
\end{equation}
We can also do a similar thing for $\Lxy$.  If we set
\begin{equation}
  \gamma(\xi) = \la \sqrt{\nu} (1-\xi^2) (\la - (1-\nu^2) \xi - \la\nu \xi^2) 
\end{equation} 
then we find that
\begin{equation} 
  \Lxy = \Gx \gy - \Gy \gx .
\end{equation}

The ring-like coordinates can be related to two pairs of polar coordinates\\ $(t,r_1,\phi,r_2,\psi)$ via 
\begin{equation} 
  r_1 = R \frac{\sqrt{1-x^2}}{x-y} \eqand r_2 = R \frac{\sqrt{y^2-1}}{x-y},\label{eqn:planep}. 
\end{equation}
Note that, in these coordinates, the flat space limit takes the standard form
\begin{equation} 
  ds^2 = -dt^2 + dr_1^2 + r_1^2 d\phi^2 + dr_2^2 + r_2^2 d\psi^2 .
\end{equation} 
The black ring has a ring-like curvature singularity at $y\rightarrow -\infty$, which is the ring $(r_1,r_2)=(0,R)$ in the polar coordinates (\ref{eqn:planep}).

\subsection{Inverse Metric} \label{sec:inv}
The inverse metric will be useful later, so we give it here for convenience, it reads
\begin{multline} \label{eqn:inv} 
  \left( \frac{\pd}{\pd s} \right)^2 
    = -\frac{\Hxy}{\Hyx} \left( \frac{\pd}{\pd t} \right)^2 +
       \frac{(x-y)^2}{R^2 \Hxy} \Bigg[ (1-\nu)^2 \left(\Gx \left( \frac{\pd}{\pd x} \right)^2 
                                                       - \Gy \left( \frac{\pd}{\pd y} \right)^2
                                                 \right) \\
      + \frac{\Axy}{\Gx \Gy} \left( \frac{\pd}{\pd \phi} -\Om_\phi \frac{\pd}{\pd t} \right)^2 
      - \frac{2\Lxy}{\Gx \Gy} \left( \frac{\pd}{\pd \phi} -\Om_\phi \frac{\pd}{\pd t} \right) \left( \frac{\pd}{\pd \psi} 
                                     -\Om_\psi \frac{\pd}{\pd t}
                              \right)\\ 
          - \frac{\Ayx}{\Gx \Gy} \left( \frac{\pd}{\pd \psi} -\Om_\psi \frac{\pd}{\pd t} \right)^2 \Bigg].
\end{multline}
Note that
\begin{equation}
  \frac{\Axy}{\Gx \Gy} = \frac{\ayb}{\Gy} +  \frac{\bxb}{\Gx}
\end{equation}
separates into $x$ and $y$ components, as do the analagous expressions for $\Ayx$ and $\Lxy$.

\subsection{Horizon}\label{sec:hor}
The metric is singular when the function $G(y)$ vanishes.  The root at
\begin{equation} 
  y = y_h \equiv \frac{-\la + \sqrt{\la^2-4\nu}}{2\nu}
\end{equation}
is a coordinate singularity corresponding to an event horizon.  Elvang \& Rodriguez \cite{Elvang:2007bi} give a prescription for changing to new coordinates that are valid across the horizon, although it is very complicated to write the transformed metric down explicitly.  In Section \ref{sec:br4coord}, we will construct an alternative set of coordinates that are valid as we cross the horizon, by looking for coordinates adapted to a particular class of null geodesics.

When $\la=2\sqrt{\nu}$, $G(y)$ has a double root at $y=y_h$ and the black ring is extremal.  In this case, Ref.\ \cite{Kunduri:2007vf} derived the near-horizon geometry, and found that it is the same as that of a boosted extremal Kerr black string.  This allows one to search for instabilities of this spacetime using the methods derived in Chapter \ref{chap:decoupling}.

\subsection{Asymptotic Flatness}\label{sec:asymp}
This spacetime is (globally) asymptotically flat, but this is not manifest in the ring-like coordinates, where asymptotic infinity corresponds to the point $(x,y)=(-1,-1)$.  To see the asymptotics explicitly, we can make a change of variables $(x,y)\mapsto (\rho,\theta)$ by setting 
\begin{equation}
  x=-1+\frac{2R^2}{\rho^2}\frac{1+\nu-\la}{1-\nu} \cos^2 \theta \eqand 
  y=-1 - \frac{2R^2}{\rho^2}\frac{1+\nu-\la}{1-\nu} \sin^2 \theta,
\end{equation}
with $R\sqrt{(1+\nu-\la)/(1-\nu)} \leq \rho < \infty$ and $0\leq \theta \leq \pi$.  Therefore, for large values of $\rho$, the metric reduces to
\begin{equation}
  ds^2 \approx -dt^2 + d\rho^2 + \rho^2 (d\theta^2 + \cos^2 \theta d\phi^2 + \sin^2 \theta d\psi^2),
\end{equation}
which is 5-dimensional Minkowski space expressed in polar coordinates, with the angular variables having the correct periodicities.  This transformation was motivated by that given in \cite{Elvang:2007bi} (although the formula given in that paper is incorrect).

\subsection{Singly Spinning Limit}\label{sec:singly}
Since the coordinates used here vary slightly from those used in most papers on singly spinning rings, e.g.\ \cite{ER:2008,ER:2001,ER:2006,Hoskisson}, it is worth showing explicitly how this reduces to the original Emparan-Reall solution.

The singly spinning limit corresponds to setting $\nu=0$.  This reduces the metric functions to the following:
\begin{equation}
  \Gx =(1-x^2)(1+\la x), \qquad H(x,y) = 1+2x\la+\la^2 \equiv H(x),
\end{equation}
\begin{equation}
  \axb = \gx = \Lxy = 0,\qquad \beta(x) = H(x) , \qquad A(x,y) = H(x) G(y)
\end{equation}
and
\begin{equation}
  \Om = \Om_\psi(y) d\psi = -CR\frac{1+y}{H(y)}d\psi, \qquad \mathrm{where} \qquad 
  C\equiv \sqrt{2\la^2\frac{(1+\la)^3}{1-\la}}.\label{eqn:Cdef}
\end{equation}
The convenience of the limits here is our main motivation for working with the particular choices of $\alpha$ and $\beta$ that we made above.

The metric reduces to
\begin{equation} \label{eqn:singlymet}
  ds^2 = -\frac{\Hy}{\Hx}\big(dt+\Om_\psi(y) d\psi\big)^2 
         + \frac{R^2 \Hx}{(x-y)^2} \left[ \frac{\Gx}{\Hx} d\phi^2 
                                         + \frac{dx^2}{\Gx} - \frac{\Gy}{\Hy} d\psi^2 - \frac{dy^2}{\Gy} \right].
\end{equation}

\subsection{Ergoregion}\label{sec:ergo}
For the singly-spinning black ring, the ergoregion was first described in \cite{ER:2001}.  It is straightforward to see that, in our notation, the ergosurface is where $H(y)$ vanishes, which occurs at \begin{equation}y = y_e \equiv -\frac{1+\la^2}{2\la}.\end{equation}  Furthermore, we have that $ y_h < y_e < -1$, for all $\la$, so the ergoregion does indeed exist, and, like the horizon, has topology $S^1 \times S^2$ (like all surfaces $y=\mathrm{const}$ for $y \neq -1$).

Things become significantly more complicated in the doubly spinning case.  The ergosurface is defined by the vanishing of $H(y,x)$, so can described (locally) as a surface $y=y_e(x)$.

Note that $H(-1,-1)=(1-\nu)(1+\nu-\la)^2 > 0$, and therefore $H(y,x) > 0$ in some neighbourhood of asymptotic infinity.  Hence, far from the ring, $\pd / \pd t$ is indeed timelike as expected.  It can also be shown that, for all $x\in[-1,1]$, $y_e(x)>y_h$, and hence the horizon is always surrounded by an ergoregion, with no intersection between the ergosurface and the horizon.  This is in contrast to the Kerr case, where they touch at the poles.  There is a clear reason for this; in Kerr the poles are the points on the horizon that are left invariant under rotations generated by the angular Killing vector, but in the black ring there are no points on the horizon left invariant under $\pd / \pd \psi$.
 
For the singly-spinning ring, $H(-1)>0$, and hence the axis $y=-1$ lies outside the ergoregion, which must therefore have ringlike topology.  However, in the doubly-spinning case, for sufficiently large $\nu$, there are some values of $x$ for which $H(-1,x)<0$, and hence the ergosurface intersects the axis and can therefore no longer have the ring-like topology $S^1\times S^2$.

What is the new topology?  Note that
\begin{equation}
  H(-1,x)=H(-1,-x)=(1-\la)^2-\nu^2 + \nu x^2 \left(1-\la^2-\nu^2+2\la\nu\right)
\end{equation}
is even as a function of $x$, and that therefore
\begin{equation}
  H(-1,1)=H(-1,-1) = (1-\nu)(1+\nu-\la)^2 > 0.
\end{equation}
Thus, for all allowed values of $\la$ and $\nu$ we have that the point at the centre of the ring lies outside of the ergoregion.  As $\nu \rightarrow 1$ (and hence $\la\rightarrow 2$), the size of the ergoregion becomes larger and larger, but there is always a region near to the centre of the ring that remains outside it.  Thus, the ergosurface topology is that of two disconnected 3-spheres, $S^3\cup S^3$.
\begin{figure}[htb]
 \begin{center}
  \begin{minipage}[c]{0.9\textwidth}
   \includegraphics[width=\textwidth]{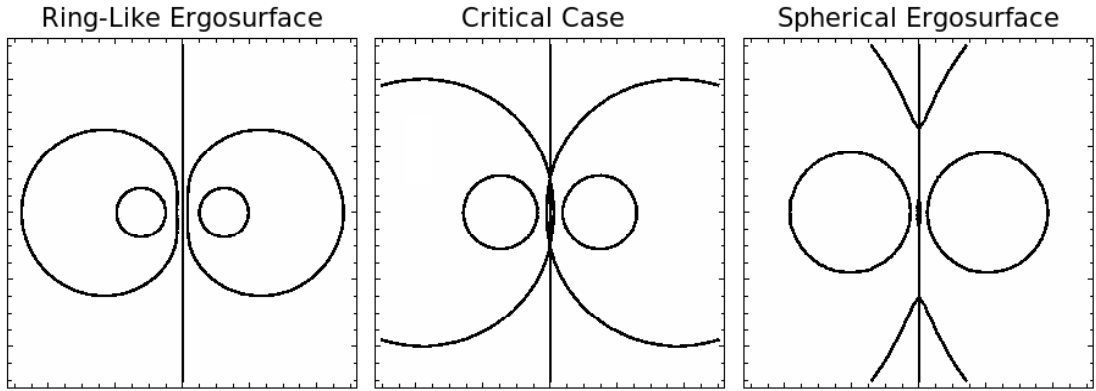}
   \caption[Two-dimensional projection of the shape of the ergoregion in the three difference cases that arise]{ \label{fig:ergoall} \small\it Two-dimensional projection of the shape of the ergoregion in the case $\nu=1/9$, for $\la = 7/9$ ($S^1\times S^2$ ergosurface), $\la = 8/9$ (critical case) and $\la = 1$ ($S^3 \cup S^3$ ergosurface).  The inner circles are the edge of the horizon, the outer lines the ergosurface and the central line the axis $y=-1$. (Plotted in $r_1$, $r_2$ coordinates.)}
  \end{minipage}
 \end{center}
\end{figure}

Note that $H(-1,x)$ is minimum at $x=0$, so to determine where in the black ring family the change of topology occurs we need to look at the case where
\begin{equation}
  H(-1,0)= 1+\la^2 - \nu^2 - 2\la = 0.
\end{equation}
This occurs when $\la=1-\nu$.  Note that we must have $\nu \leq 3-2\sqrt{2} \simeq 0.17$ for it to be possible to have this condition satisfied.  For this metric, we have that
\begin{equation}
  H(-1,x) = 4\nu^2 x^2 (1-\nu),
\end{equation}
so the ergosurface touches the $y=-1$ axis on the circle $x=0$, $y=-1$.  In the plane polar type coordinates (\ref{eqn:planep}), the locus of points where the ergosurface pinches is at $r_1 = R$, $r_2 = 0$, which makes clear that this is indeed a circle.  We will see later (\S \ref{sec:axisgeo}) that there exist stable `trapped' null geodesics orbiting around this circle.  Figure \ref{fig:ergoall} shows a 2D projection of the shape of the ergoregion in this case.

\begin{figure}[htb!]\begin{center} \leavevmode 
\input{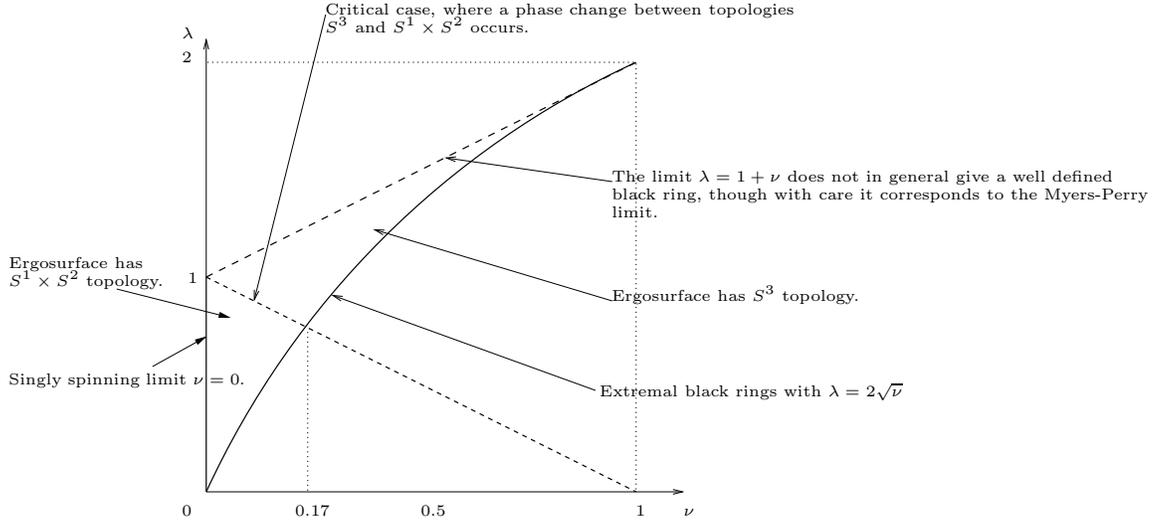}
\caption[The allowed parameter space for doubly spinning black rings.]{ \label{fig:ergophase} \small\it The allowed parameter space for doubly spinning black rings.}
\end{center}
\end{figure}

Finally, there is a nice intuitive way to think about why the ergoregion takes this form.  We can think, rather loosely, of the black ring as a Kerr black hole at each point around the $S^1$.  When the Kerr black hole is rotating rapidly (corresponding to rapid $S^2$ rotation of the black ring), its ergoregion becomes increasingly elliptical, so that eventually an observer near the centre of the ring feels frame dragging from the $S^2$ rotations on opposite sides of him simultaneously.  The effects cancel near the centre of the ring, leaving a region which does not lie in the ergoregion.  To summarise, Figure \ref{fig:ergophase} shows the parameter space for all allowed doubly-spinning black rings.

Recently, Cortier \cite{Cortier:2010cc} has provided a rigorous analysis of the ergosurface for this spacetime, confirming the results of this section.

\section{Geodesic Structure} \label{sec:br3geo}
Hoskisson \cite{Hoskisson} has studied in detail certain classes of geodesics for the singly spinning black ring.  In particular, he studies analytically families of geodesics restricted to the axes $y=-1$ and $x=\pm 1$, as well as performing numerical investigations into some more general possibilities.  Here, we concentrate on a different class of geodesics, which we can also find explicitly.  We show that, in the full doubly spinning case, the Hamilton-Jacobi equation is separable for null, zero energy geodesics.  Having demonstrated the separability of the HJ equation, we will then go on to analyse the behaviour of the geodesics that result from this.

\subsection{Conjugate momenta}
We look for geodesics by noting that they are extremal curves of the Lagrangian 
\begin{equation}
  \lag = \frac{1}{2}g_{\mu\nu}\dot{x}^\mu \dot{x}^\nu ,
\end{equation}
where a dot denotes differentiation with respect to an affine parameter $\tau$.  The conjugate momenta for this Lagrangian are
\begin{eqnarray}
  E \equiv - p_t &=& \frac{\Hyx}{\Hxy}(\dot{t}+\dot{\Om})\label{eqn:mom}\\
  \Phi \equiv p_\phi &=& -\Om_\phi E 
                         - \frac{R^2\big(-\Ayx \dot{\phi}+\Lxy \dot{\psi}\big)}{\Hyx (x-y)^2 (1-\nu)^2} \nn\\
  \Psi \equiv p_\psi &=& -\Om_\psi E 
                         - \frac{R^2\big(\Lxy \dot{\phi}+\Axy \dot{\psi}\big)}{\Hyx (x-y)^2 (1-\nu)^2} \nn\\
  p_x &=& \frac{R^2 \Hxy \dot{x}}{(x-y)^2 (1-\nu)^2 \Gx} \nonumber\\
  p_y &=& \frac{-R^2 \Hxy \dot{y}}{(x-y)^2 (1-\nu)^2 \Gy} \nonumber
\end{eqnarray}
where $\dot{\Om} \equiv \Om_\psi \dot{\psi} + \Om_\phi \dot{\phi}$.  The vector fields $\pd / \pd t$, $\pd / \pd \phi$ and $\pd / \pd \psi$ are Killing, so the conjugate momenta $-E$, $\Phi$ and $\Psi$ associated with them are conserved along any geodesics.

\subsection{The Hamilton-Jacobi Equation} \label{sec:hj}
Let $\ham (x^\mu,p_\nu)$ be the Hamiltonian for particle motion in this background, derived from the Lagrangian $\mathcal{L}(x^\mu,\dot{x}^\nu)$ in the usual way through a Legendre transformation
\begin{equation}
  \ham (x^\mu,p_\nu) \equiv p_\mu \dot{x}^\mu - \mathcal{L}(x^\mu,\dot{x}^\nu) 
                     = \frac{1}{2} g^{\mu\nu}p_\mu p_\nu.
\end{equation}
Now, consider the Hamilton-Jacobi equation
\begin{equation} \label{eqn:hj} 
  \frac{\pd S}{\pd \tau} + \mathcal{H}\left(x^\mu,\frac{\pd S}{\pd x^\nu}\right) = 0.
\end{equation}
This equation gives a useful way of encoding the geodesic structure of a system; the function $S$ contains information about all of the conjugate momenta $p_\mu = \pd S/\pd x^\mu$.  The aim of this approach is to give us an additional constant of motion.  The system is 5-dimensional, so we need 5 constants of motion in order to be able to completely integrate it.  Applying Noether's theorem to the Killing vectors $\pd / \pd t$, $\pd / \pd \psi$ and $\pd / \pd \phi$ has already given 3 of them, and we also impose the mass shell condition $g^{\mu\nu} p_\mu p_\nu = -\mu^2$ which gives a fourth.  Therefore, one more is required.

We look for additively separable solutions of the HJ equation \eqref{eqn:hj}.  Given our prior knowledge of 4 constants of motion, we make an ansatz
\begin{equation} 
  S(\tau,t,x,y,\psi,\phi) = \frac{1}{2}\mu^2 \tau - Et + \Phi \phi + \Psi \psi + S_x(x) + S_y(y),
\end{equation} where $\tau$ is an affine parameter along a geodesic, and $S_x$, $S_y$ are arbitrary functions of $x$ and $y$ respectively.  We hope that this ansatz will leave the HJ equation (\ref{eqn:hj}) in a separable form.

Inserting this ansatz into (\ref{eqn:hj}) gives, after some rearrangement,
\begin{multline}
  \Gx\dSdx-\Gy\dSdy = \frac{R^2 \Hxy}{(1-\nu)^2(x-y)^2} \left(- \mu^2 + \frac{\Hxy}{\Hyx}E^2\right) \\
           -\frac{\Hxy \Hyx}{\Axy \Ayx + \Lxy^2}\Big[\Axy(\Phi+\Om_\phi E)^2 - \Ayx(\Psi+\Om_\psi E)^2
             \qquad  \\
           - 2\Lxy (\Phi+\Om_\phi E)(\Psi+\Om_\psi E) \Big] .\label{eqn:hj1}
\end{multline}
At first glance, it appears that there is little hope of separating this.  However, it is possible to make some progress, using relations between the metric functions that are not immediately apparent from the solution as presented in \cite{Pomeransky}:
\begin{itemize}
\item Firstly, note the identity 
      \begin{equation} 
        \Axy \Ayx + \Lxy^2 \equiv \Gx \Gy \Hxy \Hyx (1-\nu)^2 .  \label{eqn:id}
      \end{equation}
      This simplifies (\ref{eqn:hj1}) to 
      \begin{multline}
        \Gx\dSdx-\Gy\dSdy= \frac{R^2 \Hxy}{(1-\nu)^2(x-y)^2}\left(-\mu^2 + \frac{\Hxy}{\Hyx} E^2\right) \\
                            - \frac{\left[\Axy(\Phi+\Om_\phi E)^2 - 2\Lxy (\Phi+\Om_\phi E)(\Psi+\Om_\psi E)  
                            - \Ayx(\Psi+\Om_\psi E)^2\right]}{\Gx\Gy(1-\nu)^2} \label{eqn:hj2}.
      \end{multline}
\item Writing 
      \begin{equation}
        \Axy = \Gx \ayb + \Gy \bxb 
      \end{equation}
      allows us to separate the $\Phi^2$ and $\Psi^2$ terms of (\ref{eqn:hj2}).
\item It is also possible to separate the $\Phi\Psi$ term using the relation
      \begin{equation}
        \Lxy = \Gx \gy-\Gx\gx.
      \end{equation}
\item It is not possible, in general, to separate the terms containing $\mu^2$, $E^2$, $E\Phi$ or $E\Psi$.
\end{itemize}

Therefore, the only separable solutions in these coordinates correspond to null ($\mu=0$), zero energy ($E=0$) geodesics, with $S_x$ and $S_y$ satisfying
\begin{multline}
    \Gx\dSdx- \frac{-\bxb \Phi^2 - 2\gx \Phi\Psi + \axb \Psi^2}{(1-\nu)^2 \Gx} \\
  = \Gy\dSdy - \frac{\ayb\Phi^2 - 2\gy \Phi\Psi - \byb \Psi^2}{(1-\nu)^2 \Gy} . \label{eqn:sep}
\end{multline}
Given this separation of variables, we can then immediately write
\begin{equation} 
  \mathrm{LHS} = \mathrm{RHS} = \frac{c}{(1-\nu)^2} 
\end{equation}
for some constant $c$.  This describes all possible null, zero energy geodesics.  $c$ is the extra constant required to allow the geodesic equations to be completely integrated in this case.  Unlike the Noether constants associated with Killing vectors it is quadratic in the momenta (see Section \ref{sec:br5sym}).  Are these geodesics physically realisable?  The answer is yes, but only in the ergoregion, where $\pd / \pd t$ is spacelike:
  Note that:
\begin{lemma}
  A null, zero energy geodesic in a black hole spacetime must be contained within the ergoregion.
\end{lemma}
\proof
Let $V$ be tangent to the geodesic, and $k$ be the (asymptotically timelike) generator of time translations.  Then, the null, zero energy condition is equivalent to saying that $V.V = 0$ and $k.V = 0$.  Given a null $V$, we can (locally) pick a basis for the tangent space of the form $\{V,n,m_i\}$ where $V.n = 1$, $m_i.m_j = \del_{ij}$, other dot products vanishing (c.f.\ Chapter \ref{chap:ghp}).

Thus, $k.V = 0$ iff $k \in \mathrm{span}(V,m_i)$ (a vector subspace of the tangent space).  Thus we can expand $k = k^0 V + k^i m_i$ and see that $k.k = k^i k^j \delta_{ij} \geq 0$, which is the definition of the ergoregion.$\Box$

It is worth emphasizing at this point that the separability of the HJ equation is a coordinate dependent phenomenon.  This is clearly illustrated by the fact that the HJ equation describing flat space geodesics is not separable in ring-like coordinates.  In fact, the general solution for flat space geodesics can be written in ring-like coordinates as
\begin{multline} 
  S(t,x,y,\phi,\psi;\tau) = K + \frac{1}{2}\mu^2 \tau - Et \\
       + \frac{R}{x-y} \left[ R_1\sqrt{1-x^2} \cos(\phi-\phi_0) + R_2 \sqrt{y^2-1} \cos(\psi-\psi_0)\right]
\end{multline} 
with $\phi_0$, $\psi_0$, $R_1$, $R_2$, $\mu^2$, $E$ and $K$ arbitrary constants.  This illustrates clearly that the failure of the Hamilton-Jacobi equation to separate for other classes of geodesics does not imply that it is impossible to find a new coordinate system in which separation occurs.

\subsection{Analysis of Paths of Ergoregion Geodesics}\label{sec:ergogeo}
Given the results of Section \ref{sec:hj}, we can study the paths of zero energy, null geodesics explicitly.  Since the zero energy, null condition is only realisable in the ergoregion, an observer moving along such a geodesic cannot pass through the ergosurface (though can fall through the horizon).

The separated Hamilton-Jacobi equation gives us that
\begin{equation}\label{eqn:ergox}
  \frac{R^4 \Hxy^2}{(x-y)^4 (1-\nu)^2} \dot{x}^2 + U(x) = 0 
\end{equation}
and
\begin{equation} \label{eqn:ergoy}
  \frac{R^4 \Hxy^2}{(x-y)^4 (1-\nu)^2} \dot{y}^2 + V(y) = 0
\end{equation} 
where
\begin{eqnarray}
  U(x) &=&  \bxb \Phi^2 + 2\gx \Phi\Psi - \axb \Psi^2 - c \Gx \\
  V(y) &=& - \ayb\Phi^2 + 2\gy \Phi\Psi + \byb \Psi^2 - c \Gy  .
\end{eqnarray}
These equations give coupled effective potential formulations for the motion, and we can use them to deduce the behaviour of this class of geodesics.  When dealing with effective potentials, it it usually useful to rearrange the equation such that one of the Noether constants (usually the energy) sits alone on the RHS, making it easy to understand how things change as that parameter varies.  Unfortunately, this is not possible in all cases here.

Note that, at least implicitly, we can use these equations to find $x$ as a function of $y$.  Dividing through, and noting that the prefactors with mixed $x$ and $y$ dependence cancel, we have that
\begin{equation}\label{eqn:dxdy}
  \left(\frac{dx}{dy}\right)^2 = \frac{U(x)}{V(y)} \quad \Rightarrow \quad 
             \int^x \frac{dx}{\sqrt{-U(x)}} = \int^y \frac{dy}{\sqrt{-V(y)}},
\end{equation} which gives us what we need.

Although these two effective potential equations are coupled to each other, the coupling arises only through the strictly positive pre-factor of the kinetic term.  Thus, the coupling has no effect on whether the potential is attractive or repulsive, or on its turning points.  Therefore, we can effectively treat the two parts independently when studying the qualitative behaviour of geodesics.

\subsubsection{Singly spinning case}\label{sec:singlygeo}
To begin with, it is easier to study these ergoregion geodesics in the singly spinning case $\nu=0$.  Here, the equations (\ref{eqn:ergox}) and (\ref{eqn:ergoy}) reduce to
\begin{equation}\label{eqn:ergoxsing}
  \dot{x}^2 + \frac{(x-y)^4 }{R^4 \Hx^2} \left[ \Phi^2\Hx-c\Gx \right] = 0 
\end{equation}
and
\begin{equation}
   \dot{y}^2 + \frac{(x-y)^4}{R^4 \Hx^2} \left[\Psi^2 \Hy - c\Gy \right] =0.
\end{equation}
Note that the ergoregion is given by $ -\frac{1}{\la} < y < - \frac{1+\la^2}{2\la} $ here, with topology $S^1 \times S^2$.  The $y$ motion is of the most immediate interest, since that governs how close to the horizon the path lies.

Care is needed when we get near to the axes $y=-1$ or $x=\pm 1$, since the angular coordinates $\psi$ or $\phi$ respectively become singular there.  However, this is a coordinate singularity, originating from the singularity at the origin in the plane polar coordinates (\ref{eqn:planep}), and hence we expect that taking limits like $y \rightarrow -1$ should be valid.  This can be confirmed in a straightforward (though messy) manner using the transformations to cartesian coordinates described in \cite{ER:2006}.

There are several cases to consider:
\paragraph{Case $c=0$:} Recall that $c$ is the separation constant from the Hamilton-Jacobi equation, so it parametrises a set of geodesic curves.  Now, we must have $c \geq 0$ to have an effective potential for $x$ that is non-positive somewhere, and hence some allowed solutions, so it is natural to begin with the bounding case $c=0$.  Note that $\Hx>0$ for all $x\in [-1,1]$, so in this case we also require $\Phi=0$ for any solution.  We must then have $\Psi\neq 0$ (else $\dot{y}=0$), and thus are left with the effective potential formulation \begin{equation}
  \dot{x}^2 = 0 \eqand \dot{y}^2 + \frac{(x-y)^4 \Psi^2\Hy}{R^4 \Hx^2} = 0.
\end{equation}
We have $\Hy <0$ everywhere inside the ergoregion, and $\Hy=0$ on the ergosurface, so the only turning point $\dot{y} = 0$ of the geodesic lies on the ergosurface.  The other coordinate $x$ is constant along these geodesics, so acts as an arbitrary constant rather than a dynamical variable in the $y$ equation, and in fact has no qualitative effects on the paths.  These solutions must correspond to geodesics that have come out of the white hole horizon in the past, move outwards away from the black ring until they just touch the ergosurface and then turn round and fall back into the black hole horizon in finite parameter time in the future.

\begin{figure}[hbt]
\begin{center}
\begin{minipage}[c]{0.9\textwidth}
\includegraphics[width=0.49\textwidth]{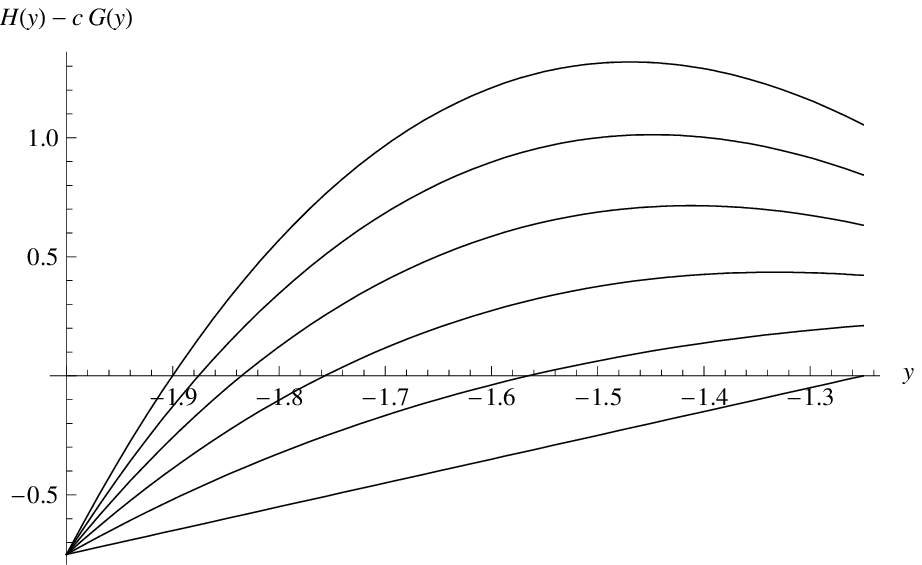}
\includegraphics[width=0.49\textwidth]{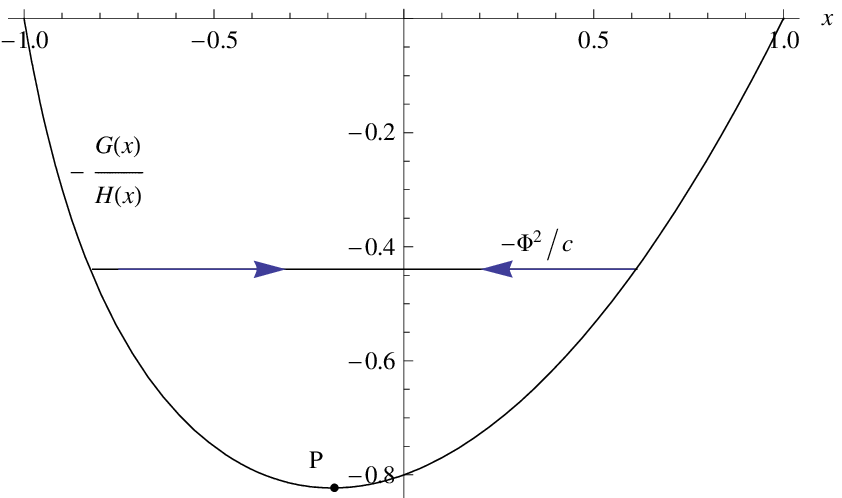}
\caption[Effective potentials for a class of zero-energy, null geodesics in the singly-spinning black ring spacetime.]{\it\small{\label{fig:singlypot}\label{fig:GoverH}\textbf{(a)} $\Hy - \bar{c} \Gy$ plotted against $y$ in the ergoregion ($-2 \leq y \leq -\frac{5}{4}$) for $\la=\tfrac{1}{2}$, $\nu=0$, for $\bar{c}=0,1,2,3,4,5$.  The potential in each case is bounded below by the $\bar{c}=0$ potential (the bottom line). \textbf{(b)} The $x$-motion effective potential $-G(x)/H(x)$ plotted against $x$.  This potential determines the allowed values of the constant $-\Phi^2/c$, an example path is plotted. (Figure has $\la=\frac{1}{2}$, $\nu=0$.)}} 
\end{minipage}
\end{center}
\end{figure}

\paragraph{Case $c>0$ and $\Phi=0$:} Here it is less easy to be explicit, but we can deduce the behaviour of these geodesics by relating them to the $c=0$ case.  The relevant equations are \begin{equation} \frac{R^4 \Hx^2}{(x-y)^4} \dot{x}^2 -  c\Gx = 0 \eqand \frac{R^4 \Hx^2}{(x-y)^4 \Psi^2 } \dot{y}^2 + \left[\Hy - \bar{c} \Gy \right] =0,\end{equation} where $\bar{c}\equiv c/\Psi^2$.  Since $\Gy<0$ outside the horizon, the effective potential for $y$-motion in the $c>0$ case is bounded below by that in the $c=0$ case, with equality only at $y=-1$ and $y=-\frac{1}{\la},$ that is at the horizon.  Thus, the geodesics in this case have the same qualitative behaviour, but stop short of the ergosurface before falling inwards again.  Figure \ref{fig:singlypot}(a) shows how the turning point of the geodesic (occurring where $\Hy - \bar{c}\Gy=0$) moves inwards as $c$ is increased.

Note that in this case, $x$ also varies, which makes integrating the motion explicitly far more difficult, though it has no real effect on the qualitative  form of the motion in $y$.  Since $c\Gx \geq 0 $ everywhere,  $x$ can take any value in $[-1,1]$.  This corresponds to the particle continually rotating around the $S^2$ part of the horizon as it moves in $y$.

\paragraph{Case $c>0$ and $\Phi>0$:}
In the singly spinning case, $\Phi$ does not enter into the effective potential for $y$, and therefore does not change the turning points in the $y$ motion.  However, the $x$ dynamics are now more interesting.  We can write the effective potential equation for $x$ as \begin{equation} \frac{R^4 \Hx}{c (x-y)^4} \dot{x}^2 - \frac{\Gx}{\Hx} = -\Phi^2/c,\end{equation} and hence see that there is a restriction on the values of $x$ that are possible.  For $\Phi^2/c=0$, any values of $x$ are allowed, but as $\Phi^2/c$ is increased, $x$ is restricted to an increasingly narrow range of values, corresponding to a centrifugal repulsion keeping the particle away from the axis $x=\pm 1$.  Rather than continuously rotating around the $S^2$, the particle follows a more complicated path, bouncing back and forth between two different extremal values of $x$.  This also gives us an upper bound on the values of $\Phi^2/c$ that are allowed, as shown by Figure \ref{fig:GoverH}(b).
There is a non-trivial fixed point in the $x$ potential (marked $P$ in Figure \ref{fig:GoverH}(b)), corresponding to an orbit at fixed $x$ when $\Phi^2/c$ takes its maximum allowed value.  It is messy to solve the cubic required to compute the exact location of the fixed point, and the corresponding maximum value of $\Phi^2/c$, and we do not do it here.

\subsubsection{Doubly spinning} \label{sec:dsgeo}
This concludes the possibilities for the singly spinning ring, and describes all of the possibilities for the behaviour of zero energy, null geodesics lying inside the ergosurface.  We now move on to the doubly spinning case.  Unfortunately, it is less easy to be explicit here, so we will limit ourselves to showing the existence of the geodesics, and discussing their properties in a couple of special cases.  The relevant effective potential equations are (\ref{eqn:ergox}) and (\ref{eqn:ergoy}).

In the previous section, we showed explicitly that the geodesics turned around before reaching the ergosurface (or in the limiting case, on the surface itself).  However, it is not strictly necessary to do this, since it can be deduced from well-known properties of geodesics.  Having found a section of a null, zero energy geodesic, we know that we can extend the geodesic indefinitely both forwards and backwards in time in a unique way, unless it hits a singularity (indeed, this is how one usually defines a singularity in a spacetime).  Furthermore, the geodesic extension of this curve must remain a null, zero energy geodesic.  Since the zero energy, null condition cannot be satisfied outside of the ergoregion, a particle travelling along such a geodesic cannot possibly pass through the ergosurface, and can only leave the ergoregion by passing through a horizon.

Now let's move on to consider some particular cases:
\paragraph{Case $\Phi=0$}
The full equations simplify significantly if we set one of the angular momenta to zero, specifically $\Phi$ (recall from the singly spinning case that there were no allowed zero-energy paths with $\Psi=0$; it is straightforward to show that the same applies here).  This leaves us with 
\begin{equation}
  U(x) = -\axb \Psi^2 - c\Gx \;\; \mathrm{and} \;\; V(y) = \byb \Psi^2 - c\Gy, 
\end{equation}
essentially leaving us with one tunable parameter $\bar{c} \equiv c / \Psi^2$.

\begin{figure}
 \begin{center}
  \begin{minipage}[c]{0.99\textwidth}
   \includegraphics[width=0.49\textwidth]{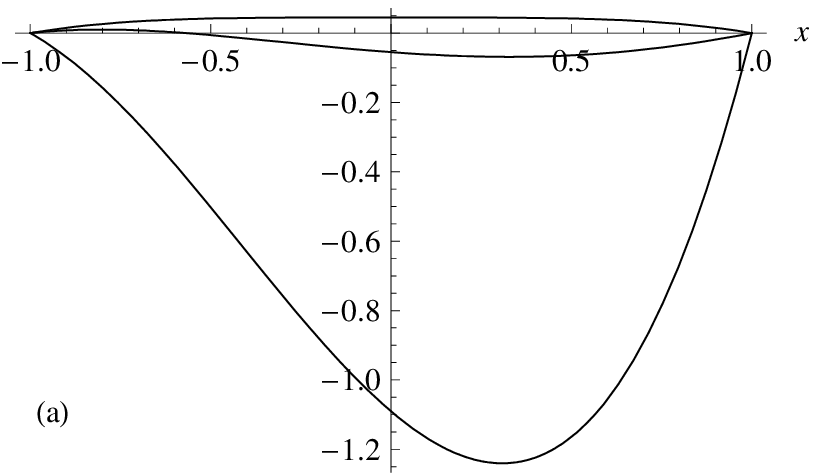}
   \includegraphics[width=0.49\textwidth]{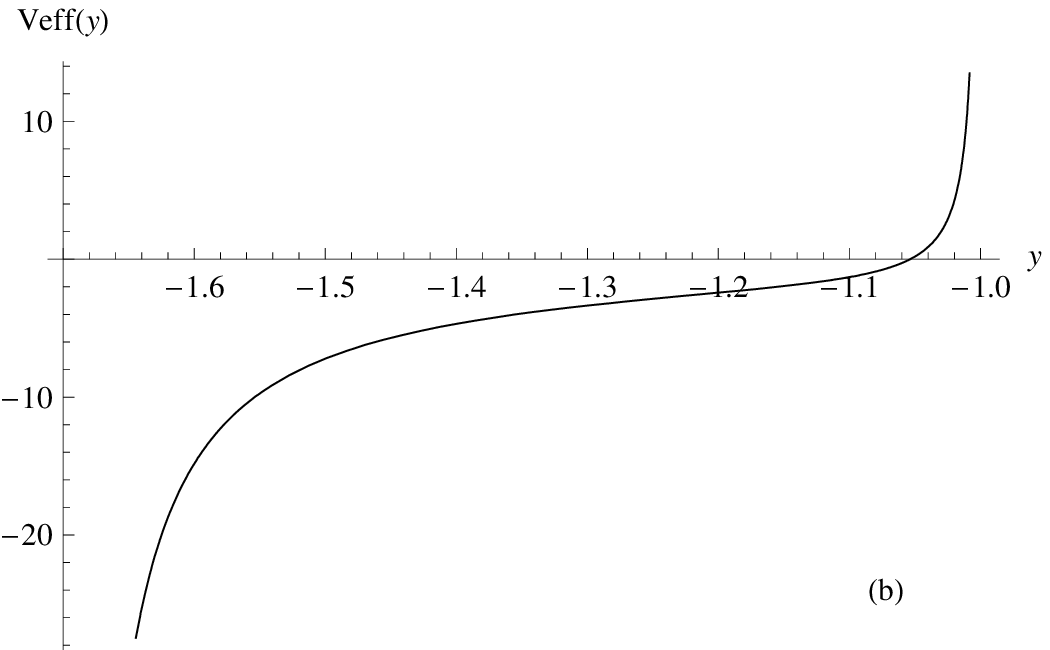}
   \caption[Effective potentials for a class of zero-energy, null geodesics in the doubly-spinning black ring spacetime.]{\it\small{\label{fig:doublyypot}\label{fig:doublyxpot}\textbf{(a)} Possible behaviours of the effective potential $U(x)$ for the doubly spinning ring in the case $\Phi=0$, for 3 different values of $\bar{c}=0,\frac{1}{10},\frac{276}{243}$.  The top curve gives no allowed geodesics, the bottom one allows all values of $x$. \textbf{(b)} The effective potential $\Veff(y) = -{\beta}(y)/G(y)$ for $y$-motion.  The horizon is located at the vertical axis on the left.  Both parts of this figure are plotted for $\la=\frac{1}{9}$, $\nu=\frac{7}{9}$, but the shape of the potentials is insensitive to changes in $\la$, $\nu$.}}
  \end{minipage}
 \end{center}
\end{figure}

Firstly, let us consider the motion in $x$.  Qualitatively there are 3 different possibilities for the potential $U(x)$ in this case, as shown in Figure \ref{fig:doublyxpot}(a).  Setting 
\begin{equation}
  \bar{c}_\pm = \frac{\nu}{1\mp\la+\nu} \left[ 2(1\pm\la)+\nu(1-\nu) \mp 3\la\nu \right],
\end{equation}
the cases are:
\begin{itemize}
  \item {\bf Case $\bar{c} < \bar{c}_-$:} Here, $U(x) >0$ for all $x$, i.e.\ there are no allowed values of $x$ and hence there can be no geodesics.  This occurs iff $U'(1) <0$, or equivalently $\bar{c} < \bar{c}_-$, and hence fixes a lower bound for $\bar{c}$.
  \item {\bf Case $\bar{c}_- <  \bar{c} < \bar{c}_+ $:} If $U'(1) > 0$, but also $U'(-1)>0$, then there are allowed geodesics, but they are restricted to a certain range in $x$, with the very `outside' of the ring excluded.
  \item {\bf Case $\bar{c} \geq \bar{c}_+$:} The $x$-range of the geodesics is entirely unrestricted, and they are free to loop all of the way around the $S^2$ of the ring.
\end{itemize}

Note that the middle case does not occur for the singly spinning ring (where $\bar{c}_+ = \bar{c}_-$), and the analysis above reduces to noting the geodesics exist only for $c\geq 0$.
For the $y$-motion, it turns out that the qualitative form of the motion is exactly the same as in the singly-spinning case.  Note that \begin{equation}V(y_h) = {\beta}(x) \Psi^2 < 0 ,\end{equation} so the potential is negative in some neighbourhood of the origin, and there is nothing (locally) to block a geodesic from crossing it.   Given this, the easiest way to study the behaviour away from the horizon is to express the potential equation as 
\begin{equation}
  \frac{R^4 \Hxy^2 \Psi^2}{(x-y)^4 (1-\nu)^2 (-\Gy)} \dot{y}^2 + V_{\mathrm{eff}}(y) = -\bar{c} 
\end{equation}
where $ \Veff(y) = -\byb/\Gy$.

To analyse the system, we need to study $\Veff(y)$ in the ergoregion.  Finding roots explicitly is hard, since it requires finding roots of a complicated quartic equation, but it can be shown (by differentiating and using the bounds on allowed values of $\la$, $\nu$ in various ways) that outside the horizon, for all values of $\la$ and $\nu$, $\Veff'(y)  > 0$ and hence there are no fixed points of the potential.  Therefore there can be no closed orbits.  As described above, we know from general principles of geodesics that all of these geodesics must turn around before getting outside of the ergoregion, so we know that $\Veff(y)$ must vanish for some $y<y_e(x)$.  However, this is only true for for a certain subset of $x$ values, and thus, there is a restriction on the allowed $x$ values near to the turning point of the geodesic.  We know that this must be consistent with the restrictions on $x$ obtained from analysing the $x$-potential.

\paragraph{General $\Phi$}
Note that $U(\pm 1) = (1-\nu)^2 (1+\nu\pm \la)^2 \Phi^2$, which is strictly positive for $|\Phi| >0$.  Therefore, the $x$ potential can no longer be categorised by finding derivatives at either end of the allowed range of $x$ values.  Instead, it is necessary to find turning points of the quartic $U(x)$ explicitly in order to find the range of $x$ values where $U(x)\leq 0$.  This is extremely messy, so we will not do it here.  However, there is a clear qualitative difference here; as soon as $|\Phi| >0$ there is a centrifugal barrier preventing these geodesics from touching the plane $x=\pm 1$.  Otherwise, the basic qualitative result is the same as in the singly spinning case; there is an upper bound on the allowed value of $\Phi^2/c$ in order to get allowed orbits of any kind.

The $y$ motion here is more complicated still, however numerical investigations suggest that, in general, no new behaviour occurs; that is all geodesics come out of the white hole and fall back into the black hole in finite proper time.

\begin{figure}[htb]
 \begin{center}
  \begin{minipage}[l]{0.5\textwidth}
   \includegraphics[width=\textwidth]{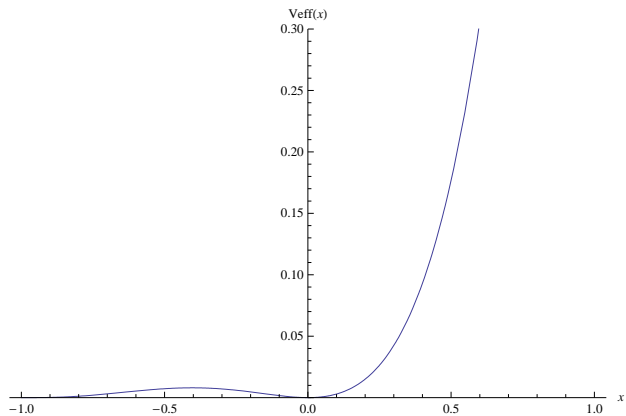}
  \end{minipage}
  \begin{minipage}[r]{0.8\textwidth}
   \caption[Effective potential for zero energy, null geodesics along the axis of a black ring with `pinched' ergosurface.]{\it\small \label{fig:critpot}The effective potential $\Veff(x)$ for zero energy, null geodesics along the axis in the critical case, where $\la=1-\nu$.  We see that the only possible orbit is a stable circular one at $x=0$. (Plot has $\nu=1/9$, $\la=8/9$, $\Phi=1$)}
  \end{minipage}
 \end{center}
\end{figure}

An exception to this occurs in the critical case $\la = 1-\nu$, where the ergoregion `pinches'.  Here, the motion in the case $\mu=E=\Psi=0$ is given by $\tfrac{1}{2} \dot{x}^2 + \Veff(x) = 0$ where
\begin{equation}
  \Veff (x) = \frac{x^2 (1+x)^4 (1-\nu)^2 \Phi^2}{ 4R^4 H(x,-1) },
\end{equation}
which means that there is a minimum at $x=0$, and hence a stable particle orbit there (see Figure \ref{fig:critpot}).  Thus, in this very special case, a lightlike particle can follow a trapped circular orbit at $(r_1,r_2)=(0,R)$, on the edge of the ergoregion.
\newpage
\subsection{Other analytically tractable geodesics}\label{sec:axisgeo}
While it is extremely unlikely to be possible to study all geodesics of this metric analytically, some progress can be made with finding geodesics that have particular symmetry.  In particular, it is possible to find geodesics lying entirely within surfaces that are fixed-point sets of the axial Killing vectors $\pd / \pd\phi$ and $\pd / \pd\psi$.  These surfaces are totally geodesic submanifolds, in that any geodesic that lies tangent to the submanifold at some point must lie entirely within the submanifold.  Typically, this introduces an extra constraint on the equations of motion, and reduces the problem to solving an ODE, the qualitative behaviour of which can be analysed via effective potential techniques.
 
In my paper \cite{ringgeo}, I derive the appropriate effective potential equations for these classes of geodesics, as well as commenting on some interesting generalities and special cases.  A full classification of all possibilities would be extremely complicated, since there is a large parameter space (any of $E$,$\la$,$\nu$,$\mu$ and one of $\Phi$ and $\Psi$ can vary), and the complexity of the potentials means that numerical graph plotting is the only reasonable approach to finding the shape of potentials in most cases.  We will not discuss any further details here.

\section{New Coordinate Systems} \label{sec:br4coord}

In order to fully understand a black hole spacetime it is necessary to construct a set of coordinates that cover the future black hole horizon.  This has been done for the singly spinning ring by Emparan \& Reall \cite{ER:2001,ER:2006}, and for the doubly spinning ring by Elvang \& Rodriguez \cite{Elvang:2007bi}.  The coordinates $(\bar{t},x,\bar{\phi},y,\bar{\psi})$ of \cite{Elvang:2007bi} are defined by setting
\begin{equation}
  d\bar{\phi} = d\phi - \frac{A}{y-y_h} dy, \quad
  d\bar{\psi} = d\psi - \frac{B}{y-y_h} dy \eqand 
  d\bar{t} = dt - \frac{C}{y-y_h} dy .\label{eqn:elvrodcoords},
\end{equation}
and attempting to find real constants $A$, $B$, $C$ such that divergences at the horizon in metric components cancel.  This works (with an additional quadratic term needed in the extremal case $\la=2\sqrt{\nu}$), and therefore proves that the horizon is regular.  However, it makes it hard to write down the transformed metric in a form that is manifestly regular at the horizon, to the extent that this has not been done in the literature.

In Section \ref{sec:ergogeo}, we found some null geodesics that cross the horizon.  Here, we construct a set of coordinates based around these geodesics, and find that these coordinates are valid across the horizon.  This provides some geometrical insight into why the choice of coordinates across the singly-spinning horizon in \cite{ER:2006} works, and also gives a more convenient set of coordinates for the doubly-spinning case than those of \cite{Elvang:2007bi}.

For convenience, we define functions $\zeta(y)$ and $\xi(x)$, related to the potentials of Section \ref{sec:ergogeo} by 
\begin{eqnarray}
  \xi(x) &\equiv & (1-\nu)^2 (-\bxb \Phi^2 - 2\gx \Phi\Psi + \axb \Psi^2 + c \Gx) = -(1-\nu)^2 U(x), \nn\\
  \zeta(y) &\equiv & (1-\nu)^2( \ayb\Phi^2 - 2\gy \Phi\Psi - \byb \Psi^2 + c \Gy) = -(1-\nu)^2 V(y).
\end{eqnarray}
Given this, the zero energy, null ergoregion geodesics of Section \ref{sec:br3geo} are described in our original set of coordinates by
\begin{eqnarray*}
  \dot{x}    &=& \pm \frac{(x-y)^2 (1-\nu)}{R^2 \Hxy}  \sqrt{\xi(x)} ,\\ 
  \dot{y}    &=& - \frac{(x-y)^2 (1-\nu)}{R^2 \Hxy} \sqrt{\zeta(y)}  ,\\
  \dot{\phi} &=& \frac{(x-y)^2}{R^2 \Hxy \Gx \Gy}\left[\Axy \Phi - \Lxy \Psi \right] ,\\
  \dot{\psi} &=& \frac{(x-y)^2}{R^2 \Hxy \Gx \Gy}\left[-\Lxy \Phi - \Ayx \Psi \right] ,\\
  \dot{t}    &=& -\Omy \dot{\psi} - \Omf \dot{\phi} \\
             &=& \frac{(x-y)^2 \la \sqrt{2(1+\nu-\la)(1+\nu+\la)}}{R \Hxy \Hyx \Gx \Gy} 
                             \bigg\{ \frac{1+y}{1-\la+\nu}(1+\la-\nu + x^2 y \nu (1-\la-\nu)  \\ 
             & & + 2\nu x (1-y))[-\Lxy \Phi - \Ayx \Psi ] + (1-x^2)y\sqrt{\nu} [ \Axy \Phi - \Lxy \Psi] \bigg\},
\end{eqnarray*}
where we have chosen signs such that $y$ is decreasing with $\tau$; that is we consider the part of a geodesic infalling across the horizon.\footnote{We could of course look at the outgoing sections of geodesics by simply changing the sign of the timelike coordinate, which we would expect to produce coordinates suitable for the white hole horizon rather than the black hole.}

Given a geodesic in this class, we might look to find a set of coordinates $(\tau,\tilde{x}^i)$ such that the geodesic is the line $\frac{d}{d\tau}(\tilde{x}^i) = 0$, where $\tau$ is an affine parameter along the geodesic (and $i=1,2,3,4$).  However, a nice feature of the original metric is the symmetry that exists between $x$ and $y$, so attempting to preserve this by transforming only three of the coordinates might well be desirable.  Our revised target will therefore be to find functions $\eta^i(x,y)$ such that
\begin{equation}
  \dot{t} - \frac{\pd \eta^t}{\pd x} \dot{x} - \frac{\pd \eta^t}{\pd y} \dot{y} 
        = \dot{\phi} - \frac{\pd \eta^\phi}{\pd x} \dot{x} - \frac{\pd \eta^\phi}{\pd y} \dot{y} 
        = \dot{\psi} - \frac{\pd \eta^\psi}{\pd x} \dot{x} - \frac{\pd \eta^\psi}{\pd y} \dot{y} =  0.
\end{equation}
Given this, we can construct the new coordinates $v=t-\eta^t$, $\tilde{\phi} = \phi - \eta^\phi$ and $\tilde{\psi} = \psi - \eta^\psi$.  These three new coordinates will be constant along the geodesics, and therefore we can expect the new coordinate system to be regular at the future horizon.  This is the most general form of coordinate change for these three coordinates that preserves the Killing vectors, that is with
\begin{equation} 
  \frac{\pd}{\pd v} = \frac{\pd}{\pd t}, \quad
  \frac{\pd}{\pd \tilde{\phi}} = \frac{\pd}{\pd \phi} \eqand
  \frac{\pd}{\pd \tilde{\psi}} = \frac{\pd}{\pd \psi}. 
\end{equation}  

\subsection{Singly-spinning case}
To see how this works, we will first apply it to the singly spinning case $\nu=0$.  Here, we have
\begin{equation}
  \zeta(y) =c{G(y)}-\Psi^2 {H(y)} \eqand \xi(x) = c{G(x)}-\Phi^2 {H(x)}, 
\end{equation}
with equations of motion 
\begin{align}
  \dot{x}    &= \pm \frac{(x-y)^2}{R^2 {H(x)}} \sqrt{{\xi(x)}}, &
  \dot{\phi} &= \frac{(x-y)^2}{R^2 {H(x)}} \left[ \frac{{H(x)}\Phi}{{G(x)}} \right], \\
  \dot{y}    &= - \frac{(x-y)^2}{R^2 {H(x)}} \sqrt{{\zeta(y)}}, &
  \dot{\psi} &= \frac{(x-y)^2}{R^2 {H(x)}} \left[ -\frac{{H(y)}\Psi}{{G(y)}} \right],\\
  \dot{t}    &= -\Omy \dot{\psi} = \frac{(x-y)^2}{R^2 {H(x)}} \left[ -\frac{CR(1+y)\Psi}{{G(y)}}\right], & & 
\end{align}
where the constant $C$ is defined by (\ref{eqn:Cdef}).

Then,
\begin{equation}
  \dot{\psi} - \frac{\pd \eta^\psi}{\pd x} \dot{x} - \frac{\pd \eta^\psi}{\pd y} \dot{y}
       = \frac{(x-y)^2}{R^2 {H(x)}} \left[ -\frac{{H(y)}}{{G(y)}}
                                           \mp \sqrt{{\xi(x)}} \frac{\pd \eta^\psi}{\pd x}  
                                           + \sqrt{{\zeta(y)}} \frac{\pd \eta^\psi}{\pd y}  \right] .
\end{equation}
If we pick
\begin{equation}
  \eta^\psi = \Psi \int_{y_0}^y \frac{ H(y') dy'}{G(y') \sqrt{\zeta(y')}}
\end{equation}
then this vanishes as required.  Similarly, picking
\begin{equation}
  \eta^\phi = \pm \Phi \int_{x_0}^x \frac{ H(x') dx'}{G(x') \sqrt{\xi(x')}} \eqand 
  \eta \equiv \eta^t = \Psi \int_{y_0}^y \frac{RC (1+y') dy' }{G(y') \sqrt{\zeta(y')}} 
\end{equation}
solves the analogous equations for $\phi$ and $t$.  Note that the lower (constant) bounds $y_0$ and $x_0$ on the integrals above are essentially arbitrary, though care must be taken to make sure that they leave well defined integrals.  A sensible choice, that is guaranteed to be well defined, is to pick $x_0 = 0$, and $y_0$ to be the turning point in the $y$ motion of the geodesic, that is $\zeta(y_0)=0$.  Note that $\pd \eta /\pd y$ and $\pd \eta^\psi /\pd y$ diverge at the horizon.  This is necessary in order to cancel the divergence at the horizon in the original coordinates, and analogous to what happens for coordinate changes across the horizon in more familiar cases.

The resulting change in the basis of 1-forms is
\begin{equation}
  dv = dt - \frac{CR(1+y)\Psi}{{G(y)} \sqrt{{\zeta(y)}}} dy, \qquad
  d\tilde{\psi} = d\psi - \frac{\Psi {H(y)}}{{G(y)} \sqrt{{\zeta(y)}}} dy, \qquad
  d\tilde{\phi} = d\phi \mp \frac{\Phi {H(x)}}{{G(x)} \sqrt{{\xi(x)}}} dx,
\end{equation}
and this puts the metric (\ref{eqn:singlymet}) into the form
\begin{multline}
  ds^2 = -\frac{{H(y)}}{{H(x)}}(dv +\Omy d\tilde{\psi})^2 + \frac{R^2 {H(x)}}{(x-y)^2} \Bigg[ \frac{c dx^2}{c{G(x)}
          -\Phi^2 {H(x)}} - \frac{c dy^2}{c{G(y)}-\Psi^2{H(y)}} \\
          \pm \frac{2\Phi d\tilde{\phi}dx }{\sqrt{c{G(x)}-\Phi^2 {H(x)}}} 
          -  \frac{2\Psi  d\tilde{\psi}dy }{\sqrt{c{G(y)}-\Psi^2 {H(y)}}}
          +  \frac{{G(x)}}{{H(x)}} d\tilde{\phi}^2 - \frac{{G(y)}}{{H(y)}} d\tilde{\psi}^2 \Bigg] .
\end{multline}
This nicely preserves the $x \leftrightarrow y$, $\phi \leftrightarrow \psi$ symmetry of the original metric.  The inverse metric is given by
{\small \begin{multline} \label{eqn:newsinginv}
  g^{\mu\nu} \frac{\pd}{\pd x^\mu}\frac{\pd}{\pd x^\nu}
      = -\frac{{H(x)}}{{H(y)}} \left( \frac{\pd}{\pd v} \right)^2 
           + \frac{(x-y)^2}{R^2 {H(x)}} \Bigg[ {G(x)} \left( \frac{\pd}{\pd x} \right)^2 
           -  {G(y)} \left( \frac{\pd}{\pd y} \right)^2\qquad\qquad \\
    \qquad\qquad\qquad\qquad\qquad\quad
          \mp \frac{ 2\Phi {H(x)}}{ \sqrt{{\xi(x)}}}  \frac{\pd}{\pd \tilde{\phi}}  \frac{\pd}{\pd x} 
         + \frac{ 2\Psi {H(y)}}{ \sqrt{{\zeta(y)}}} \left( \frac{\pd}{\pd \tilde{\psi}} 
         - \Omy \frac{\pd}{\pd v}\right)  \frac{\pd}{\pd y} \\
         + c \frac{{H(x)}}{{\xi(x)}} \left( \frac{\pd}{\pd \tilde{\phi}}  \right)^2 
         - c \frac{{H(y)}}{{\zeta(y)}} \left( \frac{\pd}{\pd \tilde{\psi}} - \Omy \frac{\pd}{\pd v} \right)^2 \Bigg] .
\end{multline}}
Since the components of both the metric, and its inverse are regular at $y=y_h$, it is now a well defined coordinate system across the horizon $G(y)=0$.  Note that, like in the original form of the metric, there is a coordinate singularity as we approach $x=\pm 1$, which has no physical significance, and is analogous to the singularity at the origin of plane polar coordinates.  There is a further subtlety here though, since we saw in \S\ref{sec:ergogeo} that for $|\Phi|>0$ the allowed range of $x$ along the geodesic is limited (since ${\xi(x)}<0$ for some $x\in [-1,1]$).  Thus, these coordinates can only cover the entire horizon when we set $\Phi=0$.

The simplest geodesics discussed in \S \ref{sec:ergogeo}  were those with $c=0=\Phi^2$, and $\Psi^2 > 0$.  This leaves us with the transformation 
\begin{equation}
  dv = dt - \frac{CR(1+y)}{{G(y)}\sqrt{-{H(y)}}} dy, \quad
  d\tilde{\psi} = d\psi + \frac{\sqrt{-{H(y)}}}{{G(y)}} dy \eqand d\tilde{\phi} = d\phi 
\end{equation}
which is precisely the coordinate change given in \cite{ER:2006}, leaving the metric in the form
\begin{equation}
  ds^2 = -\frac{{H(y)}}{{H(x)}}(dv +\Omy d\tilde{\psi})^2 + \\
  \frac{R^2 {H(x)}}{(x-y)^2} \left[\frac{dx^2}{G(x)} +  \frac{{G(x)}}{{H(x)}} d\tilde{\phi}^2 
                     - \frac{2 d\tilde{\psi}dy }{\sqrt{- {H(y)}}} - \frac{{G(y)}}{{H(y)}} d\tilde{\psi}^2 \right] .
\end{equation}
Thus, this technique has generated a family of possible coordinate transformations, including those that are already known, and attached a geometric significance to them.  Note that the coordinates are only valid out as far as the turning point $y=y_0$ of the geodesic in question, that is for the region $-\infty < y < y_0$ where $\zeta(y_0)=0$.  There is still a coordinate singularity at $x=\pm 1$, as in the original set of coordinates.

Note that, if we wished, it would be possible to make a further change of coordinates $x\mapsto \tilde{x}$ such that $\dot{\tilde{x}}=0$ along the geodesics.  However, the range of the new coordinate $\tilde{x}(x,y)$ is messy (and $y$ dependent).  Having done this, $y$ is the only coordinate varying along the geodesics, and it does so monotonically if we only consider the ingoing part of the geodesic (as we have been doing).  Thus, we can write $x=x(y)$ along the geodesic, and hence use the affine parameter $\tau$ rather than $y$ as the remaining coordinate, leaving us with the type of coordinate system originally suggested above.  We do not present any of this explicitly here, since the resulting form of the metric is extremely messy, and not obviously of any practical use.

\subsection{Doubly-spinning case}
Now we move on to the doubly spinning case.  Here, the form of the geodesic equations is more complicated, so we expect the coordinate change associated with it to be more complicated as a result.  We need to solve the PDEs
\begin{equation}
  \dot{\phi} - \frac{\pd \eta^\phi}{\pd x} \dot{x} - \frac{\pd \eta^\phi}{\pd y} \dot{y} = 0 \eqand
  \dot{\psi} - \frac{\pd \eta^\psi}{\pd x} \dot{x} - \frac{\pd \eta^\psi}{\pd y} \dot{y} = 0
\end{equation}
which can be written as
\begin{equation}
  \frac{(x-y)^2}{R^2 \Hxy} \left[ \left(\frac{\bxb \Phi+\gx \Psi}{\Gx} 
                                 \mp \sqrt{\tzx} \frac{\pd \eta^\phi}{\pd x} \right)
                                 + \left(\frac{\ayb \Phi - \gy \Psi}{\Gy} 
                                 + \sqrt{\zy} \frac{\pd \eta^\phi}{\pd y} \right)  \right] = 0
\end{equation}
and
\begin{multline}
  \frac{(x-y)^2}{R^2 \Hxy} \left[ \left(\frac{\gx\Phi - \axb \Psi}{\Gx}
                                        \mp \sqrt{\tzx} \frac{\pd \eta^\psi}{\pd x} \right)  
                                        + \left(\frac{-\gy\Phi - \byb \Psi}{\Gy} 
                                        + \sqrt{\zy} \frac{\pd \eta^\psi}{\pd y} \right)  \right] \\ = 0 .
\end{multline}
They have the obvious separable solutions
\begin{equation} \label{eqn:etaf}
  \eta^\phi = \pm \int_{x_0}^x \frac{{\beta}(x') \Phi + \gamma(x') \Psi}{G(x') \sqrt{\xi(x')}} dx' 
              + \int_{y_0}^y \frac{ -{\alpha}(y')\Phi + \gamma(y') \Psi}{G(y') \sqrt{\zeta(y')}} dy'
\end{equation}
and
\begin{equation} \label{eqn:etay}
  \eta^\psi = \pm \int_{x_0}^x \frac{ \gamma(x') \Phi - {\alpha}(x') \Psi}{G(x') \sqrt{\xi(x')}} dx' 
              + \int_{y_0}^y \frac{ \gamma(y') \Phi + {\beta}(y') \Psi}{G(y') \sqrt{\zeta(y')}} dy'
\end{equation}

However, it is less easy to solve the PDE
\begin{equation}\label{eqn:tdot}
  \dot{t} - \frac{\pd \eta}{\pd x} \dot{x} - \frac{\pd \eta}{\pd y} \dot{y} =0,
\end{equation}
since the dependence of $\Omf$ and $\Omy$ on both $x$ and $y$ means that the equation does not separate.  In order to get a new set of coordinates that is analogous to that of the singly spinning case, we might hope to be able to set $v = t - \eta^t$ where
\begin{equation}
  d\eta^t = -\Omy(x,y) d\eta^\psi - \Omf(x,y) d\eta^\phi,\label{eqn:nogood}
\end{equation}
which would give us the convenient result $dt+\Omf d\phi +\Omy d\psi = dv +\Omf d\tilde{\phi} +\Omy d\tilde{\psi}$.  Unfortunately, the right hand side of (\ref{eqn:nogood}) is not a total derivative for $\nu>0$, so this is impossible.

Instead, we might take either of two different approaches:
\begin{itemize}
  \item Look for an exact solution of (\ref{eqn:tdot}), even if it cannot be written in a convenient separable form like (\ref{eqn:etaf}), (\ref{eqn:etay}). 
  \item Give up on completely solving (\ref{eqn:tdot}), and instead just look for an $\eta$ such that $v=t-\eta$ has   
        \begin{equation} \label{eqn:tfinite}
          \dot{v} = \dot{t} - \frac{\pd \eta}{\pd x} \dot{x} - \frac{\pd \eta}{\pd y} \dot{y} < \infty
        \end{equation}
        at the horizon $y=y_h$, as we move along one of the geodesics.
\end{itemize}
We have investigated both of these possibilities.  In Appendix \ref{app:blackrings}, we see that it possible to construct an exact solution to (\ref{eqn:tdot}), but that it contains functions that can only be written down implicitly in terms of the inverse of certain functions defined by integrals.  This is clearly not desirable when trying to write down a metric of practical use for calculations, and hence we resort to looking for a new time coordinate $v$ that is merely finite at the horizon, rather than constant everywhere along the geodesic.

As described above, Elvang \& Rodriguez \cite{Elvang:2007bi} showed how to construct coordinates (\ref{eqn:elvrodcoords}) that are valid across the horizon.  It is useful to pause for a moment to understand how their change of coordinates works, since it will be useful in constructing a suitable $v$ here.

In order for the coordinate system across the horizon to be well-defined, we require that the divergence in $g_{yy}$ has been removed by this coordinate change, and that no new divergences are introduced in any of $g_{\bar{t} y}$, $g_{\bar{\phi} y}$ or $g_{\bar{\psi} y}$.  A straightforward computation shows that these conditions are equivalent to requiring that $A$,$B$,$C$ can be chosen such that, for all $x$,
\begin{eqnarray}
  C+\Omf(x,y_h) A + \Omy(x,y_h) B &=& 0 \label{eqn:hor1} \\
  A(y_h,x) A - L(x,y_h) B         &=& 0 \label{eqn:hor2} \\
  -L(x,y_h) A - A(x,y_h) B        &=& 0 \label{eqn:hor3}\\
  \lim_{y\rightarrow y_h} \Bigg[ - \frac{\Hxy}{\Gy} + \frac{A}{y-y_h} \left( \frac{\Ayx A - \Lxy B }{ \Hyx (y-y_h)} \right)
                                + \;\;\;\;\;\;\;\;\;\;\;\;\;\;\;   
                           && \nonumber \\ 
        \frac{B}{y-y_h} \left( \frac{-\Lxy A - \Axy B }{ \Hyx (y-y_h)} \right) \Bigg] &<& \infty \label{eqn:hor4}.
\end{eqnarray}

It is not immediately obvious that it is possible to satisfy these conditions simultaneously, though of course it must be if the doubly-spinning black ring is a well defined black hole spacetime.  Expanding in $x$ shows that (\ref{eqn:hor1}),(\ref{eqn:hor2}),(\ref{eqn:hor3}) have a 1-parameter family of solutions given by
\begin{equation}
  \frac{C}{B} = R\sqrt{2\left(\frac{1+\nu+\la}{1+\nu-\la} \right)} 
\end{equation}
and
\begin{equation}\label{eqn:AoverB}
  \frac{A}{B} = -\frac{\sqrt{\nu} (1+y_h^2)}{\la y_h} =\frac{\sqrt{\nu}}{\la} [\la - y_h (1-\nu)] 
              =\frac{ \gamma(y_h)}{{\beta}(y_h)} = -\frac{{\alpha}(y_h)}{\gamma(y_h)}.
\end{equation}
Putting this into (\ref{eqn:hor4}) fixes $B$, and hence $A$ and $C$.  Note that carrying out this last step explicitly is very fiddly, and its validity relies on the non-trivial fact that
\begin{equation} 
  \frac{H(x,y_h) H(y_h,x)}{ (A/B)^2 \axb + 2(A/B) \gx - \bxb} = \mathrm{constant} , 
\end{equation}
where $A/B$ is given by \eqref{eqn:AoverB}.

How does this link in to our solutions above?  We will see below that our change in coordinates makes the metric finite at the horizon, and hence it can only differ from the coordinate change of \cite{Elvang:2007bi} by a finite amount, that is as $y\rightarrow y_h$,
\begin{equation}\label{eqn:horlims}
   (y-y_h) \frac{\pd \eta^\phi}{\pd y} \rightarrow A \eqand (y-y_h) \frac{\pd \eta^\psi}{\pd y} \rightarrow B .
\end{equation}
Explicit computation confirms that this is the case.  Furthermore, we will see below that our change of coordinates $(\phi,\psi) \rightarrow (\tilde{\phi},\tilde{\psi})$ renders the $R^2/(x-y)^2$ part of the line element finite for any choice of $v$, and hence we do not need to do the fiddly computation to work the value of $B$ using (\ref{eqn:hor4}), but can merely read it off from (\ref{eqn:horlims}), that is 
\begin{equation}
   B = \lim_{y\rightarrow y_h}\left[(y-y_h) \frac{\pd \eta^\psi}{\pd y} \right] 
     = \frac{\gamma(y_h) \Phi + {\beta}(y_h)\Psi}{(1-y_h^2)\sqrt{(\la^2-4\nu)\zeta(y_h)}}.
\end{equation}
This is a significantly easier approach for getting this result.

Given this, we can immediately see that a valid change of time coordinate, to render the metric finite in the non-extremal case, is to set
\begin{equation}
  dv = dt - \frac{C}{y-y_h}dy,\quad \mathrm{where} \quad C 
     = R\sqrt{2\left(\frac{1+\nu+\la}{1+\nu-\la} \right) } \frac{\gamma(y_h) \Phi 
       + {\beta}(y_h)\Psi}{(1-y_h^2)\sqrt{(\la^2-4\nu)\zeta(y_h)}}. 
\end{equation}
This can be made slightly neater if we write 
\begin{equation}
  dv = dt - \frac{D (\gamma(y) \Phi + {\beta}(y)\Psi) }{G(y)\sqrt{\zeta(y)}}dy \qquad \mathrm{where} \qquad 
   D = R \sqrt{2\left(\frac{1+\nu+\la}{1+\nu-\la}\right) },
\end{equation}
which has the correct limit at the horizon, and will allow the new metric to be written more conveniently.

This transforms the first part of the metric via 
\begin{equation} 
  dt + \Omf d\phi + \Omy d\psi = dv + \Omf d\tilde{\phi} + \Omy d\tilde{\psi} + \tilde{\Om}_x dx 
                                 + \tilde{\Om}_y dy \equiv dv + \tilde{\Omega} 
\end{equation}
where
\begin{equation} 
  \tilde{\Om}_x = \pm \frac{ \Phi (\Omf \bxb + \Omy \gx) + \Psi (\Omf \gx - \Omy \axb)}{\Gx \sqrt{\tzx}}
\end{equation}
and
\begin{equation} 
  \tilde{\Om}_y = \frac{ \Phi (D \gy - \Omf \ayb - \Omy \gy) 
                        + \Psi (D \byb + \Omf \gy + \Omy \byb)}{\Gy \sqrt{\zy}}. \label{eqn:Omyt} 
\end{equation}
These are fairly complicated, but are suitably regular as we approach the horizon (though this regularity is not immediately manifest from looking at \eqref{eqn:Omyt}).  Furthermore, they remain valid in the extremal limit, while the original approach of \cite{Elvang:2007bi} needs additional corrections in this case.

\subsubsection{Transformed metric} \label{sec:transmet}
Given the above form for $ \tilde{\Om} $, we find that the metric can be written in the new coordinates as
{\small\begin{multline} \label{eqn:newmet} 
  ds^2 = -\frac{\Hyx}{\Hxy} ( dv+ \tilde{\Om} )^2 + \\
         \frac{R^2 \Hxy}{(x-y)^2 (1-\nu)^2} \Bigg[ \left(\frac{\tzx}{\Gx} + \frac{\Ayx\tx^2 
                            - 2\Lxy \tx \cx - \Axy \cx^2}{\Gx ^2 \Hxy\Hyx} \right) d(x,y)^2  \\
       \qquad  + 2\Bigg( - \frac{c (1-\nu)^2}{\sqrt{\zy}} dy
                  + \frac{\Ayx \tx - \Lxy \cx }{\Gx \Hxy \Hyx} d\tilde{\phi} \\
       \qquad\qquad\qquad\qquad\qquad\qquad\qquad\qquad\qquad
                  - \frac{\Lxy \tx + \Axy \cx }{\Gx \Hxy \Hyx}d\tilde{\psi} \Bigg)  d(x,y) \\
         - 2(1-\nu)^2 (\Phi d\tilde{\phi} + \Psi d\tilde{\psi}) \frac{dy}{\sqrt{\zy}} 
           + \frac{\Ayx d\tilde{\phi}^2-2\Lxy d\tilde{\phi} d\tilde{\psi}- \Axy d\tilde{\psi}^2}{\Hxy \Hyx} \Bigg] , 
\end{multline}} 
where
\begin{equation}
  d(x,y) \equiv  \left( \pm\frac{dx}{\sqrt{ \tzx}} + \frac{dy}{\sqrt{\zy}} \right),
\end{equation}
\begin{equation}
  \theta(x)  \equiv  {\beta}(x) \Phi + \gamma(x) \Psi \eqand
  \chi(x) \equiv \gamma (x) \Phi - {\alpha}(x) \Psi .
\end{equation}

In the singly spinning case we were able to maintain the $x\leftrightarrow y$ symmetry after the change of coordinates, but this turns out to not be possible here if we want to write the metric in a manner that is manifestly well defined as we cross the horizon.  As a result, this form of the metric is somewhat unpleasant.  Note that it has the following properties: \begin{itemize}
  \item The metric (and also its inverse) are regular at the horizon $y=y_h$.
  \item There is still a coordinate singularity at $x=\pm 1$.
  \item It depends on three arbitrary parameters $c$, $\Phi$ and $\Psi$, any two of which are independent.
\end{itemize}

As in the singly spinning case, we have found a family of geodesics with two free parameters (any two of $c$, $\Psi$, $\Phi$), so we are free to pick their values so as to simplify the metric in order to find something that might be more useful for practical applications.  As in the singly spinning case, $\Phi=0$ is a natural, legitimate choice, but unfortunately we can no longer set $c=0$ (see Section \ref{sec:dsgeo}).   Also, as in the singly spinning case, the coordinates have a restricted $x$ range for $\Phi \neq 0$.

The line element in the $\Phi=0$ case can be written in the form
{\small \begin{multline} \label{eqn:newmet2}
  ds^2 = -\frac{\Hyx}{\Hxy} ( dv + \tilde{\Om} )^2 \\ 
         + \frac{R^2 \Hxy}{(x-y)^2 (1-\nu)^2} \Bigg[\bar{c} \left( \frac{dx^2}{\bar{c} \Gx + \axb} 
                                                                  - \frac{dy^2}{\bar{c} \Gy - \byb} \right) 
          \qquad\qquad\qquad\qquad\qquad\\
         + \left(\frac{\axb}{\Gx} + \frac{\Ayx\gx^2 + 2\Lxy \gx \axb 
                                                - \Axy \axb^2}{\Gx ^2 \Hxy\Hyx (1-\nu)^2} \right) d(x,y)^2 \\
         + 2\left( \frac{\Ayx \tx - \Lxy \cx }{\Gx \Hxy \Hyx} d\tilde{\phi}  
                  - \frac{\Lxy \tx + \Axy \cx }{\Gx \Hxy \Hyx}d\tilde{\psi}
            \right) d(x,y) \\ 
         - \frac{ 2(1-\nu) d\tilde{\psi} dy}{\sqrt{\bar{c} \Gy - \byb}} 
         + \frac{\Ayx d\tilde{\phi}^2-2\Lxy d\tilde{\phi} d\tilde{\psi}- \Axy d\tilde{\psi^2}}{\Hxy \Hyx} \Bigg],
\end{multline}}
where now
\begin{equation}
  d(x,y) \equiv \left( \pm\frac{dx}{\sqrt{\bar{c} \Gx + \axb}} + \frac{dy}{\sqrt{\bar{c} \Gy - \byb}} \right).
\end{equation}
This now contains only the one arbitrary constant $\bar{c} \equiv c/\Psi^2$.  At first glance this looks equally complicated, but the only polynomial functions that appear in this expression are now those that appear in the original metric itself, and are far simpler, so progress has been made.

From Section \ref{sec:dsgeo} we have the condition that 
\begin{equation}
  \bar{c} \geq \frac{\nu}{1+\nu-\la} \left[ 2(1+\la) - 3\la\nu + \nu(1-\nu) \right]
\end{equation}
for these coordinates to be valid for all $x$ (with the exception of the coordinate singularity on the axis $x=\pm 1$.  We might hope that by saturating this bound we could obtain a simpler form for the metric (as occurs in the $c=0$ case for the singly spinning ring), but it is far from clear that this is the case.  Of course doing so does remove the last arbitrary constant from the metric and thus fix it entirely, as well as providing what seems like the natural doubly spinning generalisation of the singly spinning result of \cite{ER:2006}.  It would be interesting to see if a value of $\bar{c}$ could be chosen that really simplified things further here, but we have been unable to do this successfully.

It seems that no further progress can be made in our study of coordinate systems, so finally we move on to discuss whether the separability of part of the HJ equation that we have discovered can be used to say anything about hidden symmetries of the spacetime.

\section{Hidden Symmetries} \label{sec:br5sym}
If a $d$-dimensional metric has at least $d-1$ commuting Killing vectors, corresponding to $d-1$ Noether symmetries, then its associated Hamilton-Jacobi equation has separable solutions.  On the other hand, if it has fewer Killing vectors, but its HJ equation is still separable, then it is expected that this separability can be linked to a hidden phase space symmetry, related to the existence of a higher-rank Killing tensor $K$ satisfying the generalised Killing equation
\begin{equation}
  \nabla_{(\mu} K_{\nu_1 \nu_2...\nu_p)} = 0 .
\end{equation}
In most known cases, this tensor is rank-2, as in the case of the Kerr black hole discussed in Section \ref{sec:kerrkilling}.

Separability of the HJ equation for \emph{null} geodesics is a conformally invariant property of the geometry, and hence this is described by the conformally invariant generalization of the Killing equation, which in the rank-2 case reads
\begin{equation}
  \nabla_{(\mu} K_{\nu\rho)}= \omega_{(\mu} g_{\nu\rho)} \label{eqn:confkilling}
\end{equation}
for some 1-form $\omega$, given in dimension $d$ by
\begin{equation}
  \omega_\mu = \frac{2}{d+2} \left[ \nabla^\nu K_{\mu\nu} + \frac{1}{2} \nabla_\mu (\mathrm{tr} K) \right].
\end{equation}
If $K_{\mu\nu}$ solves this for a spacetime $(\Mcal,g)$, then $\La^4 K_{\mu\nu}$ solves it for the conformally related spacetime $(\Mcal,\La^2 g)$ for any suitably regular function $\La^2$.  Solutions of this equation are referred to as \emph{conformal Killing} (CK) tensors, and they have the property that $K^{\mu\nu} p_\mu p_\nu$ is conserved along any null geodesic with momentum $p_\mu$.

Note that the metric is itself a Killing tensor, with associated conserved quantity $\mu^2$, the mass of a particle following a geodesic.  Furthermore, the symmetrized outer product of any Killing vectors is also a CK tensor; if we are to use CK tensors to generate genuinely new conserved quantities we need a concept of independence:
\begin{defn}
  A rank-2 CK tensor is \emph{irreducible} (or \emph{non-trivial}) if it cannot be expressed in terms of the metric $g$ and Killing vectors $\{ k^{(i)} \}$ in the form
\begin{equation}
  K_{\mu\nu} = a(x^\rho) g_{\mu\nu} + \sum_{i,j} b^{ij} k_{(\mu}^{(i)} k_{\nu)}^{(j)},
\end{equation}
for some scalar function $a(x^\rho)$ and constants $b^{ij}$.  Two CK tensors are \emph{independent} if their difference is irreducible.
\end{defn}

A metric with $d-2$ mutually commuting Killing vectors can be written in a form where its components depend on only two coordinates, $x$ and $y$ say.  Then, if the HJ equation is separable for null geodesics, it can be written in the form 
\begin{equation}
  K^{\mu\nu}_{(1)}(x) p_\mu p_\nu = K^{\mu\nu}_{(2)}(y) p_\mu p_\nu = \mathcal{K}
\end{equation}
for some constant $\mathcal{K}$.  Both $K_{(1)}$ and $K_{(2)}$ must be CK tensors for the geometry, and they satisfy the relation
\begin{equation}
  K^{\mu\nu}_{(1)}(x) - K^{\mu\nu}_{(2)}(y) = f(x,y) g^{\mu\nu} \label{eqn:KminusK}
\end{equation}
for some function $f(x,y)$.  Therefore, they are not independent.

Does anything similar apply for the black ring metric?  We have a separable form (\ref{eqn:sep}) for the HJ equation, but only in the null, zero energy case.  We can read off tensors $K_{(1)}$ and $K_{(2)}$ from this, but do not expect them to be conformal Killing tensors, due to the $E=0$ condition.  Note that the components $K^{tt}$ and $K^{ti}$ of these tensors appear somewhat arbitrary, since they do not have any effect on the value of
\begin{equation}
  \frac{c}{(1-\nu)^2} = K^{\mu\nu} p_\mu p_\nu = K^{tt} E^2 -2 K^{ti} E p_i + K^{ij} p_i p_j = K^{ij} p_i p_j
\end{equation}
along one of the separable geodesics.  This hints at a way of understanding the symmetry that allows for this separation; dimensional reduction to remove the $K^{t\mu}$ components. This turns out to be a neat way of dealing with the zero-energy condition on these geodesics.

\subsection{Kaluza-Klein Reduction}
We perform a dimensional reduction to project out the $\pd/\pd t$ direction, via the standard Kaluza-Klein procedure.  We take an ansatz
\begin{equation}
  ds^2 = e^{\varphi/\sqrt{3}} h_{ij} dx^i dx^j + e^{-2\varphi/\sqrt{3}} (dt + \Acal_i dx^i)
\end{equation}
where $i,j,...$ range over $x,\phi,y,\psi$ and $h_{ij}$ is the metric on the 4-dimensional space.  

Note that $\pd/\pd t$ is spacelike in the ergoregion (to which our known geodesics are restricted), so the reduced metric has signature $(+,+,+,-)$, and we must restrict the ranges of our coordinates in the reduced metric so that they only correspond to this region (otherwise we would be performing a timelike reduction, which would require a slightly different analysis).  It is well known that the resulting 4-dimensional geometry solves the Einstein-Maxwell-Dilaton equations.

Comparison to the line element (\ref{eqn:brmetric}) gives
\begin{equation}
  e^{-2\varphi / \sqrt{3} } = -\frac{H(y,x)}{H(x,y)} \eqand
  \mathcal{A}_i dx^i = \Om = \Om_\phi d\phi + \Om_\psi d\psi .
\end{equation}
Given this, it is straightforward to show that the dimensionally reduced metric is given by
\begin{eqnarray} \label{eqn:4met}
  ds_4^2 &\equiv& h_{ij} dx^i dx^j \\
         &=&  \La^2(x,y) \left[ \frac{dx^2}{G(x)}-\frac{dy^2}{G(y)}+\frac{A(y,x) d\phi^2-2L(x,y) d\phi d\psi
                              - A(x,y) d\psi^2}{H(x,y) H(y,x)} \right] \nn
\end{eqnarray}
where
\begin{equation}\label{eqn:Tnat}
  \La^2(x,y) \equiv \frac{R^2 \sqrt{-H(x,y)H(y,x)}}{(x-y)^2 (1-\nu)^2}.
\end{equation}

Note that the singly-spinning black ring was originally constructed in \cite{ER:2001} by analytic continuation of an oxidised Kaluza-Klein C-metric \cite{Chamblin:1996}.  Here, we have found a Kaluza-Klein metric of a similar form to the C-metric that is linked more directly to the black ring; that is to say no analytic continuation is required.  Furthermore, this reduction is equally valid in the doubly-spinning case, for which a C-metric associated with the ring does not exist in the literature.

\subsection{Conformal Killing Tensors} \label{sec:confkill}
Note that the zero-energy geodesics in the 5-dimensional metric correspond precisely to the geodesics of the 4-dimensional metric (while those which are not zero-energy are related to charged particle orbits).  In the 5 dimensional case we know all of the zero energy, null geodesics, so this translates to knowing all of the null geodesics in the 4 dimensional metric.  Therefore, as described above, we should expect that the dimensionally reduced metric has a CK tensor, and now proceed to show that this is indeed the case.

In order to see the conformal invariance explicitly, it is nice to do the calculation with a general conformal factor $\La^2=\La^2(x,y)$ in the metric (\ref{eqn:4met}), where of course equation (\ref{eqn:Tnat}) gives the choice of $\La^2$ that actually results from the Kaluza-Klein reduction of the black ring.

We read off the forms of $K_{(1)}^{ij}$ and $K_{(2)}^{ij}$ from (\ref{eqn:sep}), which gives non-vanishing components
\begin{align}
  K_{(1)}^{xx} &= G(x) , & K_{(2)}^{yy} &= G(y), \nn\\[2mm]
  K_{(1)}^{\phi\phi} &= \frac{{\beta} (x)}{(1-\nu)^2 G(x)}  ,&
                  K_{(2)}^{\phi\phi} &= \frac{- {\alpha} (y)}{(1-\nu)^2 G(y)}  \nn\\[2mm]
  K_{(1)}^{\phi\psi} &= \frac{\gamma (x)}{(1-\nu)^2G(x)}    , &
                  K_{(2)}^{\phi\psi} &= \frac{\gamma (y)}{(1-\nu)^2 G(y)},\nn\\[2mm]
  K_{(1)}^{\psi\psi} &= \frac{-\alpha (x)}{(1-\nu)^2 G(x)}  , & 
                  K_{(2)}^{\psi\psi} &= \frac{{\beta} (y)}{(1-\nu)^2 G(y)}.
\end{align}
Now
\begin{equation}
  K_{(1)}^{ij} - K_{(2)}^{ij} = \La^2 h^{ij} ,
\end{equation}
so if one of these tensors is a conformal Killing tensor, so is the other, and they are not independent.  Given this, perhaps the natural choice of CK tensor to work with is $K \equiv  K_{(1)} + K_{(2)}$.

Differentiating, we see that $K$ satisfies the conformal Killing equation
\begin{equation}
  \nabla_{(i} K_{jk)} = \omega_{(i} h_{jk)} \quad \mathrm{where} \quad
  \omega = 2\La \left[ \frac{\pd \La}{\pd x} dx -  \frac{\pd \La}{\pd y} dy \right],
\end{equation}
and is therefore a CK tensor.  Note that $K$ is actually a Killing tensor of the geometry that has constant conformal factor $\La^2$.

With indices raised, $K^{ij}$ is not dependent on the conformal factor, and with coordinates $(x,\phi,y,\psi)$, it can be written in matrix form as
\begin{equation} \label{eqn:matK}
  \mathbf{K} =  \left( \begin{array}{cccc} G(x) & 0 & 0 & 0 \\
               0 & \frac{1}{(1-\nu)^2} \left( \frac{\bxb}{\Gx} - \frac{\ayb}{\Gy} \right) & 0 
                 & \frac{1}{(1-\nu)^2} \left( \frac{\gx}{\Gx} + \frac{\gy}{\Gy} \right)\\ 0 & 0 & G(y) & 0 \\
               0 & \frac{1}{(1-\nu)^2} \left( \frac{\gx}{\Gx} + \frac{\gy}{\Gy} \right) & 0 
                 & \frac{1}{(1-\nu)^2} \left( \frac{\byb}{\Gy} - \frac{\axb}{\Gx} \right) 
               \end{array}\right) .
\end{equation}

There is an alternative way of seeing the existence of this conformal Killing tensor.  Benenti \& Francaviglia \cite{Benenti:1979} give a canonical form for the metric of an $n$-dimensional spacetime admitting $(n-2)$ Killing vectors, and a non-trivial rank-2 Killing tensor.  The inverse metric takes the form
\begin{multline}
  g^{-1} = \frac{1}{\varphi_1(x^1) + \varphi_2(x^2)} \Bigg[ \psi_1(x^1) \left( \frac{\pd}{\pd x^1} \right)^2 
           + \psi_2(x^2) \left( \frac{\pd}{\pd x^2} \right)^2 \\
           + \left(\psi_1(x^1) \zeta_1^{\alpha\beta}(x^1) 
           + \psi_2(x^2) \zeta_2^{\alpha\beta}(x^2) \right) \left( \frac{\pd}{\pd \phi^\alpha} \right)
                                                            \left( \frac{\pd}{\pd \phi^\beta} \right) \Bigg]
\end{multline}
for some functions $\psi_a(x^a)$, $\varphi_a(x^a)$, $\zeta_a^{\alpha\beta}(x^a)$ depending on a single coordinate only, with $ \varphi_1 \psi_1^2 +  \varphi_2 \psi_2^2 \neq 0$ everywhere.  The indices $\alpha, \beta=3,...,n$ label the Killing directions $\pd/\pd \phi^\alpha$.  The rank-2 Killing tensor is given by
\begin{equation}
  K^{\alpha \beta} = \frac{1}{\varphi_1 + \varphi_2}\left( \zeta_1^{\alpha\beta} \psi_1 \varphi_2 
                                                          -\zeta_2^{\alpha\beta} \psi_2 \varphi_1 \right)  , \quad
  K^{11} = \frac{\varphi_2 \psi_1 }{\varphi_1 + \varphi_2} \eqand
  K^{22} = \frac{-\varphi_1 \psi_2 }{\varphi_1 + \varphi_2}.
\end{equation}

The inverse metric for the dimensionally reduced black ring is conformally related to a metric of this form, with $\varphi_a \equiv 1$ and we must therefore have a rank-2 conformal Killing tensor.  The form for this given corresponds precisely to our tensor $K^{ij}$, up to an arbitrary constant factor.

\subsection{Conformal Killing-Yano Tensors}\label{sec:cky}
Often, a conformal Killing (CK) tensor can be constructed from a more fundamental object, a conformal Killing-Yano (CKY) tensor, that is a 2-form $k$ satisfying the conformal Killing-Yano equation
\begin{equation}
  \nabla_{(\mu} k_{\nu) \rho} = g_{\mu\nu} \xi_\rho- \xi_{(\mu} g_{\nu)\rho} \qquad \mathrm{where} \qquad
  \xi_\nu = \frac{1}{d-1} \nabla^\mu k_{\mu\nu}.
\end{equation}
Note that if $k_{\mu\nu}$ solves it for spacetime $(M,g)$, then $\La^3 k_{\mu\nu}$ solves it for $(M,\La^2 g)$.  Given a CKY tensor $k$, $K_{\mu\nu} = k_{\mu\rho}k_\nu^{\phantom{\nu}\rho}$ is a CK tensor.  In this case, it turns out that a CKY tensor exists if and only if the ring is singly spinning.

\subsubsection{Singly Spinning Case}
In the singly spinning case, it is straightforward to directly construct an antisymmetric tensor that squares to the Killing tensor $K^{ij}$, that is a $k^{ij}$ such that $ K^{ij} = k^{ik} k^{jl} h_{kl} . $  The tensor
\begin{equation}
  k^{x\phi} = \frac{\sqrt{H(x)}}{\La(x,y)} = -k^{\phi x} \eqand 
  k^{y\psi} = \frac{\sqrt{-H(y)}}{\La(x,y)} = -k^{\psi y} , 
\end{equation}
with all other components vanishing, satisfies this.  Lowering indices, this gives us a 2-form
\begin{equation}
  k = \La^3 \left[\frac{1}{\sqrt{H(x)}} dx \wedge d\phi - \frac{1}{\sqrt{-H(y)}} dy \wedge d\psi \right] .
\end{equation}
Note that there is a second tensor with the same property, which can be obtained by taking the Hodge dual of $k$, resulting in
\begin{equation}
  \star k = \La^3 \left[\frac{1}{\sqrt{H(x)}} dx \wedge d\phi + \frac{1}{\sqrt{-H(y)}} dy \wedge d\psi \right] .
\end{equation}
By explicit calculation, it can be shown that
\begin{equation}
  \nabla_{i} k_{jk} = \nabla_{[i} k_{jk]} + 2 h_{i [j} \xi_{k]} \quad \mathrm{where} \quad
  \xi = \frac{\Gx}{\sqrt{\Hx}} \frac{\pd \La}{\pd x} d\phi + \frac{\Gy}{\sqrt{-\Hy}} \frac{\pd \La}{\pd y} d\psi 
\end{equation}
and therefore $k$ satisfies the conformal Killing-Yano equation (as does $\star k$).

It is interesting to briefly consider the case of constant $\La^2$, although this does not correspond to the actual dimensional reduction of the black ring.  Here, $k$ is a Killing-Yano tensor, and its square is a Killing tensor.  In fact something stronger can be said.  It is known \cite{Krtous:2008,Houri:2008} that any $d$-dimensional spacetime manifold with a globally defined closed CKY tensor $k$ (known as a \emph{principal} CKY tensor) can be written in a particular canonical form.

Here, taking an exterior derivative gives that 
\begin{equation}
  dk = -3 \La^2  dx \wedge dy \wedge \left[ \frac{\pd \La}{\pd y} \frac{d\phi}{\sqrt{\Hx}} 
       + \frac{\pd \La}{\pd x} \frac{d\psi}{\sqrt{-\Hy}} \right] 
\end{equation}
and hence we see that $k$ is closed for the 4-geometry with constant $\La^2$ (as is $\star k$).  Thus we have a principal CKY tensor here.  The existence of this tensor implies that the metric can be written in the known canonical form, separability of the HJ equation for all geodesics (rather than just null ones), as well as that this 4-metric is of algebraic Type D.  Since the algebraic type of a metric is a conformally invariant property, the 4-dimensional geometry must be Type D for all choices of conformal factor, and therefore the geometry that results directly from the KK reduction of the singly-spinning ring is also Type D.

\subsubsection{Doubly Spinning Case}\label{sec:doublycky}
In the doubly spinning case, it turns out that the conformal Killing tensor $K^{ij}$ is not derivable from a conformal Killing-Yano tensor.  Furthermore, this result is independent of our particular choice of CK tensor, and therefore proves that no CKY tensor can exist for the doubly-spinning ($\nu>0$) metric.  That is:
\begin{lemma}\label{lem:Kprime}
Define a symmetric rank-(2,0) tensor $K'$ by 
\begin{equation}
  K' = K + C(x^k) h^{-1} + p \left( \frac{\pd}{\pd \phi} \right)^2 
       + 2 q \left( \frac{\pd}{\pd \phi}\right) \left(\frac{\pd}{\pd \psi} \right) 
       + r \left( \frac{\pd}{\pd \psi}\right)^2 .
\end{equation}
Then $K'$ has the following properties:
\begin{enumerate}
  \item It is a conformal Killing tensor for all differentiable functions $C(x^k)$, and constants $p$, $q$, $r$.
  \item Up to arbitrary constant rescalings of $K$, it is the most general irreducible CK tensor.
  \item For $\nu>0$, and for any $C(x^a)$, $p$, $q$, $r$, there does not exist an antisymmetric tensor $k$ such that
        \begin{equation} \label{eqn:fsquare}
          K'^{ij} = k^{ik} k^{jl} h_{kl}.
        \end{equation}
\end{enumerate}
\end{lemma}
Note that if $k$ is a CKY tensor, then a $K'$ defined by (\ref{eqn:fsquare}) must be a CK tensor, and therefore the non-existence of a square-root for the most general non-trivial CK tensor proves the non-existence of a CKY tensor.  Thus, as a direct corollary of Lemma \ref{lem:Kprime}, we see that the dimensional reduction of the black ring spacetime possesses a CKY tensor if and only if the ring is singly-spinning.  When one CKY tensor exists, a second can be constructed by taking the Hodge dual, as described above.  The Lemma is proved in Appendix \ref{app:blackrings}.

\subsection{Klein-Gordon Equation}\label{sec:kg}
Often, when a spacetime possesses a Killing tensor, it is possible to find multiplicatively separable solutions of the Klein-Gordon (KG) equation.  Here, we have additive separability for geodesic motion in the null, zero energy case, so we might hope that this would translate into being able to find time-independent separable solutions to the massless KG equation for the 5-dimensional black ring.  However, the results linking the existence of a Killing tensor with the separability of the KG equation apply only in Einstein-Maxwell spaces, which our reduced 4-dimensional spacetime is not.  As a result of this, we don't expect separability of the KG equation to be possible for the black ring.  A straightforward calculation shows that this is indeed the case.  That is, taking an ansatz
\begin{equation}
  \varphi(t,x,\phi,y,\psi) = e^{-i \Phi \phi} e^{-i \Psi \psi} X(x) Y(y)
\end{equation}
does not render the massless 5-dimensional KG equation $\Box \varphi = 0$ into a separable form.

\section{Discussion and Outlook} \label{sec:br6disc}
In this chapter we have studied several aspects of the doubly-spinning black ring and noted that, although the metric is at first glance very complicated, it is possible to make progress in studying its properties analytically.  We have seen that in some senses the doubly-spinning system is more complicated, and richer, than the singly-spinning one, while other properties remain largely similar.

Some interesting questions remain.  We have not analysed in detail the paths of the axis geodesics in this chapter, since doing so is very complicated, but it might be interesting to do this and see if any new behaviour occurs that does not appear in the singly spinning case.  These results could perhaps be useful in calculations of scattering cross sections; Gooding and Frolov studied this problem in the Myers-Perry case \cite{Gooding:2008}.

We have also investigated possible links between our results, and the class of metrics described by \cite{Krtous:2008,Houri:2008}.  We have found that the 4-dimensional spacetime obtained by dimensional reduction along $\pd/\pd t$ in the ergoregion is conformal to a metric falling into this class, if, and only if, the black ring is singly spinning.  This provides a qualitative, algebraic difference between the singly spinning and doubly spinning cases.

An obvious question is to ask whether the more general, unbalanced, black ring solution \cite{Morisawa} has similar properties.  Studying the most general form of the unbalanced metric would be difficult, as it is extremely complicated, but some progress on this question can be made by looking at the limit where the black ring has rotation only in the $S^2$ direction, as derived by Figueras \cite{Figueras:2005}.  It turns out that here, no separation of the HJ equation is possible in ring-like coordinates; so this separability, and possibly the conformal Killing tensor structure associated with it, may rely in some way on the balancing condition being satisfied.  However, in the unbalanced, singly-spinning case \cite{ER:2006}, separation is possible, so the exact nature of this relationship is unclear.

\appendix
\chapter[GHP formalism with matter]{GHP formalism for spacetimes with arbitrary matter}\label{app:ghpmatter}
In this thesis, we have focussed almost entirely on Einstein spacetimes.  However, the definitions of the GHP formalism can be conveniently extended to spacetimes with arbitrary matter.

The first step in doing this it to expand the Ricci tensor $R_{\mu\nu}$ in the null frame (and hence the energy-momentum tensor $T_{\mu\nu}$).  Table \ref{tab:ricciweights} describes our notation for its components $R_{ab}$:
\begin{table}[h!]
 \begin{center}
   \begin{tabular}{|c|c|c|c|l|}
    \hline Compt. & Notation & Boost weight $b$ & Spin $s$ & Comment\\ [1mm]\hline
    $R_{00}$  & $\om$        & 2  & 0 & \\[1mm]
    $R_{0i}$  & $\psi_{i}$   & 1  & 1 & \\[1mm]
    $R_{ij}$  & $\phi_{ij}$  & 0  & 2 & $\phi_{ij} = \phi_{ji}$\\[1mm]
    $R_{01}$  & $\phi$       & 0  & 0 & $\phi\neq\phi_{ii}$\\[1mm]
    $R_{1i}$  & $\psi'_{i}$  & -1 & 1 & \\[1mm]
    $R_{11}$  & $\om'$       & -2 & 0 & \\[1mm]\hline
  \end{tabular}
  {\it\caption[Boost weight decomposition of the Ricci tensor in higher dimensions.]{\label{tab:ricciweights}\small Decomposition of the Ricci tensor in the frame basis. We use the convention that Ricci components use the lower case version of the Greek letter representing the Weyl components of the same boost weight.}}
 \end{center}
\end{table}

In four dimensions, the NP and GHP formalisms have found applications to spacetimes with various kinds of matter; but typically only where the matter is in some sense aligned with a null vector field.  Various examples, including `aligned null radiation' are discussed in \cite{exact}.

\section{Newman-Penrose equations}
Given this notation, the NP equations (see Section \ref{sec:npeqns}) read:\\
\setcounter{oldeq}{\value{equation}}
\renewcommand{\theequation}{NP\arabic{equation}m}
\setcounter{equation}{0}
{\noindent\bf Boost weight +2}
\begin{eqnarray}
  \tho \rho_{ij} - \eth_j \kap_i &=& - \rho_{ik} \rho_{kj} -\kap_i \tau'_j - \tau_i \kap_j  
                                     - \Om_{ij} - \frac{1}{d-2}\om \del_{ij},\label{NP1m}
\end{eqnarray}
{\noindent\bf Boost weight +1}
\begin{eqnarray}
  \tho \tau_i - \tho' \kap_i &=& \rho_{ij}(-\tau_j + \tau'_j)
                                 - \Ps_i + \frac{1}{d-2} \psi_i,\label{NP2m}\\[3mm]
  \eth_{[j|} \rho_{i|k]}     &=& \tau_i \rho_{[jk]} + \kap_i \rho'_{[jk]}
                                 - \frac{1}{2} \Ps_{ijk} 
                                 - \frac{1}{d-2}\psi_{[j}\del_{k]i},\label{NP3m}
\end{eqnarray}
{\noindent\bf Boost weight 0}
\begin{eqnarray}
  \tho' \rho_{ij} - \eth_j \tau_i &=& - \tau_i \tau_j - \kap_i \kap'_j 
                                      - \rho_{ik}\rho'_{kj}-\Phi_{ij}\nn\\
                                  &&  - \frac{1}{d-2} (\phi_{ij} + \phi \del_{ij}) 
                                      + \frac{\phi_{kk}+2\phi}{(d-1)(d-2)} \del_{ij},\label{NP4m}
\end{eqnarray}
\renewcommand{\theequation}{A.\arabic{equation}}
\setcounter{equation}{\value{oldeq}}
with another four equations obtained by taking the prime $'$ of these four. 

\section{Commutators}
In the case of arbitrary matter, the commutators (see Section \ref{sec:comms}) read:
\setcounter{oldeq}{\value{equation}}
\renewcommand{\theequation}{C\arabic{equation}m}
\setcounter{equation}{0}
\begin{eqnarray}
[\tho, \tho']T_{i_1...i_s} 
         &=& \Big[ (-\tau_j + \tau'_j) \eth_j \nn\\ 
         & & \quad  + b\left( -\tau_j\tau'_j + \kap_j\kap'_j + \Phi 
                            - \frac{2\phi}{d-1} + \frac{\phi_{jj}}{(d-1)(d-2)}\right)
             \Big]T_{i_1...i_s} \nonumber\\
         &&  + \sum_{r=1}^s \left(\kap_{i_r} \kap'_{j} - \kap'_{i_r} \kap_{j} 
                                  + \tau'_{i_r} \tau_{j} - \tau_{i_r} \tau'_{j} + 2\Phia_{i_r j}
                            \right) T_{i_1...j...i_s},\\[3mm]
[\tho, \eth_i]T_{k_1...k_s}
         &=& \Bigg[-(\kap_i \tho' + \tau'_i\tho +\rho_{ji}\eth_j)
             + b\left(-\tau'_j\rho_{ji} + \kap_j\rho'_{ji} 
             + \Ps_i-\frac{1}{d-2} \psi_{i}\right) \Bigg]T_{k_1...k_s} \nn\\
         & & + \sum_{r=1}^s \Big[ \kap_{k_r}\rho'_{li} - \rho_{k_r i}\tau'_l
            + \tau'_{k_r} \rho_{li} - \rho'_{k_r i} \kap_l
            - \Ps_{ilk_r} \nn\\
         & &\qquad\qquad - \frac{2}{d-2}\psi_{[l}\del_{k_r]i}\Big] T_{k_1...l...k_s},\\[3mm]
[\eth_i,\eth_j]T_{k_1...k_s}
         &=& \left(2\rho_{[ij]} \tho' + 2\rho'_{[ij]} \tho 
                   + 2b \rho_{l[i|} \rho'_{l|j]} + 2b\Phia_{ij}\right) T_{k_1...k_s}\nn\\
         && + \sum_{r=1}^s \Big[2\rho_{k_r [i|} \rho'_{l|j]} + 2\rho'_{k_r [i|} \rho_{l|j]} 
                                + \Phi_{ijk_r l} \label{C3m}\\
         && \quad\qquad + \frac{2}{d-2} (\del_{[i|k_r}\phi_{|j]l} - \del_{[i|l}\phi_{|j]k_r})
            - \frac{2(2 \phi+ \phi_{mm}) \del_{[i|k_r}\del_{|j]l}}{(d-1)(d-2)} \Big] T_{k_1...l...k_s}.\nn
\end{eqnarray}
\renewcommand{\theequation}{A.\arabic{equation}}
\setcounter{equation}{\value{oldeq}}

\section{Bianchi equations}
Including matter in the Bianchi equations is rather more complicated.  Noting that
\begin{equation}
  R_{abcd} = C_{abcd} + \frac{2}{d-2}(\eta_{a[c}R_{d]b} - \eta_{b[c}R_{d]a}) 
             - \frac{2R}{(d-1)(d-2)}\eta_{a[c}\eta_{d]b},
\end{equation}
the appropriate equations can then be obtained from (\ref{B1}-\ref{B7}) by making the following replacements:
\begin{eqnarray}
  \Om_{ij}   &\rightarrow& \Om_{ij}   + \frac{\om}{d-2}\delta_{ij},\\
  \Ps_{i}    &\rightarrow& \Ps_{i}    - \frac{\psi_i}{d-2},\\
  \Ps_{ijk}  &\rightarrow& \Ps_{ijk}  + \frac{2}{d-2}\psi_{[j}\delta_{k]i},\\
  \Phi_{ij} &\rightarrow& \Phi_{ij} +\frac{\phi_{ij}}{d-2}
                           +\frac{ (d-3)\phi-\phi_{kk}}{(d-1)(d-2)}\del_{ij},\\
  \Phi_{ijkl}&\rightarrow& \Phi_{ijkl} + \frac{2}{d-2}\left(\del_{i[k}\phi_{l]j} - \del_{j[k}\phi_{l]i}\right)
                                       - 2\del_{i[k}\del_{l]j}\frac{2\phi+\phi_{mm}}{{(d-1)(d-2)}},\\
  \Phi       &\rightarrow& \Phi - \frac{2\phi}{d-1}+\frac{\phi_{ii}}{(d-1)(d-2)},
\end{eqnarray}
together with the primed versions of the first three of these equations.  Note that before these replacements are made, we're interpreting these objects as Riemann, not Weyl, tensor components, so the various trace identities discussed in Table \ref{tab:weyl} no longer hold.  Hence the above replacements are valid only when made directly in equations (\ref{B1})-(\ref{B7}), not in contractions of these equations. When making these replacements, one can exclude any cosmological constant terms from the Ricci tensor, since these must all cancel out in the Bianchi equations.

The above equations must be supplemented by additional equations that are trivial in the case of an Einstein spacetime, but not when matter is present. These equations are equivalent to the contracted Bianchi identity
\begin{equation}
  \nabla^\mu R_{\mu\nu} = \half\nabla_\nu R.
\end{equation}
In the null basis, this equation reduces to
\begin{eqnarray}
  \tho'\om + \eth_i\psi_i - \half\tho \phi_{ii} 
            &=& -\rho'\om + (2\tau_i+\tau'_i)\psi_i
                + \rho_{ij} (\phi_{ij} - \phi\del_{ij})\nn\\
            & & + \kap_i \psi'_i, \label{BianchiMat1}\\
  \tho'\psi_i + \eth_j\phi_{ij}-\eth_i(\phi+\half\phi_{jj}) + \tho\psi'_i
            &=& -\kap'_i\om - (\rho'_{ij}+\rho'\del_{ij})\psi_j + (\tau_j+\tau'_j)(\phi_{ji}-\phi\del_{ji}) \nn\\
            & & -(\rho_{ij}+ \rho\del_{ij})\psi'_j - \kap_i\om', \label{BianchiMat2}
\end{eqnarray}
with a third equation following from (\ref{BianchiMat1})$'$.

\chapter[GHP equations for alg.\ spec.\ spacetimes]{GHP equations for algebraically special Einstein spacetimes}\label{app:ghpeqns}
\renewcommand{\theequation}{B.\arabic{equation}}
\setcounter{equation}{0}

In an algebraically special Einstein spacetime, it is shown in Chapter \ref{chap:nongeo} that there always exists a {\it geodesic} multiple WAND. If we choose $\ell$ to be this multiple WAND then we have
\begin{equation}
  \Om_{ij} = \Psi_{ijk} = \Psi_i = \kap_i = 0.
\end{equation}
This simplifies considerably many of the GHP equations. However, since we have now endowed $\ell$ with a property that is not enjoyed by $\nb$, we have broken the symmetry under the priming operation and therefore must write out all of the equations explicitly. 

In a Type D Einstein spacetime, we can choose both $\ell$ and $n$ to be geodesic multiple WANDs (see the discussion below Theorem \ref{thm:submfd}).  In this case, the priming symmetry is recovered again, and many of the equations below become unnecessary (and some of those that remain are simplified further).

\section{Newman-Penrose equations}\label{app:Ricci}
{\noindent\bf Boost weight +2}
\begin{eqnarray}
  \tho \rho_{ij} &=& - \rho_{ik} \rho_{kj} \label{Sachs},
\end{eqnarray}
{\noindent\bf Boost weight +1}
\begin{eqnarray}
  \tho \tau_i            &=& \rho_{ij}(-\tau_j + \tau'_j),\\[3mm]
  \eth_{[j|} \rho_{i|k]} &=& \tau_i \rho_{[jk]} ,
\end{eqnarray}
{\noindent\bf Boost weight 0}
\begin{eqnarray}
  \tho' \rho_{ij} - \eth_j \tau_i &=& -\tau_i \tau_j - \rho_{ik} \rho'_{kj} - \Phi_{ij} 
                                      - \frac{\La}{d-1}\del_{ij},\\[3mm]
  \tho \rho'_{ij} - \eth_j \tau'_i &=& -\tau'_i \tau'_j - \rho'_{ik} \rho_{kj} - \Phi_{ji} 
                                       - \frac{\La}{d-1}\del_{ij},
\end{eqnarray}
{\noindent\bf Boost weight -1}
\begin{eqnarray}
  \tho' \tau'_i - \tho \kap'_i &=& \rho'_{ij}(-\tau'_j + \tau_j) - \Ps'_i, \\[3mm]
  \eth_{[j|} \rho'_{i|k]}     &=& \tau'_i \rho'_{[jk]} + \kap'_i \rho_{[jk]} - \frac{1}{2} \Ps'_{ijk} ,
\end{eqnarray}
{\noindent\bf Boost weight -2}
\begin{eqnarray}
  \tho' \rho'_{ij} - \eth_j \kap'_i &=& - \rho'_{ik} \rho'_{kj} -\kap'_i \tau_j - \tau'_i \kap'_j - \Om'_{ij}.
\end{eqnarray}

\section{Bianchi equations}\label{app:Bianchi}

{\bf Boost weight +1:}
\begin{eqnarray}
  \tho \Phi_{ij}   &=& -(\Phi_{ik} + 2\Phia_{ik} + \Phi \del_{ik}) \rho_{kj}, \label{A2}\label{Bi2}\\[3mm]
  -\tho \Phi_{ijkl}   &=& 4\Phia_{ij} \rho_{[kl]} - 2\Phi_{[k|i} \rho_{j|l]} + 2\Phi_{[k|j} \rho_{i|l]} 
                       + 2\Phi_{ij[k|m} \rho_{m|l]} , \label{A4}\label{Bi3}\\[3mm]
  0                &=& 2\Phia_{[jk|}\rho_{i|l]} -2\Phi_{i[j}\rho_{kl]} + \Phi_{im [jk|}\rho_{m|l]} ,
                       \label{A5}\label{Bi4}
\end{eqnarray}
{\bf Boost weight 0:}
\begin{eqnarray}
  -2 \eth_{[j|}\Phi_{i|k]} &=& (2 \Phi_{i[j}\del_{k]l} - 2\del_{il}\Phia_{jk} - \Phi_{iljk}) \tau_l 
                                + 2(\Ps'_{[j|} \del_{il} - \Ps'_{[j|il}) \rho_{l|k]} \label{A8}\label{Bi5},\\[3mm]
  -2\eth_{[i} \Phia_{jk]} &=& 2\Ps'_{[i} \rho_{jk]} + \Ps'_{l[ij|} \rho_{l|k]}\label{A10}\label{Bi6},\\[3mm]
  -\eth_{[k|} \Phi_{ij|lm]}  &=& - \Ps'_{i[kl|} \rho_{j|m]} + \Ps'_{j[kl|} \rho_{i|m]} 
                              - 2\Ps'_{[k|ij} \rho_{|lm]}\label{A11}\label{Bi7},\\[3mm]
  - 2 \eth_{[j}\Phi_{k]i} + \tho \Ps'_{ijk}
                          &=& (2\Phi_{[j|i}\del_{k]l} + 2\del_{il} \Phia_{jk} - \Phi_{iljk})\tau'_l
                               + 2 (\Ps'_i \del_{[j|l} - \Ps'_{i[j|l}) \rho_{l|k]} \label{A9}\label{Bi8},
\end{eqnarray}
{\bf Boost weight -1:}
\begin{eqnarray}
   - \tho' \Phi_{ji} - \eth_{j}\Ps'_i + \tho \Om'_{ij}  
                           &=& (\Phis_{ik} - 3\Phia_{ik} + \Phi \del_{ik}) \rho'_{kj} 
                               + (\Ps'_{ijk}-\Ps'_i\del_{jk}) \tau_k \nonumber\\
                           &&  - 2(\Ps'_{(i}\del_{j)k} + \Ps'_{(ij)k}) \tau'_k 
                               - \Om'_{ik} \rho_{kj}\label{A13}\label{Bi9},\\[3mm]
  -\tho' \Phi_{ijkl} + 2 \eth_{[k}\Ps'_{l]ij}
                       &=& - 4\Phia_{ij} \rho'_{[kl]} -2\Phi_{i[k|}\rho'_{j|l]} + 2\Phi_{j[k|}\rho'_{i|l]} 
                                     + 2 \Phi_{ij[k|m}\rho'_{m|l]}\nonumber\\
                       && -2\Ps'_{[i|kl}\tau_{|j]} - 2\Ps'_{[k|ij} \tau_{|l]}- 2\Om'_{i[k|} \rho_{j|l]} 
                          + 2\Om'_{j[k} \rho_{i|l]}\label{A14}\label{Bi10},\\[3mm]
  -\eth_{[j|} \Ps'_{i|kl]} &=& -2\Phia_{[jk|} \rho'_{i|l]} - 2\Phi_{[j|i} \rho'_{|kl]} 
                               + \Phi_{im[jk|} \rho'_{m|l]} - 2\Om'_{i[j|} \rho_{|kl]}\label{A15}\label{Bi11},
\end{eqnarray}
{\bf Boost weight -2:}
\begin{eqnarray}
  \tho' \Ps'_{ijk} - 2 \eth_{[j}\Om'_{k]i} 
                  &=& (2\Phi_{[j|i} \del_{k]l} + 2\del_{il} \Phia_{jk} -\Phi_{iljk})\kap'_l \nonumber \\
                  && -2 (\Ps'_{[j|} \del_{il} + \Ps'_i\del_{[j|l} + \Ps'_{i[j|l} + \Ps'_{[j|il}) \rho'_{l|k]} 
                     + 2 \Om'_{i[j} \tau_{k]}\label{A16}\label{Bi12}.
\end{eqnarray}

\section{Commutators}\label{app:Comm}

\begin{eqnarray}
[\tho, \tho']T_{i_1...i_s}  &=& \left[ (-\tau_j + \tau'_j) \eth_j + 
                                   b\left(
                                      -\tau_j\tau'_j + \Phi - \frac{2\La}{d-1}
                                    \right)
                                  \right]T_{i_1...i_s} \nonumber\\
                              &&  \qquad\qquad+ \sum_{r=1}^s \left(
                                     \tau'_{i_r} \tau_{j} - \tau_{i_r} \tau'_{j} + 2\Phia_{i_r j}
                                   \right) T_{i_1...j...i_s} \label{C1s},\\[3mm]
[\tho, \eth_i]T_{k_1...k_s} &=& \Big(-(\tau'_i\tho +\rho_{ji}\eth_j)
                                      -b \tau'_j\rho_{ji}\Big) T_{k_1...k_s}\nn \\
                            & &  \qquad\qquad+\sum_{r=1}^s ( - \rho_{k_r i}\tau'_l + \tau'_{k_r} \rho_{li})
                                                       T_{k_1...l...k_s} \label{C2s}, \\[3mm]
[\tho', \eth_i]T_{k_1...k_s} &=& \Big[-(\tau_i\tho' +\rho'_{ji}\eth_j)
                                      -b \tau_j\rho'_{ji}\Big]T_{k_1...k_s} \nonumber\\
                            &&\qquad +\sum_{r=1}^s \Big[ \kap'_{k_r}\rho_{li} - \rho'_{k_r i}\tau_l 
                                                   + \tau_{k_r} \rho'_{li} - \rho'_{k_r i} \kap'_l - \Ps'_{ilk_r} 
                                             \Big] T_{k_1...l...k_s} \label{C3s}, \\[3mm]
[\eth_i,\eth_j]T_{k_1...k_s}&=& \left(
                                    2\rho_{[ij]} \tho' + 2\rho'_{[ij]} \tho + 2b \rho_{l[i|} \rho'_{l|j]} 
                                    + 2b\Phia_{ij}
                                  \right) T_{k_1...k_s}\label{C4s}\\
                             && \qquad+ \sum_{r=1}^s \Big[
                                    2\rho_{k_r [i|} \rho'_{l|j]} + 2\rho'_{k_r [i|} \rho_{l|j]} + \Phi_{ijk_r l}
                                + \frac{2\La}{d-1} \del_{[i|k_r}\del_{|j]l} \Big] T_{k_1...l...k_s}\nn . 
\end{eqnarray}

\chapter[Perturbation equations for NH geometries]{Perturbation equations for near-horizon geometries} \label{app:nhframe}
\renewcommand{\theequation}{C.\arabic{equation}}
\setcounter{equation}{0}
In this appendix, we explain the calculations required to obtain the results presented in Section \ref{sec:decoupling} for a general metric ansatz (\ref{nhgeom}) including all known near-horizon geometries.

Consider a near horizon geometry of the form (\ref{nhgeom}), with $n$ rotational Killing vectors $\pd/\pd \phi^I$, and indices $I,J,\ldots=2,\ldots n+1$ and $A,B,\ldots = n+2,\ldots d-1$.  We think of this as a fibration over $AdS_2$ of some manifold $\Hcal$ with metric
\begin{equation}
  d\hat{s}^2 = g_{IJ}(y)d\phi^I d\phi^J + g_{AB}(y) dy^A dy^B 
             = \hat{g}_{\mu\nu}d\hat{x}^\mu d\hat{x}^\nu .
\end{equation}
The rotation of the black hole is described by the constants $k^I$.  It is useful to define a (Killing) vector field $k = k^I \pard{\phi^I}$. 

In Chapter \ref{chap:decoupling} we derived decoupled equations for gravitational perturbations and test Maxwell fields in the background of Kundt spacetimes, using the higher-dimensional GHP formalism of Chapter \ref{chap:ghp}.  In this section, we show that all metrics of the form (\ref{nhgeom}) are (doubly) Kundt spacetimes, and compute the relevant equations in these particular cases.  The results obtained will be expressed in notation independent of this formalism.

We work in a null frame
\begin{align}
  \lb  &= e_0 = e^1 = \tfrac{1}{\sqrt{2}} L(y) \left(-R dT + \tfrac{dR}{R} \right),\nn\\
  \nb  &= e_1 = e^0 =  \tfrac{1}{\sqrt{2}} L(y) \left(R dT + \tfrac{dR}{R} \right) ,\nn\\
  m_\Ih  &= e_\Ih = e^\Ih = \hat{e}_{\Ih I}\left( d\phi^I - k^I R dT \right),\nn\\
  m_\Ah &= e_\Ah = e^\Ah =\hat{e}_\Ah,
\end{align}
where $\hat{e}$ are vielbeins for $\Hcal$.  Indices $\Ih,\Jh,\dots=2,\dots n+1$ are frame indices in the Killing directions, while $\Ah,\Bh,\dots = n+2,\dots d-1$ are frame indices in the non-Killing directions.

With indices raised this gives
\begin{align}
  e_0   &= \frac{1}{L\sqrt{2}} \left( \frac{1}{R}\pard{T}+k^I\pard{\phi^I}+R\pard{R} \right),\nn\\
  e_1   &= \frac{1}{L\sqrt{2}} \left(-\frac{1}{R}\pard{T}-k^I\pard{\phi^I}+R\pard{R} \right),\nn\\
  e_\Ih   &= \hat{e}_\Ih^I \pard{\phi^I }, \nn\\
  e_\Ah &= \hat{e}_\Ah.
\end{align}

Using the Cartan equations $de_a + \om_{ab}\wedge e^b = 0$ we find that the spin connection is given by
\begin{align}
  \om_{01} &= \tfrac{1}{L\sqrt{2}} (e_0 - e_1) - \tfrac{1}{2L^2}(k.\hat{e}_\Ih) e_\Ih , &
  \om_{0\Ih} &= - \tfrac{1}{2L^2}(k.\hat{e}_\Ih) e_0, \nn\\
  \om_{0\Ah} &= \tfrac{1}{L}(dL)_\Ah \, e_0 , &
  \om_{1\Ih} &= + \tfrac{1}{2L^2}(k.\hat{e}_\Ih) e_1,\nn\\
  \om_{1\Ah} &= \tfrac{1}{L}(dL)_\Ah \, e_1 , &
  \om_{\Ih\Jh} &= -\hat{e}_\Jh.\left[(e_\Ah.\nabla)\hat{e}_\Ih\right] e_\Ah,\nn\\
  \om_{\Ah\Bh} &= \hat{\om}_{\Ah\Bh}, &
  \om_{\Ah \Ih} &= 0.
\end{align}
This is sufficient to give us the GHP optical scalars for the spacetime, which read
\begin{equation}
  \kap_i = \kap'_i = 0, \qquad
  \rho_{ij} = \rho'_{ij} = 0, \qquad
  \tau_i = \frac{ k_i - d(L^2)_i}{2L^2}
\end{equation}
where $i,j\dots = 2,\dots,d-1$ are frame indices on the $d-2$ spacelike dimensions (or equivalently on $\Hcal$).  This implies that both $\lb$ and $\nb$ define geodesic, non-expanding, non-shearing, non-twisting null congruences, and hence that this is a (doubly) Kundt spacetime.  By a simple extension of Theorem \ref{thm:kundt}, it is easy to see that all doubly Kundt Einstein spacetimes are Type D.

For this metric, the GHP derivative operators, acting on a GHP scalar $T_{i_1\dots i_s}$ of boost weight $b$ and spin $s$, are
\begin{align}
  \tho T_{i_1\dots i_s}
       &= \frac{1}{L\sqrt{2}} \left( \frac{1}{R}\pard{T}+k.\pard{\phi}
                                    + R\pard{R} - b\right)T_{i_1\dots i_s},\label{eqn:tho}\\
  \tho' T_{i_1\dots i_s}
       &= \frac{1}{L\sqrt{2}} \left(-\frac{1}{R}\pard{T}-k.\pard{\phi}
                                    + R\pard{R} + b\right)T_{i_1\dots i_s},\label{eqn:thop}\\
  \eth_j T_{i_1\dots i_s} 
       &= \left(\nablah_j - \frac{b}{2L^2} k_j \right) T_{i_1\dots i_s}\label{eqn:eth}
\end{align}
where $\nablah$ is the covariant derivative on $\Hcal$.

Now consider a GHP covariant field $T_{i_1\dots i_s}$ of boost weight $b$ and spin $s$.  We are interested in the cases where $T$ is one of $\phi$, $\vphi_i$, $\Om_{ij}$, which have $(b,s)=(0,0), (1,1), (2,2)$ respectively.

Consider a separable ansatz
\begin{equation}\label{eqn:separation}
  T_{i_1\dots i_s}(T,R,\phi^I,y^A) = \chi_b(T,R) \, Y_{i_1\dots i_s}(\phi^I,y^A),
\end{equation}
where $\chi_b$ has boost weight $b$, and $Y$ has boost weight 0.  We think of $\chi_b$ as a field on $AdS_2$, and $Y$ as a tensor on $\Hcal$.  Eventually it will be useful to move away from the null frame, so let $\mu,\nu,\dots$ be coordinate indices on $\Hcal$.

Note that the GHP derivative $\eth_i$ reduces to the standard covariant derivative on $\Hcal$ when acting on boost weight zero fields such as $Y$.  Hence, given a decomposition of the form (\ref{eqn:separation}), we see that equation (\ref{eqn:eth}) reduces to
\begin{equation}
  \eth_j T_{i_1\dots i_s} = \chi_b \nablah_j Y_{i_1\dots i_s}
                            - Y_{i_1\dots i_s} \frac{b}{2L^2} k_j \chi_b.
\end{equation}

We can take Fourier expansions of the dependence of $Y$ on the coordinates $\phi^I$, of the form $Y \sim e^{i m_I \phi^I}$, which is equivalent to the statement that the Lie derivative of $Y$ with respect to $\pd/\pd \phi^I$ is given by
\begin{equation}
  (\Lcal_I Y)_{\mu_1\dots \mu_s} = i m_I Y_{\mu_1\dots \mu_s},
\end{equation}
and hence
\begin{equation}
  (\Lcal_k Y)_{\mu_1\dots \mu_s} = i k.m Y_{\mu_1\dots \mu_s},
\end{equation}
where $k.m \equiv k^I m_I$.  For the three different kinds of field, this implies that
\begin{align}
  k.\nablah Y &= i k.m Y, \\
  k.\nablah Y_\mu &= ik.m Y_\mu - (\nablah_\mu k^\nu) Y_\nu\\
  k.\nablah Y_{\mu\nu} &= ik.m Y_{\mu\nu} - 2(\nablah_{(\mu|} k^\rho) Y_{\rho|\nu)}.
\end{align}

Recall now the equation of motion $(D^2-\mu^2)\chi_b = 0$ for a charged massive scalar field $\chi$ on a unit radius $AdS_2$ space, described by the metric (\ref{ads2}), where the charged covariant derivative $D$ was defined by \eqref{eqn:ads2deriv}.  Explicitly, this equation of motion reads 
\begin{equation}
  - \frac{1}{R^2} \frac{\pd^2 \chi}{\pd T^2} - \frac{2iq}{R} \frac{\pd\chi}{\pd T}
  + \frac{\pd}{\pd R} \left( R^2 \frac{\pd\chi}{\pd R} \right) + (q^2-\mu^2) \chi = 0.
\end{equation}

Using the equations (\ref{eqn:tho},\ref{eqn:thop}), it can then be shown that
\begin{equation}
  2\tho'\tho T_{i_1\dots i_s} = \frac{1}{L^2} \left[ D^2 \chi_b + iq \chi_b \right] Y_{i_1\dots i_s}
\end{equation}
where $D^2$ is the square of the $AdS_2$ operator (\ref{eqn:ads2deriv}) and $q = ib+k.m$.  Also, we have
\begin{equation}
  \eth_j \eth_j T_{i_1\dots i_s} 
             = \frac{\chi_b}{L^2} \left[ \frac{b^2}{4L^2} k.k - b k.\nablah  
                                         + L^2 \nablah^2 \right] Y_{i_1\dots i_s}
\end{equation}
and
\begin{equation}
  -2(b+1) \tau_j \eth_j T_{i_1\dots i_s} 
               = \frac{\chi_b}{L^2}\left[ \frac{b(b+1)}{2L^2} (k.k) - (b+1) k.\nablah 
                                          + (b+1) d(L^2).\nablah \right] Y_{i_1\dots i_s}.
\end{equation}

Now consider the boost weight zero Weyl tensor components $\Phi$, $\Phi^S_{ij}$, $\Phi^A_{ij}$, $\Phi_{ijkl}$ that appear in equations \eqref{eqn:maxpert}, \eqref{eqn:scalarperts} and \eqref{eqn:gravperts}.  Recall that the NH geometry is an Einstein spacetime with Ricci tensor $R_{ab} = \La g_{ab}$.  Given this, we can use equation \eqref{eqn:hypersurfacecurv} to write
\begin{equation}
  \Phi_{ijkl} = \Rh_{ijkl} - \tfrac{2\La}{d-1} \del_{[i|k}\del_{|j]l}
\end{equation}
where $\Rh_{ijkl}$ is the Riemann tensor of $\Hcal$.  Taking traces of this with the metric on $\Hcal$ implies that
\begin{equation}
  2\Phis_{ij} =  - \Rh_{ij} + \tfrac{d-3}{d-1} \La \del_{ij} ,\qquad
  2\Phi = - \Rh + \tfrac{(d-2)(d-3)}{d-1}\La .
\end{equation}
The remaining components $\Phia_{ij}$ are not related to the curvature of $\Hcal$, but instead can be computed using equation \eqref{NP4}, giving
\begin{equation}
  2\Phia_{ij} = -2\eth_{[i} \tau_{j]} = -(d\tau)_{ij} 
              = - \left( \frac{dk}{2L^2} - \frac{(dL^2)\wedge k}{2L^4} \right)_{ij}.
\end{equation}

In the case of a scalar field, $b=s=0$, and this is enough to allow us to immediately write out equation (\ref{eqn:scalarperts}) as 
\begin{equation}
  Y \left[(D^2 - q^2)\chi_0\right] = \chi_0\left[- \nablah^\mu (L^2 \nablah_\mu Y) 
                                           - (k.m)^2 Y + M^2 L^2 Y \right]
\end{equation}
and hence we can separate variables to obtain
\begin{equation}
    (D^2 - q^2 - \la)\chi_0(T,R) =0
\end{equation}
and
\begin{equation}
 \left[ - \nablah^\mu (L(y)^2 \nablah_\mu ) - (k.m)^2 + M^2 L(y)^2  \right]Y(\phi^I,x^A) = \la\, Y(\phi^I,x^A)
\end{equation}
for some separation constant $\la$.  We can use the left hand side to define an operator $\Ocal{0}$ acting on scalar fields on $\Hcal$, whose properties are discussed in Section \ref{sec:decomposition}.

In the gravitational case $b=s=2$, and inserting the terms given above into (\ref{eqn:gravperts}) allows us to define an operator $\Ocal{2}$ by
\begin{equation}
  Y_{\mu\nu}\left[(D^2 - q^2)\chi_2\right] =  \\ \chi_2 (\Ocal{2} Y)_{\mu\nu}
\end{equation}
The operator $\Ocal{2}$ obtained in this way is given explicitly by (\ref{eqn:Ograv}).  Proving that this operator is self-adjoint with respect to the inner product (\ref{eqn:gravinner}) given is a now a case of integrating by parts.

Similarly, for electromagnetic perturbations, $b=s=1$, and inserting the terms given into (\ref{eqn:maxpert}) give us the operator (\ref{eqn:Omax}).

\chapter[MP black holes with equal angular momentum]{Myers-Perry black holes with equal angular momenta}\label{app:technical}
\renewcommand{\theequation}{D.\arabic{equation}}
\setcounter{equation}{0}
Here, we explain in detail how to obtain the results described in Section \ref{sec:nhmp}.

\section{Computing perturbation operators}
\subsection*{Structure of the near-horizon geometry}\label{app:nhmp}
Consider the near-horizon geometry of an extremal Myers-Perry black hole, described by the metric \eqref{eqn:metric}.  Given the results of Section \ref{sec:decoupling}, it suffices to study the $(d-2)$-dimensional space $\Hcal$.  We work in a frame
\begin{equation}\label{eqn:Hframe}
  e_2 = B(d\psi + \Acal), \quad
  e_\alh = r_+ \hat{e}_\alh ,
\end{equation}
where $\hat{e}_\alh$ are a real, orthonormal frame for $\CP{N}$, and $\Acal = \Acal_\alh e_\alh$.  With indices raised this gives
\begin{equation}
  e_2 = \frac{1}{B}\frac{\pd}{\pd \psi}, \quad
  e_\alh = \frac{1}{r_+} \left(\hat{e}_\alh - \Acal_\alh \frac{\pd}{\pd \psi}\right),
\end{equation}
Note that these vectors satisfy $e_i.e_j = \del_{ij}$, where $i,j,\ldots$ are frame basis indices on $\Hcal$.

The spin connection 1-forms $\om_{ij}$ associated with this basis are
\begin{equation}
  \om_{2\alh}     = \frac{B}{r_+^2} \Jcalh_{\alh\betah} e_\betah, \qquad
  \om_{\alh\betah} = -\frac{B}{r_+^2} \Jcalh_{\alh\betah} e_2 
                   + \frac{1}{r_+} \hat{\om}_{\alh\betah}.\label{eqn:curv1forms}
\end{equation}
where $\hat{\om}$ is the spin connection for $\CP{N}$, and $\Jcal = \frac{1}{2} \Jcalh_{\alh\betah}\hat{e}_\alh\wedge \hat{e}_\betah$ are the components of the complex structure for $\CP{N}$ (recall also that $E = 2L^2 / (B\Om)$).  The resulting curvature 2-forms are
\begin{equation}
  \Rcal_{2\alh}   = \frac{B^2}{r_+^4} \del_{\alh\betah} e_2 \wedge e_\betah \eqand
  \Rcal_{\alh\betah} = \frac{1}{r_+^2} \hat{\Rcal}_{\alh\betah}
                      - \frac{B^2}{r_+^4} (\Jcalh_{\alh\betah}\Jcalh_{\gammah\delh} 
                      + \Jcalh_{\alh[\gammah|}\Jcalh_{\betah|\delh]}) e_\gammah\wedge e_\delh
\end{equation}
where $\hat{\Rcal}_{\alh\betah}$ are the curvature 2-forms on $\CP{N}$.

This results in a Riemann tensor with non-vanishing components
\begin{equation}\label{eqn:riemann}
  R_{2\alh 2\betah} = \frac{B^2}{r_+^4} \del_{\alh\betah} , \qquad
  R_{\alh\betah\gammah\delh} = \frac{1}{r_+^2} \hat{R}_{\alh\betah\gammah\delh} 
                              - \frac{2B^2}{r_+^4} (\Jcalh_{\alh\betah}\Jcalh_{\gammah\delh} 
                              + \Jcalh_{\alh[\gammah|}\Jcalh_{\betah|\delh]}).
\end{equation}
where 
\begin{equation}
  \Rh_{\al\beta\gamma\del} = \gh_{\al\gamma}\gh_{\beta\del} - \gh_{\al\del}\gh_{\beta\gamma}
                                 + \Jcalh_{\al\gamma}\Jcalh_{\beta\del} - \Jcalh_{\al\del}\Jcalh_{\beta\gamma} + 2\Jcalh_{\al\beta} \Jcalh_{\gamma\del}
\end{equation}
is the Riemann tensor of $\CP{N}$.  The non-vanishing Ricci tensor components and Ricci scalar are
\begin{equation}\label{eqn:ricci}
  R_{22} = \frac{2NB^2}{r_+^4}, \qquad
  R_{\alh\betah} = \left( \frac{2(N+1)}{r_+^2} - \frac{2B^2}{r_+^4} \right)\del_{\alh\betah}, \qquad
  R = \frac{4N(N+1)}{r_+^2} - \frac{2NB^2}{r_+^4}
\end{equation}

Note that the Einstein equations for the metric (\ref{eqn:metric}) are equivalent to the following algebraic relations:
\begin{equation}\label{id:einstein}
  \La = \frac{2}{E^2} - \frac{1}{L^2} 
      = -\frac{2}{E^2} + \frac{2NB^2}{r_+^4}
      =  \frac{2(N+1)}{r_+^2} - \frac{2B^2}{r_+^4} 
\end{equation}
These are solved automatically by equations (\ref{eqn:Ldef}-\ref{eqn:Edef}), but these relations are often useful for simplifying calculations.

When $\La=0$ (or equivalently $l\rightarrow \infty$), the full spacetime is asymptotically flat, and the identities (\ref{id:einstein}) simplify to
\begin{equation} \label{eqn:asflateinstein}
  E^2 = 2L^2 = \frac{B^2}{N(N+1)^2} = \frac{r_+^2}{N(N+1)}.
\end{equation}

\subsection*{Computation of operators}
In Section \ref{sec:decoupling}, and the associated Appendix \ref{app:nhframe}, we derived equations that are covariant on $\Hcal$, with indices $\mu,\nu,\dots$.  This is convenient, in that it now allows us to evaluate these equations without using the particular basis choice that we used to derive them.

Here, $\Hcal$ can be written as a fibration over $\CP{N}$.  It will be convenient in this section to write equations in a way that is covariant over $\CP{N}$; since this will then allow us to divide components up into scalar, vector and tensor parts, depending on how they transform as fields on $\CP{N}$.  We define indices $\al,\beta,\ldots$ that are covariant on $\CP{N}$, raised and lowered with the Fubini-Study metric $\gh_{\al\beta}$ on $\CP{N}$.

For quantities transforming as vectors on $\CP{N}$, it is often useful to project into the $\mp i$ eigenspaces of $\Jcalh$ using the operator
\begin{equation}
  \Pcalh^\pm_{\al\beta} = \frac{1}{2} \left( \gh_{\al\beta} \pm i \Jcalh_{\al\beta}\right).
\end{equation}

We now look to evaluate the perturbation operators $\Ocal{b}$ in the case of this metric, using equations (\ref{eqn:Oscalar},\ref{eqn:Ograv},\ref{eqn:Omax}).  Here, $L$ is constant, so $d(L^2) = 0$ and various terms vanish.  Furthermore, (\ref{eqn:nhmpk}) implies that the vector field $k$ satisfies
\begin{equation}\label{eqn:kforms}
  k = \Om B e_2, \qquad
  k.m = \Om m, \qquad
  k.k = B^2\Om^2 ,\qquad 
  dk = \Om B \, de_2 = 2 \Om B^2 \Jcalh .
\end{equation}
Finally, we need to expand the covariant derivative on $\Hcal$ in terms of derivatives on $\CP{N}$.  It is convenient to define the following charged covariant derivative on $\CP{N}$:
\begin{equation}
   \Dcalh_\al = \hat{D}_\al - im \Acalh_\al,
\end{equation}
where $\Jcalh = \tfrac{1}{2}d\Acalh$ is the K\"ahler form on $\CP{N}$, and $\hat{D}_\al$ is the Levi-Civita connection.  Note that $\Dcalh$ satisfies
\begin{equation}
  [\Dcalh_\al,\Dcalh_\beta] = -2im \Jcalh_{\al\beta} \eqand
  \Dcalh^\pm . \Dcalh^\mp = \tfrac{1}{2} \Dcalh^2 \mp 2mN,
\end{equation}
where $\Dcalh^\pm_\al \equiv \Pcalh^{\pm\beta}_{\al}\Dcalh_\beta$.

Given this, we can expand terms of the form $\nabla^2 Y$ and $\nabla Y$ in terms of this derivative, some examples of components in the gravitational case include
\begin{equation}
  (\nablah^2 Y)_{22} = \left(\frac{1}{r_+^2} \Dcalh^2 -\frac{m^2}{B^2} - \frac{2(2N+1)B^2}{r_+^4} \right) Y_{22}
                      -\frac{4B}{r_+^3} \Jcalh^{\al\beta} \Dcalh_\al Y_{2\beta}
\end{equation}
and
\begin{equation}
  \nablah^\al Y_{2\al} = \frac{1}{r_+} \Dcalh^\al Y_{2\al}.
\end{equation}

Putting these expressions, together with equations (\ref{eqn:riemann},\ref{eqn:ricci},\ref{eqn:kforms}) into the general equations (\ref{eqn:Oscalar},\ref{eqn:Omax},\ref{eqn:Ograv}) gives us explicit expressions for the operators $\Ocal{0}$, $\Ocal{1}$ and $\Ocal{2}$ in the case of this metric.  The explicit expressions for these operators can then be simplified to those given in (\ref{eqn:2maxpert}-\ref{eqn:amaxpert}) for $\Ocal{1}$ and (\ref{eqn:22pert}-\ref{eqn:abpert}) for  $\Ocal{2}$.

\subsection*{Mode decomposition of operators}
We now move on to consider the more complicated case of electromagnetic and gravitational perturbations.  Firstly, it is useful decompose the action of the operators $\Ocal{1}$ and $\Ocal{2}$ on an arbitrary eigenvector $Y$ into components tangent, and normal to, $\CP{N}$.

The operator $\Ocal{1}$ describing Maxwell perturbation modes (defined in (\ref{eqn:Omax})) reduces to
\begin{equation}\label{eqn:2maxpert}
  (\Ocal{1} Y)_{2} = \left( -\frac{2Nm^2L^4}{r_+^4} - \frac{L^2}{r_+^2} \Dcalh^2 + 2 + 4\La L^2
                     \right) Y_2 + 2\xi^{\al\beta}\Dcalh_\beta Y_\al
\end{equation}
and
\begin{equation}\label{eqn:amaxpert}
  (\Ocal{1} Y)_\al = \left( -\frac{2Nm^2L^4}{r_+^4} - \frac{L^2}{r_+^2} \Dcalh^2 
                           + \frac{2B^2L^2}{r_+^4} + \La L^2\right) Y_\al
        + \frac{2imL^2}{r_+^2} \Jcalh_{\al}^{\;\;\beta} Y_{\beta}
        - \xi_{\al}^{\phantom{\al}\beta}\Dcalh_\beta Y_2.
\end{equation}
where
\begin{equation}
  \xi_{\al\beta} \equiv \frac{L^2}{r_+} \left(\frac{1}{E} \gh_{\al\beta}
                                   - \frac{B}{r_+^2}\Jcalh_{\al\beta} \right).
\end{equation}
Indices in these equations are raised and lowered with the metric $\gh_{\al\beta}$ on $\CP{N}$.

It is also useful to define $\DelLA$; a charged Lichnerowicz operator acting on rank-2 symmetric tensors on $\CP{N}$:
\begin{equation}\label{eqn:Lichnerowicz}
  \DelLA \Ybb_{\al\beta} 
       = -\Dcalh^2 \Om_{\al\beta} - 2 \hat{R}_{\al\gamma\beta\del} \Ybb^{\gamma\del} 
         + 4(N+1) \Ybb_{\al\beta}.
\end{equation}
This is the obvious generalization of the standard Lichnerowicz operator on $\CP{N}$, with the Laplacian $\hat{\nabla}^2$ replaced by our charged Laplacian $\Dcalh^2$ (following \cite{Kunduri:2006}).

Given this definition, the action of the operator $\Ocal{2}$ for gravitational perturbations (\ref{eqn:Ograv}) on an arbitrary 2-tensor with Fourier dependence $e^{im\psi}$ is given by:
\begin{multline}\label{eqn:22pert}
  (\Ocal{2} Y)_{22} 
     = \left(-\frac{2Nm^2L^4}{r_+^4} + 2
        - \frac{L^2}{r_+^2} \Dcalh^2 + 4(N+1)\frac{L^2B^2}{r_+^4} -\frac{4(N+1)L^2}{l^2} 
       \right) Y_{22} \\
        + 4\xi^{\al\beta} \Dcal_\beta Y_{2\al},
\end{multline}
\begin{multline}\label{eqn:2bpert}
  (\Ocal{2} Y)_{2\al} 
       = \left(-\frac{2Nm^2L^4}{r_+^4} + 2
               - \frac{L^2}{r_+^2} \Dcalh^2
               - \frac{2L^2}{E^2} + (2N+6)\frac{B^2 L^2}{r_+^4} -\frac{4(N+1)L^2}{l^2} 
         \right) Y_{2\al} \\
        + \frac{2imL^2}{r_+^2} \Jcalh_\al^{\;\;\beta}Y_{2\beta}
        - 2\xi_{\;\;\al}^{\beta} \Dcalh_\beta Y_{22}
        + 2\xi^{\beta\gamma}\Dcalh_\gamma Y_{\al\beta},
\end{multline}
\begin{multline}\label{eqn:abpert}
  (\Ocal{2} Y)_{\al\beta} 
    = \left( -\frac{2Nm^2L^4}{r_+^4} - \frac{4(N+1)L^2}{r_+^2} 
             + \frac{4B^2 L^2}{r_+^4} \right)Y_{\al\beta} 
      + \frac{L^2}{r_+^2} \DelLA Y_{\al\beta} + \frac{2imL^2}{r_+^2}[\Jcalh,Y]_{\al\beta}\\
    - \frac{4B^2L^2}{r_+^4}\left( (\Jcalh Y \Jcalh)_{\al\beta} + \del_{\al\beta} Y_{22} \right)
    -4\xi_{(\al|}^{\quad \gamma} \Dcal_{\gamma} Y_{2|\beta)} ,
\end{multline}
Note that (\ref{eqn:22pert}) is equivalent to the trace of (\ref{eqn:abpert}), given that $Y_{22} = -Y_{\al}^{\;\;\al}$.

Recall that in Chapter \ref{chap:decoupling}, we found decoupled equations for the quantities $\vphi_i$ and $\Om_{ij}$, and then in Section \ref{sec:decomposition} we separated each equation into an $AdS_2$ part and a part on $\Hcal\sim S^{2N+1}$.  In this example, we now see that there is further coupling that we want to get rid of, between equations on the different parts of $\Hcal$, namely the directions normal and tangent to $\CP{N}$.

We now look to complete the decoupling by taking a scalar-vector-tensor decomposition with respect to $\CP{N}$.  Our decomposition is equivalent to that used in the numerical studies of perturbations of the full spacetime \cite{Kunduri:2006,Murata:2008,Dias:2010eu}.  The result of this is that we can expand general perturbations in terms of scalar, vector and tensor harmonics on $\CP{N}$, and the relevant eigenvalues of the Laplacian $\Dcalh^2$ are known (see \cite{Martin:2008pf} for further details).  We describe this in detail below.  

Note that, for $N=1$, there are no vector or tensor modes.  That is, imposing either the conditions \eqref{eqn:TTdef} or the conditions \eqref{eqn:vecdef} implies that $Y_{\mu\nu}=0$.

\section{Gravitational perturbations}\label{sec:gravcalcs}
\subsection{Tensor modes}\label{sec:TTmodes}
Gravitational tensor modes are those that only have transverse, traceless parts of $\Om_{\al\beta}$ turned on, i.e.\ perturbations of the form
\begin{equation}\label{eqn:TTdef}
   Y_{22} = 0 = Y_{2\al}, \qquad 
  \gh^{\al\beta} Y_{\al\beta}=0, \qquad
  \Dcalh^{\pm\al} Y_{\al\beta} = 0. 
\end{equation}
The components of the equations (\ref{eqn:22pert},\ref{eqn:2bpert}) vanish for tensor type perturbations, and \eqref{eqn:abpert} reduces to
\begin{multline}\label{eqn:abpertTT}
  (\Ocal{2} Y)_{\al\beta}
    = \left( -\frac{2Nm^2L^4}{r_+^4} - \frac{4(N+1)L^2}{r_+^2} 
            + \frac{4B^2L^2}{r_+^4} \right)Y_{\al\beta} + \frac{L^2}{r_+^2} \DelLA Y_{\al\beta}
     + \frac{2imL^2}{r_+^2}[\Jcalh,Y]_{\al\beta}\\
    - \frac{4B^2L^2}{r_+^4}(\Jcalh Y \Jcalh)_{\al\beta},
\end{multline}
We expand $Y_{\al\beta}$ in terms of separable Fourier modes
\begin{equation}
  Y_{\al\beta} = e^{im\psi} \Ybb_{\al\beta}
\end{equation}
where $\Ybb_{\al\beta}(x)$ a tensor harmonic on $\CP{N}$,with $\Dcalh^{\al\pm} \Ybb_{\al\beta} = 0$.

As $\CP{N}$ is a complex manifold, we can split both $\Ybb$ and equation (\ref{eqn:abpertTT}) into hermitian and anti-hermitian parts, which are eigenvectors of the linear map
\begin{equation}\label{map:Jsquared}
  \Ybb_{\alpha\beta} \mapsto \Jcalh_{\al}^{\;\gamma} \Jcalh_{\beta}^{\;\del} \Ybb_{\gamma\del} 
\end{equation}
with eigenvalues $+1$ and $-1$ respectively.  In other words, we write $\Ybb_{\al\beta} = \Ybb^+_{\al\beta} + \Ybb^-_{\al\beta}$ where $(\Jcalh\Ybb^\pm \Jcalh)_{\al\beta} = \mp \Ybb^\pm_{\al\beta}$, with the upper signs corresponding to hermitian modes.

In the anti-Hermitian case, the modes can be divided further into the $\mp i$ eigenspaces of $\Jcalh$, with $\Jcalh_{\al\beta} \Ybb^{\pm}_\beta = \mp i \Ybb^{\pm}_\beta$.  Following \cite{Kunduri:2006}, we summarize this by setting $\sigma = \mp 1$ ($-$ for Hermitian, $+$ for anti-Hermitian), and $\eps = \pm 1$ for the two cases of anti-Hermitian modes, and then see that
\begin{equation}
  (\Jcalh\Ybb \Jcalh)_{\al\beta} = \sigma \Ybb_{\al\beta} \eqand [\Jcalh,\Ybb]_{\al\beta} 
                           = i\eps(1+\sigma) \Ybb_{\al\beta}.
\end{equation}

We can take $\Ybb$ to be an eigenstate of the generalized Lichnerowicz operator on $\CP{N}$ (as such eigenstates form a complete set), i.e. we assume that
\begin{equation}\label{eqn:evals}
  (\DelLA \Ybb)_{\al\beta} = \la_{\kap,m}^{\mathrm{T}} \Ybb_{\al\beta}.
\end{equation}

This eigenvalue equation has known solutions, discussed in \cite{Kunduri:2006}.  For $N=1$, there are no tensor harmonics on $\CP{1}=S^2$. For $N\geq 2$, the $m=0$ eigenvalues are given by
\begin{equation} \label{eqn:tensorevals}
  \lakz{T}= 4\kap(\kap + N) + 4(N+\sig),
\end{equation}
for non-negative integers $\kap$.\footnote{Note that the allowed range of values for $\kap$ is unknown in general, e.g.\ there may be a positive lower bound on the allowed values of $\kap$ in some dimensions, but this will not turn out to be relevant here.} 

Inserting all this into (\ref{eqn:abpertTT}) implies that
\begin{equation}
  (\Ocal{2} Y)_{\al\beta} = \la Y_{\al\beta}
\end{equation}
where 
\begin{equation}\label{eqn:TTreqn}
  \la  =  -\frac{2Nm^2L^4}{r_+^4}
            + \frac{4B^2L^2}{r_+^4}(1-\sig) 
            + (\la_{\kap,m}^{\mathrm{T}} - 4(N+1) - 2m(1+\sig))\frac{L^2}{r_+^2},
\end{equation}
In Section \ref{sec:nhmp} we gave this eigenvalue explicity in the asymptotically flat case \eqref{eqn:tensorevalsflat} and the asymptotically $AdS$ case \eqref{eqn:tensorevalsads}. 

\subsection{Vector Modes}\label{sec:gravvec}
There have currently been no studies in the literature of the stability of this black hole to vector type gravitational perturbations, which exist in dimensions $d\geq 7$.

Vector modes consist of divergence free vectors $Y_{2\al}$, along with the traceless, but not transverse, contributions to $Y_{\al\beta}$ that can be constructed from them by differentiation, that is
\begin{equation}\label{eqn:vecdef}
  Y_{22} = 0, \qquad 
  \Dcalh^{\pm\al} Y_{2\al} = 0,\qquad
  Y_{\al}^{\;\;\al} = 0.
\end{equation}
We expand these perturbations as
\begin{equation}
  Y_{2\al} = g e^{im\psi} \Ybb_\al, \qquad
  Y_{\al\beta} = e^{im\psi} \left(h^+ \Ybb_{\al\beta}^+ + h^- \Ybb_{\al\beta}^-\right)
               \equiv Y_{\al\beta}^+ + Y_{\al\beta}^-
\end{equation}
where $\Ybb_\al$ is a divergence-free vector harmonic with
\begin{equation} \label{eqn:vecevals}
  \Dcalh^2 \Ybb_\al = - \lak{V} \Ybb_\al, \qquad
  \Dcalh^{\pm\al} \Ybb_{\al} = 0, \eqand
  \Ybb_{\al\beta}^\pm \equiv \frac{-1 }{\sqrt{\lak{V}}} \Dcalh^\pm_{(\al}\Ybb_{\beta)}. 
\end{equation}

There are several different separable modes that couple to each other in this sector of perturbations.  Therefore, in order to find the relevant eigenvalues we need to consider all such modes together.  In particular, the eigenvalues of $\Ocal{2}$ will be the eigenvalues of the matrix that describes the coupling between the different components of $Y_{ij}$.

We can take $\Ybb_\al$ to be an eigenvector of the complex structure $\Jcalh$, with eigenvalue $i\eps = \mp i$, that is: $\Jcalh_{\al}^{\;\;\beta} \Ybb_\beta = -i\eps \Ybb_\al$.

Note that $\Ybb_{\al\beta}=0$ is traceless,
\begin{eqnarray}
 \Dcalh^2 Y_{\al\beta} 
    &=& - \left[\lak{V}-2(N+1) - 4m - 2(1+3\eps) 
                                       \right] Y_{\al\beta}^+ \nn\\
    & & - \left[\lak{V} -2(N+1)+ 4m - 2(1-3\eps) 
                                          \right] Y_{\al\beta}^-
\end{eqnarray}
and
\begin{equation}
  \Dcalh^{\pm\beta} Y_{\al\beta} = \frac{e^{im\psi}}{2\sqrt{\lak{V}}}
        \left[ \frac{\lak{V}}{2} \pm m(N+1\mp \eps) - (1\mp 2\eps)(N+1) \right]h^\mp \Ybb_\al.
\end{equation}

The action of $\Ocal{2}$ on $Y$ now reduces to three equations:
\begin{eqnarray}\label{eqn:2bpertvec}
  (\Ocal{2}Y)_{2\al}
      &=& \Bigg[ L^2\left(
                -\frac{2Nm^2L^2}{r_+^4} + \frac{2m\eps}{r_+^2} 
                + \frac{\lak{V}}{r_+^2}
                + \frac{2}{E^2} + (2N+6)\frac{B^2}{r_+^4}
          \right) g \nn\\
      & & \qquad + \frac{\xi^+}{\sqrt{\lak{V}}} \left( \tfrac{1}{2}\lak{V} + m(N+1-\eps)
                                               - (1-2\eps)(N+1) \right) h^- \\
      & & \qquad + \frac{\xi^-}{\sqrt{\lak{V}}} \left(\tfrac{1}{2}\lak{V} - m(N+1+\eps) 
                 - (1+2\eps)(N+1) \right) h^+  \Bigg] e^{im\psi}\Ybb_{\al}\nn
\end{eqnarray}
and
\begin{multline}\label{eqn:abpertvec}
  (\Ocal{2}Y)_{\al\beta}^\pm
    = \Bigg[ 
      L^2\bigg( -\frac{2Nm^2L^2}{r_+^4} \mp \frac{2m}{r_+^2} + \frac{2m\eps}{r_+^2}
             + \frac{4B^2}{r_+^4}(1\mp \eps) - \frac{2(N+1)}{r_+^2} 
             + \frac{\la}{r_+^2}
      \bigg) h^\pm \\
     + 4\sqrt{\lak{V}} \xi^\pm g \Bigg]e^{im\psi}\Ybb_{\al\beta}^\pm,
\end{multline}
where
\begin{equation}
  \xi^\pm \equiv  \frac{L^2}{r_+} \left(\frac{1}{E} \pm \frac{iB}{r_+^2} \right), \qquad \xi \equiv \xi^+.
\end{equation}
To obtain the latter equation, we have separated out the components proportional to $\Ybb^\pm$ by noting that they are both eigenfunctions of the map \eqref{map:Jsquared} with differing eigenvalues $\pm\eps$.

Hence we have obtained a matrix formulation of the operator $\Ocal{2}$ in this case, acting on $[g,h^+,h^-]^\mathrm{T}$.  We can think of this as describing the mixing between the sectors $Y_\al$, $Y^+_{\al\beta}$, $Y^-_{\al\beta}$:
\begin{multline}
  \Ocal{2}
    = L^2 \left( \tfrac{\lak{V}+2m\eps}{r_+^2} -\tfrac{2Nm^2L^2}{r_+^4} \right) \Id + \\
      \left( \begin{array}{ccc}
           \frac{2L^2}{E^2} +  \frac{(2N+6)L^2 B^2}{r_+^4} 
             & {\scriptstyle \tfrac{\left(\frac{1}{2}\lak{V} - m(N+1+\eps) 
                                               - (1+2\eps)(N+1) \right)\xi^*}{\sqrt{\lak{V}}} }
             & {\scriptstyle \tfrac{\left( \frac{1}{2}\lak{V} + m(N+1-\eps)
                                               - (1-2\eps)(N+1) \right)\xi}{\sqrt{\lak{V}}} }\\
           4\xi\sqrt{\lak{V}} & \frac{4B^2L^2}{r_+^4}(1-\eps) - \frac{2(N+1+m)L^2}{r_+^2} & 0 \\
           4\xi^*\sqrt{\lak{V}} & 0 & \frac{4B^2L^2}{r_+^4}(1+\eps) - \frac{2(N+1-m)L^2}{r_+^2}
        \end{array}\right)
\end{multline}

We now restrict to the case $m=0$ that is relevant to our conjecture, and find that here the matrix now reduces to
\begin{equation} \label{eqn:vecmatrix}
  \left( \begin{array}{ccc}
           \frac{\lakz{V}L^2}{r_+^2} + \frac{2}{E^2} +\frac{(2N+6)B^2}{r_+^4} 
             & {\scriptstyle \left(\tfrac{1}{2}\lakz{V}
                                   - (1+2\eps)(N+1) \right)\tfrac{\xi^*}{\sqrt{\lakz{V}}}}
             & {\scriptstyle\left(\tfrac{1}{2}\lakz{V} 
                                  - (1-2\eps)(N+1)\right)\tfrac{\xi}{\sqrt{\lakz{V}}}}\\
           4\xi\sqrt{\lakz{V}} & \frac{L^2(\lakz{V}-2(N+1))}{r_+^2} + \frac{4B^2L^2}{r_+^4}(1-\eps)
             & 0 \\
           4\xi^*\sqrt{\lakz{V}} 
             & 0 
             & \frac{L^2(\lakz{V}-2(N+1))}{r_+^2} + \frac{4B^2L^2}{r_+^4}(1+\eps)
        \end{array}\right)
\end{equation}
We can find all eigenvalues of $\Ocal{2}$ by finding the eigenvalues of this matrix.  However, to do this explicitly we need to determine the allowed eigenvalues $\lakz{V}$ of $-\Dcalh^2 = -\hat{\nabla}^2$.  Note that the eigenvalues $\la_H$ of the Hodge-de Rham Laplacian
\begin{equation}
 \Del_H = -(\star d \star d + d \star d \star)
\end{equation}
on $\CP{3}$ were given in Ref.\ \cite[Table 2]{Martin:2008pf} (determined from \cite{Ikeda:1978}).  These can be generalized to $\CP{N}$ to give
\begin{equation}
  \la_H = 4(\kap+2)(\kap+N+1) \quad \mathrm{where} \quad \kap = 0,1,2,\ldots.
\end{equation}
The eigenvalues of the standard Laplacian are related to this by the Bochner-Weitzenb\"ock identity on $\CP{N}$, which implies that
\begin{equation}
  \Del_H \Ybb_\al = -\hat{\nabla}^2\Ybb_\al + 2(N+1) \Ybb_\al
\end{equation}
where we have made use of the Ricci tensor
\begin{equation}\label{eqn:cpnricci}
  \Rh_{\al\beta} = 2(N+1)\gh_{\al\beta}
\end{equation}
of $\CP{N}$.  Hence the eigenvalues of $-\hat{\nabla}^2$are actually
\begin{equation}\label{eqn:vecla}
  \lakz{V} =  4(\kap+2)(\kap+N+1) - 2(N+1) = 4\kap (\kap + 2) + 2(N+1)(2\kap+3)
\end{equation}
where $\kap = 0,1,2,\dots$.  This gives us enough information to evaluate the eigenvalues of $\Ocal{2}$.

In the asymptotically flat case the matrix representation of $\Ocal{2}$ reduces, using the identities \eqref{eqn:asflateinstein}, to
\begin{multline}\label{eqn:O2scalar}
  \frac{\lakz{V}-2(N+1)}{r_+^2}\Id\\
      + \left( \begin{array}{ccc}
           \frac{2}{NL^2}(N+2) 
            &  \left(\tfrac{1}{2}\lakz{V}- (1+2\eps)(N+1) \right)\tfrac{\xi^*}{\sqrt{\lakz{V}}}
            & \left(\tfrac{1}{2}\lakz{V} - (1-2\eps)(N+1)\right)\tfrac{\xi}{\sqrt{\lakz{V}}}\\
           4\xi \sqrt{\lakz{V}}    & \frac{2}{NL^2}(1-\eps)  & 0 \\
           4\xi^* \sqrt{\lakz{V}}  &             0           
                                   & \frac{2}{NL^2}(1+\eps)
        \end{array}\right).
\end{multline}
The characteristic equation is then independent of $\eps$.  Inserting the allowed values \eqref{eqn:vecla} into this, we find that the eigenvalues of $\Ocal{2}$ are simple rational numbers, given by equation \eqref{eqn:gravvecevals}.

In the asymptotically $AdS$ case, it is not possible to find the eigenvalues explicitly (at least in a simple form).  However, it is reasonably straightforward to prove that all eigenvalues are positive for all $N$ and $\kap$, and hence there is no instability in this sector.  The proof of this is as follows.
\proof
Consider first the modes with $\kap=0$.  Writing $z \equiv \La L^2$ for convenience, the characteristic equation of the matrix \eqref{eqn:O2scalar} takes the form
\begin{equation}
  \Big(t - \frac{2}{N}\big(1+(N+2)z\big) \Big)Q(t) = 0
\end{equation}
where the quadratic $Q(t)$ is given by
\begin{multline}
Q(t) = \Big(t^2 -\left[6 (1+2 z)+ N (1+4 z)\right]\frac{2t}{N} \\
      + \frac{8}{N^2}\left[ N^2 z^2+ N (1+2 z)+5 N z (1+2 z)+4 (1+2 z)^2\right]\Big).
\end{multline}
Clearly there is an eigenvalue $\frac{2}{N}\big(1+(N+2)z\big)$, which is positive, as $z$ is restricted to lie in the range
\begin{equation}\label{eqn:zrange}
  -\frac{1}{N+2} < z \leq 0.
\end{equation}
(this is since $1+(N+2)z = 2N(N+1)L^2/r_+^2$).

To see that the other eigenvalues are also positive, note that
\begin{equation}
  Q(0) = \frac{8 (N^2 z^2 + 4 (1 + 2 z)^2 + N (1 + 7 z + 10 z^2)}{N^2}
\end{equation}
is increasing with $z$ in the range \eqref{eqn:zrange}, while
\begin{equation}
  Q'(0) = -\frac{2 (6 + N + 12 z + 4 N z)}{N}
\end{equation}
is decreasing over the same range.  When $z=-1/(N+2)$,
\begin{equation}
  Q(0) = \frac{8}{N+2} > 0 \eqand Q'(0) = -\frac{2(N+4)}{N+2} < 0 
\end{equation}
and hence $Q(0)>0$ and $Q'(0)<0$ for all allowed $z=\La L^2$ and $N$.  As $Q(t)$ is a quadratic, this is sufficient to prove that its roots are positive, and hence all eigenvalues of $\Ocal{2}$ are positive when $\kap=0$.

Now look to generalize this to all $\kap$, and set 
\begin{equation}
  P(t) \equiv \det (t\Id - \Ocal{2}),
\end{equation}
where $\Ocal{2}$ is the matrix representation of $\Ocal{2}$ given by equation \eqref{eqn:vecmatrix}.  Consider the quantities $P(0)$ and $P'(0)$.  When $\kap=0$, all roots of the quadratic $P(t)$ are positive, so we must have $P(0)<0$ and $P'(0)>0$.

For general $\kap\geq 0$, $P(0)$ and $P'(0)$ are respectively 6th and 5th order polynomials in $\kap$, with coefficients depending on $N$ and $z$.  If we temporarily allow $\kap$ to vary smoothly in the range $[0,\infty)$, the roots of $P(t)$ must vary continuously with $\kap$.  Hence, if we find that the conditions $P(0)<0$ and $P'(0)>0$ always remain true, then we can conclude that all roots of $P(t)$ are positive for all allowed $\kap$.  It can be shown by explicit computation that all coefficients of $\kap$ in $P(0)$ are negative and all coefficients of $\kap$ in $P'(0)$ are positive, and hence $P(0)$ is decreasing with $\kap$ and $P'(0)$ is increasing, for all $z$ and $N$.  Therefore $P(0)<0$ and $P'(0)>0$ for all (fixed) $\kap$, $z$, $N$, and this is sufficient to prove the result.$\Box$

\subsection{Scalar Modes}\label{sec:gravscalar}
Next, we consider the sector of gravitational scalar-type perturbations.  For the (non-extremal) full black hole solution, such perturbations have been previously studied by Murata \& Soda \cite{Murata:2008} (for $d=5$) and Dias \etal \cite{Dias:2010eu} (for $d=5,7,9$).

Scalar modes are the most complicated, with all possible parts of the perturbations turned on.  Starting with $Y_{22}$, contributions to $Y_{2\al}$ and $Y_{\al\beta}$ are constructed by taking derivatives.  Recall that the scalar eigenfunctions (\ref{eqn:scalarYbb}) of the charged covariant Laplacian $\Dcalh^2$ on $\CP{N}$ have eigenvalues given in (\ref{eqn:scalarevals}).  We can describe the full set of scalar perturbations as
\begin{align}
  Y_{22}          &= e^{im\psi} f \Ybb,\nn \\
  Y_{2\alpha}     &= e^{im\psi} \left[g^+ \Ybb_\al^+ + g^- \Ybb_\al^- \right],\nn\\
  Y_{\alpha\beta} 
      &= e^{im\psi}\bigg[- \tfrac{1}{\sqrt{\lak{S}}}\Big(h^{++} \Ybb_{\al\beta}^{++} 
                                                     + h^{--} \Ybb_{\al\beta}^{--} 
                                                     + h^{+-} \Ybb_{\al\beta}^{+-} \Big)
                           - \tfrac{1}{2N}f\del_{\al\beta} \Ybb \bigg],\label{eqn:scalaransatz}
\end{align}
where $\Ybb$ is the scalar eigenfunction defined in $(\ref{eqn:scalarYbb})$ and $\Ybb_\al^\pm$, $\Ybb_{\al\beta}^{\pm\pm}$, $\Ybb_{\al\beta}^{+-}$ are scalar-derived vector/tensor eigenfunctions, defined by
\begin{equation}\label{eqn:scalarderived}
  \Ybb_{\al}^\pm \equiv -\frac{\Dcalh_\al^\pm \Ybb}{\sqrt{\lak{S}}}  , \qquad
  \Ybb_{\al\beta}^{\pm\pm} \equiv \Dcalh^\pm_{(\al} \Ybb_{\beta)}^\pm 
\end{equation}
and
\begin{equation}
  \Ybb_{\al\beta}^{+-} = \Dcalh^+_{(\al} \Ybb_{\beta)}^- + \Dcalh^-_{(\al} \Ybb_{\beta)}^+ 
                           - \tfrac{\sqrt{\lak{S}}}{2N}\del_{\al\beta}\Ybb.
\end{equation}
These have the following properties:
\begin{align}
  \Jcalh_{\al}^{\;\;\beta} \Ybb^{\pm}_\beta &= \mp i \Ybb^\pm_\al, &
  \Dcalh^2 \Ybb_\al^\pm         &= -\left[\lak{S} - 2(N+1)\mp 4m \right] \Ybb_\al^\pm \nn\\
  \gh^{\al\beta} \Dcalh_\al \Ybb^\pm_{\beta} 
                                 &= \tfrac{ \lak{S} \mp 2mN}{2\sqrt{\lak{S}}} \Ybb ,&
   \Dcalh^2 \Ybb^{\pm\pm}_{\al\beta} &= -\left[\lak{S} - 4(N+3) \mp 8m \right] \Ybb^\pm_{\al\beta}, \nn
\end{align}
\begin{align}
   \Jcalh^{\al\beta} \Dcalh_\al \Ybb^\pm_{\beta} 
                                 &= \tfrac{\mp i}{2\sqrt{\lak{S}}} \left( \lak{S} \mp 2mN \right) \Ybb ,&
   \Dcalh^2 \Ybb^{+-}_{\al\beta}     &= -\left(\lak{S} - 4N \right) \Ybb^{+-}_{\al\beta},\nn
\end{align}
\begin{equation}
  (\Jcalh \Ybb^{\pm\pm}\Jcalh)_{\al\beta} = + \Ybb_{\al\beta}, \qquad
  (\Jcalh \Ybb^{+-}\Jcalh)_{\al\beta} = - \Ybb_{\al\beta},\qquad
  (\Jcalh\Ybb^{\pm\pm})_{\al\beta} = \mp i \Ybb_{\al\beta}\nn
\end{equation}
\begin{align}
  \Dcalh^\beta \Ybb_{\al\beta}^{\pm\pm} 
                &= -\tfrac{1}{2}\left( \lak{S} - 4(N+1) \mp 2m(N+2) \right)\Ybb^\pm_\al,\nn\\
  \Dcalh^\beta \Ybb_{\al\beta}^{+-}
                &= -\tfrac{N-1}{2N}\left[ (\lak{S}+2mN)\Ybb^+_\al + (\lak{S}-2mN)\Ybb^-_\al \right],\nn\\
  \Jcalh^{\beta\gamma} \Dcalh_\gamma \Ybb_{\al\beta}^{\pm\pm} 
                &= \mp\tfrac{i}{2}\left[ \lak{S} - 4(N+1) \mp 2m(N+2) \right]\Ybb^\pm_\al,\nn\\
  \Jcalh^{\beta\gamma} \Dcalh_\gamma \Ybb_{\al\beta}^{+-} 
                &= \tfrac{i(N-1)}{2N}\left[ (\lak{S}+2mN)\Ybb^+_\al - (\lak{S}-2mN)\Ybb^-_\al \right].
\end{align} 
Note that there are three exceptions to this description:
\begin{itemize}
\item For $\kap = m = 0$, $\Ybb$ is constant, and there are no scalar-derived vectors or tensors.  Here the system is described by just one equation.
\item For $\kap=1, m=0$, the functions $\Ybb^{\pm\pm}$ vanish, and there are only four relevant types of component.
\item For $N=1$ (i.e.\ in five dimensions), the function $\Ybb^\pm$ vanishes identically (as there are no traceless, symmetric type (1,1) tensors on $\CP{1}$).
\end{itemize}

Inserting the ansatz \eqref{eqn:scalaransatz} into equations (\ref{eqn:22pert}-\ref{eqn:abpert}), we obtain the following.  From (\ref{eqn:22pert}) we get
\begin{multline} \label{eqn:f1}
  (\Ocal{2}Y)_{22} = \Bigg[\left( -\frac{2Nm^2L^4}{r_+^4} 
                     + \frac{4L^2}{E^2}  
                     + \frac{\lak{S}L^2}{r_+^2} + 4(N+1)\frac{B^2L^2}{r_+^4}\right) f\\
                     + \frac{2\xi^- (\lak{S} - 2mN)g^+}{\sqrt{\lak{S}}} 
                     + \frac{2\xi^+ (\lak{S} + 2mN)g^-}{\sqrt{\lak{S}}} \Bigg] e^{im\psi}\Ybb .
\end{multline}
Splitting (\ref{eqn:2bpert}) into $\mp i$ eigenspaces of $\Jcalh$ gives two equations
\begin{multline} \label{eqn:gpm1}
  (\Ocal{2}Y)_{2\al}^\pm
           = \Bigg[ 2\sqrt{\lak{S}} \xi^\pm \left(1+\frac{1}{2N}\right) f 
                   + \frac{\xi^\pm}{\sqrt{\lak{S}}} \left( \frac{N-1}{N}\right) (\lak{S} \pm 2mN) h^{+-} \\
                   \qquad\qquad\quad + \left( -\frac{2Nm^2L^4}{r_+^4} 
                       + \frac{\left(\lak{S} - 2(N+1) \mp 2m\right)L^2}{r_+^2}
                       + \frac{2L^2}{E^2} + \frac{(2N+6) B^2L^2}{r_+^4} \right) g^\pm \\
              + \frac{\xi^\pm}{\sqrt{\lak{S}}} \left(\lak{S} - 4(N+1) \mp 2m(N+2)\right) h^{\pm\pm} 
             \Bigg] e^{im\psi}\Ybb_{\al}^\pm
\end{multline}
and from (\ref{eqn:abpert}) we obtain three equations
\begin{multline} \label{eqn:hpmpm1}
  (\Ocal{2}Y)_{\al\beta}^{\pm\pm}
      = \Bigg[ \left( -\frac{2Nm^2L^4}{r_+^4} 
                        + \frac{\left(\lak{S}-4(N+1)\mp 4m\right)L^2 }{r_+^2} \right) h^{\pm\pm} \\
                   + 4\sqrt{\lak{S}}\xi^\pm g^\pm \Bigg] e^{im\psi}\Ybb_{\al\beta}^{\pm\pm}
\end{multline}
and
\begin{multline} \label{eqn:hpm1}
  (\Ocal{2}Y)_{\al\beta}^{+-} 
           =\Bigg[ \left( -\frac{2Nm^2L^4}{r_+^4} 
                            + \frac{\left(\lak{S} - 4(N+1)\right)L^2}{r_+^2} 
                           + \frac{8B^2L^2}{r_+^4} \right) h^{+-} \\
               + 2\sqrt{\lak{S}} (\xi^- g^+ + \xi^+ g^- ) \Bigg] e^{im\psi}\Ybb_{\al\beta}^{+-},
\end{multline}
as well as again obtaining (\ref{eqn:f1}) from the trace terms.

In a similar way to the vector case, we now get a matrix representation of $\Ocal{2}$, acting on $[f,g^+,g^-,h^{++},h^{--},h^{+-}]^\mathrm{T}$.  For simplicity, we display it explicitly here only in the case $m=0$:
\begin{equation}
 \tfrac{1}{L^2} \Ocal{2} = \tfrac{\lakz{S}-4(N+1)}{r_+^2} \Id + \nn
  \qquad\qquad\qquad\qquad \qquad\qquad\qquad\qquad\qquad\qquad\qquad\qquad\qquad
\end{equation}
{\tiny
\begin{equation}\label{eqn:scalarmat}
 \left(
      \begin{array}{cccccc}
        2\La + \frac{4(N+2)B^2}{r_+^4} + \tfrac{4}{E^2}
              & \tfrac{2\xi^* \sqrt{\lakz{S}}}{L^2} & \tfrac{2\xi  \sqrt{\lakz{S}}}{L^2} 
              & 0 & 0 & 0 \\[3mm]
       \frac{2\xi}{L^2} \sqrt{\lakz{S}}  \left(1+\frac{1}{2N}\right) 
              &  \La + \frac{2}{E^2} + \frac{2(N+4) B^2}{r_+^4}   & 0 
              &{\scriptstyle\frac{(\lakz{S} - 4(N+1))\xi^*}{L^2\sqrt{\lakz{S}}}} 
              & 0 & \frac{(N-1)\xi \sqrt{\lakz{S}}}{NL^2} \\[3mm]
      \frac{2\xi^*}{L^2}\sqrt{\lakz{S}} \left(1+\frac{1}{2N}\right) & 0 
              & \La + \frac{2}{E^2}  + \frac{2(N+4) B^2}{r_+^4} 
              & 0 & {\scriptstyle\frac{(\lakz{S} - 4(N+1))\xi}{L^2\sqrt{\lakz{S}}}}
              & \frac{(N-1)\xi^* \sqrt{\lakz{S}}}{NL^2} \\[3mm]
      0 & \frac{4\xi}{L^2} \sqrt{\lakz{S}} & 0 &  0 & 0 & 0 \\[3mm]
      0 & 0 & \frac{4\xi^*}{L^2} \sqrt{\lakz{S}} & 0 & 0 & 0 \\[3mm]      
      0 & \frac{2\xi^*}{L^2}\sqrt{\lakz{S}} & \frac{2\xi}{L^2}\sqrt{\lakz{S}} 
              & 0 & 0 & \frac{8B^2}{r_+^4}
      \end{array}
  \right)
\end{equation}}
Again, although this matrix is complex, its eigenvalues are all real, and we now look to compute these explicitly, using the list of scalar eigenvalues $\lak{S}$ of $\Dcalh^2$ given by \eqref{eqn:scalarevals}.

Recall from above the there are three special cases that need to be dealt with separately.

Firstly, the case $\kap=m=0 = \la_{0,0}^\mathrm{S}$ is degenerate, in the sense that $Y_{2\al}$ and $Y_{\al\beta}$ vanish.  Hence this matrix reduces to a $1\times 1$ matrix,
\begin{equation}
  (\Ocal{2} Y)_{22} = L^2\left( \frac{4}{E^2} + 4(N+1)\frac{B^2}{r_+^4}\right) Y_{22}
\end{equation}
which has a trivially positive eigenvalue.

When $m=0,\kap=1$,  $\la_{1,0}^\mathrm{S} =4(N+1)$ and the eigenfunctions $\Ybb^{\pm\pm}$ vanish, which means that equations (\ref{eqn:hpmpm1}) have vanishing RHS, and hence the mass matrix is actually a $4\times4$ matrix, with
{\small
\begin{multline}
 \Ocal{2}= \\ L^2
 \left(
      \begin{array}{cccc}
        \frac{4(N+2)B^2}{r_+^4}+ 2\La + \frac{4}{E^2}
             & \frac{4\xi^*\sqrt{N+1}}{L^2}  
             & \frac{4\xi\sqrt{N+1}}{L^2} & 0 \\[3mm]
        \frac{4 \xi\sqrt{N+1}}{L^2} \left(1+\frac{1}{2N}\right) 
             &  \La + \frac{2}{E^2} + \frac{2(N+4) B^2}{r_+^4} & 0 
             & \frac{2(N-1)\xi\sqrt{N+1}}{NL^2} \\[3mm]
        \frac{4\xi^* \sqrt{N+1}}{L^2} \left(1+\frac{1}{2N}\right) 
             & 0 & \La + \frac{2}{E^2} + \frac{2(N+4) B^2}{r_+^4}  
             & \frac{2(N-1)\xi^*\sqrt{N+1}}{NL^2} \\[3mm]
       0 & \frac{4\xi^*\sqrt{N+1}}{L^2} & \frac{4\xi\sqrt{N+1}}{L^2} & \frac{8B^2}{r_+^4}
      \end{array}
 \right)\nn
\end{multline} }
The eigenvalues of this matrix were analysed in Sections \ref{sec:gravscalarevalsflat} and \ref{sec:gravscalarevalsads} in the asymptotically flat and asymptotically $AdS$ cases respectively, along with the eigenvalues of the $6\times 6$ matrix \eqref{eqn:scalarmat} for the case $\kap\geq 2$.

Finally, consider the case $N=1$, for which $\Ybb^{+-}$ vanishes.  This has the effect of eliminating the final row and column from the above matrices, and hence reduces the number of eigenvalues from six to five.

\section{Electromagnetic fields}
Following a similar approach to that of the gravitational case, we can obtain results for electromagnetic perturbations.

Note that we do not necessarily see all possible Maxwell perturbations with this approach, as perturbations that change $F_{ij}$ or $F$, but not $\vphi$ or $\vphi'$, cannot be analysed.  It is not clear whether there exist non-trivial perturbations with this property.\footnote{One can of course consider perturbations of $\vphi'_i$ rather than $\vphi$ by taking the prime of all equations above.  This has the effect of mapping $q \mapsto q^*$, $\chi\mapsto \chi^*$, $\eps\mapsto -\eps$ and $m\mapsto -m$, but leaves all results unchanged.}

The Maxwell perturbation modes can be divided into two categories which we will refer to as `vectors' and `scalars', according to their transformation properties on $\CP{N}$.  Vector modes are those that only have a divergence-free $\CP{N}$ part of $Y$ turned on, that is
\begin{equation}
  Y_2 = 0  \eqand \Dcalh^{\pm\al} Y_\al = 0.
\end{equation}

\subsection{Vector modes}
The simplest class of electromagnetic perturbations are the vector modes, which we can parametrize as
\begin{equation}
  Y_2 = 0, \qquad Y_\al = e^{im\psi} \Ybb_\al,
\end{equation}
where $\Ybb_\al$ are the divergence-free vector eigenfunctions of $\Dcalh^2$ defined by \eqref{eqn:vecevals} above.  The component $(\Ocal{1}Y)_2$ vanishes, and \eqref{eqn:amaxpert} reduces to
\begin{equation}
 (\Ocal{1}Y)_\al = \left[-\frac{2Nm^2L^4}{r_+^4} + \frac{\left(\la+2(N+1)+2m\eps\right)L^2}{r_+^2}\right]Y_\al.
\end{equation}
This gives the eigenvalues described in Section \ref{sec:emvectors}.

\subsection{Scalar modes}
The $\CP{N}$ scalar modes are more complicated, as for vector and scalar eigenvalues in the gravitational case.  We can expand the perturbations as
\begin{equation}
  Y_2   = e^{im\psi} f \Ybb, \qquad 
  Y_\al = e^{im\psi}\left(g^+ \Ybb^+_\al + g^- \Ybb^-_\al \right)
\end{equation}
where $\Ybb$ are the scalar eigenfunctions defined in \eqref{eqn:scalarYbb}, and $\Ybb^\pm_\al$ the scalar-derived vectors defined in \eqref{eqn:scalarderived}.

Note that for $\kap = m = 0$, when $\lak{S} = 0$, the associated eigenfunction $\Ybb(x)$ is constant, and hence $Y_\al=0$.  In this case, the operator $\Ocal{2}$ has simple eigenvalues, given by equation \eqref{eqn:emevals0}.

For $\lak{S}>0$, we follow an analagous separation procedure to that of the gravitational case, and find that the effective $AdS_2$ masses of various modes are given by eigenvalues of the matrix 
\begin{equation} \label{eqnmaxscalarmatrix}
 \Ocal{1} = \left( \begin{array}{ccc}
           \frac{\lak{S}L^2}{r_+^2} + 2 +4\La L^2
             &\tfrac{ (\lak{S}-2mN)\xi^*}{\sqrt{\lak{S}}}
             & \tfrac{(\lak{S}+2mN)\xi}{\sqrt{\lak{S}}}\\
           2\sqrt{\lak{S}}\xi & \frac{L^2(\lak{S}-2m)}{r_+^2}  & 0 \\
           2\sqrt{\lak{S}}\xi^* & 0 & \frac{L^2(\lak{S} +2m)}{r_+^2} 
        \end{array}\right).
\end{equation}
In the case $m=0$, the characteristic equation reduces to
\begin{equation}\label{eqn:maxwellcheqn}
  \left( \tfrac{L^2}{r_+^2}\lakz{S} - t\right)
  \Bigg[ t^2 - 2\left( \tfrac{L^2}{r_+^2}\lakz{S} + 1 + 2\La L^2 \right)t
         + \lakz{S} \left( \tfrac{L^4}{r_+^4} \lakz{S}  
                          + \tfrac{2L^2\left(1+2\La L^2 \right)}{r_+^2} - 4|\xi|^2\right) \Bigg] = 0,
\end{equation} 
with allowed values of $\lakz{S}$ given by $\lakz{S} = 4\kap(\kap+N)$ for $\kap = 0,1,\ldots$.  This leads to the eigenvalues listed in Section \ref{sec:emscalars}.

\chapter{Details of black ring calculations}\label{app:blackrings}
\renewcommand{\theequation}{E.\arabic{equation}}
\setcounter{equation}{0}

In this Appendix, we give further details of a couple of results from Chapter \ref{chap:blackrings}.

\section{Construction of an exact solution to (\ref{eqn:tdot})}
It is possible to solve equation (\ref{eqn:tdot}) exactly, using the method of characteristics.  However, this solution turns out to be fairly complicated, and as such is not particularly useful for constructing a new set of coordinates.  The construction of this solution is described below.  The end result is that we get functions in the new metric that, though well defined, can only be written down implicitly in terms of inverse functions, which would make the resulting metric highly inconvenient to work with.  Furthermore, regularity at the horizon is not manifest, which is the main motivation for doing this.

Note that, assuming that our separable solutions for $\eta^\phi$ and $\eta^\psi$ are the correct ones we can rewrite (\ref{eqn:tdot}) as
\begin{equation}\label{eqn:chpde}
  f(x) \frac{\pd\eta}{\pd x} + g(y) \frac{\pd\eta}{\pd y} = h(x,y) 
\end{equation}
where
\begin{eqnarray*} 
  f(x) &=& \pm \sqrt{\xi(x)} \\
  g(y) &=& -\sqrt{\zeta(y)} \\
  h(x,y) &=& \frac{\Omf (-\Axy \Phi + \Lxy \Psi) + \Omy (\Lxy \Phi + \Ayx \Psi)}{\Gx \Gy} . 
\end{eqnarray*}

To find a solution to this, we apply the method of characteristics.  Note that the characteristic curves follow the same paths in the $xy$ plane as the geodesics, with the parameter $s$ a non-affine parameter along them.  We pick an arbitrary initial surface $y=b$, and pick our initial data to be $\eta(x,b)=0$.  The non-characteristic condition for surfaces of constant $y$ is that $g(y)\neq 0$.  This fails at $y=y_0$, so we must pick $b<y_0$, and clearly the initial surface should also lie outside the horizon.  Thus, we are free to choose any arbitrary $b$ with $ y_h < b < y_0$.  The initial surface can be parametrised as $\{(a,b)\}_{a\in [-1,1]}$, and given this the characteristic curves $(x(s;a),y(s;a))$ obey the equations
\begin{equation}
  \frac{dx}{ds}(s;a) = f(x(s;a)) \eqand \frac{dx}{ds}(s;a) = g(y(s;a)),
\end{equation}
with solutions given implicitly by 
\begin{equation}
  \int_a^{x(s;a)} \frac{dx'}{f(x')} = s \; \; \mathrm{and} \;\; \int_b^{y(s;a)} \frac{dy'}{g(y')} = s . 
\end{equation}
Now define\footnote{This definition implicitly assumes the $x$ motion to be in the positive direction, the argument runs through in basically the same way with the opposite choice of sign.}
\begin{equation}
  F: [-1,1] \rightarrow \left[ 0, \int_{-1}^1 \frac{dx'}{f(x')} \right] \eqand 
  \Gamma : [y_h, b] \rightarrow \left[ 0, \int_{y_h}^b \frac{dy'}{\sqrt{\zeta(y')}} \right]  
\end{equation}
by
\begin{equation}
  F(x)  \equiv \int_{-1}^x \frac{dx'}{f(x')} = \pm \int_{-1}^x \frac{dx'}{\sqrt{\xi(x)}} \eqand 
  \Gamma(y) \equiv \int_b^y \frac{dy'}{g(y')} = \int_y^b \frac{dy'}{\sqrt{\zeta(y')}}. 
\end{equation}
Note that both $F$ and $\Gamma$ are bijective, and hence have well defined inverses.  Therefore, we can write \begin{equation}
  x(s;a) = F^{-1} (s+F(a)) \eqand y(s;a) = \Gamma^{-1} (s) .\label{eqn:flowmap} 
\end{equation}
Now, by (\ref{eqn:chpde}), 
\begin{equation}
  \frac{d}{ds} \eta(x(s;a),y(s;a)) = h(x(s;a),y(s;a)) 
\end{equation}
and integrating this gives
\begin{equation}
  \eta(x(s),y(s)) = \eta(a,b) + \int_0^s h\left[ F^{-1} (s' + F(a)), \Gamma^{-1} (s')\right] ds'. 
\end{equation}
Finally, we invert (\ref{eqn:flowmap}), change variables $ds' = d(\Gamma(y'))$ in the integral and insert our initial data $\eta(a,b)=0$ to give
\begin{equation} \label{eqn:etat}
  \eta(x,y) = \int_b^y \frac{ h\left[ F^{-1}\left(\Gamma(y')-\Gamma(y) +F(x) \right), y' \right]}{g(y')} dy'. 
\end{equation}
This is a well defined solution to the system, which reduces to the known solution for the singly spinning case if we set $\nu=0$ (which means $h(x,y)$ is a function of $y$ only).  Unfortunately, it is not of a form where it is particularly convenient for use in a coordinate system.

It appears in the transformed metric via
\begin{multline}
  dt + \Omf d\phi + \Omy d\psi = dv + \Omf d\tilde{\phi} + \Omy d\tilde{\psi} 
                                 + \left( \frac{\pd \eta}{dx} + \Omf \frac{\pd \eta^\phi}{dx}  
                                         + \Omy \frac{\pd \eta^\psi}{dx} \right) dx \\ 
                                 + \left( \frac{\pd \eta}{dy} + \Omf \frac{\pd \eta^\phi}{dy}  
                                        + \Omy \frac{\pd \eta^\psi}{dy} \right) dy \equiv dv + \tilde{\Omega}, 
\end{multline}
where this final equality defines
\begin{equation}
  \tilde{\Om} =\Omf d\tilde{\phi} + \Omy d\tilde{\psi} + \tilde{\Om}_x dx + \tilde{\Om}_y dy .
\end{equation}
Given our solution (\ref{eqn:etat}), we can write 
\begin{eqnarray}
  \frac{ \pd \eta}{\pd x} &=& \frac{1}{f(x)} \int_b^y \frac{ (\pd_1 h)(x',y') f(x')}{g(y')} dy' \eqand \\
  \frac{ \pd \eta}{\pd y} &=& \frac{ h(x,y)}{g(y)} - \frac{1}{g(y)} \int_b^y \frac{ (\pd_1 h)(x',y') f(x')}{g(y')} dy', 
\end{eqnarray}
where
\begin{equation} 
  x'(y';x,y) \equiv F^{-1}\left(\Gamma(y')-\Gamma(y) +F(x) \right). 
\end{equation}
Thus,
\begin{eqnarray} 
  \tilde{\Om}_x &=& \frac{\pd \eta}{\pd x} + \Omf \frac{\pd \eta^\phi}{\pd x} + \Omy \frac{\pd \eta^\psi}{\pd x} \nn \\
                &=& \pm \frac{1}{ \sqrt{\tzx}} \left[ \frac{\Omf \tx +  \Omy \cx }{\Gx} 
                                                     + \int_b^y \frac{ (\pd_1 h)(x',y') f(x')}{g(y')} dy' \right] \\
  \tilde{\Om}_y &=& \frac{\pd \eta}{\pd y} + \Omf \frac{\pd \eta^\phi}{\pd y} + \Omy \frac{\pd \eta^\psi}{\pd y} \nn \\
                &=& \frac{1}{ \sqrt{\zy}} \left[ \frac{\Omf \tx +  \Omy \cx }{\Gx} 
                                                - \int_b^y \frac{ (\pd_1 h)(x',y') f(x')}{g(y')} dy' \right] \\
\end{eqnarray}
where
\begin{equation}
  \theta(x) \equiv \beta(x) \Phi + \gamma(x) \Psi \eqand \chi(x) \equiv  \gamma(x) \Phi - \alpha(x) \Psi . 
\end{equation}

This form can then be inserted into the new metric (\ref{eqn:newmet}).  Note that we have not proved that this exact solution renders the metric regular at the horizon, and in fact it is not clear that it has this property.  The complicated form of the metric that we end up with here motivates us to look instead to merely solve the finiteness condition described above for the change of coordinates.

\section{Proof of Lemma \ref{lem:Kprime}}
There are three parts to this lemma, the first two of which are essentially trivial.  Property 1 follows directly from the conformal Killing equation for $K'$, and it is easy to verify that $K$, and by extension $K'$ cannot be constructed from the metric and Killing vectors and is therefore independent of the metric.  Each independent CK tensor defines a conserved quantity $K^{\mu_1...\mu_p} p_{\mu_1}...p_{\mu_p}$, along a geodesic with null momentum $p_\mu$.  We already have 3 of these conserved quantities from $\pd/\pd\phi$, $\pd/\pd\psi$, and the metric itself.  In a 4-dimensional geometry, finding the geodesics reduces to solving 4 coupled first order ODEs, so there are only 4 independent conserved quantities.  If there was another tensor that we could add to $K$ to give a more general conformal Killing tensor, then this would itself give a new independent CK tensor, and hence a new conserved quantity, which is a contradiction.  It remains, therefore, to establish the non-trivial third property; the non-existence of a `square-root' of $K'$.

The equations for the components $K'^{xx}$, $K'^{yy}$, $K'^{x\phi}$, $K'^{x\psi}$, $K'^{y\phi}$, $K'^{y\psi}$ respectively of (\ref{eqn:fsquare}) can be written in the form
\begin{eqnarray}
\label{eqn:Kxx} \frac{\Gx (1+C) }{\La^2} 
   &=& \left(\! f^{x\phi} \; f^{x\psi} \! \right) \mathbf{M} \left( \! \begin{array}{c} f^{x\phi} \\
                                                                                        f^{x\psi}\end{array} \!\right) 
       - \frac{ (f^{xy})^2}{\Gy} \\
\label{eqn:Kyy} \frac{\Gy (1-C)}{\La^2} 
   &=& \left(\! f^{y\phi} \; f^{y\psi} \! \right) \mathbf{M} \left( \! \begin{array}{c} f^{y\phi} \\
                                                                                        f^{y\psi}\end{array} \!\right)
       + \frac{ (f^{xy})^2}{\Gx} \\
\label{eqn:Kxp} f^{\phi\psi} \mathbf{M} \left( \!\begin{array}{c} f^{x\phi} \\ f^{x\psi} \end{array} \!\right) 
   &=& \frac{f^{xy}}{\Gy} \left(\!\!\begin{array}{c} f^{y\psi} \\ - f^{y\phi}\end{array} \!\! \right) \\
\label{eqn:Kyp} f^{\phi\psi} \mathbf{M} \left( \!\begin{array}{c} f^{y\phi} \\ f^{y\psi} \end{array} \!\right)
   &=& \frac{f^{xy}}{\Gx} \left( \!\! \begin{array}{c} f^{x\psi} \\ - f^{x\phi}\end{array} \!\! \right)
\end{eqnarray}
where
\begin{equation}
  \mathbf{M} \equiv \frac{1}{\La^2} \left(\! \begin{array}{cc} h_{\phi\phi} & h_{\phi\psi}\\
                                                               h_{\phi\psi} & h_{\psi\psi} \end{array} \! \right)
             = \frac{1}{\Hxy \Hyx} \left( \! \begin{array}{cc} \Ayx & -\Lxy \\ -\Lxy & -\Axy \end{array} \! \right) . 
\end{equation}

Contracting (\ref{eqn:Kxp}) with $G(y) \left( f^{x\phi} \; f^{x\psi}\right)$ and (\ref{eqn:Kxp}) with $G(x) \left( f^{y\phi} \; f^{y\psi}\right)$ gives us two new expressions for the LHS of equations (\ref{eqn:Kxx}), (\ref{eqn:Kyy}).  Substituting these in, and taking the difference of the resulting equations leaves us with
\begin{equation}
  \frac{2 f^{\phi\psi} \Gx \Gy}{\La^2} = 0 \Rightarrow f^{\phi\psi} = 0.
\end{equation}
Inserting this back into (\ref{eqn:Kxp}), (\ref{eqn:Kyp}) gives 
\begin{equation}
  f^{xy} \left( \!\begin{array}{c} f^{y\psi} \\
  - f^{y\phi}\end{array} \!\right) = 0 =  f^{xy} \left( \!\begin{array}{c} f^{x\psi} \\
  - f^{x\phi}\end{array} \!\right), 
\end{equation}
and hence we must have $f^{xy}=0$ (since otherwise we would have all other components vanishing, which leads us into an immediate contradiction).

Given these results, we then consider the components $K'^{\phi\phi}$, $K'^{\psi\psi}$ and $K'^{\phi\psi}$:
\begin{eqnarray} 
  \frac{1}{(1-\nu)^2} \left( \frac{\bxb}{\Gx} - \frac{\ayb}{\Gy} + p \right)  
    &=& \La^2 \left( \frac{(f^{x\phi})^2}{\Gx} - \frac{(f^{y\phi})^2}{\Gy} \right), \\
  \frac{1}{(1-\nu)^2} \left( \frac{\byb}{\Gy} - \frac{\axb}{\Gx} + r \right)  
    &=& \La^2 \left( \frac{(f^{x\psi})^2}{\Gx} - \frac{(f^{y\psi})^2}{\Gy} \right) , \\
  \frac{1}{(1-\nu)^2} \left( \frac{\gx}{\Gx} + \frac{\gy}{\Gy} + q \right) 
    &=& \La^2 \left( \frac{f^{x\phi} f^{x\psi}}{\Gx} - \frac{f^{y\phi} f^{y\psi}}{\Gy} \right).
\end{eqnarray}
We can use these three equations to express $(f^{y\phi})^2$, $(f^{y\psi})^2$ and $f^{y\phi} f^{y\psi}$ in terms of $(f^{x\phi})^2$, $(f^{x\psi})^2$ and $f^{x\phi} f^{x\psi}$, and then put this into (\ref{eqn:Kyy}).  Comparing this to (\ref{eqn:Kxx}) leads to a consistency condition
\begin{multline}
  \frac{\ayb \byb + \gy^2 + (-p\byb + 2q \gy + r \ayb)\Gy}{\Gy^2}\\ 
      = \frac{\axb \bxb + \gx^2 + (p\axb + 2q \gx - r \bxb)\Gx}{\Gx^2}  
\end{multline}
that is independent of $C$.  This separates $x$ and $y$, and hence can only be satisfied if both sides are constant for some choice of constants $p$,$q$,$r$.  In the singly spinning case this holds since $\alpha(\xi) = 0 = \gamma(\xi)$ for all $\xi\in (-\infty, 1]$, and we can then choose $p=r=0$ to make both sides vanish.  In the doubly spinning case, however, we are required to set 
\begin{equation}
  r=\lim_{x\rightarrow \pm 1} \frac{\axb}{\Gx}
\end{equation}
to avoid a pole in the RHS at $x=\pm 1$.  But these two limits are not the same for $\nu>0$, so we have a contradiction, which completes the proof of the Lemma.$\Box$

\backmatter
\bibliographystyle{utphys}
\bibliography{durkee}

\providecommand{\href}[2]{#2}\begingroup\raggedright\begin{thebibliography}{10%
0}

\bibitem{ringgeo}
M.~N. Durkee, ``{Geodesics and Symmetries of Doubly-Spinning Black Rings},''
  \href{http://dx.doi.org/10.1088/0264-9381/26/8/085016}{{\em Class. Quantum
  Grav..} {\bfseries 26} (2009) 085016},
\href{http://arxiv.org/abs/0812.0235}{{\ttfamily arXiv:0812.0235 [gr-qc]}}.

\bibitem{TypeII}
M.~N. Durkee, ``{Type II Einstein spacetimes in higher dimensions},''
  \href{http://dx.doi.org/10.1088/0264-9381/26/19/195010}{{\em Class. Quantum
  Grav..} {\bfseries 26} (2009) 195010},
\href{http://arxiv.org/abs/0904.4367}{{\ttfamily arXiv:0904.4367 [gr-qc]}}.

\bibitem{nongeo}
M.~N. Durkee and H.~S. Reall, ``{A higher-dimensional generalization of the
  geodesic part of the Goldberg-Sachs theorem},''
  \href{http://dx.doi.org/10.1088/0264-9381/26/24/245005}{{\em Class. Quantum
  Grav..} {\bfseries 26} (2009) 245005},
\href{http://arxiv.org/abs/0908.2771}{{\ttfamily arXiv:0908.2771 [gr-qc]}}.

\bibitem{higherghp}
M.~N. Durkee, V.~Pravda, A.~Pravdov{\' a}, and H.~S. Reall, ``{Generalization
  of the Geroch-Held-Penrose formalism to higher dimensions},''
  \href{http://dx.doi.org/10.1088/0264-9381/27/21/215010}{{\em Class. Quant.
  Grav.} {\bfseries 27} (2010) 215010},
\href{http://arxiv.org/abs/1002.4826}{{\ttfamily arXiv:1002.4826 [gr-qc]}}.

\bibitem{decoupling}
M.~N. Durkee and H.~S. Reall, ``{Perturbations of higher-dimensional
  spacetimes},'' \href{http://dx.doi.org/10.1088/0264-9381/28/3/035011}{{\em
  Class. Quantum Grav.} {\bfseries 28} (2011) 035011},
\href{http://arxiv.org/abs/1009.0015}{{\ttfamily arXiv:1009.0015 [gr-qc]}}.

\bibitem{nhperturb}
M.~Durkee and H.~S. Reall, ``{Perturbations of near-horizon geometries and
  instabilities of Myers-Perry black holes},'' {\em Accepted by PRD} (2011) ,
\href{http://arxiv.org/abs/1012.4805}{{\ttfamily arXiv:1012.4805 [hep-th]}}.

\bibitem{Einstein:1905}
A.~{Einstein}, ``{Does the inertia of a body depend upon its energy
  content?},'' {\em Annalen der Physik} {\bfseries 18} 639--641.

\bibitem{Einstein:1916}
A.~Einstein, ``{The foundation of the general theory of relativity},''
\href{http://dx.doi.org/10.1002/andp.200590044}{{\em Annalen Phys.} {\bfseries
  49} (1916) 769--822}.

\bibitem{Bruhat:1958}
Y.~{Four\`es-Bruhat}, ``{Th\'eor\`eme d'existence pour certains syst\`emes
  d'\'equations aux d\'eriv\'ees partielles non lin\'eaires},'' {\em Acta
  Math.} {\bfseries 88} (1952) 141--225.

\bibitem{Willreview}
C.~M. Will, ``The Confrontation between General Relativity and Experiment,''
  {\em Living Reviews in Relativity} {\bfseries 9} no.~3, (2006) (online),
  \href{http://arxiv.org/abs/gr-qc/0510072}{{\ttfamily arXiv:gr-qc/0510072
  [gr-qc]}}. \url{http://www.livingreviews.org/lrr-2006-3}.

\bibitem{Green:1984sg}
M.~B. Green and J.~H. Schwarz, ``{Anomaly Cancellation in Supersymmetric D=10
  Gauge Theory and Superstring Theory},''
\href{http://dx.doi.org/10.1016/0370-2693(84)91565-X}{{\em Phys. Lett.}
  {\bfseries B149} (1984) 117--122}.

\bibitem{Maldacena:1997re}
J.~M. Maldacena, ``{The large N limit of superconformal field theories and
  supergravity},'' \href{http://dx.doi.org/10.1023/A:1026654312961}{{\em Adv.
  Theor. Math. Phys.} {\bfseries 2} (1998) 231--252},
\href{http://arxiv.org/abs/hep-th/9711200}{{\ttfamily arXiv:hep-th/9711200}}.

\bibitem{Witten:1998qj}
E.~Witten, ``{Anti-de Sitter space and holography},'' {\em Adv. Theor. Math.
  Phys.} {\bfseries 2} (1998) 253--291,
\href{http://arxiv.org/abs/hep-th/9802150}{{\ttfamily arXiv:hep-th/9802150}}.

\bibitem{Aharony:1999ti}
O.~Aharony, S.~S. Gubser, J.~M. Maldacena, H.~Ooguri, and Y.~Oz, ``{Large N
  field theories, string theory and gravity},''
  \href{http://dx.doi.org/10.1016/S0370-1573(99)00083-6}{{\em Phys. Rept.}
  {\bfseries 323} (2000) 183--386},
\href{http://arxiv.org/abs/hep-th/9905111}{{\ttfamily arXiv:hep-th/9905111}}.

\bibitem{Witten:1998zw}
E.~Witten, ``{Anti-de Sitter space, thermal phase transition, and confinement
  in gauge theories},'' {\em Adv. Theor. Math. Phys.} {\bfseries 2} (1998)
  505--532,
\href{http://arxiv.org/abs/hep-th/9803131}{{\ttfamily arXiv:hep-th/9803131}}.

\bibitem{ER:2008}
R.~{Emparan} and H.~S. {Reall}, ``{Black Holes in Higher Dimensions},'' {\em
  Living Rev. Rel.} {\bfseries 11} (2008) 6,
\href{http://arxiv.org/abs/0801.3471}{{\ttfamily arXiv:0801.3471 [hep-th]}}.

\bibitem{wald}
R.~M. {Wald}, {\em {General Relativity}}.
\newblock University of Chicago Press, 1984.

\bibitem{Michell:1784}
J.~Michell, ``On the Means of Discovering the Distance, Magnitude, etc. of the
  Fixed Stars, in Consequence of the Diminution of the Velocity of Their Light,
  in Case Such a Diminution Should be Found to Take Place in any of Them, and
  Such Other Data Should be Procured from Observations, as Would be Farther
  Necessary for That Purpose.,''
  \href{http://dx.doi.org/10.1098/rstl.1784.0008}{{\em Phil. Trans. R. Soc.
  Lond.} {\bfseries 74} (1784) 25--57}.

\bibitem{HawkingEllis}
S.~W. {Hawking} and G.~F.~R. {Ellis}, {\em {The Large Scale Structure of
  Space-time}}.
\newblock Cambridge University Press, 1973.

\bibitem{Ghez:2008ms}
A.~M. Ghez, S.~Salim, N.~N. Weinberg, J.~R. Lu, T.~Do, J.~K. Dunn, K.~Matthews,
  M.~R. Morris, S.~Yelda, E.~E. Becklin, T.~Kremenek, M.~Milosavljevic, and
  J.~Naiman, ``{Measuring Distance and Properties of the Milky Way's Central
  Supermassive Black Hole with Stellar Orbits},''
  \href{http://dx.doi.org/10.1086/592738}{{\em Astrophys. J.} {\bfseries 689}
  (2008) 1044--1062},
\href{http://arxiv.org/abs/0808.2870}{{\ttfamily arXiv:0808.2870 [astro-ph]}}.

\bibitem{schwarzchild}
K.~Schwarzchild, ``{On the gravitational field of a point mass in Einstein's
  theory},'' {\em Reimer, Berlin , S.} (1916) .

\bibitem{Finkelstein}
D.~Finkelstein, ``Past-Future Asymmetry of the Gravitational Field of a Point
  Particle,'' \href{http://dx.doi.org/10.1103/PhysRev.110.965}{{\em Phys. Rev.}
  {\bfseries 110} no.~4, (May, 1958) 965--967}.

\bibitem{Kruskal}
M.~D. Kruskal, ``Maximal Extension of Schwarzschild Metric,''
  \href{http://dx.doi.org/10.1103/PhysRev.119.1743}{{\em Phys. Rev.} {\bfseries
  119} no.~5, (Sep, 1960) 1743--1745}.

\bibitem{Kerr}
R.~P. {Kerr}, ``{Gravitational Field of a Spinning Mass as an Example of
  Algebraically Special Metrics},''
  \href{http://dx.doi.org/10.1103/PhysRevLett.11.237}{{\em Phys. Rev. Lett.}
  {\bfseries 11} no.~5, (Sep, 1963) 237--238}.

\bibitem{Carter:1968b}
B.~Carter, ``{Hamilton-Jacobi and Schrodinger separable solutions of Einstein's
  equations},''
{\em Commun. Math. Phys.} {\bfseries 10} (1968) 280.

\bibitem{Walker:1970}
M.~{Walker} and R.~{Penrose}, ``{On Quadratic First Integrals of the Geodesic
  Equations for Type [22] Spacetimes},''
\href{http://dx.doi.org/10.1007/BF01649445}{{\em Commun. Math. Phys.}
  {\bfseries 18} (1970) 265--274}.

\bibitem{exact}
H.~Stephani, D.~Kramer, M.~MacCallum, C.~Hoenselaers, and E.~Herlt, {\em {Exact
  solutions of Einstein's field equations}}.
\newblock Camb. Univ. Press, 2003.

\bibitem{Whiting:1988vc}
B.~F. Whiting, ``{Mode stability of the Kerr black hole},''
\href{http://dx.doi.org/10.1063/1.528308}{{\em J. Math. Phys.} {\bfseries 30}
  (1989) 1301}.

\bibitem{Teukolsky}
S.~A. {Teukolsky}, ``{Perturbations of a rotating black hole. 1. Fundamental
  equations for gravitational electromagnetic and neutrino field
  perturbations},''
\href{http://dx.doi.org/10.1086/152444}{{\em Astrophys. J.} {\bfseries 185}
  (1973) 635--647}.

\bibitem{Teukolsky:1972}
S.~A. Teukolsky, ``{Rotating black holes - separable wave equations for
  gravitational and electromagnetic perturbations},''
\href{http://dx.doi.org/10.1103/PhysRevLett.29.1114}{{\em Phys. Rev. Lett.}
  {\bfseries 29} (1972) 1114--1118}.

\bibitem{stewperts}
J.~M. {Stewart} and M.~{Walker}, ``{Perturbations of spacetimes in general
  relativity},''
{\em Proc. Roy. Soc. Lond.} {\bfseries A341} (1974) 49--74.

\bibitem{Press:1973zz}
W.~H. Press and S.~A. Teukolsky, ``{Perturbations of a Rotating Black Hole. II.
  Dynamical Stability of the Kerr Metric},''
\href{http://dx.doi.org/10.1086/152445}{{\em Astrophys. J.} {\bfseries 185}
  (1973) 649--674}.

\bibitem{Stewart:1975vg}
J.~M. Stewart, ``{Stability of Kerr's spacetime},''
{\em Proc. Roy. Soc. Lond.} {\bfseries A344} (1975) 65--79.

\bibitem{Christodoulou:1993uv}
D.~Christodoulou and S.~Klainerman, {\em {The Global nonlinear stability of the
  Minkowski space}}.
\newblock Princeton University Press, 1993.

\bibitem{Dafermos:2010hd}
M.~Dafermos and I.~Rodnianski, ``{The black hole stability problem for linear
  scalar perturbations},''
\href{http://arxiv.org/abs/1010.5137}{{\ttfamily arXiv:1010.5137 [gr-qc]}}.

\bibitem{Chrusciel:2010fn}
P.~T. Chrusciel, G.~J. Galloway, and D.~Pollack, ``{Mathematical general
  relativity: a sampler},'' \href{http://arxiv.org/abs/1004.1016}{{\ttfamily
  arXiv:1004.1016 [gr-qc]}}.

\bibitem{Hollands:2003ie}
S.~Hollands and A.~Ishibashi, ``{Asymptotic flatness and Bondi energy in higher
  dimensional gravity},'' \href{http://dx.doi.org/10.1063/1.1829152}{{\em
  J.Math.Phys.} {\bfseries 46} (2005) 022503},
  \href{http://arxiv.org/abs/gr-qc/0304054}{{\ttfamily arXiv:gr-qc/0304054
  [gr-qc]}}.

\bibitem{Tanabe:2009xb}
K.~Tanabe, N.~Tanahashi, and T.~Shiromizu, ``{Asymptotic flatness at spatial
  infinity in higher dimensions},''
  \href{http://dx.doi.org/10.1063/1.3166141}{{\em J.Math.Phys.} {\bfseries 50}
  (2009) 072502}, \href{http://arxiv.org/abs/0902.1583}{{\ttfamily
  arXiv:0902.1583 [gr-qc]}}.

\bibitem{Hawking:1972}
S.~W. {Hawking}, ``{Black Holes in General Relativity},''
{\em Commun. Math. Phys.} {\bfseries 25} (1972) 152--166.

\bibitem{Chrusciel:1994tr}
P.~T. Chrusciel and R.~M. Wald, ``{On the topology of stationary black
  holes},'' \href{http://dx.doi.org/10.1088/0264-9381/11/12/001}{{\em
  Class.Quant.Grav.} {\bfseries 11} (1994) L147--L152},
  \href{http://arxiv.org/abs/gr-qc/9410004}{{\ttfamily arXiv:gr-qc/9410004
  [gr-qc]}}.

\bibitem{Galloway:2005}
G.~J. Galloway and R.~Schoen, ``{A generalization of Hawking's black hole
  topology theorem to higher dimensions},'' {\em Commun. Math. Phys.}
  {\bfseries 266} (2006) 571--576,
\href{http://arxiv.org/abs/gr-qc/0509107}{{\ttfamily arXiv:gr-qc/0509107}}.

\bibitem{Hollands:2010qy}
S.~Hollands, J.~Holland, and A.~Ishibashi, ``{Further restrictions on the
  topology of stationary black holes in five dimensions},''
\href{http://arxiv.org/abs/1002.0490}{{\ttfamily arXiv:1002.0490 [gr-qc]}}.

\bibitem{Hollands:2006rj}
S.~Hollands, A.~Ishibashi, and R.~M. Wald, ``{A Higher Dimensional Stationary
  Rotating Black Hole Must be Axisymmetric},''
  \href{http://dx.doi.org/10.1007/s00220-007-0216-4}{{\em Commun. Math. Phys.}
  {\bfseries 271} (2007) 699--722},
\href{http://arxiv.org/abs/gr-qc/0605106}{{\ttfamily arXiv:gr-qc/0605106}}.

\bibitem{Moncrief:2008}
V.~Moncrief and J.~Isenberg, ``{Symmetries of Higher Dimensional Black
  Holes},'' \href{http://dx.doi.org/10.1088/0264-9381/25/19/195015}{{\em Class.
  Quant. Grav.} {\bfseries 25} (2008) 195015},
\href{http://arxiv.org/abs/0805.1451}{{\ttfamily arXiv:0805.1451 [gr-qc]}}.

\bibitem{Friedrich:1998wq}
H.~Friedrich, I.~Racz, and R.~M. Wald, ``{On the Rigidity Theorem for
  Spacetimes with a Stationary Event Horizon or a Compact Cauchy Horizon},''
  \href{http://dx.doi.org/10.1007/s002200050662}{{\em Commun. Math. Phys.}
  {\bfseries 204} (1999) 691--707},
\href{http://arxiv.org/abs/gr-qc/9811021}{{\ttfamily arXiv:gr-qc/9811021}}.

\bibitem{Alexakis:2009gi}
S.~Alexakis, A.~D. Ionescu, and S.~Klainerman, ``{Hawking's local rigidity
  theorem without analyticity},''
\href{http://arxiv.org/abs/0902.1173}{{\ttfamily arXiv:0902.1173 [gr-qc]}}.

\bibitem{Israel:1967wq}
W.~Israel, ``{Event horizons in static vacuum space-times},''
\href{http://dx.doi.org/10.1103/PhysRev.164.1776}{{\em Phys. Rev.} {\bfseries
  164} (1967) 1776--1779}.

\bibitem{Carter:1971zc}
B.~Carter, ``{Axisymmetric Black Hole Has Only Two Degrees of Freedom},''
\href{http://dx.doi.org/10.1103/PhysRevLett.26.331}{{\em Phys. Rev. Lett.}
  {\bfseries 26} (1971) 331--333}.

\bibitem{uniquenessbook}
M.~{Heusler}, {\em {Black Hole Uniqueness Theorems}}.
\newblock Cambridge University Press, 2003.

\bibitem{ArkaniHamed:1998rs}
N.~Arkani-Hamed, S.~Dimopoulos, and G.~R. Dvali, ``{The hierarchy problem and
  new dimensions at a millimeter},''
  \href{http://dx.doi.org/10.1016/S0370-2693(98)00466-3}{{\em Phys. Lett.}
  {\bfseries B429} (1998) 263--272},
\href{http://arxiv.org/abs/hep-ph/9803315}{{\ttfamily arXiv:hep-ph/9803315}}.

\bibitem{Hawking:1974sw}
S.~W. Hawking, ``{Particle Creation by Black Holes},''
\href{http://dx.doi.org/10.1007/BF02345020}{{\em Commun. Math. Phys.}
  {\bfseries 43} (1975) 199--220}.

\bibitem{Emparan:2000rs}
R.~Emparan, G.~T. Horowitz, and R.~C. Myers, ``{Black holes radiate mainly on
  the brane},'' \href{http://dx.doi.org/10.1103/PhysRevLett.85.499}{{\em Phys.
  Rev. Lett.} {\bfseries 85} (2000) 499--502},
\href{http://arxiv.org/abs/hep-th/0003118}{{\ttfamily arXiv:hep-th/0003118}}.

\bibitem{Policastro:2001yc}
G.~Policastro, D.~T. Son, and A.~O. Starinets, ``{The shear viscosity of
  strongly coupled N = 4 supersymmetric Yang-Mills plasma},''
  \href{http://dx.doi.org/10.1103/PhysRevLett.87.081601}{{\em Phys. Rev. Lett.}
  {\bfseries 87} (2001) 081601},
\href{http://arxiv.org/abs/hep-th/0104066}{{\ttfamily arXiv:hep-th/0104066}}.

\bibitem{Buchel:2003tz}
A.~Buchel and J.~T. Liu, ``{Universality of the shear viscosity in
  supergravity},'' \href{http://dx.doi.org/10.1103/PhysRevLett.93.090602}{{\em
  Phys. Rev. Lett.} {\bfseries 93} (2004) 090602},
\href{http://arxiv.org/abs/hep-th/0311175}{{\ttfamily arXiv:hep-th/0311175}}.

\bibitem{Son:2007vk}
D.~T. Son and A.~O. Starinets, ``{Viscosity, Black Holes, and Quantum Field
  Theory},''
  \href{http://dx.doi.org/10.1146/annurev.nucl.57.090506.123120}{{\em Ann. Rev.
  Nucl. Part. Sci.} {\bfseries 57} (2007) 95--118},
\href{http://arxiv.org/abs/0704.0240}{{\ttfamily arXiv:0704.0240 [hep-th]}}.

\bibitem{Rangamani:2009xk}
M.~Rangamani, ``{Gravity and Hydrodynamics: Lectures on the fluid-gravity
  correspondence},''
  \href{http://dx.doi.org/10.1088/0264-9381/26/22/224003}{{\em
  Class.Quant.Grav.} {\bfseries 26} (2009) 224003},
  \href{http://arxiv.org/abs/0905.4352}{{\ttfamily arXiv:0905.4352 [hep-th]}}.

\bibitem{Herzog:2009xv}
C.~P. Herzog, ``{Lectures on Holographic Superfluidity and
  Superconductivity},''
  \href{http://dx.doi.org/10.1088/1751-8113/42/34/343001}{{\em J. Phys.}
  {\bfseries A42} (2009) 343001},
\href{http://arxiv.org/abs/0904.1975}{{\ttfamily arXiv:0904.1975 [hep-th]}}.

\bibitem{Janik:2010we}
R.~A. Janik, ``{The dynamics of quark-gluon plasma and AdS/CFT},''
\href{http://arxiv.org/abs/1003.3291}{{\ttfamily arXiv:1003.3291 [hep-th]}}.

\bibitem{Hartnoll:2008vx}
S.~A. Hartnoll, C.~P. Herzog, and G.~T. Horowitz, ``{Building a Holographic
  Superconductor},''
  \href{http://dx.doi.org/10.1103/PhysRevLett.101.031601}{{\em Phys.Rev.Lett.}
  {\bfseries 101} (2008) 031601},
  \href{http://arxiv.org/abs/0803.3295}{{\ttfamily arXiv:0803.3295 [hep-th]}}.

\bibitem{Horowitz:2010gk}
G.~T. Horowitz, ``{Introduction to Holographic Superconductors},''
\href{http://arxiv.org/abs/1002.1722}{{\ttfamily arXiv:1002.1722 [hep-th]}}.

\bibitem{Tangherlini}
F.~R. Tangherlini, ``{Schwarzschild field in n dimensions and the
  dimensionality of space problem},''
\href{http://dx.doi.org/10.1007/BF02784569}{{\em Nuovo Cim.} {\bfseries 27}
  (1963) 636--651}.

\bibitem{mp}
R.~C. {Myers} and M.~J. {Perry}, ``{Black Holes in Higher Dimensional
  Space-Times},''
\href{http://dx.doi.org/10.1016/0003-4916(86)90186-7}{{\em Ann. Phys.}
  {\bfseries 172} (1986) 304}.

\bibitem{Frolov:2007nt}
V.~P. Frolov and D.~Kubiz{\v n}\'ak, ``{`Hidden' symmetries of higher
  dimensional rotating black holes},''
  \href{http://dx.doi.org/10.1103/PhysRevLett.98.011101}{{\em Phys. Rev. Lett.}
  {\bfseries 98} (2007) 011101},
\href{http://arxiv.org/abs/gr-qc/0605058}{{\ttfamily arXiv:gr-qc/0605058}}.

\bibitem{Page:2006ka}
D.~N. Page, D.~Kubiz{\v n}\'ak, M.~Vasudevan, and P.~Krtou{\v s}, ``{Complete
  Integrability of Geodesic Motion in General Kerr- NUT-AdS Spacetimes},''
  \href{http://dx.doi.org/10.1103/PhysRevLett.98.061102}{{\em Phys. Rev. Lett.}
  {\bfseries 98} (2007) 061102},
\href{http://arxiv.org/abs/hep-th/0611083}{{\ttfamily arXiv:hep-th/0611083}}.

\bibitem{Hawking:1998kw}
S.~W. Hawking, C.~J. Hunter, and M.~M. Taylor-Robinson, ``{Rotation and the
  AdS/CFT correspondence},''
  \href{http://dx.doi.org/10.1103/PhysRevD.59.064005}{{\em Phys. Rev.}
  {\bfseries D59} (1999) 064005},
\href{http://arxiv.org/abs/hep-th/9811056}{{\ttfamily arXiv:hep-th/9811056}}.

\bibitem{Gibbons:2004}
G.~W. Gibbons, H.~L\"{u}, D.~N. Page, and C.~N. Pope, ``{The general Kerr-de
  Sitter metrics in all dimensions},''
  \href{http://dx.doi.org/10.1016/j.geomphys.2004.05.001}{{\em J. Geom. Phys.}
  {\bfseries 53} (2005) 49--73},
\href{http://arxiv.org/abs/hep-th/0404008}{{\ttfamily arXiv:hep-th/0404008}}.

\bibitem{Gibbons:2004js}
G.~W. Gibbons, H.~L\"{u}, D.~N. Page, and C.~N. Pope, ``{Rotating black holes
  in higher dimensions with a cosmological constant},''
  \href{http://dx.doi.org/10.1103/PhysRevLett.93.171102}{{\em Phys. Rev. Lett.}
  {\bfseries 93} (2004) 171102},
\href{http://arxiv.org/abs/hep-th/0409155}{{\ttfamily arXiv:hep-th/0409155}}.

\bibitem{ER:2001}
R.~{Emparan} and H.~S. {Reall}, ``A rotating black ring in five dimensions,''
  {\em Phys. Rev. Lett.} {\bfseries 88} (2002) 101101,
\href{http://arxiv.org/abs/hep-th/0110260}{{\ttfamily arXiv:hep-th/0110260}}.

\bibitem{ER:2006}
R.~{Emparan} and H.~S. {Reall}, ``{Black Rings},'' {\em Class. Quantum Grav..}
  {\bfseries 23} (2006) R169,
\href{http://arxiv.org/abs/hep-th/0608012}{{\ttfamily hep-th/0608012}}.

\bibitem{Pomeransky}
A.~A. {Pomeransky} and R.~A. {Sen'kov}, ``{Black ring with two angular
  momenta},''
\href{http://arxiv.org/abs/hep-th/0612005}{{\ttfamily arXiv:hep-th/0612005}}.

\bibitem{Elvang:2007sat}
H.~{Elvang} and P.~{Figueras}, ``{Black Saturn},'' {\em JHEP} {\bfseries 05}
  (2007) 050,
\href{http://arxiv.org/abs/hep-th/0701035}{{\ttfamily arXiv:hep-th/0701035}}.

\bibitem{Izumi:2007}
K.~Izumi, ``{Orthogonal black di-ring solution},''
  \href{http://dx.doi.org/10.1143/PTP.119.757}{{\em Prog. Theor. Phys.}
  {\bfseries 119} (2008) 757--774},
\href{http://arxiv.org/abs/0712.0902}{{\ttfamily arXiv:0712.0902 [hep-th]}}.

\bibitem{Evslin:2007}
J.~Evslin and C.~Krishnan, ``{The Black Di-Ring: An Inverse Scattering
  Construction},'' \href{http://dx.doi.org/10.1088/0264-9381/26/12/125018}{{\em
  Class. Quant. Grav.} {\bfseries 26} (2009) 125018},
\href{http://arxiv.org/abs/0706.1231}{{\ttfamily arXiv:0706.1231 [hep-th]}}.

\bibitem{Elvang:2007bi}
H.~{Elvang} and M.~J. {Rodriguez}, ``{Bicycling Black Rings},''
  \href{http://dx.doi.org/10.1088/1126-6708/2008/04/045}{{\em JHEP} {\bfseries
  04} (2008) 045},
\href{http://arxiv.org/abs/0712.2425}{{\ttfamily arXiv:0712.2425 [hep-th]}}.

\bibitem{ER:weyl}
R.~Emparan and H.~S. Reall, ``{Generalized Weyl solutions},''
  \href{http://dx.doi.org/10.1103/PhysRevD.65.084025}{{\em Phys. Rev.}
  {\bfseries D65} (2002) 084025},
\href{http://arxiv.org/abs/hep-th/0110258}{{\ttfamily arXiv:hep-th/0110258}}.

\bibitem{Caldarelli:2008}
M.~M. Caldarelli, R.~Emparan, and M.~J. Rodriguez, ``{Black Rings in
  (Anti)-deSitter space},''
  \href{http://dx.doi.org/10.1088/1126-6708/2008/11/011}{{\em JHEP} {\bfseries
  0811} (2008) 011},
\href{http://arxiv.org/abs/0806.1954}{{\ttfamily arXiv:0806.1954 [hep-th]}}.

\bibitem{Reall:2002bh}
H.~S. Reall, ``{Higher dimensional black holes and supersymmetry},''
  \href{http://dx.doi.org/10.1103/PhysRevD.68.024024}{{\em Phys. Rev.}
  {\bfseries D68} (2003) 024024},
\href{http://arxiv.org/abs/hep-th/0211290}{{\ttfamily arXiv:hep-th/0211290}}.

\bibitem{Chrusciel:2005pa}
P.~T. Chrusciel, H.~S. Reall, and P.~Tod, ``{On non-existence of static vacuum
  black holes with degenerate components of the event horizon},''
  \href{http://dx.doi.org/10.1088/0264-9381/23/2/018}{{\em Class. Quant. Grav.}
  {\bfseries 23} (2006) 549--554},
\href{http://arxiv.org/abs/gr-qc/0512041}{{\ttfamily arXiv:gr-qc/0512041}}.

\bibitem{Hajicek:1974}
P.~H\'aji{\v e}k, ``Three remarks on axisymmetric stationary horizons,'' {\em
  Communications in Mathematical Physics} {\bfseries 36} (1974) 305--320.
  \url{http://dx.doi.org/10.1007/BF01646202}.

\bibitem{Kunduri:2008rs}
H.~K. Kunduri and J.~Lucietti, ``{A classification of near-horizon geometries
  of extremal vacuum black holes},''
  \href{http://dx.doi.org/10.1063/1.3190480}{{\em J. Math. Phys.} {\bfseries
  50} (2009) 082502},
\href{http://arxiv.org/abs/0806.2051}{{\ttfamily arXiv:0806.2051 [hep-th]}}.

\bibitem{Hollands:2009}
S.~Hollands and A.~Ishibashi, ``{All vacuum near horizon geometries in
  arbitrary dimensions},''
  \href{http://dx.doi.org/10.1007/s00023-010-0022-y}{{\em Annales Henri
  Poincare} {\bfseries 10} (2010) 1537--1557},
\href{http://arxiv.org/abs/0909.3462}{{\ttfamily arXiv:0909.3462 [gr-qc]}}.

\bibitem{Holland:2010bd}
J.~Holland, ``{Non-existence of toroidal cohomogeneity-1 near horizon
  geometries},''
\href{http://arxiv.org/abs/1008.0520}{{\ttfamily arXiv:1008.0520 [gr-qc]}}.

\bibitem{Moncrief:1983}
V.~{Moncrief} and J.~{Isenberg}, ``{Symmetries of cosmological Cauchy
  horizons},'' {\em Commun. Math. Phys.} {\bfseries 89} (1983) 387–413.

\bibitem{Kundt:1961}
W.~{Kundt}, ``{The plane-fronted gravitational waves},'' {\em Zeitschrift für
  Physik A Hadrons and Nuclei} {\bfseries 163} (1961) 77--86.
  \url{http://dx.doi.org/10.1007/BF01328918}.

\bibitem{Podolsky:2008ec}
J.~Podolsk\'y and M.~{\v Z}ofka, ``{General Kundt spacetimes in higher
  dimensions},'' \href{http://dx.doi.org/10.1088/0264-9381/26/10/105008}{{\em
  Class. Quantum Grav..} {\bfseries 26} (2009) 105008},
\href{http://arxiv.org/abs/0812.4928}{{\ttfamily arXiv:0812.4928 [gr-qc]}}.

\bibitem{Coley:2009ut}
A.~Coley, S.~Hervik, G.~O. Papadopoulos, and N.~Pelavas, ``{Kundt
  Spacetimes},''
\href{http://arxiv.org/abs/0901.0394}{{\ttfamily arXiv:0901.0394 [gr-qc]}}.

\bibitem{Bardeen:1999px}
J.~M. Bardeen and G.~T. Horowitz, ``{The extreme Kerr throat geometry: A vacuum
  analog of AdS(2) x S(2)},''
  \href{http://dx.doi.org/10.1103/PhysRevD.60.104030}{{\em Phys. Rev.}
  {\bfseries D60} (1999) 104030},
\href{http://arxiv.org/abs/hep-th/9905099}{{\ttfamily arXiv:hep-th/9905099}}.

\bibitem{Kunduri:2007vf}
H.~K. {Kunduri}, J.~{Lucietti}, and H.~S. {Reall}, ``{Near-horizon symmetries
  of extremal black holes},''
  \href{http://dx.doi.org/10.1088/0264-9381/24/16/012}{{\em Class. Quantum
  Grav..} {\bfseries 24} (2007) 4169--4190},
\href{http://arxiv.org/abs/0705.4214}{{\ttfamily arXiv:0705.4214 [hep-th]}}.

\bibitem{Figueras:2008qh}
P.~Figueras, H.~K. Kunduri, J.~Lucietti, and M.~Rangamani, ``{Extremal vacuum
  black holes in higher dimensions},''
  \href{http://dx.doi.org/10.1103/PhysRevD.78.044042}{{\em Phys. Rev.}
  {\bfseries D78} (2008) 044042},
\href{http://arxiv.org/abs/0803.2998}{{\ttfamily arXiv:0803.2998 [hep-th]}}.

\bibitem{Chow:2008dp}
D.~D.~K. Chow, M.~Cvetic, H.~L\"{u}, and C.~N. Pope, ``{Extremal Black Hole/CFT
  Correspondence in (Gauged) Supergravities},''
  \href{http://dx.doi.org/10.1103/PhysRevD.79.084018}{{\em Phys. Rev.}
  {\bfseries D79} (2009) 084018},
\href{http://arxiv.org/abs/0812.2918}{{\ttfamily arXiv:0812.2918 [hep-th]}}.

\bibitem{kerrcft}
M.~Guica, T.~Hartman, W.~Song, and A.~Strominger, ``{The Kerr/CFT
  Correspondence},'' \href{http://dx.doi.org/10.1103/PhysRevD.80.124008}{{\em
  Phys. Rev.} {\bfseries D80} (2009) 124008},
\href{http://arxiv.org/abs/0809.4266}{{\ttfamily arXiv:0809.4266 [hep-th]}}.

\bibitem{popekerrcft}
H.~L\"{u}, J.~Mei, and C.~N. Pope, ``{Kerr/CFT Correspondence in Diverse
  Dimensions},'' \href{http://dx.doi.org/10.1088/1126-6708/2009/04/054}{{\em
  JHEP} {\bfseries 04} (2009) 054},
\href{http://arxiv.org/abs/0811.2225}{{\ttfamily arXiv:0811.2225 [hep-th]}}.

\bibitem{GL}
R.~{Gregory} and R.~{Laflamme}, ``{Black strings and p-branes are unstable},''
  \href{http://dx.doi.org/10.1103/PhysRevLett.70.2837}{{\em Phys. Rev. Lett.}
  {\bfseries 70} (1993) 2837--2840},
\href{http://arxiv.org/abs/hep-th/9301052}{{\ttfamily arXiv:hep-th/9301052}}.

\bibitem{Lehner:2010pn}
L.~Lehner and F.~Pretorius, ``{Black Strings, Low Viscosity Fluids, and
  Violation of Cosmic Censorship},''
  \href{http://dx.doi.org/10.1103/PhysRevLett.105.101102}{{\em Phys.Rev.Lett.}
  {\bfseries 105} (2010) 101102},
  \href{http://arxiv.org/abs/1006.5960}{{\ttfamily arXiv:1006.5960 [hep-th]}}.

\bibitem{Emparan:2003sy}
R.~{Emparan} and R.~C. {Myers}, ``{Instability of ultra-spinning black
  holes},'' {\em JHEP} {\bfseries 09} (2003) 025,
\href{http://arxiv.org/abs/hep-th/0308056}{{\ttfamily arXiv:hep-th/0308056}}.

\bibitem{Ishibashi:2003}
A.~Ishibashi and H.~Kodama, ``{Stability of higher-dimensional Schwarzschild
  black holes},'' \href{http://dx.doi.org/10.1143/PTP.110.901}{{\em Prog.
  Theor. Phys.} {\bfseries 110} (2003) 901--919},
\href{http://arxiv.org/abs/hep-th/0305185}{{\ttfamily arXiv:hep-th/0305185}}.

\bibitem{Bekenstein:1973ur}
J.~D. Bekenstein, ``{Black holes and entropy},''
\href{http://dx.doi.org/10.1103/PhysRevD.7.2333}{{\em Phys. Rev.} {\bfseries
  D7} (1973) 2333--2346}.

\bibitem{Elvang:2007hg}
H.~Elvang, R.~Emparan, and P.~Figueras, ``{Phases of Five-Dimensional Black
  Holes},'' \href{http://dx.doi.org/10.1088/1126-6708/2007/05/056}{{\em JHEP}
  {\bfseries 05} (2007) 056},
\href{http://arxiv.org/abs/hep-th/0702111}{{\ttfamily arXiv:hep-th/0702111}}.

\bibitem{Emparan:2007wm}
R.~Emparan, T.~Harmark, V.~Niarchos, N.~A. Obers, and M.~J. Rodriguez, ``{The
  Phase Structure of Higher-Dimensional Black Rings and Black Holes},''
  \href{http://dx.doi.org/10.1088/1126-6708/2007/10/110}{{\em JHEP} {\bfseries
  10} (2007) 110},
\href{http://arxiv.org/abs/0708.2181}{{\ttfamily arXiv:0708.2181 [hep-th]}}.

\bibitem{Emparan:2009vd}
R.~Emparan, T.~Harmark, V.~Niarchos, and N.~A. Obers, ``{New Horizons for Black
  Holes and Branes},'' \href{http://dx.doi.org/10.1007/JHEP04(2010)046}{{\em
  JHEP} {\bfseries 04} (2010) 046},
\href{http://arxiv.org/abs/0912.2352}{{\ttfamily arXiv:0912.2352 [hep-th]}}.

\bibitem{Emparan:2010sx}
R.~Emparan and P.~Figueras, ``{Multi-black rings and the phase diagram of
  higher- dimensional black holes},''
  \href{http://dx.doi.org/10.1007/JHEP11(2010)022}{{\em JHEP} {\bfseries 11}
  (2010) 022},
\href{http://arxiv.org/abs/1008.3243}{{\ttfamily arXiv:1008.3243 [hep-th]}}.

\bibitem{Gubser:2000ec}
S.~S. Gubser and I.~Mitra, ``{Instability of charged black holes in Anti-de
  Sitter space},'' \href{http://arxiv.org/abs/hep-th/0009126}{{\ttfamily
  arXiv:hep-th/0009126 [hep-th]}}.

\bibitem{Gubser:2000mm}
S.~S. Gubser and I.~Mitra, ``{The Evolution of unstable black holes in anti-de
  Sitter space},'' {\em JHEP} {\bfseries 0108} (2001) 018,
  \href{http://arxiv.org/abs/hep-th/0011127}{{\ttfamily arXiv:hep-th/0011127
  [hep-th]}}.

\bibitem{Reall:2001ag}
H.~S. Reall, ``{Classical and thermodynamic stability of black branes},''
  \href{http://dx.doi.org/10.1103/PhysRevD.64.044005}{{\em Phys.Rev.}
  {\bfseries D64} (2001) 044005},
  \href{http://arxiv.org/abs/hep-th/0104071}{{\ttfamily arXiv:hep-th/0104071
  [hep-th]}}.

\bibitem{Monteiro:2009tc}
R.~Monteiro, M.~J. Perry, and J.~E. Santos, ``{Thermodynamic instability of
  rotating black holes},''
  \href{http://dx.doi.org/10.1103/PhysRevD.80.024041}{{\em Phys. Rev.}
  {\bfseries D80} (2009) 024041},
\href{http://arxiv.org/abs/0903.3256}{{\ttfamily arXiv:0903.3256 [gr-qc]}}.

\bibitem{Monteiro:2009ke}
R.~Monteiro, M.~J. Perry, and J.~E. Santos, ``{Semiclassical instabilities of
  Kerr-AdS black holes},''
  \href{http://dx.doi.org/10.1103/PhysRevD.81.024001}{{\em Phys.Rev.}
  {\bfseries D81} (2010) 024001},
  \href{http://arxiv.org/abs/0905.2334}{{\ttfamily arXiv:0905.2334 [gr-qc]}}.

\bibitem{Dias:2009iu}
O.~J.~C. Dias, P.~Figueras, R.~Monteiro, J.~E. Santos, and R.~Emparan,
  ``{Instability and new phases of higher-dimensional rotating black holes},''
  \href{http://dx.doi.org/10.1103/PhysRevD.80.111701}{{\em Phys. Rev.}
  {\bfseries D80} (2009) 111701},
\href{http://arxiv.org/abs/0907.2248}{{\ttfamily arXiv:0907.2248 [hep-th]}}.

\bibitem{Dias:2010maa}
O.~J.~C. Dias, P.~Figueras, R.~Monteiro, and J.~E. Santos, ``{Ultraspinning
  instability of rotating black holes},''
  \href{http://dx.doi.org/10.1103/PhysRevD.82.104025}{{\em Phys. Rev.}
  {\bfseries D82} (2010) 104025},
\href{http://arxiv.org/abs/1006.1904}{{\ttfamily arXiv:1006.1904 [hep-th]}}.

\bibitem{Dias:2010eu}
O.~J.~C. Dias, P.~Figueras, R.~Monteiro, H.~S. Reall, and J.~E. Santos, ``{An
  instability of higher-dimensional rotating black holes},''
  \href{http://dx.doi.org/10.1007/JHEP05(2010)076}{{\em JHEP} {\bfseries 05}
  (2010) 076},
\href{http://arxiv.org/abs/1001.4527}{{\ttfamily arXiv:1001.4527 [hep-th]}}.

\bibitem{Shibata:2009ad}
M.~Shibata and H.~Yoshino, ``{Nonaxisymmetric instability of rapidly rotating
  black hole in five dimensions},''
  \href{http://dx.doi.org/10.1103/PhysRevD.81.021501}{{\em Phys. Rev.}
  {\bfseries D81} (2010) 021501},
\href{http://arxiv.org/abs/0912.3606}{{\ttfamily arXiv:0912.3606 [gr-qc]}}.

\bibitem{Shibata:2010wz}
M.~Shibata and H.~Yoshino, ``{Bar-mode instability of rapidly spinning black
  hole in higher dimensions: Numerical simulation in general relativity},''
  \href{http://dx.doi.org/10.1103/PhysRevD.81.104035}{{\em Phys.Rev.}
  {\bfseries D81} (2010) 104035},
  \href{http://arxiv.org/abs/1004.4970}{{\ttfamily arXiv:1004.4970 [gr-qc]}}.

\bibitem{Elvang:2006dd}
H.~Elvang, R.~Emparan, and A.~Virmani, ``{Dynamics and stability of black
  rings},'' \href{http://dx.doi.org/10.1088/1126-6708/2006/12/074}{{\em JHEP}
  {\bfseries 12} (2006) 074},
\href{http://arxiv.org/abs/hep-th/0608076}{{\ttfamily arXiv:hep-th/0608076}}.

\bibitem{Dias:2006zv}
O.~J.~C. Dias, ``{Superradiant instability of large radius doubly spinning
  black rings},'' \href{http://dx.doi.org/10.1103/PhysRevD.73.124035}{{\em
  Phys. Rev.} {\bfseries D73} (2006) 124035},
\href{http://arxiv.org/abs/hep-th/0602064}{{\ttfamily arXiv:hep-th/0602064}}.

\bibitem{ghp}
R.~{Geroch}, A.~{Held}, and R.~{Penrose}, ``{A space-time calculus based on
  pairs of null directions},'' \href{http://dx.doi.org/10.1063/1.1666410}{{\em
  Journal of Mathematical Physics} {\bfseries 14} no.~7, (1973) 874--881}.
  \url{http://link.aip.org/link/?JMP/14/874/1}.

\bibitem{Goldberg}
J.~N. {Goldberg} and R.~K. {Sachs}, ``{A Theorem on Petrov Types},'' {\em Acta
  Phys. Pol.} {\bfseries 22} no.~13, (1962) .

\bibitem{Petrov:1954}
A.~Petrov, ``{Classification of spaces defined by gravitational fields},'' {\em
  Uch. Zapiski Kazan Gos. Univ.} {\bfseries 144} (1954) 55.

\bibitem{stewart}
J.~{Stewart}, {\em {Advanced general relativity}}.
\newblock Cambridge University Press, 1991.

\bibitem{cmpp}
A.~Coley, R.~Milson, V.~Pravda, and A.~Pravdov\'a, ``{Classification of the
  Weyl tensor in higher-dimensions},'' {\em Class. Quantum Grav..} {\bfseries
  21} (2004) L35--L42,
\href{http://arxiv.org/abs/gr-qc/0401008}{{\ttfamily arXiv:gr-qc/0401008}}.

\bibitem{Alignment}
R.~Milson, A.~Coley, V.~Pravda, and A.~Pravdov\'a, ``{Alignment and
  algebraically special tensors in Lorentzian geometry},''
  \href{http://dx.doi.org/10.1142/S0219887805000491}{{\em Int. J. Geom. Meth.
  Mod. Phys.} {\bfseries 2} (2005) 41--61},
\href{http://arxiv.org/abs/gr-qc/0401010}{{\ttfamily arXiv:gr-qc/0401010}}.

\bibitem{desmet}
P.-J. De~Smet, ``{Black holes on cylinders are not algebraically special},''
  \href{http://dx.doi.org/10.1088/0264-9381/19/19/307}{{\em Class. Quantum
  Grav..} {\bfseries 19} (2002) 4877--4896},
\href{http://arxiv.org/abs/hep-th/0206106}{{\ttfamily arXiv:hep-th/0206106}}.

\bibitem{np}
E.~{Newman} and R.~{Penrose}, ``{An Approach to Gravitational Radiation by a
  Method of Spin Coefficients},''
  \href{http://dx.doi.org/10.1063/1.1724257}{{\em Journal of Mathematical
  Physics} {\bfseries 3} no.~3, (1962) 566--578}.
  \url{http://link.aip.org/link/?JMP/3/566/1}.

\bibitem{Bianchi}
V.~Pravda, A.~Pravdov\'a, A.~Coley, and R.~Milson, ``{Bianchi identities in
  higher dimensions},''
  \href{http://dx.doi.org/10.1088/0264-9381/21/12/007}{{\em Class. Quantum
  Grav..} {\bfseries 21} (2004) 2873--2898},
\href{http://arxiv.org/abs/gr-qc/0401013}{{\ttfamily arXiv:gr-qc/0401013}}.

\bibitem{Ricci}
M.~Ortaggio, V.~Pravda, and A.~Pravdov\'a, ``{Ricci identities in higher
  dimensions},'' \href{http://dx.doi.org/10.1088/0264-9381/24/6/018}{{\em
  Class. Quantum Grav..} {\bfseries 24} (2007) 1657--1664},
\href{http://arxiv.org/abs/gr-qc/0701150}{{\ttfamily arXiv:gr-qc/0701150}}.

\bibitem{Coley:2004hu}
A.~Coley, R.~Milson, V.~Pravda, and A.~Pravdov\'a, ``{Vanishing scalar
  invariant spacetimes in higher dimensions},''
  \href{http://dx.doi.org/10.1088/0264-9381/21/23/014}{{\em Class. Quantum
  Grav..} {\bfseries 21} (2004) 5519--5542},
\href{http://arxiv.org/abs/gr-qc/0410070}{{\ttfamily arXiv:gr-qc/0410070}}.

\bibitem{Coley:disc}
A.~Coley and S.~Hervik, ``{Algebraic classification of spacetimes using
  discriminating scalar curvature invariants},''
\href{http://arxiv.org/abs/1011.2175}{{\ttfamily arXiv:1011.2175 [gr-qc]}}.

\bibitem{Coley:2009eb}
A.~Coley, S.~Hervik, and N.~Pelavas, ``{Spacetimes characterized by their
  scalar curvature invariants},''
  \href{http://dx.doi.org/10.1088/0264-9381/26/2/025013}{{\em Class. Quant.
  Grav.} {\bfseries 26} (2009) 025013},
\href{http://arxiv.org/abs/0901.0791}{{\ttfamily arXiv:0901.0791 [gr-qc]}}.

\bibitem{bivectors}
A.~Coley and S.~Hervik, ``{Higher dimensional bivectors and classification of
  the Weyl operator},''
  \href{http://dx.doi.org/10.1088/0264-9381/27/1/015002}{{\em Class. Quantum
  Grav..} {\bfseries 27} (2010) 015002},
\href{http://arxiv.org/abs/0909.1160}{{\ttfamily arXiv:0909.1160 [gr-qc]}}.

\bibitem{brwands}
V.~Pravda and A.~Pravdov\'a, ``{WANDs of the black ring},''
  \href{http://dx.doi.org/10.1007/s10714-005-0110-3}{{\em Gen. Rel. Grav.}
  {\bfseries 37} (2005) 1277--1287},
\href{http://arxiv.org/abs/gr-qc/0501003}{{\ttfamily arXiv:gr-qc/0501003}}.

\bibitem{Mahdi}
M.~{Godazgar} and H.~S. {Reall}, ``{Algebraically special axisymmetric
  solutions of the higher-dimensional vacuum Einstein equation},''
  \href{http://dx.doi.org/10.1088/0264-9381/26/16/165009}{{\em Class. Quantum
  Grav..} {\bfseries 26} (2009) 165009},
\href{http://arxiv.org/abs/0904.4368}{{\ttfamily arXiv:0904.4368 [gr-qc]}}.

\bibitem{Ortaggio:2009sb}
M.~Ortaggio, ``{Bel-Debever criteria for the classification of the Weyl tensors
  in higher dimensions},''
  \href{http://dx.doi.org/10.1088/0264-9381/26/19/195015}{{\em Class. Quant.
  Grav.} {\bfseries 26} (2009) 195015},
\href{http://arxiv.org/abs/0906.3818}{{\ttfamily arXiv:0906.3818 [gr-qc]}}.

\bibitem{TypeD}
V.~Pravda, A.~Pravdov\'a, and M.~Ortaggio, ``{Type D Einstein spacetimes in
  higher dimensions},''
  \href{http://dx.doi.org/10.1088/0264-9381/24/17/009}{{\em Class. Quantum
  Grav..} {\bfseries 24} (2007) 4407--4428},
\href{http://arxiv.org/abs/0704.0435}{{\ttfamily arXiv:0704.0435 [gr-qc]}}.

\bibitem{kerrschild}
M.~Ortaggio, V.~Pravda, and A.~Pravdov\'a, ``{Higher dimensional Kerr-Schild
  spacetimes},'' \href{http://dx.doi.org/10.1088/0264-9381/26/2/025008}{{\em
  Class. Quantum Grav..} {\bfseries 26} (2009) 025008},
\href{http://arxiv.org/abs/0808.2165}{{\ttfamily arXiv:0808.2165 [gr-qc]}}.

\bibitem{Godazgar:2010ks}
M.~Godazgar, ``{Spinor classification of the Weyl tensor in five dimensions},''
  \href{http://dx.doi.org/10.1088/0264-9381/27/24/245013}{{\em Class. Quant.
  Grav.} {\bfseries 27} (2010) 245013},
\href{http://arxiv.org/abs/1008.2955}{{\ttfamily arXiv:1008.2955 [gr-qc]}}.

\bibitem{DeSmet:2003kt}
P.-J. De~Smet, ``{The Petrov type of the five-dimensional Myers-Perry
  metric},'' \href{http://dx.doi.org/10.1023/B:GERG.0000022586.06313.fc}{{\em
  Gen. Rel. Grav.} {\bfseries 36} (2004) 1501--1504},
\href{http://arxiv.org/abs/gr-qc/0312021}{{\ttfamily arXiv:gr-qc/0312021}}.

\bibitem{BMPV}
J.~C. Breckenridge, R.~C. Myers, A.~W. Peet, and C.~Vafa, ``{D-branes and
  spinning black holes},''
  \href{http://dx.doi.org/10.1016/S0370-2693(96)01460-8}{{\em Phys. Lett.}
  {\bfseries B391} (1997) 93--98},
\href{http://arxiv.org/abs/hep-th/9602065}{{\ttfamily arXiv:hep-th/9602065}}.

\bibitem{DeSmet:2004if}
P.-J. De~Smet, ``{The Petrov type of the BMPV metric},''
  \href{http://dx.doi.org/10.1007/s10714-005-0013-3}{{\em Gen. Rel. Grav.}
  {\bfseries 37} (2005) 237--242},
\href{http://arxiv.org/abs/gr-qc/0401033}{{\ttfamily arXiv:gr-qc/0401033}}.

\bibitem{Hughston:1973}
L.~P. Hughston and P.~Sommers, ``Spacetimes with killing tensors,'' {\em
  Communications in Mathematical Physics} {\bfseries 32} (1973) 147--152.
  \url{http://dx.doi.org/10.1007/BF01645652}. 10.1007/BF01645652.

\bibitem{Collinson:1976}
C.~D. Collinson, ``On the relationship between Killing tensors and Killing-Yano
  tensors,'' {\em International Journal of Theoretical Physics} {\bfseries 15}
  (1976) 311--314. \url{http://dx.doi.org/10.1007/BF01807593}.
  10.1007/BF01807593.

\bibitem{Stephani:1978}
H.~Stephani, ``A note on Killing tensors,'' {\em General Relativity and
  Gravitation} {\bfseries 9} (1978) 789--792.
  \url{http://dx.doi.org/10.1007/BF00760867}. 10.1007/BF00760867.

\bibitem{Krtous:2008tb}
P.~Krtou{\v s}, V.~P. Frolov, and D.~Kubiz{\v n}\'ak, ``{Hidden Symmetries of
  Higher Dimensional Black Holes and Uniqueness of the Kerr-NUT-(A)dS
  spacetime},'' \href{http://dx.doi.org/10.1103/PhysRevD.78.064022}{{\em
  Phys.Rev.} {\bfseries D78} (2008) 064022},
\href{http://arxiv.org/abs/0804.4705}{{\ttfamily arXiv:0804.4705 [hep-th]}}.

\bibitem{Kinnersley}
W.~Kinnersley, ``{Type D Vacuum Metrics},''
\href{http://dx.doi.org/10.1063/1.1664958}{{\em J. Math. Phys.} {\bfseries 10}
  (1969) 1195--1203}.

\bibitem{Mason:2010zzc}
L.~Mason and A.~Taghavi-Chabert, ``{Killing-Yano tensors and multi-Hermitian
  structures},''
\href{http://dx.doi.org/10.1016/j.geomphys.2010.02.008}{{\em J. Geom. Phys.}
  {\bfseries 60} (2010) 907--923}.

\bibitem{Krtous:2008}
P.~{Krtou{\v s}}, V.~P. {Frolov}, and D.~{Kubiz{\v n}\'ak}, ``{Hidden
  Symmetries of Higher Dimensional Black Holes and Uniqueness of the
  Kerr-NUT-(A)dS spacetime},''
  \href{http://dx.doi.org/10.1103/PhysRevD.78.064022}{{\em Phys. Rev.}
  {\bfseries D78} (2008) 064022},
\href{http://arxiv.org/abs/0804.4705}{{\ttfamily arXiv:0804.4705 [hep-th]}}.

\bibitem{Houri:2008}
T.~{Houri}, T.~{Oota}, and Y.~{Yasui}, ``{Closed conformal Killing-Yano tensor
  and uniqueness of generalized Kerr-NUT-de Sitter spacetime},''
  \href{http://dx.doi.org/10.1088/0264-9381/26/4/045015}{{\em Class. Quantum
  Grav..} {\bfseries 26} (2009) 045015},
\href{http://arxiv.org/abs/0805.3877}{{\ttfamily arXiv:0805.3877 [hep-th]}}.

\bibitem{GomezLobo:2009ct}
A.~Garc\'ia-Parrado G\'omez-Lobo and J.~M. Mart\'in-Garc\'ia, ``{Spinor
  calculus on 5-dimensional spacetimes},''
  \href{http://dx.doi.org/10.1063/1.3256124}{{\em J. Math. Phys.} {\bfseries
  50} (2009) 122504},
\href{http://arxiv.org/abs/0905.2846}{{\ttfamily arXiv:0905.2846 [gr-qc]}}.

\bibitem{VSI}
A.~Coley, R.~Milson, V.~Pravda, and A.~Pravdov\'a, ``{Vanishing scalar
  invariant spacetimes in higher dimensions},''
  \href{http://dx.doi.org/10.1088/0264-9381/21/23/014}{{\em Class. Quantum
  Grav..} {\bfseries 21} (2004) 5519--5542},
\href{http://arxiv.org/abs/gr-qc/0410070}{{\ttfamily arXiv:gr-qc/0410070}}.

\bibitem{Pravdova:2008gp}
A.~Pravdov\'a and V.~Pravda, ``{Newman-Penrose formalism in higher dimensions:
  vacuum spacetimes with a non-twisting multiple WAND},''
  \href{http://dx.doi.org/10.1088/0264-9381/25/23/235008}{{\em Class. Quantum
  Grav..} {\bfseries 25} (2008) 235008},
\href{http://arxiv.org/abs/0806.2423}{{\ttfamily arXiv:0806.2423 [gr-qc]}}.

\bibitem{Pravdova:2005ey}
A.~Pravdov\'a, V.~Pravda, and A.~Coley, ``{A note on the peeling theorem in
  higher dimensions},''
  \href{http://dx.doi.org/10.1088/0264-9381/22/13/001}{{\em Class. Quantum
  Grav..} {\bfseries 22} (2005) 2535--2538},
\href{http://arxiv.org/abs/gr-qc/0505026}{{\ttfamily arXiv:gr-qc/0505026}}.

\bibitem{Ortaggio:2009zt}
M.~Ortaggio, V.~Pravda, and A.~Pravdova, ``{On asymptotically flat
  algebraically special spacetimes in higher dimensions},''
  \href{http://dx.doi.org/10.1103/PhysRevD.80.084041}{{\em Phys.Rev.}
  {\bfseries D80} (2009) 084041},
  \href{http://arxiv.org/abs/0907.1780}{{\ttfamily arXiv:0907.1780 [gr-qc]}}.

\bibitem{odonnell}
P.~{O'Donnell}, {\em {Introduction to 2-spinors in General Relativity}}.
\newblock World Scientific, 2003.

\bibitem{penrind}
R.~{Penrose} and W.~{Rindler}, {\em {Spinors and space-time (Vol 1. Two-spinor
  calculus and relativistic fields, Vol 2. Spinor and twistor methods in
  spacetime geometry)}}.
\newblock Cambridge University Press, 1984, 1986.

\bibitem{Ortaggio:2007}
M.~Ortaggio, ``{Higher dimensional spacetimes with a geodesic, shearfree,
  twistfree and expanding null congruence},'' {\em Proceedings of the 17th
  SIGRAV Conference, Turin, Italy, 4-7 Sep 2006} {\bfseries .} (2007) ,
\href{http://arxiv.org/abs/gr-qc/0701036}{{\ttfamily arXiv:gr-qc/0701036}}.

\bibitem{Milson:align}
R.~Milson, ``{Alignment and the classification of Lorentz-signature tensors},''
  in {\em Symmetry and Perturbation Theory: Proceedings of the International
  Conference on SPT2004, Cala Genone, Italy 30 May 6 June 2004}, pp.~215--222.
\newblock World Scientific, 2004.
\newblock
\href{http://arxiv.org/abs/gr-qc/0411036}{{\ttfamily arXiv:gr-qc/0411036}}.
\newblock

\bibitem{Frolov}
V.~P. Frolov and D.~Stojkovic, ``{Particle and light motion in a space-time of
  a five- dimensional rotating black hole},''
  \href{http://dx.doi.org/10.1103/PhysRevD.68.064011}{{\em Phys. Rev.}
  {\bfseries D68} (2003) 064011},
\href{http://arxiv.org/abs/gr-qc/0301016}{{\ttfamily arXiv:gr-qc/0301016}}.

\bibitem{witten1}
E.~Witten, ``{Instability of the Kaluza-Klein Vacuum},''
\href{http://dx.doi.org/10.1016/0550-3213(82)90007-4}{{\em Nucl. Phys.}
  {\bfseries B195} (1982) 481}.

\bibitem{adssoliton}
G.~T. {Horowitz} and R.~C. {Myers}, ``{AdS-CFT correspondence and a new
  positive energy conjecture for general relativity},''
  \href{http://dx.doi.org/10.1103/PhysRevD.59.026005}{{\em Phys. Rev. D}
  {\bfseries 59} no.~2, (Dec, 1998) 026005}.

\bibitem{Ortaggio:2009bz}
M.~Ortaggio, V.~Pravda, and A.~Pravdov\'a, ``{On Kerr-Schild spacetimes in
  higher dimensions},''
\href{http://arxiv.org/abs/0901.1561}{{\ttfamily arXiv:0901.1561 [gr-qc]}}.

\bibitem{RobTraut}
J.~Podolsk\'y and M.~Ortaggio, ``{Robinson-Trautman spacetimes in higher
  dimensions},'' \href{http://dx.doi.org/10.1088/0264-9381/23/20/002}{{\em
  Class. Quantum Grav..} {\bfseries 23} (2006) 5785--5797},
\href{http://arxiv.org/abs/gr-qc/0605136}{{\ttfamily arXiv:gr-qc/0605136}}.

\bibitem{TaghaviChabert:2010bm}
A.~Taghavi-Chabert, ``{Optical structures, algebraically special spacetimes,
  and the Goldberg-Sachs theorem in five dimensions},''
\href{http://arxiv.org/abs/1011.6168}{{\ttfamily arXiv:1011.6168 [gr-qc]}}.

\bibitem{lode}
A.~Coley, S.~Hervik, M.~Ortaggio, and L.~Wylleman {\em Talk given by L.\
  Wylleman, October 2010} .

\bibitem{Teukolsky:1973}
S.~A. Teukolsky, ``{Perturbations of a rotating black hole. 1. Fundamental
  equations for gravitational electromagnetic and neutrino field
  perturbations},''
\href{http://dx.doi.org/10.1086/152444}{{\em Astrophys. J.} {\bfseries 185}
  (1973) 635--647}.

\bibitem{Cadabra1}
K.~Peeters, ``{A field-theory motivated approach to symbolic computer
  algebra},'' \href{http://dx.doi.org/10.1016/j.cpc.2007.01.003}{{\em Comput.
  Phys. Commun.} {\bfseries 176} (2007) 550--558},
\href{http://arxiv.org/abs/cs/0608005}{{\ttfamily arXiv:cs/0608005}}.

\bibitem{Cadabra2}
K.~Peeters, ``{Introducing Cadabra: A symbolic computer algebra system for
  field theory problems},''
\href{http://arxiv.org/abs/hep-th/0701238}{{\ttfamily arXiv:hep-th/0701238}}.

\bibitem{waldtypeDpert}
R.~M. Wald, ``{On perturbations of a Kerr black hole},''
  \href{http://dx.doi.org/10.1063/1.1666203}{{\em Journal of Mathematical
  Physics} {\bfseries 14} no.~10, (1973) 1453--1461}.
  \url{http://link.aip.org/link/?JMP/14/1453/1}.

\bibitem{Wald:1978vm}
R.~M. Wald, ``Construction of Solutions of Gravitational, Electromagnetic, or
  Other Perturbation Equations from Solutions of Decoupled Equations,''
\href{http://dx.doi.org/10.1103/PhysRevLett.41.203}{{\em Phys. Rev. Lett.}
  {\bfseries 41} no.~4, (Jul, 1978) 203--206}.

\bibitem{BF}
P.~Breitenlohner and D.~Z. Freedman, ``{Stability in Gauged Extended
  Supergravity},''
\href{http://dx.doi.org/10.1016/0003-4916(82)90116-6}{{\em Ann. Phys.}
  {\bfseries 144} (1982) 249}.

\bibitem{Hartnoll:2008kx}
S.~A. Hartnoll, C.~P. Herzog, and G.~T. Horowitz, ``{Holographic
  Superconductors},''
  \href{http://dx.doi.org/10.1088/1126-6708/2008/12/015}{{\em JHEP} {\bfseries
  12} (2008) 015},
\href{http://arxiv.org/abs/0810.1563}{{\ttfamily arXiv:0810.1563 [hep-th]}}.

\bibitem{Amsel:2009ev}
A.~J. Amsel, G.~T. Horowitz, D.~Marolf, and M.~M. Roberts, ``{No Dynamics in
  the Extremal Kerr Throat},''
  \href{http://dx.doi.org/10.1088/1126-6708/2009/09/044}{{\em JHEP} {\bfseries
  09} (2009) 044},
\href{http://arxiv.org/abs/0906.2376}{{\ttfamily arXiv:0906.2376 [hep-th]}}.

\bibitem{Dias:2009ex}
O.~J.~C. Dias, H.~S. Reall, and J.~E. Santos, ``{Kerr-CFT and gravitational
  perturbations},'' \href{http://dx.doi.org/10.1088/1126-6708/2009/08/101}{{\em
  JHEP} {\bfseries 08} (2009) 101},
\href{http://arxiv.org/abs/0906.2380}{{\ttfamily arXiv:0906.2380 [hep-th]}}.

\bibitem{Kunduri:2006}
H.~K. Kunduri, J.~Lucietti, and H.~S. Reall, ``{Gravitational perturbations of
  higher dimensional rotating black holes: Tensor Perturbations},''
  \href{http://dx.doi.org/10.1103/PhysRevD.74.084021}{{\em Phys. Rev.}
  {\bfseries D74} (2006) 084021},
\href{http://arxiv.org/abs/hep-th/0606076}{{\ttfamily arXiv:hep-th/0606076}}.

\bibitem{Murata:2008}
K.~Murata and J.~Soda, ``{Stability of Five-dimensional Myers-Perry Black Holes
  with Equal Angular Momenta},''
  \href{http://dx.doi.org/10.1143/PTP.120.561}{{\em Prog. Theor. Phys.}
  {\bfseries 120} (2008) 561--579},
\href{http://arxiv.org/abs/0803.1371}{{\ttfamily arXiv:0803.1371 [hep-th]}}.

\bibitem{Strominger:1998yg}
A.~Strominger, ``{$AdS_2$ quantum gravity and string theory},'' {\em JHEP}
  {\bfseries 01} (1999) 007,
\href{http://arxiv.org/abs/hep-th/9809027}{{\ttfamily arXiv:hep-th/9809027}}.

\bibitem{Ishibashi:2004wx}
A.~Ishibashi and R.~M. Wald, ``{Dynamics in non-globally hyperbolic static
  spacetimes. III: anti-de Sitter spacetime},''
  \href{http://dx.doi.org/10.1088/0264-9381/21/12/012}{{\em Class. Quant.
  Grav.} {\bfseries 21} (2004) 2981--3014},
\href{http://arxiv.org/abs/hep-th/0402184}{{\ttfamily arXiv:hep-th/0402184}}.

\bibitem{Hoxha:2000}
P.~Hoxha, R.~R. Martinez-Acosta, and C.~N. Pope, ``{Kaluza-Klein consistency,
  Killing vectors, and Kaehler spaces},''
  \href{http://dx.doi.org/10.1088/0264-9381/17/20/305}{{\em Class. Quantum
  Grav..} {\bfseries 17} (2000) 4207--4240},
\href{http://arxiv.org/abs/hep-th/0005172}{{\ttfamily arXiv:hep-th/0005172}}.

\bibitem{Dias:2010ma}
O.~J.~C. Dias, R.~Monteiro, H.~S. Reall, and J.~E. Santos, ``{A scalar field
  condensation instability of rotating anti- de Sitter black holes},''
  \href{http://dx.doi.org/10.1007/JHEP11(2010)036}{{\em JHEP} {\bfseries 11}
  (2010) 036},
\href{http://arxiv.org/abs/1007.3745}{{\ttfamily arXiv:1007.3745 [hep-th]}}.

\bibitem{Martin:2008pf}
J.~E. Martin and H.~S. Reall, ``{On the stability and spectrum of
  non-supersymmetric AdS(5) solutions of M-theory compactified on
  Kahler-Einstein spaces},''
  \href{http://dx.doi.org/10.1088/1126-6708/2009/03/002}{{\em JHEP} {\bfseries
  03} (2009) 002},
\href{http://arxiv.org/abs/0810.2707}{{\ttfamily arXiv:0810.2707 [hep-th]}}.

\bibitem{Denef:2009tp}
F.~Denef and S.~A. Hartnoll, ``{Landscape of superconducting membranes},''
  \href{http://dx.doi.org/10.1103/PhysRevD.79.126008}{{\em Phys. Rev.}
  {\bfseries D79} (2009) 126008},
\href{http://arxiv.org/abs/0901.1160}{{\ttfamily arXiv:0901.1160 [hep-th]}}.

\bibitem{chandra}
S.~Chandrasekhar, {\em {The mathematical theory of black holes}}.
\newblock Oxford University Press, 1992.

\bibitem{Morisawa}
Y.~{Morisawa}, S.~{Tomizawa}, and Y.~{Yasui}, ``{Boundary Value Problem for
  Black Rings},'' {\em Phys. Rev.} {\bfseries D77} (2008) 064019,
\href{http://arxiv.org/abs/0710.4600}{{\ttfamily arXiv:0710.4600 [hep-th]}}.

\bibitem{Chrusciel:2009}
P.~T. Chrusciel, J.~Cortier, and A.~G.-P. Gomez-Lobo, ``{On the global
  structure of the Pomeransky-Senkov black holes},''
\href{http://arxiv.org/abs/0911.0802}{{\ttfamily arXiv:0911.0802 [gr-qc]}}.

\bibitem{Chrusciel:2010jg}
P.~T. Chrusciel and S.~J. Szybka, ``{Stable causality of the Pomeransky-Senkov
  black holes},''
\href{http://arxiv.org/abs/1010.0213}{{\ttfamily arXiv:1010.0213 [hep-th]}}.

\bibitem{Elvang:2008erg}
H.~Elvang, P.~Figueras, G.~T. Horowitz, V.~E. Hubeny, and M.~Rangamani, ``{On
  Universality of Ergoregion Mergers},''
  \href{http://dx.doi.org/10.1088/0264-9381/26/8/085011}{{\em
  Class.Quant.Grav.} {\bfseries 26} (2009) 085011},
\href{http://arxiv.org/abs/0810.2778}{{\ttfamily arXiv:0810.2778 [gr-qc]}}.

\bibitem{Hoskisson}
J.~{Hoskisson}, ``{Particle motion in the rotating black ring metric},''
  \href{http://dx.doi.org/10.1103/PhysRevD.78.064039}{{\em Phys. Rev.}
  {\bfseries D78} no.~6, (2008) 064039},
  \href{http://arxiv.org/abs/0705.0117}{{\ttfamily 0705.0117}}.
  \url{http://link.aps.org/abstract/PRD/v78/e064039}.

\bibitem{Cortier:2010cc}
J.~Cortier, ``{On the structure of the ergosurface of Pomeransky-Senkov black
  rings},''
\href{http://arxiv.org/abs/1012.3594}{{\ttfamily arXiv:1012.3594 [hep-th]}}.

\bibitem{Chamblin:1996}
A.~{Chamblin} and R.~Emparan, ``{Bubbles in Kaluza-Klein theories with space-
  or time-like internal dimensions},''
  \href{http://dx.doi.org/10.1103/PhysRevD.55.754}{{\em Phys. Rev.} {\bfseries
  D55} (1997) 754--765},
\href{http://arxiv.org/abs/hep-th/9607236}{{\ttfamily arXiv:hep-th/9607236}}.

\bibitem{Benenti:1979}
S.~{Benenti} and M.~{Francaviglia}, ``{Remarks on Certain Separability
  Structures and Their Applications to General Relativity},'' {\em Gen. Rel.
  Grav.} {\bfseries 10} no.~1, (1979) 79--92.

\bibitem{Gooding:2008}
C.~Gooding and A.~V. Frolov, ``{Five-Dimensional Black Hole Capture
  Cross-Sections},'' \href{http://dx.doi.org/10.1103/PhysRevD.77.104026}{{\em
  Phys. Rev.} {\bfseries D77} (2008) 104026},
\href{http://arxiv.org/abs/0803.1031}{{\ttfamily arXiv:0803.1031 [gr-qc]}}.

\bibitem{Figueras:2005}
P.~{Figueras}, ``{A black ring with a rotating 2-sphere},'' {\em JHEP}
  {\bfseries 07} (2005) 039,
\href{http://arxiv.org/abs/hep-th/0505244}{{\ttfamily arXiv:hep-th/0505244}}.

\bibitem{Ikeda:1978}
{Ikeda, A. and Taniguchi, Y.}, ``{Spectra and eigenforms of the laplacian on
  $S^n$ and $P^n(C)$},'' {\em Osaka J. Math} {\bfseries 15} (1978) 515.

\end{thebibliography}\endgroup


\providecommand{\href}[2]{#2}\begingroup\raggedright\endgroup

\end{document}